\documentclass[11pt,oneside,reqno]{article} 

\usepackage[USenglish]{babel} 

\usepackage[utf8]{inputenc} 

\usepackage[a4paper,top=30mm,bottom=20mm,left=27.5mm,right=27.5mm]{geometry} 
\usepackage{lscape} 

\usepackage{setspace} 
\usepackage{amsmath} 
\usepackage{mathtools} 
\usepackage{array} 
\usepackage{relsize} 

\usepackage{tocloft} 
\usepackage{titlesec} 
    \titleformat*{\section}{\bfseries} 
    \titleformat*{\subsection}{\bfseries} 
    \titleformat*{\subsubsection}{\normalsize\bfseries} 
    \titleformat*{\paragraph}{\normalsize\bfseries} 
    \titleformat{name=\section}{\normalfont\bfseries}{\thesection}{0.7em}{\hspace{0em}}{} 
    \titleformat{name=\subsection}{\normalfont\bfseries}{\thesubsection}{0.7em}{\hspace{0em}}{} 
    \titleformat{name=\subsubsection}{\normalfont\normalsize\bfseries}{\thesubsubsection}{0.7em}{\hspace{0em}}{} 
    \titleformat{name=\paragraph}[runin]{\normalfont\normalsize\bfseries}{\theparagraph}{0.7em}{\hspace{0em}}{} 

\usepackage{titletoc}  

\usepackage[hang,flushmargin,symbol]{footmisc} 
    \renewcommand*{\thefootnote}{\fnsymbol{footnote}} 

\usepackage{lipsum} 

\usepackage{amssymb} 
\usepackage{marvosym} 
\usepackage{ifsym} 
\usepackage{wasysym} 
\usepackage{eurosym} 
\usepackage{textcomp} 
\usepackage{MnSymbol} 

\usepackage{caption} 
\usepackage{subcaption} 
\usepackage[capposition=top]{floatrow} 
\usepackage{float} 
\usepackage{afterpage} 

\usepackage{xcolor} 
\usepackage{soul} 

\usepackage{tabularx} 
    \newcolumntype{L}[1]{>{\raggedright\arraybackslash}p{#1}} 
    \newcolumntype{C}[1]{>{\centering\arraybackslash}p{#1}} 
    \newcolumntype{R}[1]{>{\raggedleft\arraybackslash}p{#1}} 
\usepackage{multirow} 
\usepackage{hhline} 
\usepackage{rotating} 
\usepackage[flushleft]{threeparttable} 
\usepackage{diagbox} 
\usepackage{colortbl} 
\usepackage{arydshln} 
\usepackage{booktabs}
    \newcommand{\tabitem}{~~\llap{--\!}~~} 

\usepackage{pgfplots} 
    \pgfplotsset{compat=newest}
    \usepgfplotslibrary{statistics} 
    \usepgfplotslibrary{external} 
    \usepgfplotslibrary{fillbetween} 
    \pgfkeys{/pgf/number format/.cd,1000 sep={}} 
\usepackage{tikz} 
    \usetikzlibrary{decorations.pathreplacing,decorations.markings, patterns, intersections} 

\usepackage{enumitem} 

\usepackage{graphicx} 
\usepackage{pdfpages} 

\usepackage{comment} 

\usepackage{bm} 

\usepackage[backend=bibtex,natbib=true,style=authoryear,doi=false,isbn=false,url=false,eprint=false,giveninits=true,uniquename=false,dashed=false,maxnames=3,maxbibnames=30]{biblatex} 

\usepackage{csquotes} 

    \DeclareNameAlias{sortname}{family-given} 
    \DeclareNameAlias{default}{family-given} 

    \DeclareFieldFormat[article,inbook,incollection,inproceedings,patent,thesis,unpublished]{title}{#1\isdot}

    \renewbibmacro{in:}{\ifentrytype{article}{}{\printtext{\bibstring{in}\intitlepunct}}}

    \renewbibmacro*{volume+number+eid}{%
      \printfield{volume}%
      \setunit*{\addnbthinspace}
      \printfield{number}%
      \setunit{\addcomma\space}%
      \printfield{eid}}
    \DeclareFieldFormat[article]{number}{\mkbibparens{#1}}

    \usepackage{xpatch}
    \def\dopatchbibdrivereditorcomma#1{%
      \xpatchbibdriver{#1}
        {\usebibmacro{maintitle+booktitle}%
         \newunit\newblock}
        {\usebibmacro{maintitle+booktitle}%
         \setunit{\addcomma\space}\newblock}
        {}
        {\typeout{failed to patch driver for type #1}}}
    \forcsvlist{\dopatchbibdrivereditorcomma}{inbook,incollection,inproceedings}

    \renewbibmacro*{byeditor+others}{%
      \ifnameundef{editor}
        {}
        {\usebibmacro{byeditor+othersstrg}%
         \setunit{\addspace}%
         \printnames[byeditor]{editor}%
         \clearname{editor}%
         \addcomma}%
      \usebibmacro{byeditorx}%
      \usebibmacro{bytranslator+others}}

    \makeatletter
    \pretocmd{\blx@head@bibintoc}{\phantomsection}{}{\ddt}
    \makeatother

    \AtBeginBibliography{
    }

    \DeclareCiteCommand{\citeyear}
        {}
        {\bibhyperref{\printdate}}
        {\multicitedelim}
        {}
    \DeclareCiteCommand{\citeyearpar}
        {}
        {\mkbibparens{\bibhyperref{\printdate}}}
        {\multicitedelim}
        {}

\usepackage[colorlinks=true,linkcolor=blue,citecolor=blue,urlcolor=black,linkbordercolor=cyan,citebordercolor=cyan,urlbordercolor=cyan,hyperfootnotes=true]{hyperref} 

\pgfplotsset{
boxplot/every average/.style={solid,/tikz/mark=square}, 
boxplot/every box/.style={solid}, 
boxplot/every whisker/.style={solid}, 
boxplot/every median/.style={solid}, 
}

\pgfplotsset{error bar legend/.style={%
    /pgfplots/legend image code/.prefix code={%
      \pgfkeysgetvalue{/pgfplots/error bars/error mark}{\pgfplotserrorbarsmark}%
      \draw[%
        /pgfplots/every error bar,
        mark=\pgfplotserrorbarsmark,
        /pgfplots/error bars/error mark options,
        sharp plot,
        ##1
      ] plot coordinates {(0.3cm, -0.15cm) (0.3cm, 0.15cm)};%
    }
  }
}

\begin{document}

\hypersetup{pageanchor=false} 
\begin{titlepage}
\begin{center}


\Large

\textbf{Minimum Wages in Concentrated Labor Markets}

\normalsize

\vspace*{1.75cm}

\href{https://scholar.google.de/citations?hl=de&user=xCBmJSEAAAAJ}{\textcolor{black}{\textbf{Martin Popp}}} \\
\href{https://iab.de/en/employee/?id=12380200}{\textcolor{black}{Institute for Employment Research (IAB)\vphantom{/}}} \\[-0.2cm]
\href{https://www.iza.org/person/29456/martin-popp}{\textcolor{black}{Institute of Labor Economics (IZA)\vphantom{/}}} \\[-0.2cm]
\href{https://www.laser.fau.de/associate.php?assoc_id=390}{\textcolor{black}{Labor and Socio-Economic Research Center (LASER)\vphantom{/}}}  \\[-0.2cm]
\textcolor{black}{martin.popp@iab.de}

\vspace*{0.6cm}

arXiv Preprint\footnote[1]{Corresponding Address: Martin Popp, Institute for Employment Research (IAB), Regensburger Straße 100, 90478 Nuremberg, Germany, Email: martin.popp@iab.de, Phone: $\plus$49 (0)911/179-3697. Acknowledgements: I am grateful to the Joint Graduate Program of IAB and FAU Erlangen-Nuremberg (GradAB) for financial support of my research. I thank Miren Azkarate-Askasua, Lutz Bellmann, Mario Bossler, David Card, Matthias Collischon, Martin Friedrich, Nicole Gürtzgen, Anna Herget, Katrin Hohmeyer, Etienne Lal\'{e}, Ioana Marinescu, Alan Manning, Andreas Peichl, Michael Oberfichtner, Duncan Roth, Ronja Röttger, and Philipp vom Berge for helpful discussions and suggestions. I am grateful to Emily Breuer and Franka Vetter for excellent research assistance. Earlier versions of this paper were presented at the 12th Ph.D.\ Workshop ``Perspectives on (Un-)Employment'' (IAB), the 4th Workshop on Labor Statistics (IZA), the 26th SOLE Conference (U Philadelphia), the 25th Spring Meeting of Young Economists (U Bologna), the 86th Midwest Economics Association Annual Meeting (MEA), the Scottish Economic Society Annual Meeting 2022 (U Glasgow), the 9th Conference on Social and Economic Data (RatSWD), the 15th Workshop on Labor Economics (IAAEU), the Workshop on Imperfect Competition in the Labor Market (IAB), the 35th EALE Conference (U Prague/CERGE-EI), the 16th Workshop on Microeconomics (U Lueneburg), the Statistical Week 2024 (OTH Regensburg) as well as in seminars in Nuremberg (IAB/FAU) and Uppsala (IFAU).}  \\
\today
\\[1.2cm]

\begin{abstract} 
\onehalfspacing

\noindent Economists increasingly refer to monopsony power to reconcile the absence of negative employment effects of minimum wages with theory. However, systematic evidence for the monopsony argument is scarce. In this paper, I perform a comprehensive test of this argument by using labor market concentration as a proxy for monopsony power. Labor market concentration turns out substantial in Germany. Absent wage floors, higher concentration reduces wages and employment, reflecting monopsonistic conduct of firms. Sectoral minimum wages lead to negative employment effects in slightly concentrated or more competitive labor markets. This effect weakens with increasing concentration and, ultimately, becomes positive in highly concentrated or monopsonistic markets. Overall, the results lend empirical support to the monopsony argument, implying that conventional minimum wage effects on employment conceal heterogeneity across market forms. 

\end{abstract}

\vspace*{0.5cm}
\end{center}
\small



\end{titlepage}
\hypersetup{pageanchor=true} 


\renewcommand*{\thefootnote}{\arabic{footnote}}
\setcounter{footnote}{0}

\clearpage

\normalsize


\section{Introduction}
\label{sec:1}

Although minimum wage regulations are in effect in many economies, the benefits and costs of such an intervention remain a controversial topic. For instance, there is an ongoing debate in the U.S.\ about raising the federal minimum wage from \$7.25 to \$15 per hour. While proponents of minimum wages seek to raise the pay of low-wage workers, opponents warn that this policy harms employment. Building on the notion of competitive labor markets, economists have long argued that firms respond to higher minimum wages by reducing employment. In contrast, the empirical minimum wage literature shows inconclusive results: whereas some studies find negative employment effects (e.g., \citealp{NeumarkWascher1992}; \citealp{CurrieFallick1996}; \citealp{ClemensWither2019}; \citealp{BosslerGerner2020}), a plethora of studies is not in line with conventional wisdom and reports employment effects of minimum wages to be close to zero (e.g., \citealp{CardKrueger1994}; \citealp{DubeEtAl2010}; \citealp{AllegrettoEtAl2011}; \citealp{CengizEtAl2019}; \citealp{HarasztosiLindner2019}; \citealp{DustmannEtAl2022}). Many of these studies refer to firms' monopsony power to reconcile the absence of negative employment effects of minimum wages with theory. When commanding monopsony power, firms mark down wages below workers' marginal productivity. In such a setting, an adequate minimum wage can counteract monopsony power without having adverse consequences on employment. However, systematic empirical support for the monopsony argument is scarce.

In this paper, I calculate measures of labor market concentration to perform a comprehensive empirical test of the monopsony argument. Labor market concentration gives rise to monopsony power for three reasons: First, firms with higher market shares have a greater downward leverage on market wages. Second, individuals in highly concentrated labor markets face a lower effective arrival rate of job offers, rendering labor supply to the single firm less sensitively to wages. Third, labor market concentration facilitates anti-competitive collusion between firms which further restricts personnel turnover. I make use of these channels and produce fine-grained measures of labor market concentration to gauge the degree of firms' monopsony power in the labor market. To this end, I leverage administrative records on the near-universe of workers in the German labor market between 1999 and 2017.

The test of the monopsony argument refers to sixteen low-wage sectors for which the German Ministry of Labor and Social Affairs has declared sector-specific minimum wages. Before minimum wages have been legislated in these sectors, I examine whether wages and employment turn out lower in more concentrated labor markets, reflecting monopsonistic conduct of firms in the absence of wage floors. As this hypothesis is supported by the data, I scrutinize next whether minimum wages can counteract firms' monopsony power without harming employment, the so-called ``monopsony argument''. Given a significant bite of these sectoral wage floors, the analysis endorses the monopsony argument when two conditions are met: On the one hand, slightly concentrated (i.e., more competitive) labor markets must experience negative minimum wage effects on employment. On the other hand, this negative effect must weaken for increasing levels of labor market concentration. Moreover, provided that minimum wages are set close to market-clearing rates, minimum wage effects on employment can even turn out positive in highly concentrated (i.e., more monopsonistic) markets.

The near-universal coverage of workers in the German administrative social security records allows me to calculate reliable measures of labor market concentration, the classical source of monopsony power, for each local labor market in Germany. In this study, I follow common practice and define labor markets as pair-wise combinations of industries and commuting zones. For the baseline delineation with 4-digit industries, I document high levels of concentration in 51.8 percent of all labor markets and 12.9 percent of workers, operationalized by a market-level Herfindahl-Hirschman Index (HHI) above the threshold of 0.2 from E.U.\ antitrust policy. The average labor market features roughly three firms (HHI=0.342), and this number has been relatively stable throughout the last two decade. The high levels of concentration indicate that firms in the German labor market command considerable monopsony power. While this finding mirrors evidence that the labor supply elasticity to the single firm is quite inelastic (e.g., \citealp{HirschEtAl2010}), it also manifests in substantial wage markdowns, which I calculate from survey information on a subset of firms. Importantly, labor market concentration positively correlates with higher markdowns, bolstering my practice of using labor market concentration as a proxy for monopsony power.

Building on the concentration indices, I document that both wages and employment turn out lower in more concentrated labor markets -- as put forward by monopsony theory. Specifically, I select all those observations of firms in the respective minimum wage sectors before wage floors came into force. For these observations, I perform OLS and instrumental variables (IV) regressions of firms' average wages and employment on market-level HHI values, based on yearly panel information from 1999 onwards. For the IV regressions, I construct a so-called ``leave-one-out'' instrument for HHI by calculating the average of the log inverse number of firms in the same industry for all other commuting zones. Importantly, this instrument harnesses variation from national changes in labor market concentration over time and, thus, rules out potentially endogenous variation in HHI that originates from the very same labor market. Although OLS and IV findings are qualitatively alike, the IV estimates turn out to be larger in absolute terms. The baseline IV specifications indicate that an increase in labor market concentration by 100 percent reduces average daily wages by 2.0 percent and employment by 3.4 percent, thus lending support to monopsonistic conduct of firms in concentrated labor markets. The results are robust to many alternative specifications, labor market definitions and concentration measures. In particular, the effects become more negative for a flow-adjusted HHI formulation that incorporates jobs in related industries as additional outside options of workers.

Next, I analyze whether minimum wages represent an effective policy to counteract monopsonistic conduct of firms. Specifically, I explore whether firms' responses to minimum wages vary by the underlying market form. To this end, I regress firms' average wages and employment on sectoral minimum wage levels and an interaction term with labor market concentration that serves to inspect the moderating role of monopsony power. Before turning to employment effects, I verify whether sectoral minimum wages in Germany effectively increased the pay of the workforce. I find that, on average, a 10 percent increase in sectoral minimum wages raises firms' average daily wages by 0.8 percent in atomistic labor markets with zero HHI where wages are supposed to align with marginal productivity. In line with the evidence on monopsonistic conduct, this positive wage effect becomes more pronounced with higher levels of labor market concentration, under which workers' wages are increasingly set below marginal productivity. Eventually, in the polar case of a monopsony (i.e., markets where HHI equals 1), a 10 percent minimum wage increase is associated with a rise in average wages by 3.3 percent.

Given this bite, I report that, on average, a 10 percent increase in sectoral minimum wages reduces firms' employment by 2.3 percent in zero-HHI markets, corroborating the notion that minimum wage policies harm employment in slightly concentrated (i.e., more competitive) markets. However, in line with the monopsony argument, the negative employment effects gradually disappear in more concentrated (i.e., more monopsonistic) labor markets. The elasticity is zero around HHI values of 0.2, which reflects a market with five equally-sized employers and represents the threshold to separate moderately from highly concentrated product markets in E.U.\ antitrust policy. For higher HHI values, the minimum wage elasticity of employment turns out positive. Ultimately, a 10 percent minimum wage increase is associated with a rise in employment by 9.0 percent when HHI equals 1. In these highly concentrated labor markets, the minimum wage effect on employment conceals important heterogeneity because the underlying relationship is non-monotonous: at first, the positive employment effect amplifies with the underlying bite but falls rapidly when the minimum wage is set close to the median wage of the firm in these low-wage sectors. In general, the results are robust to using many alternative model specifications, different concentration measures, and broader and narrower labor market definitions. Crucially, I also find empirical support for the monopsony argument when using markdowns -- the most comprehensive measure of monopsony power -- as moderating variable which, however, is only available for a subsample of firms in the data. Moreover, additional checks rule out omitted variable bias from product market power or productivity advantages, thus endorsing monopsony power as the true moderator of minimum wage effects. Finally, the results carry over to the aggregate level and a complementary event-study analysis highlighting the effects of the first-time introduction of minimum wages in the respective sectors.

Absent structural assumptions, the joint pattern of estimated wage and employment effects provides systematic support for the monopsony argument. Accordingly, when reasonably set and rigorously enforced, minimum wages constitute an effective measure to counteract negative consequences of monopsony power without detrimental or, in extreme cases of labor market concentration, even positive effects on employment.

In a final step, I divide the estimated minimum wage elasticities of employment by the respective minimum wage elasticities of wages to trace out the underlying own-wage elasticities of employment. Along the HHI distribution, I compare these values with elasticities from the German minimum wage literature. In low-HHI markets, I arrive at significantly negative elasticities, which are diametrically opposed to positive elasticities in high-HHI markets. At the same, my elasticities in the middle of the HHI distribution resemble the majority of elasticities from the literature which turn out either close to zero or insignificant. This pattern emphasizes that the frequent finding of close-to-zero employment effects in the minimum wage literature is (partly) stemming from the practice that conventional minimum wage studies pool together opposite effects across market forms and, therefore, report average elasticities that conceal heterogeneity by labor market concentration.

This paper contributes to the understanding of monopsonistic labor markets and minimum wage effects in several ways. First, I provide first evidence on labor market concentration in the German labor market. The concentration of workers on employers constitutes a key source for monopsony power that employers may exert over their employees. In his pioneering work, \citet{Bunting1962} was the first to calculate measures of labor market concentration for U.S.\ local labor markets. In recent years, the literature on labor market concentration experienced a sudden revival \citep{AzarEtAl2017} spurred by the recognition of a falling labor share in the U.S. \citep{AutorEtAl2020} and the increased availability of data from registers or online job boards. Building on social security register data, my concentration indices for the German labor market mirror international evidence which is unanimous that labor market concentration is substantial, such as for France \citep{MarinescuEtAl2021}, Norway \citep{DodiniEtAl2024}, or Portugal \citep{MartinsMelo2024}. For instance, \citet{AzarEtAl2020} harness data on U.S.\ online job vacancies and show that around 60 percent of labor markets are highly concentrated. By contrast, evidence on the development of labor market concentration over time is more mixed. Stable concentration in the German labor market contrasts with contemporaneous evidence from the U.S., where local labor markets became less concentrated \citep{Rinz2022}, and the U.K., which featured a bell-shaped path in labor market concentration \citep{AbelEtAl2020}.

Second, this study comprehensively examines firms' monopsonistic conduct -- in terms of not only wages but also employment. \citet{Robinson1933} established the theory of a single profit-maximizing employer who cuts employment along a positively sloped labor supply curve to the firm to mark down workers' wages. \citet{Manning2003a} popularized the estimation of dynamic monopsony models to quantify the degree of monopsony power in the labor market. These semi-structural models commonly estimate rather inelastic labor supply elasticities to the single firm, indicating a high level of wage-setting power of firms \citep{SokolovaSorensen2021}. Similar results are obtained by another strand of the literature that regresses firm-specific employment on plausibly exogenous variation in wage rates (e.g., \citealp{Falch2010}; \citealp{StaigerEtAl2010}). Except for the markets of nurses and teachers (e.g., \citealp{LuizerThornton1986}), direct empirical evidence on whether firms in fact exercise their monopsony power to push down wages was hardly available until the mid-2010s. \citet{Webber2015} shows that firms pay lower wages when labor supply is less wage-elastic. Meanwhile, a rapidly growing literature reports negative effects of labor market concentration on wages \citep{AzarEtAl2022,BassaniniEtAl2023,BassaniniEtAl2024,BenmelechEtAl2022,DodiniEtAl2024,MartinsMelo2024,PragerSchmitt2021,QiuSojourner2023,Rinz2022,SchubertEtAl2022,Thoresson2024}. In addition, \citet{MarinescuEtAl2021} show that higher concentration comes along with fewer hires in the labor market. By contrast, and building on micro-level data on firms, this study is the first to simultaneously show that firms in concentrated labor markets not only suppress wages but that they also do so by curtailing their employment -- the key rationale of monopsony theory.

Third, under specific conditions, the German Federal Ministry of Labor and Social Affairs has the right to declare collectively agreed wages to be universally binding. Since 1997, such minimum wages have come into effect in an increasing number of sectors and are regularly adjusted in the process of collective bargaining. As of 2015, more than 12 percent of all workers in Germany were employed in these minimum wages sectors. Several studies have investigated the effects of these minimum wages on single targeted sectors, such as main construction \citep{KoenigMoeller2009,Rattenhuber2014,VomBergeFrings2020}, electrical trade \citep{BoockmannEtAl2013, Frings2013}, or roofing \citep{AretzEtAl2013,GregoryZierahn2022}. The consensus among these studies is that the sectoral minimum wages effectively raised wages, especially in East Germany, but disemployment effects are hard to find in the sectors under study \citep{Moeller2012,FitzenbergerDoerr2016}. To date, however, a comprehensive analysis is missing. This study is the first to complement the literature with an analysis across sectors, providing holistic evidence on the labor market effects of sectoral minimum wages in Germany.

Fourth, my results provide systematic empirical support for the monopsony argument: whereas minimum wages are detrimental to jobs in more competitive labor markets, they do hardly affect or even stimulate employment in more monopsonistic labor markets. The international and German minimum wage literature frequently arrive at near-zero employment effects \citep{CardKrueger1995,NeumarkWascher2008,CaliendoEtAl2019,DubeZipperer2024}, and many of these articles rely on the notion of monopsony power to reconcile the absence of disemployment effects with economic theory (e.g., \citealp{CardKrueger1994}; \citealp{Moeller2012}). From an empirical point of view, however, systematic evidence in favor of the monopsony argument is limited. \citet{AzarEtAl2024} use vacancy-based concentration measures to study the potentially moderating role of monopsony power on minimum wage effects in three occupations from the U.S.\ general merchandise sector (and the fast food industry as a robustness check). In a similar fashion, \citet{Munguia2020} differentiates minimum wage effects on U.S.\ teenage employment by measures of employment-based concentration. At the market level, both studies find that labor market concentration positively moderates employment effects of minimum wages in U.S.\ labor markets.\footnote{Using an alternative approach with Japanese manufacturing data, \citet{OkudairaEtAl2019} show that minimum wage employment effects disappear when firms' pre-treatment gap between the value of the marginal product of labor and the wage rate is sufficiently large.} By contrast, this study is the first to provide a systematic examination of the monopsony argument for a European labor market. In doing so, and unlike the two aforementioned studies, this paper is not limited to a narrow set of workers. Rather, it extends to overall employment across the near-universe of German firms from sixteen heterogeneous low-wage sectors.

Specifically, this study extends the work by \citet{AzarEtAl2024}, which constitutes the most comprehensive inquiry into the monopsony argument, in five important respects, thus providing more comprehensive evidence on the interplay between minimum wages and labor market power. First, I ascertain that higher labor market concentration can indeed be interpreted as more monopsony power by revealing a positive association between HHI values and wage markdowns. Second, the monopsony argument presupposes that workers are negatively affected by firms' monopsonistic conduct in the first place. To substantiate that the minimum wages are in fact counteracting detrimental consequences of monopsony power, I additionally show that higher labor market concentration comes along with lower wage and employment levels before minimum wages were introduced in the sectors. Third, this study is the first to demonstrate that labor market concentration not only weakens the negative minimum wage effects on employment but that it also strengthens the positive effects on wages, implying that minimum wages bite harder in more concentrated markets. Fourth, this study is also unique in empirically documenting that the relationship between minimum wages and employment is non-monotonous in monopsonistic markets: the positive employment effect in high-HHI markets disappears when the level of the minimum wage is overly high. Fifth, the monopsony argument not only holds for increases in minimum wage levels but is also valid for the first-time introduction of the minimum wages in the respective sectors.

The remainder of the paper has the following structure: Section \ref{sec:2} theoretically establishes the moderating role of monopsony power on the employment effects of minimum wages. Section \ref{sec:3} outlines the institutional setting of sectoral minimum wages in the German labor market. Section \ref{sec:4} describes the data. In Section \ref{sec:5}, I provide descriptive evidence on labor market concentration and its correlates with markdowns. Section \ref{sec:6} estimates the effect of labor market concentration on wages and employment absent minimum wage regulations. In Section \ref{sec:7}, I test the monopsony argument by investigating whether the effects of minimum wages on wages and employment vary by the underlying market form. The discussion in Section \ref{sec:8} elaborates on whether monopsony power can rationalize the widespread finding of near-zero employment effects of minimum wages. I conclude in Section \ref{sec:9}.

\section{Theoretical Background}
\label{sec:2}

In the following section, I discuss the effects of minimum wages in the presence of monopsony power from a theoretical perspective to clarify the hypotheses to be tested in the empirical part of this paper. In doing so, I largely focus on oligopsony models, in which monopsony power is originating from a finite set of employers. For more details on the standard Cournot oligopsony model, I refer the reader to Appendix \ref{sec:A}.

In the standard model of perfect competition, a large number of atomistic firms face an infinitely elastic labor supply curve that implies constant marginal cost of labor (MCL) at market wage $w^{C}$. In equilibrium, each profit-maximizing firm optimizes employment $L$ such that the marginal revenue product of labor (MRPL) equals the wage rate. Thus, firms' markdowns -- the ratio between MRPL and wages -- are unity. In default of other channels of adjustment, the introduction or increase of a binding minimum wage $w^{min}$ raises marginal cost and, thus, makes firms lay off the least productive units of labor. As a result, minimum wages initiate unambiguously negative effects on employment along the intensive or the extensive margin \citep{Stigler1946}. During the early 1990s, however, novel research designs delivered near-zero employment effects \citep{Card1992a,Card1992b,CardKrueger1994} and, thus, challenged the common belief among economists that minimum wages are detrimental to employment \citep{BrownEtAl1982}. The minimum wage literature offers several explanations to reconcile the absence of disemployment effects with the notion of competitive labor markets \citep{Schmitt2015,Clemens2021}. Indeed, any specification in terms of heads (instead of hours) will underestimate the disemployment effect to the extent that minimum wage responses occur along the intensive margin. In addition, firms can pass on the higher wage costs to customers in the form of price increases \citep{Lemos2008}. Moreover, with inadequate enforcement, studies will report small employment effects simply because firms did not fully comply with the minimum wage regulation \citep{AshenfelterSmith1979}. Moreover, in an efficiency wage model, minimum wages carry the advantage that they enhance workers' productivity \citep{RebitzerTaylor1995}.

Importantly, the prevalence of monopsony power -- the argument most frequently put forward to rationalize the absence of negative minimum wage effects -- departs from perfect competition \citep{CardKrueger1995}. The textbook monopsony model from \citet{Robinson1933} postulates a single firm that is the only buyer in the labor market. Such a profit-maximizing monopsonist equates MRPL with MCL to choose the optimal wage-employment combination along the labor supply curve to the firm. The firm derives its monopsony power from the fact that the labor supply to the firm $\mu$ is less than perfectly elastic, implying that an infinitesimally small wage cut does not make all workers leave the firm. In case the firm pays all workers the same wage, such an upward-sloping labor supply curve gives rise to increasing marginal cost of labor. Unlike in the competitive model, the monopsonist cannot recruit workers at the ongoing wage rate but, instead, must offer higher wages to both new hires and incumbent workers. Consequently, the monopsonist will find it optimal to reduce employment $L^{M}$ vis-\`{a}-vis the competitive level to mark down the wage rate $w^{M}$ below workers' marginal productivity. This markdown negatively depends on the wage elasticity of labor supply to the firm which, in the polar case of a single employer, coincides with the wage elasticity of labor supply to the market, $\mu$: $ \frac{MRPL}{w^{M}} = \frac{\mu\,+\,1}{\mu} $ \citep{AshenfelterEtAl2010}.

In monopsonistic labor markets, minimum wages entails less negative employment effects than in perfectly competitive markets and, for moderately-set wage floors, can even stimulate employment \citep{Stigler1946,Lester1947}. Specifically, the employment effect differs between three regimes: First, in the unconstrained regime below the monopsonistic wage $w^{M}$, minimum wages do not bind. Second, in the supply-determined regime, the minimum wage lies in the interval between the monopsonistic wage $w^{M}$ and the market-clearing wage $w^{C}$. Such a regulation renders marginal cost of labor constant because hiring at the minimum wage no longer involves higher wages for incumbents (who also earn the minimum wage). The reduced marginal cost will make the monopsonist increase employment until the minimum wage meets the labor supply curve (after which MCL coincides with pre-minimum-wage levels). Put differently, the minimum wage eliminates the firm's incentive to take advantage of inelastic labor supply at the bottom part of the wage distribution. Third, in the demand-determined regime, the minimum wage exceeds the market-clearing wage $w^{C}$, and the minimum wage materializes along the negatively sloped labor demand curve. Thus, the negative employment effects along the demand-determined regime may (partly) offset the positive effects along the supply-determined regime, resulting in a non-monotonous relationship.

The case of a few employers in the labor market, a so-called oligopsony, constitutes a more general market structure in the midst of the two polar settings described above. In the standard Cournot oligopsony model \citep{BoalRansom1997}, $J$ firms with possibly different revenue functions choose their labor input $L_{j}$ while taking each other's quantities as given. The market wage $w^{O}$ is determined by the intersection of the labor supply curve and joint employment of these firms. The key insight of the model is that firms with large employment shares have a greater downward leverage on wages. In the optimum, high-productivity firms feature higher markdowns than low-productivity firms. Using this relationship, the employment-weighted markdown in the Cournot oligopsony becomes:
\begin{equation}
\label{eq:1}
\frac{\,\,MRPL\,\,}{w^{O}} =  \frac{\,\,\mu\,+\,H\!H\!I\,\,}{\mu}
\end{equation}
The markdown not only negatively depends on the wage elasticity of labor supply to the market but also increases with the market's employment-based Herfindahl-Hirschman Index. Specifically, by rendering labor supply to the firm less elastic, higher labor market concentration confers more monopsony power upon employers, resulting in increasingly marked-down wages through lower employment. The Cournot oligopsony model nests the standard models of perfect competition, $H\!H\!I=0$, and monopsony, $H\!H\!I=1$, as limiting cases.

In the Cournot oligopsony model, the range of the supply-determined regime becomes increasingly smaller for lower employment concentration. Nevertheless, \citet{AzarEtAl2024} derive that an adequately-set minimum wage between the oligopsonistic wage $w^{O}$ and the market-clearing wage $w^{C}$ stimulates employment in the Cournot oligopsony model.\footnote{The supply-determined regime fully disappears in the polar case of perfect competition (HHI=0). In such a setting, there is no leeway for monopsonistic conduct of firms. Thus, minimum wages are either non-binding or reduce employment along the labor demand curve.}

The previous elaborations highlight that monopsony power moderates employment effects of minimum wages in oligopsony models. In these pioneering models, firms derive their wage-setting power from strategic interactions among a finite number of firms, which is why labor market concentration is often termed as the ``classical'' source of monopsony power \citep{Manning2003b}. However, the key prediction that monopsony power dampens the negative employment responses to minimum wages emerges also from the two other hosts of monopsony models which primarily rely on so-called ``modern'' sources, namely job differentiation and search frictions \citep{AzarMarinescu2024}. First, in job differentiation models, monopsony power arises because worker have heterogeneous preferences over jobs that differ in (non-wage) amenities and, thus, their labor supply to the firm becomes finitely elastic even when labor markets are atomistic (e.g., \citealp{CardEtAl2018}; \citealp{LamadonEtAl2022}). In a model of horizontal job differentiation, \citet{BhaskarTo1999} and \citet{Walsh2003} allow for free entry and exit of firms and derive that minimum wages just above the monopsonistic wage rate stimulate employment. Second, in search-and-matching models, search frictions -- manifesting in a finite arrival rate of job offers -- limit workers' access to outside options, allowing firms to mark down wages despite being infinitesimally small (e.g., \citealp{Postel-VinayRobin2002}; \citealp{JaroschEtAl2024}). In a random search model with on-the-job search, \citet{BurdettMortensen1998} demonstrate that minimum wages spur employment when set below workers' marginal productivity.

Taken together, minimum wages will cause less negative effects on employment to the extent that the market is monopsonistic \citep{BhaskarEtAl2002}. Building on the Cournot oligopsony model, I lever concentration indices, the classical source of monopsony power, to directly test these hypotheses based on variation in sectoral minimum wages from Germany. Moreover, for a narrow subsample of firms, I perform the analysis for the more comprehensive measure of markdowns, which additionally reflects job differentiation and search frictions as modern sources of monopsony power.

\section{Institutional Background}
\label{sec:3}

For many decades, high coverage rates of collective agreements ensured effective wage floors in Germany, stifling any discussion about the introduction of a nation-wide minimum wage. But, in the last 25 years, Germany experienced a sharp decline in collective bargaining and rising inequality at the bottom of the wage distribution \citep{DustmannEtAl2009}. Against this backdrop, Germany became one of the last E.U.\ countries to impose a nation-wide minimum wage in 2015. Beginning in 1997, however, minimum wages had already come into force in a large variety of sectors.\footnote{Throughout this study, I distinguish between the concepts of ``sector'' and ``industry''. I define an ``industry'' as an entry within the 5-digit German Classification of Economic Activities (WZ). On the contrary, I use the term ``sector'' to describe a (sub-)set of industries that share the same collective bargaining agreement and, consequently, the same minimum wage regulation.} These sectoral minimum wages were not set by government but, instead, originated from sector-wise collective bargaining agreements (CBA) that the Federal Ministry of Labor and Social Affairs declared universally binding. Every few years, these sectoral minimum wages were subject to adjustments in the underlying CBAs, offering rich variation to study the minimum wage effects over time. In the following, I will briefly describe the main characteristics of sectoral minimum wages in the German labor market. In Appendix \ref{sec:B}, I provide additional details on the institutional background.

When a CBA is declared universally binding, the negotiated minimum working conditions no longer apply only to the social partners, but to the entire sector. Crucially, the sectoral minimum wages apply to all firms in the sector, including those firms that decided against a membership in the respective employer association. In principle, the regulation covers all employees in these firms, irrespective of their occupation or union membership. However, several sectors grant exemptions to white-collar workers or apprentices.

To date, twenty sectors have imposed minimum wages throughout Germany. In this paper, I study sixteen of these sectors that can be identified in the German version of the NACE industry classification. In 2015, these sectors covered about 380,000 firms and 5.2 million workers, representing 12.2 percent of the workforce in Germany. The Kaitz Indices in the sectors typically range between 65 and 85 percent, indicating that the minimum wages bit strongly into the wage distribution. However, these relatively high values are largely attributable to the fact that these minimum wages refer to low-wage sectors rather than the economy as a whole. Some of the sectoral minimum wages are set uniformly, while others differentiate between East and West Germany, or the 16 federal states. The Kaitz Indices generally prove to be higher in East Germany although, in many sectors, the social partners negotiated lower minimum wages than in West Germany. The Ministry of Labor and Social Affairs is obliged to publish sectoral minimum wages in the German Federal Bulletin (Bundesanzeiger) from which I extract sector-specific variation over the years 1999 to 2017.

\section{Data}
\label{sec:4}

I assemble several datasets on the German labor market to examine the monopsony argument. The primary sources for my main analyses comprise administrative information from the German social security registers as well as survey information from the IAB Establishment Panel. To perform certain robustness checks, I further leverage secondary data sources which, along with descriptive statistics, are described in Appendix \ref{sec:C}.

Calculating unbiased measures of labor market concentration necessitates full information on the population of workers and their employers. In this study, I leverage administrative records on the near-universe of workers in Germany to calculate indices of labor market concentration. Specifically, I use information available in the Integrated Employment Biographies (IEB) from the Institute of Employment Research (IAB) in Germany \citep{MuellerWolter2020}. The IEB collect job notifications on the entirety of workers in Germany who are liable to social security contributions.\footnote{As a consequence, the IEB data do not include civil servants, self-employed persons and family workers who are exempt from social security payments.} Amongst many other variables, the data assemble longitudinal information on employment spells, earnings (up to the annual censoring limit), type of contract, time-consistent 5-digit industry affiliation, place of work, and an establishment (or firm) identifier. The term ``establishment'' refers to a regionally and economically delimited unit of production in which employees work.\footnote{In this study, I use the terms ``establishment" and ``firm" interchangeably. In contrast, the term ``company'' combines all establishments from the same employer. An establishment may consist of one or more branch offices or workplaces. In principle, branch offices of one company which belong to the same industry and the same municipality are given a joint establishment number. However, it is not possible to differentiate between branch offices from the same establishment in the data. Furthermore, no information is available as to which establishments belong to the same company.} In principle, IEB information is available from 1975 (West Germany) and 1993 (East Germany) onward. Due to a major structural break in the data in the late 1990s, I restrict the analysis to the years 1999-2017. For June 30 of these years, I calculate a broad range of different measures of labor market concentration.

I combine the resulting concentration indices with information from the Establishment History Panel (BHP) to study wage and employment effects by market form. The BHP is an annual panel dataset, which consolidates IEB notifications on workers' jobs at the firm level as of June 30 each year \citep{GanzerEtAl2020}. As a result, the BHP covers the universe of German firms with at least one employee subject to social security contributions. The dataset offers information on firm size and workforce composition, such as the number of regular full-time, regular part-time, and marginal part-time workers. After imputing right-censored wages using a two-step procedure \citep{CardEtAl2013}, the BHP provides averages of individual gross wages in each firm. For lack of information on individual hours worked, these averages refer to daily wages of regular full-time workers (who are supposed to work a similar number of hours). Based on detailed 5-digit industry codes, I select all firm-year observations that refer to one of the sixteen minimum wage sectors between 1999 and 2017.

Given the sector affiliation, I enrich the data with minimum wages levels that the Ministry of Labor and Social Affairs declared universally binding in the German Federal Bulletin. Finally, I augment the dataset with an aggregate time series on sector-specific shares of firms that abide by collective bargaining agreements. To this end, I make use of information from the IAB Establishment Panel, which is an annual representative survey on German firms \citep{EllguthEtAl2014}. The final dataset comprises 6,865,711 firm-year observations. 43.4 percent of these observations refer to years before a minimum wage was legislated for the first time in the respective sectors, and the remaining 56.6 percent refer to the years thereafter. The panel covers 930,823 firms that appear on average 7.4 times between 1999 and 2017. These firms employ 3.8-5.3 million workers per year, which totals 84.3 million worker-year observations over the period of study.

\section{Labor Market Concentration}
\label{sec:5}

In the following section, I provide first evidence on labor market concentration in Germany and inspect its correlation with markdowns in the labor market and markups in the product market. Appendix \ref{sec:D} delivers information on the statistical properties of the concentration indices, the delineation of the underlying markets, and further descriptive evidence on concentration in the German labor market. For more details on the flow-adjusted measure of labor market concentration, I refer the reader to Appendix \ref{sec:E}. Appendix \ref{sec:F} deals with the estimation of markdowns and markups and its correlates with labor market concentration.

\paragraph{Sources of Monopsony Power.} Labor market concentration constitutes the classical source of monopsony power. In the Cournot oligopsony model, labor market concentration (in terms of employment HHI) directly manifests in higher markdowns because firms with higher market shares can exert greater downward leverage on market wages, as implied by Equation (\ref{eq:1}). Besides, labor supply to the firm may become more inelastic with higher labor market concentration for two additional reasons, resulting in even higher markdowns: First, high labor market concentration facilitates anti-competitive collusion among firms \citep{BoalRansom1997}. For instance, large firms can arrange mutual non-poaching agreements to effectively restrict personnel turnover between close competitors \citep{KruegerAshenfelter2022}. Second, by lowering the effective arrival rate of jobs, higher labor market concentration (in terms of HHI) exacerbates workers' search frictions, resulting in lower wages through fewer outside options \citep{JaroschEtAl2024}.

\paragraph{Measurement.} I follow standard practice and construct measures of absolute labor market concentration on the basis of the Herfindahl-Hirschman Index (HHI), which equals the sum of squared market shares \citep{Hirschman1945,Herfindahl1950}. Specifically, I calculate labor market concentration as
\begin{equation}
\label{eq:2}
H\!H\!I_{mt} = \sum_{j=1}^{J} s_{jmt}^{2}
\end{equation}
where $ s_{jmt} = \frac{L_{jmt}}{\sum_{j=1}^{J} L_{jmt}} $ represents the share of firm $j$ in employment of labor market $m$ in year $t$. HHI values range from 0 to 1, and higher values signal greater concentration. In many countries, the HHI offers a guideline for antitrust policy. The E.U.\ Commission \citeyearpar{EC2004} scrutinizes the intensity of product market power by means of three intervals for HHI values: low (0.0-0.1), medium (0.1-0.2) and high levels of concentration (0.2-1.0).\footnote{Note that these intervals were defined in an ad-hoc manner for the evaluation of product markets in the European Union. Nevertheless, I interpret labor market concentration in light of these thresholds because, against the backdrop of limited awareness of monopsony power in past decades, analogous thresholds for labor markets have not (yet) been released by E.U.\ antitrust authorities. In the U.S., similar thresholds have been legislated and, since recently, additionally refer to labor markets \citep{FTC2023}.}

The use of concentration indices necessitates an appropriate definition of labor markets. Following the literature, I operationalize labor markets on the basis of observable firm characteristics, namely industry and workplace (e.g., \citealp{BergerEtAl2022}; \citealp{Rinz2022}). As baseline specification, I define labor markets $m$ as combinations of 4-digit NACE industries $i$ and commuting zones $z$. In terms of industry, the 4-digit classification entails $I=615$ industry classes, and 91 of these entries relate to (subsets of the) minimum wage sectors. In terms of region, I employ the graph-theoretical method from \citet{KroppSchwengler2016} to merge 401 administrative districts (3-digit NUTS regions) to more adequate functional regions. Based on commuting patterns, the optimization yields $Z=51$ commuting zones with strong interactions within but few connections between zones. By construction, concentration indices implicitly treat labor markets as discrete segments between which workers cannot move \citep{Manning2021}. I address this issue in later robustness checks and show that the regression results are robust to broader or narrower definitions of labor markets. To more rigorously address the role of outside options in wage determination (e.g., \citealp{CaldwellDanieli2022}; \citealp{SchubertEtAl2022}), I demonstrate that the results also hold for a flow-adjusted version of the HHI along the lines of \citet{Arnold2021}. Unlike the standard HHI, the flow-adjusted HHI takes into account jobs in other but related industries as additional outside options while the weights of these outside options are determined on the basis of empirically observable transition probabilities.

\paragraph{Baseline Concentration.} The histogram in Figure \ref{fig:1} illustrates the distribution of labor market concentration. Given its baseline version for employment in pairs of 4-digit NACE industries and commuting zones, the average HHI equals 0.342 which, by taking the reciprocal, is equivalent to 2.9 equally-sized firms in the labor market. At the 25th percentile, the median, and the 75th percentile, the equivalent number of competitors is 14.5 (HHI=0.069), 4.6 (HHI=0.217), and 1.9 (HHI=0.526). A large fraction of labor markets features near-zero HHI values, resembling the notion of atomistic labor markets. For higher HHI values, the density becomes increasingly smaller. However, a spike appears at the upper end of the distribution, indicating a considerable portion of labor markets with a single employer (11.7 percent). Overall, 15.8 of the labor markets feature medium levels of concentration, operationalized by the HHI interval between 0.10 and 0.20 from E.U.\ antitrust policy. 51.8 percent of the labor markets exceed the threshold of 0.20, mirroring high levels of concentration. By and large, in the 91 industries that belong to the minimum wage sectors (HHI=0.314), labor market concentration does not markedly differ from the universe of 615 4-digit industries.

\tikzset{external/export =false} 

\begin{figure}[!ht]
\centering
\caption{Distribution of Baseline HHI}
\label{fig:1}
\scalebox{0.80}{
\begin{tikzpicture}
\pgfplotsset{set layers}
\begin{axis}[area style, xlabel=Herfindahl-Hirschman Index, ylabel=Density, xtick align=outside, xtick pos = left, height=14cm, width=14cm, xmin=-0.05, xmax=1.05, ymin = -0.01, ymax=0.16, ytick={0,0.03,0.06,0.09,0.12,0.15},xtick={0,0.2,0.4,0.6,0.8,1}, grid=major, set layers, major grid style = {dotted, /pgfplots/on layer=axis background}, legend pos = north east, y tick label style={/pgf/number format/.cd,fixed,fixed zerofill, precision=2,/tikz/.cd}, x tick label style={/pgf/number format/.cd,fixed,fixed zerofill, precision=1,/tikz/.cd}]
\draw[dashed, color=blue] (0.069,0) -- (0.069,0.13) node[above, color=blue]{P25=0.069};
\draw[dashed, color=blue] (0.217,0) -- (0.217,0.11) node[above, color=blue]{P50=0.217};
\draw[dashed, color=blue] (0.342,0) -- (0.342,0.09) node[above, color=blue]{Mean=0.342};
\draw[dashed, color=blue] (0.526,0) -- (0.526,0.07) node[above, color=blue]{P75=0.526};
\addplot+[ybar interval, mark=no, color=black, fill=gray!5!white] plot coordinates { (0,.11292495) (.025,.08659977)};
\addplot+[ybar interval, mark=no, color=black, fill=gray!5!white] plot coordinates { (.025,.08659977) (.05,.06551824)};
\addplot+[ybar interval, mark=no, color=black, fill=gray!5!white] plot coordinates { (.05,.06551824) (.075,.05505403)};
\addplot+[ybar interval, mark=no, color=black, fill=gray!5!white] plot coordinates { (.075,.05505403) (.1,.04678795)};
\addplot+[ybar interval, mark=no, color=black, fill=gray!20!white] plot coordinates { (.1,.04678795) (.125,.04169323)};
\addplot+[ybar interval, mark=no, color=black, fill=gray!20!white] plot coordinates { (.125,.04169323) (.15000001,.03697815)};
\addplot+[ybar interval, mark=no, color=black, fill=gray!20!white] plot coordinates { (.15000001,.03697815) (.175,.0333139)};
\addplot+[ybar interval, mark=no, color=black, fill=gray!20!white] plot coordinates { (.175,.0333139) (.2,.03181631)};
\addplot+[ybar interval, mark=no, color=black, fill=gray!60!white] plot coordinates { (.2,.03181631) (.22499999,.02627903)};
\addplot+[ybar interval, mark=no, color=black, fill=gray!60!white] plot coordinates { (.22499999,.02627903) (.25,.02591197)};
\addplot+[ybar interval, mark=no, color=black, fill=gray!60!white] plot coordinates { (.25,.02591197) (.27500001,.02395714)};
\addplot+[ybar interval, mark=no, color=black, fill=gray!60!white] plot coordinates { (.27500001,.02395714) (.30000001,.02183662)};
\addplot+[ybar interval, mark=no, color=black, fill=gray!60!white] plot coordinates { (.30000001,.02183662) (.32499999,.02213445)};
\addplot+[ybar interval, mark=no, color=black, fill=gray!60!white] plot coordinates { (.32499999,.02213445) (.34999999,.01869043)};
\addplot+[ybar interval, mark=no, color=black, fill=gray!60!white] plot coordinates { (.34999999,.01869043) (.375,.01966575)};
\addplot+[ybar interval, mark=no, color=black, fill=gray!60!white] plot coordinates { (.375,.01966575) (.40000001,.01455425)};
\addplot+[ybar interval, mark=no, color=black, fill=gray!60!white] plot coordinates { (.40000001,.01455425) (.42500001,.01518349)};
\addplot+[ybar interval, mark=no, color=black, fill=gray!60!white] plot coordinates { (.42500001,.01518349) (.44999999,.01243162)};
\addplot+[ybar interval, mark=no, color=black, fill=gray!60!white] plot coordinates { (.44999999,.01243162) (.47499999,.01091936)};
\addplot+[ybar interval, mark=no, color=black, fill=gray!60!white] plot coordinates { (.47499999,.01091936) (.5,.02749765)};
\addplot+[ybar interval, mark=no, color=black, fill=gray!60!white] plot coordinates { (.5,.02749765) (.52499998,.01135773)};
\addplot+[ybar interval, mark=no, color=black, fill=gray!60!white] plot coordinates { (.52499998,.01135773) (.55000001,.01572253)};
\addplot+[ybar interval, mark=no, color=black, fill=gray!60!white] plot coordinates { (.55000001,.01572253) (.57499999,.00935046)};
\addplot+[ybar interval, mark=no, color=black, fill=gray!60!white] plot coordinates { (.57499999,.00935046) (.60000002,.00760538)};
\addplot+[ybar interval, mark=no, color=black, fill=gray!60!white] plot coordinates { (.60000002,.00760538) (.625,.01061313)};
\addplot+[ybar interval, mark=no, color=black, fill=gray!60!white] plot coordinates { (.625,.01061313) (.64999998,.00679576)};
\addplot+[ybar interval, mark=no, color=black, fill=gray!60!white] plot coordinates { (.64999998,.00679576) (.67500001,.00812974)};
\addplot+[ybar interval, mark=no, color=black, fill=gray!60!white] plot coordinates { (.67500001,.00812974) (.69999999,.00719008)};
\addplot+[ybar interval, mark=no, color=black, fill=gray!60!white] plot coordinates { (.69999999,.00719008) (.72500002,.00521008)};
\addplot+[ybar interval, mark=no, color=black, fill=gray!60!white] plot coordinates { (.72500002,.00521008) (.75,.00612038)};
\addplot+[ybar interval, mark=no, color=black, fill=gray!60!white] plot coordinates { (.75,.00612038) (.77499998,.00563796)};
\addplot+[ybar interval, mark=no, color=black, fill=gray!60!white] plot coordinates { (.77499998,.00563796) (.80000001,.00604277)};
\addplot+[ybar interval, mark=no, color=black, fill=gray!60!white] plot coordinates { (.80000001,.00604277) (.82499999,.00518701)};
\addplot+[ybar interval, mark=no, color=black, fill=gray!60!white] plot coordinates { (.82499999,.00518701) (.85000002,.00472976)};
\addplot+[ybar interval, mark=no, color=black, fill=gray!60!white] plot coordinates { (.85000002,.00472976) (.875,.00518072)};
\addplot+[ybar interval, mark=no, color=black, fill=gray!60!white] plot coordinates { (.875,.00518072) (.89999998,.004864)};
\addplot+[ybar interval, mark=no, color=black, fill=gray!60!white] plot coordinates { (.89999998,.004864) (.92500001,.00448856)};
\addplot+[ybar interval, mark=no, color=black, fill=gray!60!white] plot coordinates { (.92500001,.00448856) (.94999999,.00505697)};
\addplot+[ybar interval, mark=no, color=black, fill=gray!60!white] plot coordinates { (.94999999,.00505697) (.97500002,.12096869)};
\addplot+[ybar interval, mark=no, color=black, fill=gray!60!white] plot coordinates { (.97500002,.12096869) (1,.0001)};
\legend{{~Low (32.4 Percent)},,,,{~Medium (15.8 Percent)},,,,{~High (51.8 Percent)}}
\end{axis}
\end{tikzpicture}
}
\floatfoot{\footnotesize\textsc{Note. ---} The figure illustrates the distribution of labor market concentration in Germany. Labor market concentration refers to employment-based HHI values for pair-wise combinations of 4-digit NACE industries and commuting zones, and is tracked with annual frequency. HHI = Herfindahl-Hirschman Index. NACE = Statistical Nomenclature of Economic Activities in the European Community. Source: IEB, 1999-2017.}
\end{figure}

\paragraph{Relationship with Markdowns.} The practice of using concentration indices as a measure of market power -- known as the ``structure-conduct-performance'' literature -- implicitly relies on the Cournot model in which a higher HHI value implies a lower degree of competition in the product or labor market. \citet{BerryEtAl2019} and \citet{Syverson2019} argue that caution about this practice is warranted because higher concentration could in principle also be positively correlated with the competitiveness of the market. For instance, in the Burdett-Mortensen (\citeyear{BurdettMortensen1998}) model, an increase in labor market competition (approximated by a higher arrival rate of job offers) is reflected in higher HHI values \citep{Manning2021}.

To explore whether labor market concentration can serve as a proxy for monopsony power in the German setup, I use the IAB Establishment Panel to determine yearly markdowns for a subset of the firms in the BHP data. To determine these markdowns, I follow the literature on the ``production-function approach'' \citep{Hall1988,DeLoecker2011,DobbelaereMairesse2013} and take advantage of the fact that wedges between output elasticities and revenue shares reflect market power in input and output markets. Across firms from all sectors between 1999 and 2017, the firm-level markdowns averaged 1.456 which, by taking the reciprocal, implies that workers are paid about 69 percent of their marginal revenue product. When filtering for firms that belong to the subset of minimum wage sectors, the average markdown turns out lower, implying that workers obtain 81 percent of their MRPL. Simple regressions underline that there is a positive relationship between firms' markdowns and HHI values: a 100 percent HHI increase (e.g., a reduction in the number of equally-sized employers from 10 to 5 firms) raises firms' markdowns on average by 0.037 (see Appendix Table \ref{tab:F2}). For the subset of firms in minimum wage sectors, this effect turns out even more pronounced. Further regressions show that a minimum wage increase by 1 Euro significantly reduces markdowns by 0.025, providing tentative evidence that the minimum wage policies limit employers' wage-setting power. By and large, the presented evidence corroborates that higher labor market concentration confers more monopsony power upon employers in the German labor market -- as put forward by the ``structure-conduct-performance'' literature. In the following, I largely resort to the more broadly obtainable measures of labor market concentration and interpret higher values as evidence for more monopsony power.

\paragraph{Relationship with Markups.} To ease the later interpretation of the multivariate results through the lens of monopsony theory, it is important to rule out any interplay between labor and product markets beforehand. In general, industry-based measures of local labor market concentration might also reflect product market concentration, notably when most of the products are sold locally \citep{Manning2021}. However, certain implications of firms' wage-setting power in the labor market could, in principle, also arise from price-setting power in the product market. In particular, monopolists reduce their output and, like monopsonists, employ fewer workers than in the absence of market power.

I examine the relationship between labor market concentration and product market power using estimated price-cost markups. These firm-level markups are recovered in course of the markdown estimation. Indeed, there is a positive, albeit weak, relationship between labor market concentration and markups (see Appendix Table \ref{tab:F4}). Reassuringly, however, the relationship collapses when looking at minimum wage sectors only, mitigating the concern that labor market concentration could reflect product market power in the later multivariate analyses. Furthermore, the relationship turns even negative when excluding service sectors whose product markets are more local by nature. Thus, in the later analysis, I will corroborate the robustness of the multivariate results under exclusion of the service sectors.

\section{Effects of Labor Market Concentration}
\label{sec:6}

In a next step, I explore whether -- ahead of the minimum wage regulations -- higher labor market concentration results ceteris paribus in lower wages and employment, as postulated by the standard Cournot oligopsony model. Such a negative interrelation is an important piece of evidence for the monopsony argument because, in the absence of monopsonistic conduct, there is no reason why minimum wage effects should vary between slightly and highly concentrated labor markets. In Appendix \ref{sec:G}, I provide the full set of sensitivity and heterogeneity analyses, along with further checks to facilitate the causal interpretation of the effects.

\paragraph{Empirical Model.}

To study the effect of labor market concentration on the outcome variables of interest, I estimate the following fixed-effects model
\begin{equation}
\label{eq:3}
ln\,Y_{jizt} \,=\,  \theta \cdot ln\,H\!H\!I_{izt}  \,+\, \delta_{j}  \,+\, \zeta_{zt} \,+\, \zeta_{it}  \,+\, \varepsilon_{jizt}
\end{equation}
where $Y$ refers to either average daily wages or employment of regular full-time workers of firm $j$ in year $t$, $H\!H\!I$ is the corresponding Herfindahl-Hirschman Index, $\delta_{j}$, $\zeta_{zt}$, and $\zeta_{it}$ are firm, commuting-zone-by-year, and industry-by-year fixed effects, and $\varepsilon$ is an error term. For these analyses, I keep only those observations that refer to years before minimum wage regulations came into effect in the respective sector.\footnote{As a consequence, the sectors of main construction, electrical trade and roofing do not enter the regressions as sectoral minimum wages were implemented already before 1999.} Thus, the period of analysis begins in 1999 and ends in 2014, which is the year before the last sectoral minimum wage was introduced. In a first specification, I condition only on firm fixed effects to capture unobserved time-invariant heterogeneity across employers. Thus, the identification of elasticity estimates $\hat{\theta}$ stems from variation within firms over time. In a second regression, I add year fixed effects to account for time-specific events common to all firms (e.g., the business cycle). I cluster standard errors at the labor market level.

The main threat of identification are time-varying variables that correlate with labor market concentration and exert a direct impact on wages or employment. Specifically, labor demand or labor supply shocks may bias the estimation. For instance, a positive productivity shock will make incumbent firms increase wages and employment along the labor supply curve while new entrants simultaneously lower the HHI, creating a downward bias. A rigorous way to address these shocks would be to control for labor-market-by-year fixed effects but this would leave no variation in concentration which is exactly defined at this level. In a third specification, I separately condition on commuting-zone-by-year fixed effects and 2-digit-industry-by-year fixed effects to absorb the local dimension and the broader industrial dimension of these shocks.

Nevertheless, the question remains whether demand or supply shocks specific to a certain labor market (i.e., at the 4-digit industry level within a commuting zone) may skew the results. Moreover, reverse causality is a first-order concern which might arise from the feedback effect of employment responses on the employment-based HHI \citep{MarinescuEtAl2021}. To rule out these issues of endogeneity, I follow an instrumental variable strategy that builds on \citet{Hausman1996} and \citet{Nevo2001} in a fourth and baseline specification. Specifically, I instrument any HHI value in a certain commuting zone $c$ by the average of the log inverse number of firms in all other commuting zones for the same industry and time period:
\begin{equation}
\label{eq:4}
\overline{ln\,I\!N\!F}^{\,-c}_{it} \,=\,  \frac{\sum_{z \neq c\vphantom{/}} \, ln\,I\!N\!F_{izt}}{Z-1} \,=\, \frac{\sum_{z \neq c\vphantom{/}} \, ln\,J^{-1}_{izt}}{50}
\end{equation}
In a similar context, such an instrument has been implemented by \citet{AzarEtAl2022}, \citet{BassaniniEtAl2024}, \citet{QiuSojourner2023}, or \citet{Rinz2022}. Favorably, the ``leave-one-out'' property of this instrument delivers variation in local labor market concentration that is driven by national forces in the respective industry and, thus, not by shifts in the industry in that certain commuting zone. For example, mergers (or divestitures) of national companies reduce (increase) the number of firms in all local markets where these companies are operating without being correlated with idiosyncratic shocks in these markets. By relying on the number of firms in other labor markets, the instrument protects against a direct effect of labor supply or labor demand shocks on the market-level HHI. Nevertheless, the instrument is not fully exogenous when these shocks are correlated across commuting zones, for instance, due to common technology shocks. To assess the magnitude of the potential bias, I apply the plausibly exogenous regression method \citep{ConleyEtAl2012} to derive bounds for the causal effect of labor market concentration on the outcomes of interest. Using this method, I will later show that my estimates are robust to large potential violations of exogeneity of the instrument.

\paragraph{Effects of Labor Market Concentration on Wages.} In line with monopsony theory, the results indicate that, prior to minimum wage regulations, higher labor market concentration is associated with significantly lower wages. Following Equation (\ref{eq:3}), Table \ref{tab:1} displays the effects of labor market concentration on the average daily wages of regular full-time workers. The first regression with mere firm fixed effects implies that an HHI increase by 100 percent, for instance, a decrease in the equivalent number of employers from 10 to 5, is ceteris paribus associated with a reduction in daily wages by 2.5 percent. In Column (2), the effect almost halves when adding year fixed effects. In Column (3), when controlling for commuting-zone-by-year and 2-digit-industry-by-year fixed effects, this effect weakens further to 0.8 percent, highlighting the role of labor supply and labor demand shocks. To allow for a more causal interpretation, the baseline specification in Column (4) performs an IV estimation with the leave-one-out instrument, as given by Equation (\ref{eq:4}). Favorably, the first-stage regression of HHI on the instrument shows the expected positive sign. The respective F statistic is 387.0 which validates the relevance of the instrument. In the second stage, the IV estimate turns out negative as well but larger in absolute terms: an HHI increase by 100 percent reduces average daily wages in the firm by 2.0 percent. Importantly, this effect size is in close proximity to IV results from the related literature.\footnote{My elasticity of -0.020 resembles estimates from the following studies: -0.020 by \citet{MarinescuEtAl2021} and -0.024 by \citet{BassaniniEtAl2023} for France, -0.014 by \citet{MartinsMelo2024} for Portugal, -0.025 by \citet{SchubertEtAl2022} for the U.S., and -0.018 by \citet{AbelEtAl2020} for the U.K. In addition, \citet{BassaniniEtAl2024} find that the elasticity of wages with respect to HHI is strikingly similar in Denmark (-0.029), France (-0.022), Germany (-0.019), and Portugal (-0.025). Moreover, my estimate is quantitatively different but still in the ballpark of those estimates reported by \citet{BenmelechEtAl2022} and \citet{Rinz2022} on the U.S. labor market.} To put my results into context, going from the 25th percentile (HHI=0.069) to the 75th percentile (HHI=0.526) of the HHI distribution is associated with a decline in average daily wages by 13.2 percent. Across specifications, all elasticities are significantly different from zero at 1 percent levels.

\afterpage{

\begin{table}[!ht]
\centering
\scalebox{0.90}{
\begin{threeparttable}
\caption{Effects of Labor Market Concentration on Wages}
\label{tab:1}
\begin{tabular}{L{4cm}C{2.7cm}C{2.7cm}C{2.7cm}C{2.7cm}} \hline
&&&& \\[-0.3cm]
\multirow{4.4}{*}{} & \multirow{4.4}{*}{\shortstack{(1) \\ Log \\ Mean Daily\vphantom{/} \\ Wages }} & \multirow{4.4}{*}{\shortstack{(2) \\ Log \\ Mean Daily\vphantom{/} \\ Wages }} & \multirow{4.4}{*}{\shortstack{(3) \\ Log \\ Mean Daily\vphantom{/} \\ Wages }} & \multirow{4.4}{*}{\shortstack{(4) \\ Log \\ Mean Daily\vphantom{/} \\ Wages }} \\
&&&& \\
&&&& \\
&&&& \\[0.3cm] \hline
&&&& \\[-0.3cm]
\multirow{2.4}{*}{Log HHI} &  \multirow{2.4}{*}{\shortstack{\hphantom{***}-0.025***\hphantom{-}  \\ (0.007)}}  & \multirow{2.4}{*}{\shortstack{\hphantom{***}-0.014***\hphantom{-}  \\ (0.004)}} & \multirow{2.4}{*}{\shortstack{\hphantom{***}-0.008***\hphantom{-}  \\ (0.002)}} & \multirow{2.4}{*}{\shortstack{\hphantom{***}-0.020***\hphantom{-}  \\ (0.004)}}   \\
&&&& \\[0.2cm] 
\multirow{2.4}{*}{\shortstack{Fixed Effects}}  & \multirow{2.4}{*}{Firm} &  \multirow{2.4}{*}{\shortstack{Firm \\ Year}} & \multirow{2.4}{*}{\shortstack{Firm \\ Year $\times$ CZ \\ NACE-2 $\times$ Year}} &  \multirow{2.4}{*}{\shortstack{Firm \\ CZ $\times$ Year \\ NACE-2 $\times$ Year}} \\
&&&& \\[0.4cm] \hline
&&&& \\[-0.2cm]
Instrument  & None & None & None & $ \overline{\text{Log}\,\,\text{INF}} \,\,\backslash\, \{c\} $  \\
\multirow{3.4}{*}{\shortstack[l]{Labor Market \\ Definition \\ (Object)}}  & \multirow{3.4}{*}{\shortstack{NACE-4 \\ $\times$ CZ \\ (Employment)}} & \multirow{3.4}{*}{\shortstack{NACE-4 \\ $\times$ CZ \\ (Employment)}} & \multirow{3.4}{*}{\shortstack{NACE-4 \\ $\times$ CZ \\ (Employment)}} & \multirow{3.4}{*}{\shortstack{NACE-4 \\ $\times$ CZ \\ (Employment)}} \\
&&&& \\
&&&& \\[0.2cm] 
Observations &  2,139,426       & 2,139,426       & 2,139,400       & 2,139,400         \\
\multirow{2.4}{*}{\shortstack{First-Stage Effect}} &   &  &  & \multirow{2.4}{*}{\shortstack{\hphantom{***}0.820***  \\ (0.042)}}   \\
&&&& \\[0.2cm] 
F: Instrument &  &  &  & ~387.0 \\[0.2cm]  \hline
\end{tabular}
\begin{tablenotes}[para]
\footnotesize\textsc{Note. ---} The table displays fixed effects regressions of log daily wages (in terms of firm-level averages of regular full-time workers) on log labor market concentration (in terms of HHI). The instrumental variable refers to the leave-one-out average of the log inverse number of firms in all other commuting zones but for the same industry and time period. Standard errors (in parentheses) are clustered at the labor market level. CZ = Commuting Zone. HHI = Herfindahl-Hirschman Index. INF = Inverse Number of Firms. NACE-X = X-Digit Statistical Nomenclature of Economic Activities in the European Community. * = p$<$0.10. ** = p$<$0.05. *** = p$<$0.01. Sources: IEB $\plus$ BHP, 1999-2014.
\end{tablenotes}
\end{threeparttable}
}
\end{table}

}

\paragraph{Effects of Labor Market Concentration on Employment.} To see whether the negative effect on wages in more concentrated labor markets also comes along with fewer workers, Table \ref{tab:2} estimates the effect of labor market concentration on employment of regular full-time workers, using the very same specification of Equation (\ref{eq:3}). Indeed, all four regressions point towards a negative employment effect, indicating that more concentrated labor markets employ significantly less workers vis-\`{a}-vis less concentrated labor markets. The baseline IV estimate in Column (4) is substantial: an HHI increase by 100 percent makes firms reduce their employment of regular full-time workers by 3.4 percent.\footnote{Similarly, \citet{MarinescuEtAl2021} find that a 100 percent increase in labor market concentration reduces the number of new hires by 3.2 percent.} In other words, an increase in HHI from the 25th to the 75th percentile comes along with an employment reduction by 22.5 percent. All reported effects are significantly different from zero.

\afterpage{

\begin{table}[!ht]
\centering
\scalebox{0.90}{
\begin{threeparttable}
\caption{Effects of Labor Market Concentration on Employment}
\label{tab:2}
\begin{tabular}{L{4cm}C{2.7cm}C{2.7cm}C{2.7cm}C{2.7cm}} \hline
&&&& \\[-0.3cm]
\multirow{4.4}{*}{} & \multirow{4.4}{*}{\shortstack{(1) \\ Log \\ Regular FT\vphantom{/} \\ Employment }} & \multirow{4.4}{*}{\shortstack{(2) \\ Log \\ Regular FT\vphantom{/} \\ Employment }} & \multirow{4.4}{*}{\shortstack{(3) \\ Log \\ Regular FT\vphantom{/} \\ Employment }} & \multirow{4.4}{*}{\shortstack{(4) \\ Log \\ Regular FT\vphantom{/} \\ Employment }} \\
&&&& \\
&&&& \\
&&&& \\[0.3cm] \hline
&&&& \\[-0.3cm]
\multirow{2.4}{*}{Log HHI} &  \multirow{2.4}{*}{\shortstack{\hphantom{***}-0.050***\hphantom{-}  \\ (0.012)}}  & \multirow{2.4}{*}{\shortstack{\hphantom{***}-0.055***\hphantom{-}  \\ (0.011)}} & \multirow{2.4}{*}{\shortstack{\hphantom{***}-0.020***\hphantom{-}  \\ (0.005)}} & \multirow{2.4}{*}{\shortstack{\hphantom{***}-0.034***\hphantom{-}  \\ (0.011)}}   \\
&&&& \\[0.2cm] 
\multirow{2.4}{*}{\shortstack{Fixed Effects}}  & \multirow{2.4}{*}{Firm} &  \multirow{2.4}{*}{\shortstack{Firm \\ Year}} & \multirow{2.4}{*}{\shortstack{Firm \\ CZ $\times$ Year \\ NACE-2 $\times$ Year}} &  \multirow{2.4}{*}{\shortstack{Firm \\ CZ $\times$ Year \\ NACE-2 $\times$ Year}} \\
&&&& \\[0.4cm] \hline
&&&& \\[-0.2cm]
Instrument  & None & None & None & $ \overline{\text{Log}\,\,\text{INF}} \,\,\backslash\, \{c\} $  \\
\multirow{3.4}{*}{\shortstack[l]{Labor Market \\ Definition \\ (Object)}}  & \multirow{3.4}{*}{\shortstack{NACE-4 \\ $\times$ CZ \\ (Employment)}} & \multirow{3.4}{*}{\shortstack{NACE-4 \\ $\times$ CZ \\ (Employment)}} & \multirow{3.4}{*}{\shortstack{NACE-4 \\ $\times$ CZ \\ (Employment)}} & \multirow{3.4}{*}{\shortstack{NACE-4 \\ $\times$ CZ \\ (Employment)}} \\
&&&& \\
&&&& \\[0.2cm] 
Observations &  2,139,426       & 2,139,426       & 2,139,400       & 2,139,400         \\
\multirow{2.4}{*}{\shortstack{First-Stage Effect}} &   &  &  & \multirow{2.4}{*}{\shortstack{\hphantom{***}0.820***  \\ (0.042)}}   \\
&&&& \\[0.2cm] 
F: Instrument &  &  &  & ~387.0 \\[0.2cm]  \hline
\end{tabular}
\begin{tablenotes}[para]
\footnotesize\textsc{Note. ---} The table displays fixed effects regressions of log employment (in terms of regular full-time workers per firm) on log labor market concentration (in terms of HHI). The instrumental variable refers to the leave-one-out average of the log inverse number of firms in all other commuting zones but for the same industry and time period. Standard errors (in parentheses) are clustered at the labor market level. CZ = Commuting Zone. FT = Full-Time. HHI = Herfindahl-Hirschman Index. INF = Inverse Number of Firms. NACE-X = X-Digit Statistical Nomenclature of Economic Activities in the European Community. * = p$<$0.10. ** = p$<$0.05. *** = p$<$0.01. Sources: IEB $\plus$ BHP, 1999-2014.
\end{tablenotes}
\end{threeparttable}
}
\end{table}

}

\paragraph{Sensitivity and Heterogeneity.} To test the sensitivity of the baseline effects, I perform additional IV regressions with alternative specifications, labor market definitions, and concentration indices. In these checks, most of the effects remain in close vicinity to the baseline, while some turn out even more pronounced. Importantly, the negative effects of labor market concentration on wages and employment manifest both in the West and the East German labor market. Labor market concentration enables firms to push down large portions of the wage distribution in the minimum wage sectors, with greater monopsonistic conduct at the top than at the bottom end of the distribution. Importantly, I also arrive at a negative effect of labor market concentration on overall employment, which is the sum of regular full-time, regular part-time, and marginal part-time workers.

\paragraph{Causal Interpretation.} Acknowledging that the instrument might not be fully exogenous in the presence of common supply or demand shocks, I follow Conley et al.'s \citeyearpar{ConleyEtAl2012} plausibly exogenous regression technique to derive upper and lower bounds for the causal effect of labor market concentration on wages and employment (see Appendix Table \ref{tab:G11}). Under the assumption that the instrument's direct effect in the second stage ranges between zero (exogeneity) and the reduced-form effect, the bounds for the second-stage effect of labor market concentration range between -0.027 and 0.007 for wages, and between -0.51 and 0.017 for employment. Building on these intervals, I show that the negative effects of labor market concentration remain even for large potential violations of exogeneity of the instrument, namely 68.9 percent (wages) and 50.7 percent (employment) of the reduced-form effect.

As discussed in Section \ref{sec:5}, industry-based measures of local labor market concentration might pick up product market concentration unless the latter has a national rather than a local dimension. Reassuringly, the analysis in Appendix \ref{sec:F} demonstrates that labor market concentration positively correlates with markdowns and that such a positive relationship with price-cost markups is not evident, ruling out confounding impact of product market power. As an additional check, I re-run the regressions while disregarding the service sectors where product markets tend to have more of a local nature and thus might overlap with labor markets. In line with the absence of a positive relationship between labor market concentration and markups in these sectors, the HHI effects on wages and employment hardly change (see Appendix Table \ref{tab:G12}).

By and large, I document robust evidence that firms employ fewer workers at lower wage levels when operating in more concentrated labor markets. These findings corroborates the predictions of monopsony theory in a sense that employers with labor market power suppress wages along a positively-sloped labor supply curve to the firm.

\section{Minimum Wage Effects}
\label{sec:7}

Building on the evidence for monopsonistic conduct before minimum wages were introduced, I examine next whether the introduction and increase in sectoral minimum wages counteracted monopsony power without having negative effects on employment. Specifically, I scrutinize whether sectoral minimum wages exhibit less adverse employment effects in more monopsonistic labor markets where concentration tends to be higher. In Appendix \ref{sec:H}, I provide full details on the regression results, robustness checks, and heterogeneity analyses.

\paragraph{Empirical Model.} The empirical analysis of minimum wage effects proceeds in two steps. In a first step, I test whether higher sectoral minimum wages effectively raise wage levels. An empirical bite confirmation is an essential prerequisite because causal effects on employment will not manifest unless minimum wages are binding. Provided that there is a significant bite, I examine the minimum wage effects on employment in a second step. In both cases, the baseline version of the regression model takes the following log-linear form
\begin{equation}
\label{eq:5}
ln\,Y_{jsrizt} \,=\,  \alpha \cdot ln\,w^{min}_{srt}\, + \, \beta \cdot ln\,w^{min}_{srt} \cdot \overline{H\!H\!I}_{iz}  \, + \,  X^{T}_{st} \,  \gamma \,+\, \delta_{j}  \,+\, \zeta_{zt}  \,+\, \varepsilon_{jsrizt}
\end{equation}
where the outcome variable $Y$ refers to either average daily wages or employment in firm $j$ for year $t$, $w^{min}$ is the prevailing minimum wage in sector $s$ and federal state $r$, $H\!H\!I$ is the respective Herfindahl-Hirschman Index, $X$ is the set of control variables, $\delta_{j}$ and $\zeta_{zt}$ are firm and commuting-zone-by-year fixed effects, and $\epsilon$ is the error term.\footnote{The notation in (\ref{eq:5}) highlights that minimum wages differ regionally in some sectors. If so, minimum wages are set differently for West Germany and East Germany, or by federal state.} Specifically, I regress the logarithm of the outcome variable on the logarithm of the current minimum wage level in the sector. To infer a potentially moderating role of labor market concentration, I augment this specification with an interaction effect between the minimum wage levels and HHI for the labor market in which the firm is operating. Thus, the estimate for the minimum wage elasticity on wages or employment, as a function of HHI, reads: $\hat{\eta}^{\,Y}_{w^{min}} \,=\, \hat{\alpha} + \hat{\beta} \cdot H\!H\!I $. Standard errors are clustered at the sector-by-federal-state level.

As in Section \ref{sec:6}, I successively include firm fixed effects, year fixed effects, and commuting-zone-by-year fixed effects in a first, second, and third specification of (\ref{eq:5}) to absorb unobserved time-invariant heterogeneity between employers and local supply or demand shocks.\footnote{Note that, unlike in Equation (\ref{eq:2}), I cannot control for industry-by-year fixed effects in the minimum wage regressions because minimum wages are essentially set at this level.} \citet{HaucapEtAl2011} and \citet{BachmannEtAl2014} discuss strategic behavior of unions and employer associations under the possibility of coverage extension for CBAs in Germany. Hence, in a fourth and baseline specification, I address potential endogeneity of the minimum wage level using a set of sectoral control variables: the sectoral share of firms subject to CBAs and the logarithm of sectoral employment. In the spirit of \citet{AllegrettoEtAl2011}, I further control for sector-specific linear time trends to grasp heterogeneous paths of growth in the outcome variable.

Across all four specifications, I construct the interaction term as a product of the sectoral minimum wage level and the firm-specific average of HHI across available years. The use of averaged HHI values is useful for two reasons: First, my paramount objective is to examine minimum wage effects within firms at fixed values of labor market concentration. However, in a fixed-effects panel design, demeaning an interaction effect of two time-varying regressors yields a combination of mutual between- and within-unit interdependencies of both variables \citep{GiesselmannSchmidtCatran2022}. In other words, the interaction term may not only capture the effect of minimum wages at different HHI values but also the effect of HHI at different minimum wage levels. Yet, averaging over HHI will ensure that the interaction effect captures solely the moderating effect of level differences in HHI on within-unit variation in minimum wages. Second, the use of average HHI values alleviates reverse causality between employment responses and the employment-based HHI. On the downside, however, averaging over HHI will still give rise to endogeneity in case minimum wage changes correlate with HHI levels. To gauge the magnitude of this potential bias, I regress HHI values on sectoral minimum wage levels to inspect the correlation between both variables, conditional on the covariates from the baseline specification. Reassuringly, the variables do not exhibit any correlation at all (p=0.90), invalidating the concern that averaging over HHI might induce endogeneity (see Appendix Table \ref{tab:H1}). Nevertheless, in later robustness checks, I show that the results are virtually unchanged when interacting minimum wage levels with predetermined HHI values.\footnote{As a more rigorous alternative to HHI averages, predetermined HHI values eliminate any reverse causality between minimum wage effects and HHI values. As a drawback, however, predetermined HHI values provide a less representative picture of the underlying labor market concentration over the period of study.}

\paragraph{Minimum Wage Effects on Wages.} Before turning to employment effects, it is necessary to verify whether the minimum wages actually raised wages. Building on Equation (\ref{tab:5}), Table \ref{tab:3} displays the estimated minimum wage effects on wages by market form. Across all four specifications, the regressions demonstrate that increases in sectoral minimum wages were binding, regardless of the underlying market form.\footnote{In default of a bite, near-zero minimum wage effects on employment would have a fundamentally different implication relative to a setting with binding minimum wages.} Column (4) relates to the baseline specification with firm fixed effects, commuting-zone-by-year fixed effects, and sectoral control variables. In atomistic labor markets with zero HHI, a 10 percent rise in minimum wages leads ceteris paribus to an increase in average daily wages by 0.8 percent. Interestingly, the minimum wage effects on wages turn out more positive in more concentrated labor markets, seemingly reflecting the evidence on monopsonistic conduct before the minimum wage introduction in the sectors. Specifically, in the polar case of a monopsonistic labor market with HHI equal to 1, a rise in minimum wages by 10 percent results in a wage increase by 3.3 percent. Both the main effects and the interaction effects are significantly different from zero.

\begin{table}[!ht]
\centering
\scalebox{0.90}{
\begin{threeparttable}
\caption{Minimum Wage Effects on Wages}
\label{tab:3}
\begin{tabular}{L{4cm}C{2.7cm}C{2.7cm}C{2.7cm}C{2.7cm}} \hline
&&&& \\[-0.3cm]
\multirow{4}{*}{} & \multirow{4.4}{*}{\shortstack{(1) \\ Log \\ Mean Daily\vphantom{/} \\ Wages }} & \multirow{4.4}{*}{\shortstack{(2) \\ Log \\ Mean Daily\vphantom{/} \\ Wages }} & \multirow{4.4}{*}{\shortstack{(3) \\ Log \\ Mean Daily\vphantom{/} \\ Wages }} & \multirow{4.4}{*}{\shortstack{(4) \\ Log \\ Mean Daily\vphantom{/} \\ Wages }} \\
&&&& \\
&&&& \\
&&&& \\[0.3cm] \hline
&&&& \\[-0.3cm]
\multirow{2.4}{*}{Log Minimum Wage} &  \multirow{2.4}{*}{\shortstack{\hphantom{***}0.689***  \\ (0.023)}}  & \multirow{2.4}{*}{\shortstack{\hphantom{***}0.077***  \\ (0.025)}} & \multirow{2.4}{*}{\shortstack{\hphantom{*}0.045*  \\ (0.025)}} & \multirow{2.4}{*}{\shortstack{\hphantom{***}0.083***  \\ (0.019)}}   \\
                                                                                                                                                                                                                    &&&&  \\
\multirow{2.4}{*}{\shortstack[l]{Log Minimum Wage \\ $\times$ $\overline{\text{HHI}}$}} &  \multirow{2.4}{*}{\shortstack{\hphantom{***}0.733*** \\ (0.142)}}  & \multirow{2.4}{*}{\shortstack{\hphantom{***}0.382*** \\ (0.109)}} & \multirow{2.4}{*}{\shortstack{\hphantom{***}0.263*** \\ (0.096)}} & \multirow{2.4}{*}{\shortstack{\hphantom{***}0.243*** \\ (0.084)}}    \\
&&&& \\[0.2cm] 
Control Variables  & No & No  & No & Yes \\
\multirow{2.4}{*}{\shortstack{Fixed Effects}}  & \multirow{2.4}{*}{Firm} &  \multirow{2.4}{*}{\shortstack{Firm \\ Year}} & \multirow{2.4}{*}{\shortstack{Firm \\ CZ $\times$ Year}} &  \multirow{2.4}{*}{\shortstack{Firm \\ CZ $\times$ Year}} \\
&&&& \\[0.2cm] \hline
&&&& \\[-0.4cm]
\multirow{3.4}{*}{\shortstack[l]{Labor Market \\ Definition \\ (Object)}}  & \multirow{3.4}{*}{\shortstack{NACE-4 \\ $\times$ CZ \\ (Employment)}} & \multirow{3.4}{*}{\shortstack{NACE-4 \\ $\times$ CZ \\ (Employment)}} & \multirow{3.4}{*}{\shortstack{NACE-4 \\ $\times$ CZ \\ (Employment)}} & \multirow{3.4}{*}{\shortstack{NACE-4 \\ $\times$ CZ \\ (Employment)}} \\
&&&& \\
&&&& \\[0.2cm] 
Observations &  2,700,155    & 2,700,155    & 2,700,155    & 2,700,155      \\[0.2cm]
Adjusted R$^2$ &  0.799    & 0.810    &   0.811    & 0.811      \\[0.2cm] \hline
\end{tabular}
\begin{tablenotes}[para]
\footnotesize\textsc{Note. ---} The table displays fixed effects regressions of log daily wages (in terms of firm-level averages of regular full-time workers) on log sectoral minimum wage levels as well as their interaction effect with labor market concentration (measured as average HHI over time). The set of control variables includes log sectoral employment, the sectoral share of firms subject to a collective bargaining agreement, and sector-specific linear time trends. Standard errors (in parentheses) are clustered at the sector-by-federal-state level. CZ = Commuting Zone. HHI = Herfindahl-Hirschman Index. NACE-4 = 4-Digit Statistical Nomenclature of Economic Activities in the European Community. * = p$<$0.10. ** = p$<$0.05. *** = p$<$0.01. Sources: IEB $\plus$ BHP $\plus$ IAB Establishment Panel, 1999-2017.
\end{tablenotes}
\end{threeparttable}
}
\end{table}

\paragraph{Heterogeneity of Wage Effects.} In terms of East-West differences, the main effect is always significantly positive but turns out more pronounced in East Germany where the wage level generally tends to be lower. The interaction effects turn out to be positive in both regions but only the coefficient for West Germany remains significant. First and foremost, minimum wages push up wages at the bottom of the wage distribution. In line, the minimum wage effects on higher percentiles of the firm-specific wage distribution turn out increasingly smaller: in zero-HHI markets, a minimum wage increase by 10 percent significantly raises the 25th percentile, the median, and the 75th percentile of the wage distribution by 1.4, 0.9, and 0.5 percent. For each of the three percentiles, increases in sectoral minimum wages bite harder in more concentrated labor markets, although only the interaction effects at the median and 75th percentile are significantly positive.


\begin{table}[!ht]
\centering
\scalebox{0.90}{
\begin{threeparttable}
\caption{Minimum Wage Effects on Employment}
\label{tab:4}
\begin{tabular}{L{4cm}C{2.7cm}C{2.7cm}C{2.7cm}C{2.7cm}} \hline
&&&& \\[-0.3cm]
\multirow{4}{*}{} & \multirow{4.4}{*}{\shortstack{(1) \\ Log \\ Regular FT\vphantom{/} \\ Employment }} & \multirow{4.4}{*}{\shortstack{(2) \\ Log \\ Regular FT\vphantom{/} \\ Employment }} & \multirow{4.4}{*}{\shortstack{(3) \\ Log \\ Regular FT\vphantom{/} \\ Employment }} & \multirow{4.4}{*}{\shortstack{(4) \\ Log \\ Regular FT\vphantom{/} \\ Employment }} \\
&&&&  \\
&&&&  \\
&&&&  \\[0.3cm] \hline
&&&& \\[-0.3cm]
\multirow{2.4}{*}{Log Minimum Wage} &  \multirow{2.4}{*}{\shortstack{\hphantom{***}-0.375***\hphantom{-} \\ (0.060)}}  & \multirow{2.4}{*}{\shortstack{\hphantom{***}-0.326***\hphantom{-} \\ (0.042)}} & \multirow{2.4}{*}{\shortstack{\hphantom{***}-0.310***\hphantom{-} \\ (0.029)}} & \multirow{2.4}{*}{\shortstack{\hphantom{***}-0.232***\hphantom{-} \\ (0.034)}}   \\
&&&&  \\
\multirow{2.4}{*}{\shortstack[l]{Log Minimum Wage \\ $\times$ $\overline{\text{HHI}}$}} &  \multirow{2.4}{*}{\shortstack{\hphantom{***}1.081*** \\ (0.350)}}  & \multirow{2.4}{*}{\shortstack{\hphantom{**}0.679** \\ (0.333)}} & \multirow{2.4}{*}{\shortstack{\hphantom{***}1.161*** \\ (0.286)}} & \multirow{2.4}{*}{\shortstack{\hphantom{***}1.130*** \\ (0.299)}}    \\
&&&&  \\[0.2cm] 
Control Variables  & No & No  & No & Yes  \\
\multirow{2.4}{*}{\shortstack{Fixed Effects}}  & \multirow{2.4}{*}{Firm} & \multirow{2.4}{*}{\shortstack{Firm \\ Year}}  & \multirow{2.4}{*}{\shortstack{Firm \\ CZ $\times$ Year}}  & \multirow{2.4}{*}{\shortstack{Firm \\ CZ $\times$ Year}} \\
&&&& \\[0.2cm] \hline
&&&& \\[-0.4cm]
\multirow{3.4}{*}{\shortstack[l]{Labor Market \\ Definition \\ (Object)}}  & \multirow{3.4}{*}{\shortstack{NACE-4 \\ $\times$ CZ \\ (Employment)}} & \multirow{3.4}{*}{\shortstack{NACE-4 \\ $\times$ CZ \\ (Employment)}} & \multirow{3.4}{*}{\shortstack{NACE-4 \\ $\times$ CZ \\ (Employment)}}  & \multirow{3.4}{*}{\shortstack{NACE-4 \\ $\times$ CZ \\ (Employment)}} \\
&&&&  \\
&&&&  \\[0.2cm] 
Observations &  2,700,155    & 2,700,155    & 2,700,155    & 2,700,155      \\[0.2cm]
Adjusted R$^2$ &  0.874    & 0.876    &   0.877    & 0.877      \\[0.2cm] \hline
\end{tabular}
\begin{tablenotes}[para]
\footnotesize\textsc{Note. ---}The table displays fixed effects regressions of log employment (in terms of regular full-time workers per firm) on log sectoral minimum wage levels as well as their interaction effect with labor market concentration (measured as average HHI over time). The set of control variables includes log sectoral employment, the sectoral share of firms subject to a collective bargaining agreement, and sector-specific linear time trends. Standard errors (in parentheses) are clustered at the sector-by-federal-state level. CZ = Commuting Zone. FT = Full-Time. HHI = Herfindahl-Hirschman Index. NACE-4 = 4-Digit Statistical Nomenclature of Economic Activities in the European Community. * = p$<$0.10. ** = p$<$0.05. *** = p$<$0.01. Sources: IEB $\plus$ BHP $\plus$ IAB Establishment Panel, 1999-2017.
\end{tablenotes}
\end{threeparttable}
}
\end{table}


\paragraph{Minimum Wage Effects on Employment.} Given the underlying bite of sectoral minimum wages, I proceed with studying the minimum wage effects on employment. Table \ref{tab:4} displays the estimated minimum wage effects on employment, using an analogous version of Equation (\ref{eq:5}). The results buttress the monopsony argument. All specifications arrive at significantly negative minimum wage effects of employment in zero-HHI markets, corroborating the notion of perfectly competitive markets in a sense that binding wage floors harm employment. However, the HHI interaction effects turn out significantly positive, highlighting that labor market concentration moderates employment responses: the negative employment effects disappear for increasing levels of concentration and, eventually, become positive in highly concentrated labor markets -- as put forward by the monopsony argument. Column (4) refers to the baseline specification with firm fixed effects, commuting-zone-by-year fixed effects, and sectoral controls. In zero-HHI markets, a 10 percent increase in sectoral minimum wages reduces regular full-time employment by 2.3 percent. This negative effect weakens for higher HHI values and reaches zero around the HHI threshold of 0.2 from E.U.\ antitrust policy, which mirrors a market with five equally-sized firms. Above this threshold, higher minimum wages have positive effects and, for the polar case of HHI=1, a 10 percent increase in the minimum wages raises employment by 9.0 percent (see Appendix Figure \ref{fig:H1}).

\paragraph{Non-Linearities.} I perform multiple checks to examine the sensitivity and heterogeneity of the estimated employment effects. As my analysis takes place at the firm level and the firm-weighted HHI distribution is not uniform, the identification of the linear interaction effect may not reflect all parts of the HHI range equally. To ascertain that the positive HHI gradient is not just a result of the linearity assumption, I construct interaction effects between minimum wage levels and indicator variables that separate low- from high-HHI labor markets. Following E.U.\ antitrust thresholds, I estimate main and interactions effects for divisions of the HHI range into two, three, four, and five domains (see Appendix Table \ref{tab:H4}). Across these specifications, the estimates exhibit the same pattern as before: negative employment effects in slightly concentrated markets that weaken for higher HHI domains and, ultimately, become positive in highly concentrated labor markets. In this respect, Figure \ref{fig:2} illustrates the estimated pattern of elasticities for the most detailed division into five HHI domains.

\begin{figure}[!ht]
\centering
\caption{Minimum Wage Elasticity of Employment by HHI Categories}
\label{fig:2}
\scalebox{0.80}{
\begin{tikzpicture}
\begin{axis}[xlabel=Herfindahl-Hirschman Index, ylabel=Minimum Wage Elasticity of Employment, ymax=0.85, ymin=-0.60, height=14cm, width=14cm, grid=major, grid style=dotted, bar width=1.75cm,enlarge x limits={abs=1.5cm}, xtick={1,2,3,4,5},xticklabels={\shortstack{Very Low\vphantom{/}\\ 0.00-0.05}, \shortstack{Low\vphantom{/}\\ 0.05-0.10}, \shortstack{Medium\vphantom{/}\\ 0.10-0.20}, \shortstack{High\vphantom{/}\\ 0.20-0.40}, \shortstack{Very High\vphantom{/}\\ 0.40-1.00} }, ytick={-0.50,-0.25,0,0.25,0.50,0.75}, legend pos = north west, y tick label style={/pgf/number format/.cd,fixed,fixed zerofill, precision=2,/tikz/.cd}]
\addplot[ybar, area legend, color=black, fill=white] coordinates {(1,0)}; 
\addplot[ybar, area legend, color=black, fill=gray!2.5!white] coordinates {(1,-.22544844)};
\addplot[ybar, area legend, color=black, fill=gray!7.5!white] coordinates {(2,-.10963407)};
\addplot[ybar, area legend, color=black, fill=gray!15!white] coordinates {(3,.07344028)};
\addplot[ybar, area legend, color=black, fill=gray!30!white] coordinates {(4,-.0211995)};
\addplot[ybar, area legend, color=black, fill=gray!70!white] coordinates {(5,.32978871)};
\addplot[only marks,mark=square*, color=black, error bar legend, error bars/.cd, y dir=both, y explicit] coordinates { (1,-.22544844)+-(0,.05491288) (2,-.10963407)+-(0,.16845787) (3,.07344028)+-(0,.1544805) (4,-.0211995)+-(0,.15409707) (5,.32978871)+-(0,.30758283) };
\legend{~Employment Elasticity,,,,,,~90\% Confidence Interval}
\end{axis}
\end{tikzpicture}
}
\floatfoot{\footnotesize\textsc{Note. ---} The figure illustrates estimated minimum wage elasticities of employment. Estimates stem from fixed effects regressions of log employment (in terms of regular full-time workers per firm) on log sectoral minimum wag levels as well as their categorial interaction effects of labor market concentration (measured as average HHI over time). Each point estimates features a 90 percent confidence interval. Sources: IEB $\plus$ BHP $\plus$ IAB Establishment Panel, 1999-2017.}
\end{figure}

\paragraph{Sensitivity and Heterogeneity of Employment Effects.} The empirical results also withstand further scrutiny. Figure \ref{fig:3} compares the baseline minimum wage elasticities with those from sensitivity and heterogeneity analyses. Specifically, I perform sensitivity checks for alternative specifications, labor market definitions, and concentration measures. In general, the results are robust to various modifications of the baseline equation, thus lending further support to the monopsony argument. Importantly, the use of time-varying or predetermined HHI values as moderating variable does not lead to marked changes in the elasticity estimates, mirroring the insignificant correlation between HHI and sectoral minimum wage levels (see Appendix Table \ref{tab:H1}).

\afterpage{

\begin{figure}[!ht]
\centering
\caption{Sensitivity and Heterogeneity of Employment Effects}
\label{fig:3}
\scalebox{0.80}{
\begin{tikzpicture}
\begin{axis}[xlabel=Minimum Wage Elasticity of Employment, ylabel=Specification, height=25cm, width=13cm, grid=major, grid style=dotted, xmin=-1, xmax=2.6, ymin=-1, ymax=25, xtick={-0.8,-0.4,0,0.4,0.8,1.2,1.6,2.0,2.4}, extra x tick style={grid=none}, extra x ticks={0}, legend pos= north east, x tick label style={/pgf/number format/.cd,fixed,fixed zerofill, precision=1,/tikz/.cd},y tick label style = {font=\footnotesize}, ytick={1,2,3,4.5,5.5,7,8,9,10,11,12.5,13.5,14.5,15.5,16.5,17.5,19,20,21,22,23.5}, yticklabels={{Marginal Employment\vphantom{/}},{Regular Part-Time Employment\vphantom{/}},{Overall Employment\vphantom{/}},{East Germany\vphantom{/}},{West Germany\vphantom{/}},{Market Share\vphantom{/}},{Exponential Index\vphantom{/}},{Inverse Number of Firms\vphantom{/}},{1-Firm Concentration Ratio\vphantom{/}},{Rosenbluth Index\vphantom{/}},{HHI Based on Wage Bill\vphantom{/}},{HHI Based on Hires\vphantom{/}},{NUTS-3 Regions\vphantom{/}},{Flow-Adjusted NACE-4 Industries\vphantom{/}},{NACE-5 Industries\vphantom{/}},{NACE-3 Industries\vphantom{/}},{With Implicit Minimum Wage\vphantom{/}},{Without Linear Time Trends\vphantom{/}},{Predetermined HHI\vphantom{/}},{Time-Varying HHI\vphantom{/}},{\textbf{Baseline}\vphantom{/}},}]
\addplot[only marks,mark=square*,mark options={fill=white},color=gray, error bar legend, error bars/.cd, x dir=both, x explicit] coordinates {  (-.06894696,1.15)+-(.04163761,0) (.35710642,2)+-(.10417295,0) (-.06763798,3)+-(.04086132,0) (-.3378047,4.5)+-(.11063001,0) (-.18094733,5.5)+-(.06225513,0) (-.22535387,7)+-(.0540387,0) (-.22591412,8)+-(.0546672,0) (-.2215984,9)+-(.05389221,0) (-.25584477,10)+-(.06049171,0) (-.22681507,11)+-(.0547457,0) (-.23704757,12.5)+-(.05649062,0) (-.20794007,13.65)+-(.05419266,0) (-.24922261,14.5)+-(.05968681,0) (-.23298889,15.5)+-(.05596974,0) (-.22974494,16.5)+-(.05572745,0) (-.22047998,17.65)+-(.05431269,0) (-.24709602,19)+-(.04521461,0) (-.30934268,20)+-(.04753721,0) (-.23189548,21)+-(.05551267,0) (-.22665191,22)+-(.05486524,0) (-.23160227,23.5)+-(.05570547,0) };
\addplot[only marks,mark=square*,mark options={fill=black}, color=black, error bar legend, error bars/.cd, x dir=both, x explicit] coordinates {  (.27190158,.85000002)+-(.39458624,0) (1.7425697,2)+-(.60487378,0) (.93711489,3)+-(.44354185,0) (.91127497,4.5)+-(.76977336,0) (.85565376,5.5)+-(.61153316,0) (1.5135553,7)+-(.51530951,0) (1.0713435,8)+-(.53277415,0) (1.2278509,9)+-(.53616053,0) (.60008872,10)+-(.38478595,0) (.96958166,11)+-(.48907629,0) (.91025287,12.5)+-(.43593985,0) (.21755578,13.35)+-(.40428865,0) (.37427628,14.5)+-(.21375318,0) (1.3586199,15.5)+-(.67847896,0) (.63107532,16.5)+-(.37563151,0) (.23652096,17.35)+-(.72862029,0) (.57245302,19)+-(.25428355,0) (.8461464,20)+-(.46065167,0) (.84215969,21)+-(.43534595,0) (.488736,22)+-(.3918007,0) (.89809322,23.5)+-(.48252922,0) };
\addplot[loosely dashed] coordinates {(0,-1.5) (0,26.5)};
\legend{~HHI=0~,~HHI=1~}
\end{axis}
\end{tikzpicture}
}  \vspace*{-0.25cm}
\floatfoot{\footnotesize\textsc{Note. ---} The figure illustrates estimated minimum wage elasticities of employment for a variety of sensitivity and heterogeneity analyses. The baseline estimation regresses log employment (in terms of regular full-time workers per firm) on log sectoral minimum wage levels, an interaction effect with labor market concentration (measured as average HHI over time), firm fixed effects, commuting-zone-by-year fixed effects, and a set of sectoral control variables). Labor markets are combinations of 4-digit NACE industries and commuting zones. Hollow squares refer to point elasticities for HHI=0 whereas solid squares represent elasticities for HHI=1. Each point estimate features a 90 percent confidence interval.  HHI = Herfindahl-Hirschman Index. NACE-X = X-Digit Statistical Nomenclature of Economic Activities in the European Community. NUTS-X = X-Digit Statistical Nomenclature of Territorial Units. Sources: IEB $\plus$ BHP $\plus$ IAB Establishment Panel, 1999-2017.}
\end{figure}

}

In line with a stronger bite, the negative employment effects in low-HHI markets turn out more pronounced in East than in West Germany. Importantly, my findings in favor of the monopsony argument carry over to overall employment, which comprises regular full-time, regular part-time and marginal part-time workers. Interestingly, the group of regular part-time workers exhibits positive employment effects not only in highly concentrated but, to a lesser degree, also in slightly concentrated labor markets. Composition effects provide a natural explanation for the increase in regular part-time workers in low-HHI markets: On the one hand, negative employment responses along the hours margin transform regular full-time workers into regular part-time workers. On the other hand, with higher minimum wage levels, the monthly income of marginal part-time workers increasingly exceeds the threshold above which marginal part-time workers automatically turn into regular part-time workers.

\paragraph{Non-Monotonous Relationship.} In monopsonistic labor markets, the minimum wage effect on employment differs between three regimes of bindingness: zero (or small) effects in the unconstrained regime (i.e., the minimum wage does not bite), positive effects in the supply-determined regime (i.e., the minimum wage counteracts monopsony conduct), and negative effects in the demand-determined regime (i.e., firms lay off the least productive workers). To check whether reported minimum wage effects on employment obscure heterogeneity by the bite of the sectoral minimum wage, I re-run the baseline regression on a full set of interaction terms between log sectoral minimum wage levels, labor market concentration (measured as average HHI), and indicator variables for five quintile groups of bindingness of the sectoral minimum wage in a firm (measured as average Kaitz Index, see Appendix Table \ref{tab:H10}).\footnote{The firm-specific Kaitz Index is the ratio of the sectoral minimum wage to the median hourly wage rate of regular full-time workers in a firm. For lack of information on individual working hours, I approximate hourly wage rates by dividing weekly earnings of regular full-time workers by 40 working hours per week. Based on their average Kaitz Index, each firm is assigned one of five quintile groups: 0-0.68 (first group), 0.68-0.79 (second group), 0.79-0.92 (third group), 0.92-1.15 (fourth group), and higher than 1.15 (fifth group). The assumption of a 40-hour week introduces some measurement error which is why the absolute value of the Kaitz Indices should not be taken at face value.}

\afterpage{
\begin{landscape}

\vspace*{\fill}

\begin{figure}[!ht]
\centering
\caption{Minimum Wage Elasticity of Employment by Bindingness}
\label{fig:4}
\vspace*{0.5cm}
\begin{subfigure}{0.475\textwidth}
\centering
\caption{HHI=0}
\scalebox{0.85}{
\begin{tikzpicture}
\begin{axis}[xlabel=Quintile of Kaitz Index, ylabel=Minimum Wage Elasticity of Employment, ymax=2.5, ymin=-2.5, height=12cm, width=14cm, grid=major, grid style=dotted, bar width=1.75cm,enlarge x limits={abs=1.5cm}, xtick={1,2,3,4,5},xticklabels={First, Second, Third, Fourth, Fifth}, ytick={-2,-1,0,1,2}, legend pos = south west, y tick label style={/pgf/number format/.cd,fixed,fixed zerofill, precision=0,/tikz/.cd}]
\addplot[ybar, area legend, color=black, fill=white] coordinates {(1,-.23036627)};
\addplot[ybar, area legend, color=black, fill=white] coordinates {(2,-.1419978)};
\addplot[ybar, area legend, color=black, fill=white] coordinates {(3,-.13130076)};
\addplot[ybar, area legend, color=black, fill=white] coordinates {(4,-.32395279)};
\addplot[ybar, area legend, color=black, fill=white] coordinates {(5,-.53199631)};
\addplot[only marks,mark=square*, color=black, error bar legend, error bars/.cd, y dir=both, y explicit] coordinates {   (1,-.23036627)+-(0,.07722519) (2,-.1419978)+-(0,.06014347) (3,-.13130076)+-(0,.06999312) (4,-.32395279)+-(0,.08558264) (5,-.53199631)+-(0,.14565301) };
\legend{~Employment Elasticity,,,,,~90\% Confidence Interval}
\end{axis}
\end{tikzpicture}
}
\end{subfigure}
\hfill
\begin{subfigure}{0.475\textwidth}
\centering
\caption{HHI=1}
\scalebox{0.85}{
\begin{tikzpicture}
\begin{axis}[xlabel=Quintile of Kaitz Index, ylabel=Minimum Wage Elasticity of Employment, ymax=2.5, ymin=-2.5, height=12cm, width=14cm, grid=major, grid style=dotted, bar width=1.75cm,enlarge x limits={abs=1.5cm}, xtick={1,2,3,4,5},xticklabels={First, Second, Third, Fourth, Fifth}, ytick={-2,-1,0,1,2}, legend pos = south west, y tick label style={/pgf/number format/.cd,fixed,fixed zerofill, precision=0,/tikz/.cd}]
\addplot[ybar, area legend, color=black, fill=gray!70!white] coordinates {(1,.76296145)};
\addplot[ybar, area legend, color=black, fill=gray!70!white] coordinates {(2,1.4048133)};
\addplot[ybar, area legend, color=black, fill=gray!70!white] coordinates {(3,1.2972683)};
\addplot[ybar, area legend, color=black, fill=gray!70!white] coordinates {(4,.74260968)};
\addplot[ybar, area legend, color=black, fill=gray!70!white] coordinates {(5,-1.0234675)};
\addplot[only marks,mark=square*, color=black, error bar legend, error bars/.cd, y dir=both, y explicit] coordinates {  (1,.76296145)+-(0,.55190134) (2,1.4048133)+-(0,.61759931) (3,1.2972683)+-(0,.73665428) (4,.74260968)+-(0,1.1244383) (5,-1.0234675)+-(0,.92613113) };
\legend{~Employment Elasticity,,,,,~90\% Confidence Interval}
\end{axis}
\end{tikzpicture}
}
\end{subfigure}
\floatfoot{\footnotesize\textsc{Note. ---} The figures display estimated minimum wage elasticities of employment by quintile groups of the underlying Kaitz Index. Estimates stem from fixed effects regressions of log employment (in terms of regular full-time workers per firm) on a full set of interaction terms between log sectoral minimum wage levels, labor market concentration (measured as average HHI over time), and five quintile groups of minimum wage bindingness (measured as average Kaitz Index). The Kaitz Index is calculated as the ratio of the sectoral minimum wage to the median hourly wage rate of regular full-time workers per firm. The bars illustrate point elasticities for different quintile groups of the Kaitz Index at the polar values of the HHI distribution. Each point estimates features a 90 percent confidence interval. Sources: IEB $\plus$ BHP $\plus$ IAB Establishment Panel, 1999-2017.}
\end{figure}

\vspace*{\fill}

\end{landscape}
}

Figure \ref{fig:4} visualizes the resulting minimum wage elasticities of employment by quintile groups of bindingness, separately for markets with HHI=0 and HHI=1. The negative employment effect in low-HHI markets gradually becomes more negative when the bite intensifies: in the fifth quintile group of the Kaitz Index, the minimum wage elasticity of employment is about twice as negative as in the first quintile group (Panel a). In line with monopsony theory, the minimum wage elasticity of employment in labor markets, where HHI equals 1, is a non-monotonous function of the underlying bite (Panel b): At first, the positive employment effect grows with the bite, counteracting monopsony power along the labor supply curve. But, beginning with the fourth quintile group (i.e., when the minimum wage is set close to the median wage of the firm in these low-wage sectors), the positive minimum wage effect falls rapidly. The minimum wage elasticity of employment in the fifth quintile group of bindingness not only turns out smaller than the values in the first, second, and third quintile group, it is also significantly negative. Overall, the non-monotonous relationship implies that, given a higher bindingness, firms' adjustment increasingly takes place along the negatively sloped labor demand curve since minimum wages gradually exceed workers' marginal productivity. Furthermore, note that the estimated minimum wage elasticity of employment for HHI=1 exceeds the respective elasticity for HHI=0 within each of the bottom three quintile groups of the Kaitz Index (i.e., when conditioning on the absolute bite before a potential non-monotonicity kicks in). This finding rules out a possible concern that disemployment effect in low-HHI markets arise simply because low-HHI markets feature higher minimum wage levels than high HHI-markets.

\paragraph{Causal Interpretation.}  Before addressing potential confounders, I interact sectoral minimum wage levels directly with markdowns. Markdowns constitute the most comprehensive measure of monopsony power and reflect not only classical but also modern sources of monopsony power (like job differentiation or search frictions), but they can be linked to the BHP data only for the subsample of firms in the IAB Establishment Panel. Using this so-called ``Linked-Employer-Employee Dataset of the IAB'' (LIAB), Column (1) in Table \ref{tab:5} shows that the monopsony argument is also valid when taking markdowns as moderators: The minimum wage effect on employment turns out significantly negative when the markdown equals 1 (i.e., when workers are paid their marginal product). For increasing markdowns (i.e., when workers' wages are marked down relative to their marginal product), the employment effect becomes less negative and, ultimately, turns positive.\footnote{Without loss of generality, I have reformulated average markdowns as deviations from a value of 1 such that the main effect in Column (1) of Table \ref{tab:5} reflects the minimum wage effect on employment in a perfectly competitive market where wages equal marginal productivity.}

The baseline analysis demonstrates that the effect of minimum wages on employment is a positive function of labor market concentration. However, it is unclear whether labor market concentration -- as the classical source of monopsony power -- causally determines minimum wage responses or just constitutes a proxy for an omitted variable that is the factual moderator of the minimum wage effect on employment. In the latter case, however, labor market concentration would still provide correlative information on the true moderator and, thus, can guide policy-makers to assess the impact of higher minimum wages. Next, I scrutinize the two main candidates for such an omitted variable bias: product market power and productivity advantages.

As mentioned in Section \ref{sec:5}, product market concentration might inadvertently enter industry-based measures of local labor market concentration. If labor market concentration was not the causal moderator of the minimum wage effects, firms with labor market power could also pass on higher minimum wages more easily in the form of higher prices to consumers (e.g., \citealt{HarasztosiLindner2019}). If so, the positive HHI interaction term in the employment regression could not only be attributed to labor market but also product market power. Incompatible with this hypothesis, however, price-cost markup regressions do not show a positive relationship with labor market concentration for the minimum wage sectors (see Appendix Table \ref{tab:F4}). To further invalidate this argument, Column (2) of Table \ref{tab:5} highlights that higher markdowns continue to dampen the negative employment effects when additionally controlling for a moderating role of markups in the LIAB sample. Moreover, Column (3) in Table \ref{tab:5} shows that the baseline employment regressions are robust when filtering out service sectors whose product markets have more of a local nature, supporting the notion that it is labor market power that causally moderates the minimum wage effects.

In general, higher minimum wages will make more productive firms reduce employment less strongly than highly productive firms. Thus, if productivity positively correlates with labor market concentration, the HHI interaction term will be upward-biased. However, the frequent finding that labor market concentration turns out higher in rural areas, where firms also tend to be less productive, will rather result in a downward bias \citep{AzarEtAl2024}. To rule out such an omitted variable bias, I augment the baseline specification with an additional interaction effect between minimum wage levels and two different proxies for firm productivity. First, I proxy for productivity by dividing survey information on annual revenues by headcount employment for firms in the LIAB subsample. Second, I approximate productivity by firm fixed effects from log-linear wage regressions in the tradition of Abowd, Kramarz, and Margolis (\citeyear{AbowdEtAl1999}, hereafter `AKM'). These AKM effects reflect a relative wage premium paid to all workers in the firm, conditional on time-variant worker characteristics and fixed effects.\footnote{I retrieve AKM firm fixed effects for log daily wages of regular full-time workers from \citet{BellmannEtAl2020}.} In Columns (4) and (5) of Table \ref{tab:5}, the interaction effects with log productivity turn out positive: an increase in productivity by 1 percent raises the minimum wage elasticity of employment by 0.00309 (revenue per worker) or 0.00624 (AKM), respectively. Despite the moderating impact of productivity, the interaction effect with labor market concentration remains unchanged, thus lending further credence to the monopsony argument.

\afterpage{
\begin{landscape}

\begin{table}[!ht]
\centering
\scalebox{0.90}{
\begin{threeparttable}
\caption{Further Robustness Checks}
\label{tab:5}
\begin{tabular}{L{4cm}C{2.7cm}C{2.7cm}C{2.7cm}C{2.7cm}C{2.7cm}C{2.7cm}C{2.7cm}} \hline
&&&&&&& \\[-0.3cm]
\multirow{4}{*}{} & \multirow{4.4}{*}{\shortstack{(1) \\ Log \\ Regular FT\vphantom{/} \\ Employment }} & \multirow{4.4}{*}{\shortstack{(2) \\ Log \\ Regular FT\vphantom{/} \\ Employment }} & \multirow{4.4}{*}{\shortstack{(3) \\ Log \\ Regular FT\vphantom{/} \\ Employment }} & \multirow{4.4}{*}{\shortstack{(4) \\ Log \\ Regular FT\vphantom{/} \\ Employment }} & \multirow{4.4}{*}{\shortstack{(5) \\ Log \\ Regular FT\vphantom{/} \\ Employment }}  & \multirow{4.4}{*}{\shortstack{(6) \\ Firm\vphantom{/} \\ Closure\vphantom{/} }}   & \multirow{4.4}{*}{\shortstack{(7) \\ Log \\ Regular FT\vphantom{/} \\ Employment }} \\
&&&&&&&      \\
&&&&&&&      \\
&&&&&&&      \\[0.3cm] \hline
&&&&&&& \\[-0.3cm]
\multirow{2.4}{*}{Log Minimum Wage} &  \multirow{2.4}{*}{\shortstack{\hphantom{**}-0.356**\hphantom{-} \\ (0.150)}}  & \multirow{2.4}{*}{\shortstack{\hphantom{*}-0.268*\hphantom{-} \\ (0.161)}} & \multirow{2.4}{*}{\shortstack{\hphantom{***}-0.083***\hphantom{-} \\ (0.029)}} & \multirow{2.4}{*}{\shortstack{\hphantom{**}-0.240**\hphantom{-} \\ (0.118)}} & \multirow{2.4}{*}{\shortstack{\hphantom{***}-0.238***\hphantom{-} \\ (0.029)}}  & \multirow{2.4}{*}{\shortstack{\hphantom{***}0.125*** \\ (0.017)}}  & \multirow{2.4}{*}{\shortstack{\hphantom{*}-0.228*\hphantom{-} \\ (0.119)}}     \\
&&&&&&&  \\
\multirow{2.4}{*}{\shortstack[l]{Log Minimum Wage \\ $\times$ $\overline{\text{HHI}}$}} &    &     & \multirow{2.4}{*}{\shortstack{\hphantom{***}1.663*** \\ (0.354)}}   & \multirow{2.4}{*}{\shortstack{\hphantom{*}1.456* \\ (0.810)}}  & \multirow{2.4}{*}{\shortstack{\hphantom{***}1.155*** \\ (0.297)}} & \multirow{2.4}{*}{\shortstack{0.019 \\ (0.132)}} & \multirow{2.4}{*}{\shortstack{\hphantom{***}1.869*** \\ (0.423)}}    \\
&&&&&&&  \\
\multirow{2.4}{*}{\shortstack[l]{Log Minimum Wage \\ $\times$ $\overline{\text{Markdown}}$}} &  \multirow{2.4}{*}{\shortstack{\hphantom{***}0.379*** \\ (0.126)}} & \multirow{2.4}{*}{\shortstack{\hphantom{***}0.413*** \\ (0.152)}} &         &         &      &       &  \\
&&&&&&&  \\
\multirow{2.4}{*}{\shortstack[l]{Log Minimum Wage \\ $\times$ $\overline{\text{Markup}}$}} &    & \multirow{2.4}{*}{\shortstack{0.521 \\ (0.528)}}     &         &        &           &         &              \\
&&&&&&&  \\
\multirow{2.4}{*}{\shortstack[l]{Log Minimum Wage \\ $\times$ $\overline{\text{Log (Revenue/Worker)}}$}} &     &      &       & \multirow{2.4}{*}{\shortstack{\hphantom{*}0.309* \\ (0.175)}}   &             &       &        \\
&&&&&&&  \\
\multirow{2.4}{*}{\shortstack[l]{Log Minimum Wage \\ $\times$ $\overline{\text{Log AKM}}$}} &  &     &      &      &    \multirow{2.4}{*}{\shortstack{\hphantom{***}0.624*** \\ (0.140)}}      &                &          \\
&&&&&&&  \\[0.2cm] 
Control Variables  & Yes & Yes  & Yes & Yes & Yes & Yes & Yes   \\
\multirow{2.4}{*}{\shortstack{Fixed Effects}}  & \multirow{2.4}{*}{\shortstack{Firm \\ CZ $\times$ Year}} & \multirow{2.4}{*}{\shortstack{Firm \\ CZ $\times$ Year}}  & \multirow{2.4}{*}{\shortstack{Firm \\ CZ $\times$ Year}} & \multirow{2.4}{*}{\shortstack{Firm \\ CZ $\times$ Year}} & \multirow{2.4}{*}{\shortstack{Firm \\ CZ $\times$ Year}} & \multirow{2.4}{*}{\shortstack{Firm \\ CZ $\times$ Year}} & \multirow{2.4}{*}{\shortstack{Sector $\times$ Market\\ CZ $\times$ Year}} \\
&&&&&&& \\[0.2cm] \hline
&&&&&&& \\[-0.2cm]
Sample & LIAB & LIAB  & BHP~\textbackslash~Services & LIAB & BHP & BHP & BHP   \\
\multirow{2.4}{*}{Unit of Observation}  & \multirow{2.4}{*}{\shortstack{Firm \\ $\times$ Year}} & \multirow{2.4}{*}{\shortstack{Firm \\ $\times$ Year}} & \multirow{2.4}{*}{\shortstack{Firm \\ $\times$ Year}} & \multirow{2.4}{*}{\shortstack{Firm \\ $\times$ Year}} & \multirow{2.4}{*}{\shortstack{Firm \\ $\times$ Year}} & \multirow{2.4}{*}{\shortstack{Firm \\ $\times$ Year}} & \multirow{2.4}{*}{\shortstack{Sector $\times$ Market \\ $\times$ Year}}  \\
&&&&&&& \\[0.2cm] 
Observations &  10,869       & 10,394       & 2,135,562    & 16,329       & 2,585,412    & 2,700,155    & ~44,417         \\[0.2cm]
Adjusted R$^2$ &  0.951    & 0.951    &   0.860 &   0.954 &   0.872 &   0.157 &   ~0.962        \\[0.2cm] \hline
\end{tabular}
\begin{tablenotes}[para]
\footnotesize\textsc{Note. ---} The table displays fixed effects regressions of log employment (in terms of regular full-time workers) and an indicator for firm closure on log sectoral minimum wage levels as well as their interaction effect with labor market concentration and alternative moderators. The set of control variables includes log sectoral employment, the sectoral share of firms subject to a collective bargaining agreement, and sector-specific linear time trends. For better interpretation, markdowns and markups are centered at 1 while log revenue per worker and log AKM effects are centered around their mean. Labor markets are combinations of 4-digit NACE industries and commuting zones. Standard errors (in parentheses) are clustered at the sector-by-federal-state level. AKM = AKM Firm Fixed Effect. CZ = Commuting Zone. HHI = Herfindahl-Hirschman Index. NACE = Statistical Nomenclature of Economic Activities in the European Community. FT = Full-Time. * = p$<$0.10. ** = p$<$0.05. *** = p$<$0.01. Sources: IEB $\plus$ BHP $\plus$ IAB Establishment Panel $\plus$ LIAB, 1999-2017.
\end{tablenotes}
\end{threeparttable}
}
\end{table}

\end{landscape}
}

\paragraph{Firm Closures.} From a theoretical point of view, minimum wages do not only make employers adjust employment of incumbent firms but may also cause firms to close \citep{Williamson1968}. Abstracting from efficiency wage considerations, higher minimum wages are supposed to lower profits of treated firms. As a consequence, some of these firms are forced to quit the market. Following standard practice (e.g., \citealp{FacklerEtAl2013}), I identify a closure in the year when a firm appears for the last time in the data.\footnote{I use of the BHP wave from 2018 to separate factual from artificial closures at the edge of 2017.} Prior to minimum wage legislation, the probability of closure in the relevant sectors is 6.6 percent. After the minimum wage introduction in the respective sectors, this probability climbs to 9.0 percent. Column (6) of Table \ref{tab:5} displays the results from a linear probability model where the binary outcome of firm closure is regressed on the covariates from Equation (\ref{eq:5}). The main effect is significantly positive: an increase in the minimum wage by 1 percent raises the probability of closure by 0.13 percentage points in zero-HHI markets. In contrast, the HHI interaction term is not significantly different from zero. To compare market forms, however, it is necessary to normalize the coefficients by the underlying minimum wage elasticity of wages (see Column (4) in Table \ref{tab:3}): an effective wage increase by 1 percent will raise the probability of firm closure by 1.5 (=0.125/0.083) percentage points in zero-HHI markets but only by 0.4 (=(0.125+0.019)/(0.083+0.243)) percentage points in markets where HHI equals 1. A natural explanation for this difference are rents which turn out larger in more monopsonistic markets than in more competitive markets. My finding that higher sectoral minimum wages make firms exit the market complements analogous evidence on the introduction of the 2015 nation-wide minimum wage in the German labor market \citep{DustmannEtAl2022}.

\paragraph{Analysis at the Aggregate Level.} By construction, the variation in employment from closing or opening firms does not contribute to the baseline regression at the firm level. An elegant way of capturing minimum wage effects on employment of both incumbent, opening, and closing firms is to perform the analysis at an aggregate level. In Column (7) of Table \ref{tab:5}, I regress employment in pair-wise combinations of labor markets and the sixteen minimum wage sectors on the covariates from Equation (\ref{eq:5}). After collapsing the data accordingly, the monopsony argument turns out valid not only at the micro but also at the aggregate level. The minimum wage elasticity of employment in low-HHI markets is similarly negative as in Column (4) of Table \ref{tab:4}, seemingly because the positive effect on firm closures is absorbed by reallocation effects between employers at the aggregate level. In high-HHI markets, the minimum wage elasticity of aggregate employment stays significantly positive.

\paragraph{Minimum Wage Introduction in the Sectors.} In the previous analyses, the log-linear regressions deliver the effects of subsequent hikes in sectoral minimum wage levels over time. However, the first-time introduction of minimum wages in these sectors provides another valuable setting to trace out potentially varying effects by monopsony power. In the following analysis, I complement my concentration indices with quarterly firm data from the ``Administrative Worker Flow Panel'' (AWFP) and narrow the analysis to the time before and after the minimum wage introduction in the respective sectors.\footnote{Like Section \ref{sec:6}, the sectors of main construction, electrical trade, and roofing do not contribute to this analysis since the minimum wage in these sectors was introduced before the period of analysis 1999-2017.} Specifically, I estimate a timing-based event study model \citep{Miller2023} with only ever-treated units where earlier or later treated firms in the minimum wage sectors serve as control groups for each other. To portray dynamic effects with sufficient temporal frequency, calendar time $t$ in the event-study analysis refers to quarters. Let the event date $E_{s}$ be the specific quarter during which the first minimum wage came into effect in the sector $s$ and let event time $e$ denote to the number of quarters since this event took place. For outcome variable $Y$, the timing-based event-study model with treatment heterogeneity by average HHI then is
\begin{equation}
\label{eq:6}
ln\,Y_{jsizt} \,=\,  \sum_{e=-10}^{12} \phi_{e} \cdot D_{ste} \, + \, \sum_{e=-10}^{12} \psi_{e} \cdot D_{ste} \cdot \overline{H\!H\!I}_{iz} \,+\, \delta_{j}  \,+\, \zeta_{zt}  \,+\, \varepsilon_{jsizt}
\end{equation}
where $D_{ste}$ is a binary variable that takes value 1 if $E_{s}=t-e$, and 0 otherwise. For zero HHI, the coefficients $\phi_{e}$ capture the dynamic effects after the minimum wage introduction as they accumulate over time since the event ($e\geq0$). In addition, the coefficients before the event took place ($\phi_{e}$ for $e<0$) serve as placebo effects and may signal anticipation effects or omission of confounding variables. For HHI=1, the respective interaction term must be added for each event time: $\phi_{e} + \psi_{e}$. To align with my baseline regression model (\ref{eq:5}), the event-study analysis further includes firm fixed effects, $\delta_{j}$, and commuting-zone-by-year fixed effects, $\zeta_{zt}$. I narrow the event window to 10 lead quarters and 12 lag quarters, creating a relatively well balanced panel in event time. Following standard practice, I define the treatment effects relative to the immediate pre-treatment quarter: $\phi_{-1}=\psi_{-1}=0$. As event, calendar time, and event time are perfectly multicollinear, treatment effects in timing-based event-study models are only identified up to a linear trend \citep{BorusyakEtAl2024}. To circumvent this problem, I follow \citet{SchmidheinySiegloch2023} and pool treatment effects ten, eleven, and twelve quarters after the minimum wage introduction: $\phi_{10}=\phi_{11}=\phi_{12}$ and $\psi_{10}=\psi_{11}=\psi_{12}$. In a last step, after having estimated Equation (\ref{eq:6}), I correct the estimated dynamic treatment paths for linear pre-trends.

\begin{figure}[!t]
\centering
\caption{Employment Effects of Minimum Wage Introduction}
\label{fig:5}
\centering
\scalebox{0.70}{
\begin{tikzpicture}
\begin{axis}[xlabel={Quarters since Minimum Wage Introduction in the Sector}, ylabel={De-Trended Semi-Elasticity}, xmin=-11, xmax=11, ymin=-0.175, ymax=0.35, height=13.5cm, width=20cm, ymajorgrids, xtick={-10,-9,-8,-7,-6,-5,-4,-3,-2,-1,0,1,2,3,4,5,6,7,8,9,10}, xticklabels={-10,-9,-8,-7,-6,-5,-4,-3,-2,-1,0,1,2,3,4,5,6,7,8,9,{10-12}}, ytick={-0.1,0,0.1,0.2,0.3}, grid style = dotted, y tick label style={/pgf/number format/.cd,fixed,fixed zerofill, precision=1,/tikz/.cd}, x tick label style={/pgf/number format/.cd,fixed,fixed zerofill, precision=0,/tikz/.cd},legend pos= north west,scaled ticks=false,label style= {font=\Large}]

\addplot[mark=none,solid, color=gray, line width=0.25mm] coordinates {   (-10.15,.00558731) (-9.1499996,-.01800549) (-8.1499996,-.01194107) (-7.1500001,.00566238) (-6.1500001,.02337247) (-5.1500001,.00543362) (-4.1500001,.00120599) (-3.1500001,-.00356944) (-2.1500001,.00829545) (-1,-.01604122) (-.15000001,-.03312502) (.85000002,-.03945868) (1.85,-.0389297) (2.8499999,-.06690445) (3.8499999,-.07747626) (4.8499999,-.08595946) (5.8499999,-.0738515) (6.8499999,-.09013408) (7.8499999,-.10216576) (8.8500004,-.102389) (9.8500004,-.09910894) };
\addplot[only marks,mark=square*,mark options={fill=white}, color=gray, error bar legend, error bars/.cd, y dir=both, y explicit] coordinates {  (-10.15,.00558731)+-(0,.03745208) (-9.1499996,-.01800549)+-(0,.03429854) (-8.1499996,-.01194107)+-(0,.03177611) (-7.1500001,.00566238)+-(0,.02718772) (-6.1500001,.02337247)+-(0,.01890295) (-5.1500001,.00543362)+-(0,.01503882) (-4.1500001,.00120599)+-(0,.01905522) (-3.1500001,-.00356944)+-(0,.01514914) (-2.1500001,.00829545)+-(0,.0055433) (-1,-.01604122)+-(0,0) (-.15000001,-.03312502)+-(0,.01099693) (.85000002,-.03945868)+-(0,.01226174) (1.85,-.0389297)+-(0,.01360469) (2.8499999,-.06690445)+-(0,.01670393) (3.8499999,-.07747626)+-(0,.01785427) (4.8499999,-.08595946)+-(0,.02317155) (5.8499999,-.0738515)+-(0,.02730051) (6.8499999,-.09013408)+-(0,.03003832) (7.8499999,-.10216576)+-(0,.03102732) (8.8500004,-.102389)+-(0,.03847995) (9.8500004,-.09910894)+-(0,.04505306)  };

\addplot[mark=none,solid, color=black, line width=0.25mm] coordinates {   (-9.8500004,.00864938) (-8.8500004,.00868161) (-7.8499999,-.01928205) (-6.8499999,.01408627) (-5.8499999,.01181377) (-4.8499999,-.04123276) (-3.8499999,-.0078094) (-2.8499999,.01147929) (-1.85,.00949129) (-1,.0041226) (.15000001,.00203389) (1.15,.03064233) (2.1500001,.04329395) (3.1500001,.03540642) (4.1500001,.08822814) (5.1500001,.13810694) (6.1500001,.16900153) (7.1500001,.171075) (8.1499996,.17340197) (9.1499996,.2123903) (10.15,.25149369) };
\addplot[only marks,mark=square*,mark options={fill=black}, color=black, error bar legend, error bars/.cd, y dir=both, y explicit] coordinates {  (-9.8500004,.00864938)+-(0,.05558933) (-8.8500004,.00868161)+-(0,.0529115) (-7.8499999,-.01928205)+-(0,.04481434) (-6.8499999,.01408627)+-(0,.0424733) (-5.8499999,.01181377)+-(0,.04477877) (-4.8499999,-.04123276)+-(0,.0314899) (-3.8499999,-.0078094)+-(0,.02923113) (-2.8499999,.01147929)+-(0,.04119318) (-1.85,.00949129)+-(0,.03877616) (-1,.0041226)+-(0,0) (.15000001,.00203389)+-(0,.0316027) (1.15,.03064233)+-(0,.04760542) (2.1500001,.04329395)+-(0,.05547652) (3.1500001,.03540642)+-(0,.04255236) (4.1500001,.08822814)+-(0,.05099977) (5.1500001,.13810694)+-(0,.06260066) (6.1500001,.16900153)+-(0,.06777494) (7.1500001,.171075)+-(0,.05553345) (8.1499996,.17340197)+-(0,.06814297) (9.1499996,.2123903)+-(0,.07678175) (10.15,.25149369)+-(0,.07031971)  };

\addplot[loosely dashed, color=blue] coordinates {(-0.5,1) (-0.5,-1)};

\legend{,~HHI=0~,,~HHI=1~,}
\end{axis}
\end{tikzpicture}
}
\vspace{-0.25cm}
\floatfoot{\footnotesize\textsc{Note. ---} The figure illustrates estimated semi-elasticities from the introduction of minimum wages in the sectors. The timing-based event-study model regresses log employment (in terms of regular full-time workers per firm) on interactions between event-time dummy variables and labor market concentration (measured as average HHI over time), firm fixed effects, and commuting-zone-by-quarter fixed effects. The coefficients were corrected for linear pre-trends. The markers illustrate point estimates at polar HHI levels. Each point estimate features a 90 percent confidence interval. Standard errors (in parentheses) are clustered at the sector-by-federal-state level.  HHI = Herfindahl- Hirschman Index. Sources: IEB + AWFP, 1999-2017.}
\end{figure}

Favorably, the event-study results for the minimum wage introduction in the sectors corroborate the fixed effects results for minimum wage hikes. In a first step, I estimate dynamic treatment effects for firms' average daily wages of regular full-time workers. After the minimum wage introduction in the sectors, average daily wages begin to rise and the overall wage growth accumulates to around 2 percent after 10-12 quarters in zero-HHI markets (see Appendix Figure \ref{fig:H2}). In line with monopsonistic conduct before the minimum wage introduction, the bite turns out significantly larger with 6 percent in markets where HHI equals 1. Given this bite pattern, Figure \ref{fig:5} visualizes the dynamic treatment effects for employment of regular full-time workers before and after the event. For either of the polar HHI values, the pre-treatment effects turn out to be close-to-zero and largely insignificant. After the treatment, employment falls significantly in low-HHI markets whereas it increases significantly in high-HHI markets, corroborating the monopsony argument. Note that the first and second hike in the minimum wage level occurs on average 3.7 and 9.5 quarters after the minimum wage introduction in the sectors. As a consequence, the opposite development of employment accelerates during the post-event period, culminating in a reduction by around 10 percent in zero-HHI markets and an increase by more than 20 percent in markets where HHI equals 1.

\section{Discussion}
\label{sec:8}

In the following section, I put my findings into wider perspective and compare the reported effect size with available evidence from the German minimum wage literature. I refer the reader to Appendix \ref{sec:I} for further details on this comparison.

In the past, many studies have attributed the absence of negative minimum wage effects on employment to the prevalence of monopsony power (e.g., \citealp{CardKrueger1995}; \citealp{Moeller2012}). To date, however, the monopsony argument has been put forward mainly on the basis of theoretical considerations rather than systematic empirical evidence. My results provide direct empirical support to this line of argument by demonstrating that labor market concentration, the classical source for monopsony power, modulates the effect of minimum wages on employment. While, in line with perfect competition, minimum wages significantly reduce employment in slightly concentrated labor markets, I report zero and positive effects for moderately and highly concentrated labor markets in which firms enjoy more wage-setting power. The positive sign of the average employment effect in highly concentrated labor markets implies that wage floors were not raised far above market-clearing levels. In the German setup, this condition is plausibly met as sectoral minimum wages stem from collective agreements that necessitate approval from the respective employer association. My reasoning that minimum wages successfully counteracted firms' monopsony power is backed by four additional pieces of evidence: First, before minimum wages were implemented in the sectors, there is evidence for monopsonistic conduct of firms: holding other things equal, wages and employment turn out lower in more concentrated labor markets (see Section \ref{sec:6}). Second, the positive sign of the average employment effect in highly concentrated markets is governed by a non-monotonous relationship with the underlying bindingness of the minimum wage (see Figure \ref{fig:4}). Third, markdowns -- the most comprehensive measure of monopsony power -- positively moderate the minimum wage effects on employment (see Table \ref{tab:5}). Fourth, markdowns shrink with increasing sectoral minimum wage levels, suggesting less leeway for monopsonistic conduct of firms in the presence of a wage floor (see Appendix Table \ref{tab:F2}).

In light of my results, an adequate empirical design should take into account that minimum wage effects depend on the prevalence of monopsony power in the labor market. However, minimum wage studies generally pool information across different labor markets and, thus, might conceal heterogeneity by market form. If so, these studies arrive at closer-to-zero employment effects by averaging opposite effects across market forms. To make my results and estimates from the literature comparable, it is necessary to interpret employment responses to minimum wages in relation to the magnitude of the underlying bite \citep{DubeZipperer2024}. To this end, I calculate own-wage elasticities of employment based on minimum wage variation by dividing the minimum wage elasticity of employment by the respective minimum wage elasticity of wages, separately for different HHI levels:\footnote{To be precise, I derive minimum wage elasticities by dividing the effects in Column (4) of Table \ref{tab:3} by the effects in Column (4) of Table \ref{tab:4} at different values of labor market concentration.}
\begin{equation}
\label{eq:7}
\hat{\eta}^{\,L}_{w} \,(H\!H\!I) = \frac{\hat{\eta}^{\,L}_{w^{min}} (H\!H\!I) }{ \hat{\eta}^{\,w}_{w^{min}} (H\!H\!I) }
\end{equation} 

In Figure \ref{fig:6}, I contrast the resulting own-wage elasticities of employment with analogous but pooled estimates from the entirety of the German minimum wage literature. Most of the available own-wage elasticities of employment from the literature are not significantly different from zero. In contrast, my low-HHI estimates are located at the lower end of the distribution of point estimates and feature significantly negative values. For moderate HHI values, my elasticities are insignificant, resembling the consensus of close-to-zero effects in the literature. My high-HHI estimates are situated at the upper end of the distribution and feature significantly positive elasticities.


\afterpage{

\begin{figure}[!ht]
\centering
\caption{Own-Wage Elasticity of Employment Based on Minimum Wage Variation}
\label{fig:6}
\scalebox{0.80}{
\begin{tikzpicture}
\begin{axis}[ xlabel=Own-Wage Elasticity of Employment, ylabel=Study, height=23.25cm, width=11.5cm, grid=major, grid style=dotted, xmin=-6.5, xmax=6.5, ymin=0, ymax=24, xtick={-6,-4,-2,2,4,6}, extra x tick style={grid=none}, extra x ticks={0}, legend pos= north east, x tick label style={/pgf/number format/.cd,fixed,fixed zerofill, precision=0,/tikz/.cd},y tick label style = {font=\small}, ytick={1,2,3,4,5,6,7,8,9,10,11,12,13,14,15,16,17,18,19,20,21,22,23}, yticklabels={{\textbf{This Paper: HHI=1.00}\vphantom{/}},{\citet{KoenigMoeller2009}: West\vphantom{/}},{\citet{BoockmannEtAl2011a}: East\vphantom{/}},{\textbf{This Paper: HHI=0.40}\vphantom{/}},{\citet{BoockmannEtAl2011b}: East\vphantom{/}},{\citet{AhlfeldtEtAl2018}\vphantom{/}},{\citet{DustmannEtAl2022}: Regions\vphantom{/}},{\citet{Rattenhuber2014}: East\vphantom{/}},{\citet{Kunaschk2024}\vphantom{/}},{\textbf{This Paper: HHI=0.20}\vphantom{/}},{\citet{BoockmannEtAl2011c}: 2007 MW\vphantom{/}},{\citet{BosslerEtAl2024}\vphantom{/}},{\citet{BosslerSchank2023}\vphantom{/}},{\citet{BosslerGerner2020}\vphantom{/}},{\citet{BosslerEtAl2022}: 2015 MW\vphantom{/}},{\citet{DustmannEtAl2022}: Firms\vphantom{/}},{\textbf{This Paper: HHI=0.10}\vphantom{/}},{\citet{MoellerEtAl2011}: East\vphantom{/}},{\citet{BoockmannEtAl2011a}: West\vphantom{/}},{\citet{VomBergeFrings2020}: East\vphantom{/}},{\citet{KoenigMoeller2009}: East\vphantom{/}},{\textbf{This Paper: HHI=0.05}\vphantom{/}},{\textbf{This Paper: HHI=0.00}\vphantom{/}},}]
\addplot[only marks,mark=square*,color=black, error bar legend, error bars/.cd, x dir=both, x explicit] coordinates {  (2.7560141,1)+-(1.1807408,0) (1.2246759,4)+-(.71619236,0) (-.04316535,10)+-(.51126856,0) (-1.1101317,17)+-(.4546746,0) (-1.8492439,22)+-(.49222779,0) (-2.8062806,23)+-(.73936605,0) };
\addplot[only marks,mark=square*, mark options={fill=white}, color=black, error bar legend, error bars/.cd, x dir=both, x explicit] coordinates {  (2.2,2)+-(2.7201755,0) (1.5037168,3)+-(1.7227291,0) (.5,5)+-(2.1119332,0) (.12244898,6)+-(.10369049,0) (.03,7)+-(.1974,0) (.02469136,8)+-(.08243902,0) (.0177,9)+-(.054285,0) (-.16666667,11)+-(.32598919,0) (-.16666667,12)+-(.15392451,0) (-.19830029,13)+-(.3357451,0) (-.278,14)+-(.34873998,0) (-.30232558,15)+-(.32717124,0) (-.31,16)+-(.06580001,0) (-1.4366198,18)+-(3.4612279,0) (-1.7183124,19)+-(4.0736532,0) (-1.7238095,20)+-(3.0174115,0) (-1.8181819,21)+-(1.7034539,0) };
\addplot[loosely dashed] coordinates {(0,-1.5) (0,24.5)};
\legend{~This Paper,~Literature}
\end{axis}
\end{tikzpicture}
}  
\floatfoot{\footnotesize\textsc{Note.} --- The figure provides an overview of estimated own-wage elasticities of employment based on minimum wage variation in the German labor market. Solid squares represent own-wage elasticities of employment that originate from this study (at varying HHI levels) whereas hollow squares depict elasticities from the literature. Each point estimate features a 90 percent confidence interval. Regarding this study, I calculate own-wage elasticities of employment at selected HHI levels by dividing the minimum wage elasticity of employment from Column (4) in Table \ref{tab:4} by the underlying minimum wage elasticity of wages from Column (4) in Table \ref{tab:3}. The respective standard errors stem from a Bootstrap algorithm with 50 replications. When the own-wage elasticity of employment is not explicitly reported in the literature, I divide the minimum wage elasticity of employment by the minimum wage elasticity of wages by myself. If the standard error of the own-wage elasticity of employment is not reported in the literature, I apply the Delta method to reported coefficients and their standard errors and assume that the covariance between the wage and employment effects is zero. HHI = Herfindahl-Hirschman Index. Sources: IEB $\plus$ BHP $\plus$ IAB Establishment Panel, 1999-2017.}
\end{figure}

}

Taken together, the observed pattern lends empirical support to the hypothesis that monopsony power is contributing to the frequent finding of close-to-zero minimum wage effects on employment: pooling negative effects in more competitive markets with positive effects in more monopsonistic markets yields average employment effects in the midst of both extremes. The exact level of this average effect hinges on the unit of observation underlying the estimate. While regressions at the labor market level assign equal weight to each labor market, regressions at the firm level weight labor markets with many firms (i.e., with low HHI and, thus, more negative employment effects) more strongly. Consequently, the average own-wage elasticity of employment across labor markets approaches zero more closely than the average elasticities across firms (see Appendix Table \ref{tab:I1}).\footnote{Moreover, aggregate employment is genuinely less wage-elastic than employment of single firms to the extent that displaced workers take up a new job in another firm in the same labor market due to reallocation effects \citep{BeaudryEtAl2018,BosslerPopp2024}.} In line, \citet{DuetschEtAL2024} review the German minimum wage literature and conclude, that employment effects generally turn out more negative at the firm than at the regional level.

\section{Conclusion}
\label{sec:9}

For many years, minimum wages have sparked contentious debates in both scientific and political spheres. While opponents warn that higher minimum wages hurt jobs, proponents argue that such a policy would not only raise wages but, in monopsonistic labor markets, have no adverse or positive employment. In the last two decades, a growing volume of work has reported close-to-zero employment effects of minimum wages, lending tentative support to the monopsony argument. Unfortunately, empirical evidence that systematically attributes slightly negative, zero, or positive employment effects to the presence of monopsony power is rare, and as yet not available for the German labor market. In this paper, I largely follow the classical notion of thin labor markets and use detailed measures of labor market concentration to operationalize monopsony power in Germany. Building on these proxies, I inspect the validity of the monopsony argument by examining whether labor market concentration is moderating the impact of higher sectoral minimum wages on employment.

Taking advantage of administrative data on the near-universe of workers, I provide first evidence on labor market concentration in Germany, an important but hitherto unavailable friction parameter. These concentration measures, which positively correlate with markdowns, challenge the classical view that labor markets exhibit an atomistic market structure. To begin with, I explore whether there is in fact evidence for market failure before the minimum wages were introduced. Indeed, my results highlight that firms successfully reduce wages by lowering employment in more concentrated labor markets, which is indicative of monopsonistic conduct of firms and calling for policy intervention.

My core analysis aims to inject more empirical substance into the minimum wage debate. I show that sectoral minimum wages effectively raised the pay of the workforce. Importantly, the bite turns out larger in more concentrated labor markets, corroborating the evidence on monopsonistic conduct in absence of a wage floor. Given this bite, I systematically document the validity of the monopsony argument. The introduction or increase of sectoral minimum wages generally harms employment in slightly concentrated or more competitive labor markets. Crucially, however, these disemployment effect gradually disappear for increasing labor market concentration and, in highly concentrated or more monopsonistic markets, the average employment effect turns out even positive. Nevertheless, when the minimum wage approaches firms' median wage in these low-wage sectors, the positive employment effects in highly concentrated labor markets quickly disappears. All in all, the estimated effect pattern advocates a nuanced view, namely that the minimum wage effects on wages and employment systematically differ by the underlying market form.

This study bears important policy implications. Antitrust policy in Germany and the E.U.\ focuses on consumer welfare by preventing distortions in product markets \citep{ArakiEtAl2023}. However, my analysis demonstrates that many firms also enjoy considerable labor market power, with detrimental consequences for wages and employment. Hence, antitrust authorities should scrutinize mergers, collusive practices, and other non-competitive behavior also in view of a potential supremacy of employers over workers. For this purpose, \citet{NaiduEtAl2018} have developed antitrust remedies against labor market power. Alternatively, the existence of powerful worker unions may counteract monopsony power (e.g., \citealp{DodiniEtAl2021}). If both antitrust policy and worker unions fail to correct for monopsony power in the first place, this study highlights that an adequately-set minimum wage constitutes a welfare-enhancing policy when monopsony power is widespread, reinforcing both wages and employment at the lower end of the wage distribution. In particular, my concentration measures can provide standardized guidance to policy-makers whether or not certain labor markets would benefit from a minimum wage regulation. However, when the minimum wage is set universally, the benefits of counteracting monopsony power in more concentrated labor markets might be outweighed by job loss in less concentrated labor markets. Moreover, this paper has demonstrated that an excessively high wage floor may hurt employment even when labor markets are highly concentrated.

Prior work has largely ignored that minimum wage effects vary systematically between competitive and monopsonistic labor markets, frequently reporting near-zero employment effects in the midst of both extremes. Future research should acknowledge the underlying heterogeneity and increasingly test for the role of monopsony power or a proxy thereof. Apart from labor market concentration as the classical source, future work should envisage other moderators that capture ``modern'' sources of monopsony power more directly -- factors like job differentiation or search frictions that may additionally moderate minimum wage effects. Although this study provides tentative evidence that markdowns, the most comprehensive measure of monopsony power, are mitigating negative minimum wage effects on employment, additional work on this subject would prove insightful.

\clearpage

\printbibliography[heading=bibintoc] 

\clearpage


\clearpage
\begin{appendix}
\begin{refsection} 

\renewcommand\thetable{\thesection\arabic{table}} 
\renewcommand\thefigure{\thesection\arabic{figure}} 

\pdfbookmark[0]{Appendix}{appendix} 

\begin{center}

\vspace*{1cm}

\Large
Appendix \\[1cm]
\textbf{Minimum Wages in Concentrated Labor Markets} \\[1cm]

Martin Popp

\normalsize

\vspace*{2cm}

\Large
\textbf{Content}
\normalsize

\startcontents[sections] 
\printcontents[sections]{l}{1}{\setcounter{tocdepth}{2}} 

\end{center}

\counterwithin{equation}{section}

\clearpage


\section{Minimum Wages in the Cournot Oligopsony Model}
\label{sec:A}
\setcounter{table}{0} 
\setcounter{figure}{0} 

Suppose that $J$ firms simultaneously maximize profits under the following employment-setting game while taking each other's quantities as given \citep{BoalRansom1997}:
\begin{equation}
\label{eq:A1}
\max_{L_{j}} \,\, P \cdot Q(L_{j}) \,-\, W( L_{j} + L_{-j} ) \cdot L_{j}
\end{equation}
where $L_{j}$ is the firm's own quantity of employment, $L_{-j}$ is the quantity of employment of all other firms in the market, $Q$ is the produced quantity of a single homogeneous good, and $P$ is the given product price for this good. The model entails a uniform market wage $W$ that positively depends on total employment in the labor market: $L = L_{j} + L_{-j}$. The first-order condition for each firm $j$ with respect to $L_{j}$ is
\begin{equation}
\label{eq:A2}
M\!RPL_{j} \, \equiv \, P \,\cdot\, \frac{\partial Q(L_{j})}{\partial L} = W \,+\,  \frac{\partial W(L)}{\partial L_{j}} \, L_{j}
\end{equation}
where $M\!RPL_{j}$ is the firm's marginal revenue product of labor (MRPL).

\paragraph{The (Inverse) Wage Elasticity of Labor Supply to the Firm.} I begin with deriving markdowns as a function of the (inverse) wage elasticity of labor supply (LS) to the single firm. Dividing Equation (\ref{eq:A2}) by $W$ results in the following expression for firms-specific markdowns
\begin{equation}
\label{eq:A3}
\frac{\,M\!RPL_{j}\,}{W} \,=\, 1 \,+\, \underbrace{\frac{\partial W(L)}{\partial L_{j}} \cdot \frac{L_{j}}{W}}_{\mu^{-1}_{j}}  \,=\, 1 \,+\, \mu_{j}^{-1}
\end{equation}
where the wage elasticity of labor supply to the single firm $j$ is denoted by $\mu_{j}=\frac{\partial L_{j}}{\partial W(L)} \frac{W}{L_{j}}$. According to Equation (\ref{eq:A3}), the firm-specific markdown increases when the inverse wage elasticity of labor supply to the firm increases (i.e., labor supply to the firm becomes less wage-sensitive). The employment-weighted average of these firm-level markdowns is:
\begin{equation}
\label{eq:A4}
\frac{\,M\!RPL\,}{W} \, \equiv \, \mathlarger{\mathlarger{\mathlarger{\sum}}}_{j=1}^{J} \,\, \frac{L_{j}}{L} \cdot \frac{\,M\!RPL_{j}\,}{W} \, = \, \mathlarger{\mathlarger{\mathlarger{\sum}}}_{j=1}^{J} \,\, s_{j} \cdot \Big( 1 \,+\, \mu_{j}^{-1} \Big)
\end{equation}
where the employment share of firm $j$ is given by $s_{j}=\frac{L_{j}}{L}$. As employment shares sum up to 1, further rearrangement yields:
\begin{equation}
\label{eq:A5}
\frac{\,M\!RPL\,}{W}  \, = \, \underbrace{\big(\sum\nolimits_{j=1}^{J} \,\, s_{j}\big)}_{=\,1} \,+\, \underbrace{\big(\sum\nolimits_{j=1}^{J} \,\, s_{j} \cdot \mu_{j}^{-1}\big)}_{\equiv \, \bar{\mu}^{-1}} \, = \, 1 \,+\, \frac{1}{\bar{\mu}}  \,=\,  \frac{\,\bar{\mu} \,+\, 1\,}{\bar{\mu}}
\end{equation}
Thus, the average markdown in the market positively depends on the employment-weighted average of the inverse wage elasticities of labor supply to the single firm, $\bar{\mu}^{-1}$. Put differently, when the market wage reacts generally more sensitively to single firms' employment decisions, the firms may exert a greater downward leverage on wages, and the market wage falls increasingly short of the average marginal revenue product of labor. The underlying intuition is that, in markets with an on average higher inverse wage elasticity of labor supply to the single firm, the firms can more easily enforce lower wages by lowering their employment, resulting in higher markdowns. In terms of the reciprocal $\bar{\mu}$, the average markdown in the market is a decreasing function of the underlying wage elasticities of labor supply to the single firms.

\paragraph{The (Inverse) Wage Elasticity of Labor Supply to the Market.} In a next step, I derive markdowns as a function of the (inverse) wage elasticity of labor supply to the market. Dividing both sides of Equation (\ref{eq:A2}) by $W$, using $\frac{\partial W(L)}{\partial L_{j}}=\frac{\partial W(L)}{\partial L}$, and rearranging leads to a slightly different expression of firm-specific markdowns
\begin{equation}
\label{eq:A6}
\frac{\,M\!RPL_{j}\,}{W} \, = \, 1 \,+\, \underbrace{\frac{\partial W(L)}{\partial L} \cdot \frac{L}{W}}_{\mu^{-1}} \cdot \underbrace{\frac{\,\,L_{j}\,\,}{L}}_{s_{j}}  \,=\, 1 \,+\, s_{j} \, \mu^{-1} 
\end{equation}
where the wage elasticity of labor supply to the market is denoted by $\mu=\frac{\partial L}{\partial W(L)} \frac{W}{L}$. In particular, the firm-specific markdown increases when i) the inverse wage elasticity of labor supply to the market increases (i.e., labor supply to the market becomes less wage-sensitive) and ii) the firm's market share increases. Using Equation (\ref{eq:A6}), the employment-weighted average of these firm-level markdowns reads:
\begin{equation}
\label{eq:A7}
\frac{\,M\!RPL\,}{W} \equiv \, \mathlarger{\mathlarger{\mathlarger{\sum}}}_{j=1}^{J} \,\, \frac{L_{j}}{L} \cdot \frac{\,M\!RPL_{j}\,}{W} =  \mathlarger{\mathlarger{\mathlarger{\sum}}}_{j=1}^{J} \,\, s_{j} \cdot  \Big( 1 \,+\, s_{j} \, \mu^{-1} \Big)
\end{equation}
Since the wage elasticity of labor supply to the market is constant across firms and employment shares sum up to 1, the equation further simplifies to:
\begin{equation}
\label{eq:A8}
\frac{\,M\!RPL\,}{W} \, = \, \underbrace{\big(\sum\nolimits_{j=1}^{J} \,\, s_{j}\big)}_{=\,1} \,\,+\,\,\, \mu^{-1} \,\underbrace{\big(\sum\nolimits_{j=1}^{J} s^{2}_{j}\,\big)}_{=\,H\!H\!I}  \, = \,  1 \,+\, \frac{H\!H\!I}{\mu}  \, = \, \frac{\, \mu \,+\, H\!H\!I\,}{\mu}
\end{equation}
In this formulation, the average employment-weighted markdown depends on two forces. First, wages are increasingly marked down relative to the marginal revenue product when the wage elasticity of labor supply to the market decreases. The reason is that a relatively low wage sensitivity of workers allows firms to reduce the market wage without triggering a sharp decline in labor supply. Second, the markdown also becomes larger when the employment-based Herfindahl-Hirschman Index turns out higher (i.e., employment in the market is more concentrated). To facilitate intuition, combining Equation (\ref{eq:A5}) with (\ref{eq:A8}) delivers the following relationship between average wage elasticity of labor supply to the firm, the wage elasticity of labor supply to the market, and the employment-based Herfindahl-Hirschman Index:
\begin{equation}
\label{eq:A9}
\bar{\mu}  \, = \, \frac{\mu}{\,H\!H\!I\,}
\end{equation}
Importantly, with higher employment concentration, the exogenously given wage elasticity of labor supply to the market translates on average into increasingly smaller (i.e., less elastic) wage elasticities of labor supply to the single firms. Vice versa, the inverse wage elasticities of labor supply to the single firms become increasingly larger, implying that the average sensitivity of the market wage to the single firms' employment decisions increases. By construction, a higher employment-based HHI value reflects that the average firm in the market features a larger employment share. With higher average market shares, relative employment reductions of single firms exert a larger downward leverage on market wages. In other words, higher market shares render firms' marginal cost of labor (MCL) curves steeper since their increased labor demand will result in increasingly higher market wages. Therefore, in the profit-maximizing optimum, the oligopsonists will reduce their employment vis-\'{a}-vis the perfectly competitive market (where marginal cost of labor are constant), resulting in lower wage rates and higher markdowns. Taken together, by means of a higher downward leverage on wages manifesting in lower aggregate employment, higher employment concentration results ceteris paribus in a lower market wage and higher markdowns, conditional on the wage elasticity of labor supply to the market. Note that, in the symmetric case (i.e., when firms have identical marginal revenue product functions), the Herfindahl-Hirschman Index collapses to the inverse number of employers in the labor market because all firms have identical employment shares.

\paragraph{The Polar Cases of Perfect Competition and Monopsony.} Notably, the Cournot oligopsony model nests the standard models of perfect competition, $H\!H\!I$ = 0, and monopsony, $H\!H\!I$ = 1, as limiting cases. Figure \ref{fig:A1} displays both market forms graphically. For HHI=0, the atomistic market structure renders labor supply to the single firm infinitely elastic, regardless of the underlying wage elasticity of labor supply to the market. Consequently, firms are wage takers and the market wage follows the marginal revenue product of labor: $ \frac{MRPL}{w^{C}} = 1 $. At the other extreme, for HHI=1, the monopsonist controls the entire market and the market wage is a mere function of the wage elasticity of labor supply to the single firm (which, in this case, is identical to the labor supply elasticity to the market): $ \frac{MRPL}{w^{M}} = \frac{\mu\,+\,1}{\mu} $. To achieve this markdown, the monopsonist curtails employment, $L^{M}$, compared to total employment of wage-taking employers in a perfectly competitive market, $L^{C}$.

\begin{figure}[!ht]
\centering
\caption{The Polar Cases of Market Forms}
\label{fig:A1}

\huge
\begin{center}

\scalebox{0.7}{

\begin{tikzpicture}[line width=0.8pt, every node/.style={color=black}, dot/.style={circle,fill=black,minimum size=6pt,inner sep=0pt,outer sep=-1pt}]

\draw[<->,thick, xshift=-1.3cm, line width=2pt,rounded corners=0.1cm]  (0,13) node(yline)[above]{$\bm{w}$} -- (0,1) -- (13,1) node(xline)[right]{$\bm{L}$};
\draw[xshift=-1.3cm] (1.3,11.7) -- (11.7,4.3) node[below]{ \shortstack{ \textbf{LD}  \\ \textbf{MRPL}  } };
\draw[xshift=-1.3cm] (1.3,1.3) -- (11.7,8.7) node[above]{\textbf{LS}};

\draw[xshift=-1.3cm] (1.3,2.2262) -- (7.952949,11.7) node[above]{\textbf{MCL}};

\draw[dash pattern = on 6pt off 3pt,xshift=-1.3cm] (0,6.5) -- (8.602528,6.5) -- (8.602528,1);
\draw[xshift=-1.3cm] (8.602528,1) -- (8.602528,1) node[below]{$L^{C}$};
\draw[xshift=-1.3cm] (0,6.5) -- (0,6.5) node[left]{$w^{C}$};


\draw[dash pattern = on 6pt off 3pt,xshift=-1.3cm]  (5.735018727,4.458333334) -- (0,4.458333334) node[left]{$w^{M}$};
\draw[dash pattern = on 6pt off 3pt,xshift=-1.3cm]  (5.735018727,8.54166667) -- (5.735018727,1) node[below]{$L^{M}$};

\draw[dash pattern = on 6pt off 3pt,xshift=-1.3cm]   (5.735018727,8.54166667) -- (0,8.54166667)  node[left]{$\overline{w}$};

\draw[dotted, xshift=-1.3cm, thick, color=blue] (11.7,7.52083335) -- (0,7.52083335) node[left, color=blue]{$w^{min}$};
\draw[dash pattern = on 6pt off 3pt,xshift=-1.3cm, color=blue] (7.1687773385,7.52083335) -- (7.1687773385,1) node[below, color=blue]{$L^{min}$};


\node[dot,xshift=-1.3cm] at (7.1687773385,7.52083335) {};
\node[dot,xshift=-1.3cm] at (5.735018727,4.458333334) {};
\node[dot,xshift=-1.3cm] at (5.735018727,8.54166667) {};
\node[dot,xshift=-1.3cm] at (8.602528,6.5) {};

\end{tikzpicture}
}
\end{center}
\normalsize
\floatfoot{\footnotesize\textsc{Note. ---} The figure compares labor market outcomes between the polar cases of perfectly competition and monopsony. Moreover, the diagram shows the effects from an exemplary minimum wage above but close to the equilibrium wage rate (i.e., in the range of the demand-determined regime). Source: Own illustration.}
\end{figure}

\paragraph{The Role of Minimum Wages.} In oligopsonistic labor markets, minimum wages always entail less negative employment effects than in perfectly competitive labor markets and, for moderately-set wage floors, can even stimulate employment \citep{Stigler1946,Lester1947,AzarEtAl2024}. To facilitate this argument, suppose that there is the polar case of a single monopsonist, $H\!H\!I$ = 1. When raising minimum wages from zero, the minimum wage effect on employment progresses through three different regimes: First, in the unconstrained regime below the monopsonistic wage $w^{C}$, minimum wages do not bite into the wage distribution and, thus, do not exert any employment effects. Second, in the supply-determined regime, the minimum wage lies in the interval between the monopsonistic wage $w^{M}$ and the market-clearing wage $w^{C}$. The wage floor renders marginal cost of labor constant because hiring at the minimum wage no longer involves higher wages for incumbents (who also earn the minimum wage). The reduced marginal cost makes the monopsonist increase employment until the minimum wage meets the labor supply curve (after which MCL coincides with pre-minimum-wage levels). Hence, the minimum wage eliminates the firm's incentive to take advantage of inelastic labor supply at the bottom part of the wage distribution. Third, in the demand-determined regime, the minimum wage exceeds the market-clearing wage $w^{C}$ and the minimum wage materializes along the negatively sloped labor demand curve. As a consequence, firms will lay of the least productive workers. Taken together, there is a non-monotonous relationship between minimum wages and employment effects in monopsonistic markets: at first, a higher minimum wage raises employment by counteracting monopsonistic conduct along the labor supply curve but, when set increasingly above the market-clearing level, the positive effect on employment gradually disappears. If the minimum wage is set beyond the marginal revenue product of labor absent the minimum wage, $\overline{w}$, the overall minimum wage effect on employment will turn negative.

In the general case of a Cournot oligopsony model, the range of the supply-determined regime becomes increasingly smaller for lower employment concentration. Nevertheless, \citep{AzarEtAl2024} derive that an adequately-set minimum wage between the oligopsonistic wage $w^{O}$ and the market-clearing wage $w^{C}$ stimulates employment in the Cournot oligopsony model. Figure \ref{fig:A2} visualizes the non-monotonous relationship between minimum wages and employment effects for the more restrictive case of a Cournot oligopsony with symmetric firms. Under the symmetry assumption, all firms have identical marginal revenue product functions and will therefore react uniformly to the minimum wage. If, however, firms are not symmetric, low-productivity firms will feature more negative employment effects than high-productivity firms. Nonetheless, when the minimum wage is set below the market-clearing wage $w^{C}$, the aggregate minimum wage effect on employment is still positive. In the polar case of a perfectly competitive market, $H\!H\!I$ = 1, the supply-determined regime disappears completely. In such a market, wages are not marked down relative to marginal productivity and, absent other margins of adjustment, minimum wages are either non-binding or reduce employment along the labor demand curve. To sum up, minimum wages will cause less negative effects on employment to the extent that the labor market is monopsonistic.

\clearpage
\vspace*{\fill}

\begin{figure}[!ht]
\centering
\caption{Stylized Minimum Wage Effects in Cournot Oligopsony Model}
\label{fig:A2}
\vspace*{0.5cm}
\begin{subfigure}{1\textwidth}
\centering
\caption{Minimum Wage Effects on Wages}
\scalebox{0.85}{
\begin{tikzpicture}

\begin{axis}[height=10cm, width=14cm, grid=both, grid style = dotted,ytick={0},minor ytick={-2,-1,0,1,2,3,4,5,6}, xtick={3,5.5,8,12.363636}, xlabel=Minimum Wage, ylabel=$\Delta$ Wage Rate, ymin=-3, ymax=7, xmin=0, xmax=15.363636, legend pos = south west, y tick label style={/pgf/number format/.cd,fixed,fixed zerofill, precision=0,/tikz/.cd}, x tick label style={/pgf/number format/.cd,fixed,fixed zerofill, precision=1,/tikz/.cd}, legend style={font=\footnotesize}, reverse legend, xticklabels={$w^{M}$,$w^{O}$,$w^{C}$,$\overline{w}$}]

\addplot[mark=none, solid, color=lightgray, line width=1mm] coordinates { (0,0) (8,0) (16,3) };
\addplot[mark=none, solid, color=gray, line width=1mm] coordinates { (0,0) (5.5,0) (16,4) };
\addplot[mark=none, solid, color=black, line width=1mm] coordinates { (0,0) (3,0) (16,5) };

\legend{$H\!H\!I=0$, $0<H\!H\!I<1$, $H\!H\!I=1$}
\end{axis}
\end{tikzpicture}
}
\end{subfigure}
\vskip \baselineskip \vspace*{0.5cm}
\begin{subfigure}{1\textwidth}
\centering
\caption{Minimum Wage Effects on Employment}
\scalebox{0.85}{
\begin{tikzpicture}

\begin{axis}[height=10cm, width=14cm, grid=both, grid style = dotted,ytick={0},minor ytick={-4,-3,-2,-1,0,1,2,3,4}, xtick={3,5.5,8,12.363636}, xlabel=Minimum Wage, ylabel=$\Delta$ Employment, ymin=-5, ymax=5, xmin=0, xmax=15.363636, legend pos = south west, y tick label style={/pgf/number format/.cd,fixed,fixed zerofill, precision=0,/tikz/.cd}, x tick label style={/pgf/number format/.cd,fixed,fixed zerofill, precision=1,/tikz/.cd}, legend style={font=\footnotesize}, reverse legend, xticklabels={$w^{M}$,$w^{O}$,$w^{C}$,$\overline{w}$}]

\addplot[mark=none, solid, color=lightgray, line width=1mm] coordinates { (0,0) (8,0) (16,-5.5) };
\addplot[mark=none, solid, color=gray, line width=1mm] coordinates { (0,0) (5.5,0) (8,1.5) (16,-4)  };
\addplot[mark=none, solid, color=black, line width=1mm] coordinates { (0,0) (3,0) (8,3) (16,-2.5)};

\legend{$H\!H\!I=0$, $0<H\!H\!I<1$, $H\!H\!I=1$}
\end{axis}
\end{tikzpicture}
}
\end{subfigure}

\floatfoot{\footnotesize\textsc{Note. ---} The figures visualize minimum wage effects on wages and employment in a Cournot oligopsony model of symmetric firms. The x axis refers to absolute levels of the minimum wage, and the y axis displays absolute deviations from the outcome variable absent wage floors. The linearity of employment effects derives from the assumption that both the marginal revenue product of labor and labor supply to the market are first-order polynomials of the wage rate. Due to the symmetry assumptions, firms feature identical marginal revenue product functions and, consequently, identical market shares and markdowns. Thus, all firms in the market feature the same minimum wage effects on wages and employment. Source: Own illustration.}
\end{figure}

\vspace*{\fill}
\clearpage

\section{Institutional Background: Further Evidence}
\label{sec:B}
\setcounter{table}{0} 
\setcounter{figure}{0} 

When a collective bargaining agreement (CBA) is declared universally binding, the negotiated minimum working conditions no longer apply only to the social partners, but to the entire sector. Crucially, the sectoral minimum wages apply to all firms in the sector, including those firms that decided against a membership in the respective employer association. Legally, a firm belongs to a certain sector if it predominantly engages in this sector, namely with at least 50 percent of its business activity. In principle, the regulation covers all employees in these firms, irrespective of their occupation or union membership. However, several sectors grant exemptions to white-collar workers or apprentices.

\paragraph{Legislation of Sectoral Minimum Wages.} Table \ref{tab:B1} provides an overview about the minimum wage legislation in Germany. Three independent pieces of legislation lay down the requirements for the imposition of sectoral minimum wages. First, since 1949, the Collective Bargaining Law (TVG) stipulates the general conditions for universal bindingness of a collective bargaining agreement (CBA). At request of either the worker union or the employer association, the Ministry of Labor and Social Affairs, under certain conditions, may declare a CBA to be universally binding. The conditions of a declaration of universal bindingness are quite restrictive: First, an implementation necessitates a majority vote (i.e., four yes votes) from a so-called Bargaining Committee (Tarifausschuss). The Bargaining Committee has equal representation and consists of six members, three of whom are appointed from each of the national umbrella organizations of workers and employers. Second, the implementation must be in public interest. Third, until 2014, an implementation also required the CBA coverage rate of firms in the sector to exceed 50 percent. By and large, given the restrictive conditions, minimum wages based on TVG could attain long-term relevance for only few sectors, such as electrical trade or chimney sweeping.

\afterpage{
\begin{landscape}

\vspace*{0.25cm}

\begin{table}[!ht]
\centering

\renewcommand{\arraystretch}{1.3}
\scalebox{0.85}{

\begin{threeparttable}
\caption{Minimum Wage Legislation in Germany}
\label{tab:B1}
\vspace{-0.1cm}
\begin{tabular}[c]{L{4.5cm}C{8cm}C{6.25cm}C{9cm}} \hline

\multirow{3}{*}{}  & \multirow{3}{*}{\shortstack{Prerequisites for \vphantom{/} \\ Declaration of Universal Bindingness \vphantom{/} }} & \multirow{3}{*}{Upon Approval by ...\vphantom{/}} & \multirow{3}{*}{Possible Scope\vphantom{/}} \\
&&& \\
&&& \\[0.1cm] \hline		

\multirow{4}{*}{\shortstack[l]{\textbf{TVG (1949, 1969, 2014)}\vphantom{/}\\[0.1cm] valid: 04/1949-hitherto \vphantom{/}}}  &  \multirow{4}{*}{\shortstack{\tabitem MW regulation in existing CBA\vphantom{/} \\ \tabitem request from a CBA partner\vphantom{/}\\  \tabitem in public interest \vphantom{/}  \\ \tabitem CBA firm coverage $>$ 50\% (until 08/2014) \vphantom{/} }} &  \multirow{4}{*}{\shortstack{\tabitem Bargaining Committee\vphantom{/}\\ \tabitem Ministry of Labor and Social Affairs \vphantom{/}}} & \multirow{4}{*}{\tabitem sectoral\vphantom{/}}  \\
&&& \\
&&& \\
&&& \\ 

\multirow{3}{*}{\shortstack[l]{\textbf{AEntG (1998, 2007)}\vphantom{/}\\[0.1cm] valid: 01/1999-04/2009 \vphantom{/}}}    &  \multirow{3}{*}{\shortstack{\tabitem MW regulation in existing CBA\vphantom{/} \\ \tabitem request from a CBA partner\vphantom{/}}} &  \multirow{3}{*}{\tabitem Ministry of Labor and Social Affairs\vphantom{/}} & \multirow{3}{*}{\shortstack{\tabitem sector-wise in construction industry\vphantom{/}\\ \tabitem commercial cleaning (since 07/2007)\vphantom{/} }}   \\
&&& \\
&&& \\ 

\multirow{3}{*}{\shortstack[l]{\textbf{AEntG (2009, 2014)}\vphantom{/}\\[0.1cm] valid: 04/2009-hitherto \vphantom{/}}}   &  \multirow{3}{*}{\shortstack{\tabitem MW regulation in existing CBA\vphantom{/} \\ \tabitem joint request from CBA partners\vphantom{/} \\ \tabitem in public interest\vphantom{/}}} &  \multirow{3}{*}{\shortstack{\tabitem Ministry of Labor and Social Affairs\vphantom{/} \\  (\tabitem Bargaining Committee ) \vphantom{/} \\ (\tabitem Federal Government ) \vphantom{/}}} & \multirow{3}{*}{\shortstack{\tabitem sector-wise in nine industries (until 08/2014)\vphantom{/}\\ \tabitem sector-wise (since 08/2014)\vphantom{/}}}   \\
&&& \\
&&& \\ 

\multirow{4}{*}{\shortstack[l]{\textbf{AÜG (2011, 2014)}\vphantom{/}\\ valid: 12/2011-hitherto \vphantom{/}}}  &  \multirow{4}{*}{\shortstack{\tabitem MW regulation in existing CBA\vphantom{/} \\ \tabitem joint request from CBA partners\vphantom{/} \\ \tabitem MW beneficial for social security system\vphantom{/} \\ \tabitem in public interest (since 08/2014)\vphantom{/}}} &  \multirow{4}{*}{\tabitem Ministry of Labor and Social Affairs\vphantom{/}} & \multirow{4}{*}{\tabitem temporary work\vphantom{/}}   \\
&&& \\
&&& \\
&&& \\ 
&&& \\[-0.4cm]

\multirow{2}{*}{\shortstack[l]{\textbf{MiLoG (2014)}\vphantom{/}\\[0.1cm] valid: 01/2015-hitherto \vphantom{/}}}  &  \multirow{2}{*}{none} &  \multirow{2}{*}{\tabitem MW Commission\vphantom{/}} & \multirow{2}{*}{\tabitem nation-wide\vphantom{/}}   \\

&&& \\ \hline

\end{tabular}

\begin{tablenotes}[para]
\footnotesize\textsc{Note. ---} The table describes the pieces of legislation that regulate the imposition of minimum wages in Germany. With respect to TVG (1949, 1969, 2014) and AEntG (2009, 2014), the Bargaining Committee consists of six members, three of whom are appointed from each of the national umbrella organizations of employees and employers. For the AEntG (2009, 2014), initial declaration requests for a specific collective bargaining agreement in a certain sector need approval of both the Ministry of Labor and Social Affairs and the Bargaining Committee (at least four yes-votes from six members or, alternatively, tacit approval after three months) or the Federal Government (if there are only two or three yes-votes in the Bargaining Committee). On the contrary, follow-up requests for subsequent collective agreements in the same sector merely require approval of the Ministry of Labor and Social Affairs. The nine industries listed in AEntG (2009) as candidate sectors are: construction, commercial cleaning, postal services, security, specialized hard coal mining, industrial laundries, waste removal, public training services from SGB II/III, and nursing care. According to MiLoG (2014), the Minimum Wage Commission consists of a chairperson, three employee and three employer representatives and two non-voting advisory members from the scientific community. AEntG = Arbeitnehmerentsendegesetz. AÜG = Arbeitnehmerüberlassungsgesetz. CBA = Collective Bargaining Agreement. MiLoG = Mindestlohngesetz. MW = Minimum Wage. SGB II/III = Sozialgesetzbuch II/III. TVG = Tarifvertragsgesetz. Sources: AEntG $\plus$ AÜG $\plus$ MiLoG $\plus$ TVG.
\end{tablenotes}

\end{threeparttable}
}
\end{table}

\end{landscape}
}

Second, in 1996, the German parliament adopted the Posting of Workers Law (AEntG) to curtail wage competition in the construction sector \citep{Eichhorst2005}. Accordingly, foreign workers who are posted to German construction sites from abroad are subject to the same minimum working conditions of collective agreements that have been declared generally binding. When the AEntG was created, however, universal bindingness of a CBA could only be achieved under the restrictive conditions of TVG. Therefore, in 1998, the AEntG was amended to grant the Ministry of Labor and Social Affiars the right to autonomously declare CBAs universally binding provided that there is a request from either of the two social partners. Subsequently, minimum wages based on AEntG came into effect for the sectors of main construction, roofing, and painting and varnishing. Advocates of minimum wages recognized that the AEntG opened up a legal possibility to enforce minimum working conditions without explicit consent from employers. In the following years, the scope of the law steadily expanded to include other non-construction sectors where protection against foreign competition played only a subordinate role. Consequently, minimum wages based on AEntG were introduced in the sectors of commercial cleaning, industrial laundries, specialized hard coal mining, waste removal, nursing care, security, and public training services. In 2009, the AEntG was modified such that the conditions under which CBAs can be declared universally binding became more restrictive. Since then, the social partners must submit a joint request to the Ministry of Labor and Social Affairs. Moreover, the first-time declaration of universal bindingness for a certain sector necessitates approval from both the Ministry of Labor and Social Affairs and the Bargaining Committee (at least four yes votes from six members or, alternatively, tacit approval after three months) or the Federal Government (if there are only two or three yes votes in the Bargaining Committee). On the contrary, follow-up requests for subsequent CBA declarations in the same sector merely require approval of the Ministry of Labor and Social Affairs. Eventually, the possible ambit of AEntG was opened to all sectors in 2014 provided that a declaration of universal bindingness is in public interest, leading to minimum wages in the sectors of hairdressing, slaughtering and meat processing, textile and clothing, agriculture, forestry and gardening, and money and value services.

Third, the Temporary Work Law (AÜG) was revised in 2011 to prevent misuse of temporary work. Among others, the reform permitted the imposition of a minimum wage for temporary workers. To this end, the social partners have to submit joint request for their CBA. The Ministry of Labor and Social Affairs can declare the universal bindingness in case the decree serves to ensure the financial stability of the social security system. With short interruptions, minimum wages apply in the temporary work sector since 2012.

The introduction of a nation-wide minimum wage in 2015 also had direct impact on sectoral minimum wages. First, the Minimum Wage Law (MiLoG) contains a three-year exemption for sectors that abide by minimum wages from AEntG or AÜG. However, this rule of transition encouraged several sectors to impose minimum wages that undercut the statutory minimum wage of 8.50 Euro per hour, namely hairdressing, slaughtering and meat processing, textile and clothing, and agriculture, forestry and gardening. Second, the statutory minimum wage rendered sectoral wage floors useless in sectors where there was no consensus for a higher wage floor, such as industrial laundries or waste removal. In 2017, the nation-wide minimum wage was raised to 8.84 Euro per hour.

\paragraph{Minimum Wage Sectors.} In this study, I focus on those twenty sectors that feature universally binding sectoral minimum wages throughout all federal states of Germany. Hence, sectors with wage floors for only local entities are not part of the analysis. Table \ref{tab:B2} provides an overview for sixteen of these sectors that can be identified in the German ``WZ'' Classification of Economic Activities \citep{Destatis2008} and, thus, enter the multivariate analysis. Specifically, I make use of the 1993, 2003 and 2008 versions of the 5-digit WZ Classification to determine the sector affiliation of firms for the years 1999-2002, 2003-2007, and 2008-2017. The WZ Classification derives its four leading digits from the Statistical Classification of Economic Activities in the European Community (NACE). Unfortunately, the 5-digit WZ Classification is not sufficiently granular to identify firms from the following minimum wage sectors: industrial laundries, specialized hard coal mining, public training services, and money and value services. In 2015, the 16 remaining sectors covered about 380,000 firms and 5.2 million workers, representing 12.2 percent of headcount employment in Germany. In terms of employment, the largest minimum wage sectors are the sectors of nursing care, commercial cleaning, temporary work, and main construction.

The Kaitz Indices in the sectors typically range between 65 and 85 percent, indicating that the minimum wages bit strongly into the wage distribution. However, these relatively high values are largely attributable to the fact that these minimum wages were introduced in low-wage sectors rather than the economy as a whole. The largest Kaitz Indices are reported for the sectors of hairdressing, roofing, and commercial cleaning, and the lowest for waste removal, and textile and clothing.

\paragraph{Development of Minimum Wage Levels.} Figure \ref{tab:B1} visualizes the development in sectoral minimum wages between 1999 and 2017. Every few years, collective bargaining agreements are renegotiated between worker unions and employer associations in the sectors. Due to follow-up declarations of these CBAs, sectoral minimum wages are subject to regular adjustment, thus offering rich variation to study the minimum wage effects over time. Some of the sectoral minimum wages are set uniformly, while others differentiate between East and West Germany, or the 16 federal states. In many sectors, the social partners negotiated lower minimum wages for East Germany compared to West Germany. Sectoral minimum wages expire with the end of the underlying CBA. Hence, transitional periods without a minimum wage regulation can occur until the Ministry of Labor and Social Affairs declares a follow-up CBA universally binding. In such cases, the wage floor falls back to the level of the nation-wide minimum wage since 2015. Some sectors differentiate minimum wages by qualification or task. In such cases, I select the lowest wage floor.

\begin{landscape}

\begin{table}[!ht]
\centering
\renewcommand{\arraystretch}{1.475}
\scalebox{0.8}{
\begin{threeparttable}
\caption{Sectoral Minimum Wages}
\label{tab:B2}
\vspace{-0.1cm}
\begin{tabular}[c]{L{6.5cm}C{2.25cm}C{4cm}C{2.5cm}C{2cm}C{3cm}C{6cm}} \hline
\multirow{2.4}{*}{} & \multirow{2.4}{*}{\shortstack{Introduction}}   & \multirow{2.4}{*}{Legislative Basis\vphantom{/} }    & \multirow{2.4}{*}{Workers}  & \multirow{2.4}{*}{Firms}  & \multirow{2.4}{*}{\shortstack{Kaitz Index\vphantom{/} \\ (West/East)\vphantom{/} }} & \multirow{2.4}{*}{2008 NACE-5 Classification:\vphantom{/}} \\
&&&&&& \\ \hline
\multirow{2}{*}{Main Construction} &  \multirow{2}{*}{01/1997}   &  \multirow{2}{*}{TVG/AEntG}  & \multirow{2}{*}{793,738  } & \multirow{2}{*}{82,166   } &  \multirow{2}{*}{0.64/0.84} & \multirow{2}{*}{\shortstack{ 4120.1-4299.0/4312.0/4329.1 \\ 4331.0/4391.2/4399.2/4399.9}}             \\[-0.2cm]
&&&&&&  \\ 
\multirow{1}{*}{Electrical Trade} &  \multirow{1}{*}{06/1997}   &  \multirow{1}{*}{TVG/MiLoG}  & \multirow{1}{*}{246,196  } & \multirow{1}{*}{29,760   } &  \multirow{1}{*}{0.65/0.80} & \multirow{1}{*}{4321.0} \\ 
\multirow{1}{*}{Roofing} &  \multirow{1}{*}{10/1997}   &  \multirow{1}{*}{TVG/AEntG}  & \multirow{1}{*}{89,830   } & \multirow{1}{*}{12,995   } &  \multirow{1}{*}{0.73/0.97} &  \multirow{1}{*}{4391.1} \\ 
\multirow{1}{*}{Painting \& Varnishing} &  \multirow{1}{*}{12/2003}   &  \multirow{1}{*}{AEntG}  & \multirow{1}{*}{141,125  } & \multirow{1}{*}{23,706   } &  \multirow{1}{*}{0.66/0.88} & \multirow{1}{*}{4334.1} \\ 
\multirow{1}{*}{Commercial Cleaning} &  \multirow{1}{*}{04/2004}   &  \multirow{1}{*}{TVG/AEntG/MiLoG}  & \multirow{1}{*}{919,563  } & \multirow{1}{*}{29,628   } &  \multirow{1}{*}{0.83/0.89} & \multirow{1}{*}{8121.0/8122.9-8129.9} \\ 
\multirow{1}{*}{Waste Removal} &  \multirow{1}{*}{01/2010}   &  \multirow{1}{*}{AEntG/MiLoG}  & \multirow{1}{*}{177,647  } & \multirow{1}{*}{6,872    } &  \multirow{1}{*}{0.51/0.72} & \multirow{1}{*}{3811.0-3900.0} \\ 
\multirow{1}{*}{Nursing Care} &  \multirow{1}{*}{08/2010}   &  \multirow{1}{*}{AEntG}  & \multirow{1}{*}{939,660  } & \multirow{1}{*}{21,372   } &  \multirow{1}{*}{0.62/0.74} & \multirow{1}{*}{8710.0/8810.1}  \\ 
\multirow{1}{*}{Security} &  \multirow{1}{*}{06/2011}   &  \multirow{1}{*}{AEntG/MiLoG}  & \multirow{1}{*}{183,058  } & \multirow{1}{*}{4,684    } &  \multirow{1}{*}{0.63/0.80} & \multirow{1}{*}{8010.0/8020.0} \\ 
\multirow{1}{*}{Temporary Work} &  \multirow{1}{*}{01/2012}   &  \multirow{1}{*}{AÜG/MiLoG}  & \multirow{1}{*}{829,579  } & \multirow{1}{*}{12,355   } &  \multirow{1}{*}{0.81/0.88} & \multirow{1}{*}{7820.0/7830.0}   \\ 
\multirow{1}{*}{Scaffolding} &  \multirow{1}{*}{08/2013}   &  \multirow{1}{*}{AEntG/MiLoG} & \multirow{1}{*}{28,935   } & \multirow{1}{*}{2,862    } &  \multirow{1}{*}{0.70/0.86} & \multirow{1}{*}{4399.1} \\ 
\multirow{1}{*}{Stonemasonry} &  \multirow{1}{*}{10/2013}   &  \multirow{1}{*}{AEntG/MiLoG}  & \multirow{1}{*}{25,620   } & \multirow{1}{*}{4,351    } &  \multirow{1}{*}{0.73/0.93} & \multirow{1}{*}{2370.0}   \\ 
\multirow{1}{*}{Hairdressing} &  \multirow{1}{*}{11/2013}   &  \multirow{1}{*}{TVG/AEntG/MiLoG}  & \multirow{1}{*}{193,559  } & \multirow{1}{*}{48,976   } &  \multirow{1}{*}{0.93/0.99} & \multirow{1}{*}{9602.1}   \\ 
\multirow{1}{*}{Chimney Sweeping} &  \multirow{1}{*}{04/2014}   &  \multirow{1}{*}{TVG}  & \multirow{1}{*}{16,728   } & \multirow{1}{*}{7,428    } &  \multirow{1}{*}{0.71/0.86} & \multirow{1}{*}{8122.1}   \\ 
\multirow{1}{*}{Slaughtering \& Meat Processing} &  \multirow{1}{*}{08/2014}   &  \multirow{1}{*}{AEntG}  & \multirow{1}{*}{182,504  } & \multirow{1}{*}{9,549    } &  \multirow{1}{*}{0.67/0.84} & \multirow{1}{*}{1011.0-1013.0}   \\ 
\multirow{1}{*}{Textile \& Clothing} &  \multirow{1}{*}{01/2015}   &  \multirow{1}{*}{AEntG/MiLoG}  & \multirow{1}{*}{120,099  } & \multirow{1}{*}{6,398    } &  \multirow{1}{*}{0.53/0.73} & \multirow{1}{*}{1310.0-1439.0}   \\ 
\multirow{1}{*}{Agriculture, Forestry \& Gardening} &  \multirow{1}{*}{01/2015}   &  \multirow{1}{*}{AEntG}  & \multirow{1}{*}{342,451  } & \multirow{1}{*}{79,056   } &  \multirow{1}{*}{0.68/0.71} & \multirow{1}{*}{0111.0-0240.0/0312.0-0322.0}   \\ \hline
\end{tabular}
\begin{tablenotes}[para]
\footnotesize\textsc{Note. ---} The table provides an overview about sectoral minimum wages in Germany. The number of workers, the number of firms, and the Kaitz Indices always refer to June 30, 2015, except for the sectors of security (2013) and stonemasonry (2014) which had no valid sectoral minimum wage on this date. The reported Kaitz Index is the ratio of the sectoral minimum wage to the median hourly wage rate of regular full-time workers per sector, assuming 40 working hours per week. If sectoral minimum wages vary by federal state, qualification, or task, I select the lowest wage floor. The 5-digit industry codes from the 2008 version of the German NACE Classification serve to identify the sectoral affiliation of firms between 2008 and 2017. For the years 1999-2002 and 2003-2007, I make use of earlier 1993 NACE-5 and 2003 NACE-5 codes to determine the minimum wage sectors: main construction (1993/2003 NACE-5: 4511.2-4511.4/4521.1-4521.7/4522.2-4522.3/4523.2-4525.3/4525.5-4525.6/4532.0/4541.0), electrical trade (1993/2003 NACE-5: 4531.0), roofing (1993/2003 NACE-5: 4522.1), painting and varnishing (1993/2003 NACE-5: 4544.1), commercial cleaning (1993/2003 NACE-5: 7470.1/7470.3-4), waste removal (1993 NACE-5: 3710.1-3720.5/9000.3-9000.9, 2003 NACE-5: 3710.1-3720.5/9002.1-9003.0), nursing care (1993/2003 NACE-5: 8531.5/8532.6), security (1993/2003 NACE-5: 7460.2), temporary work (1993/2003 NACE-5: 7450.2), scaffolding (1993/2003 NACE-5: 4525.4), stonemasonry (1993/2003 NACE-5: 2670.1-2), hairdressing (1993 NACE-5: 9302.1-3, 2003 NACE-5: 9302.5), chimney sweeping (1993/2003 NACE-5: 7470.2), slaughtering \& meat processing (1993/2003 NACE-5: 1511.1-1513.0), textile \& clothing (1993 NACE-5: 1711.0-1830.0, 2003 NACE-5: 1711.0-1830.0), and agriculture, forestry \& gardening (1993/2003 NACE-5: 0111.1-0141.1/0142.0-0502.0). For reasons of parsimony, this table does not include the following minimum wage sectors that cannot be identified by means of available industry codes: industrial laundries, specialized hard coal mining, public training services, and money and value services. AEntG = Arbeitnehmerentsendegesetz. AÜG = Arbeitnehmerüberlassungsgesetz.  MiLoG = Mindestlohngesetz. NACE-5 = 5-Digit Statistical Nomenclature of Economic Activities in the European Community. TVG = Tarifvertragsgesetz. Sources: AEntG $\plus$ AÜG  $\plus$ IEB, 1999-2017 $\plus$ Bundesanzeiger $\plus$ MiLoG $\plus$ TVG.
\end{tablenotes}
\end{threeparttable}
}
\end{table}

\renewcommand{\arraystretch}{1.5}

\clearpage

\begin{figure}[!ht]
\centering
\caption{Variation in Sectoral Minimum Wages}
\label{fig:B1}
\begin{subfigure}{0.475\textwidth}
\centering
\caption{Main Construction}
\scalebox{0.825}{
\begin{tikzpicture}

\begin{axis}[ytick={6,7,8,9,10,11,12,13,14}, height=7cm, width=14cm, grid=both, grid style = dotted, xtick={2000.45,2005.45,2010.45,2015.45},minor xtick={1999.45,2001.45,2002.45,2003.45,2004.45,2006.45,2007.45,2008.45,2009.45,2011.45,2012.45,2013.45,2014.45,2016.45,2017.45}, xlabel=Year, ylabel=Minimum Wage, ymin=5.75, ymax=14.25, xmin=1999.00, xmax=2018.00, legend pos = north west, y tick label style={/pgf/number format/.cd,fixed,fixed zerofill, precision=0,/tikz/.cd}, x tick label style={/pgf/number format/.cd,fixed,fixed zerofill, precision=0,/tikz/.cd}, legend style={font=\footnotesize}]

\addplot[mark=none, solid, color=black, line width=1mm] coordinates { (1997.00,8.68) (1997.66,8.68) (1997.66,8.17) (1999.66,8.17) (1999.66,9.44) (2000.66,9.44) (2000.66,9.63) (2001.66,9.63) (2001.66,9.79) (2002.66,9.79) (2002.66,10.12) (2003.66,10.12) (2003.66,10.36)  (2003.83,10.36) (2003.83,10.36) (2004.66,10.36) (2004.66,10.36) (2005.66,10.36) (2005.66,10.20) (2006.66,10.20)  (2006.66,10.30) (2007.66,10.30) (2007.66,10.40) (2008.66,10.40) (2008.66,10.70) (2009.66,10.70) (2009.66,10.80) (2010.66,10.80) (2010.66,10.90) (2011.91,10.90) (2011.91,11.00) (2012.00,11.00) (2012.00,11.05) (2013.00,11.05) (2013.00,11.05) (2014.00,11.05) (2014.00,11.10) (2015.00,11.10) (2015.00,11.15) (2016.00,11.15) (2016.00,11.25) (2017.00,11.25) };

\addplot[mark=none, solid, color=lightgray, line width=1mm] coordinates { (1997.00,7.98) (1997.66,7.98) (1997.66,7.73) (1999.66,7.73) (1999.66,8.31) (2000.66,8.31) (2000.66,8.47) (2001.66,8.47) (2001.66,8.61) (2002.66,8.61) (2002.66,8.75) (2003.66,8.75) (2003.66,8.95) (2003.83,8.95) (2003.83,8.95) (2004.66,8.95) (2004.66,8.95) (2005.66,8.95) (2005.66,8.80) (2006.66,8.80) (2006.66,8.90) (2007.66,8.90) (2007.66,9.00) (2008.66,9.00) (2008.66,9.00) (2009.66,9.00) (2009.66,9.25) (2010.66,9.25) (2010.66,9.50) (2011.91,9.50) (2011.91,9.75) (2012.00,9.75) (2012.00,10.00) (2013.00,10.00) (2013.00,10.25) (2014.00,10.25) (2014.00,10.50) (2015.00,10.50) (2015.00,10.75) (2016.00,10.75)  (2016.00,11.05) (2017.00,11.05) };

\addplot[mark=none, solid, color=gray, line width=1mm] coordinates {(2017.00,11.30) (2018.00,11.30)  }; 

\legend{~West Germany \& Berlin,~East Germany,~Germany,}
\end{axis}
\end{tikzpicture}
}
\end{subfigure}
\hfill
\begin{subfigure}{0.475\textwidth}
\centering
\caption{Electrical Trade}
\scalebox{0.825}{
\begin{tikzpicture}

\begin{axis}[ytick={6,7,8,9,10,11,12,13,14}, height=7cm, width=14cm, grid=both, grid style = dotted, xtick={2000.45,2005.45,2010.45,2015.45},minor xtick={1999.45,2001.45,2002.45,2003.45,2004.45,2006.45,2007.45,2008.45,2009.45,2011.45,2012.45,2013.45,2014.45,2016.45,2017.45}, xlabel=Year, ylabel=Minimum Wage, ymin=5.75, ymax=14.25, xmin=1999.00, xmax=2018.00, legend pos = north west, y tick label style={/pgf/number format/.cd,fixed,fixed zerofill, precision=0,/tikz/.cd}, x tick label style={/pgf/number format/.cd,fixed,fixed zerofill, precision=0,/tikz/.cd}, legend style={font=\footnotesize}]

\addplot[mark=none, solid, color=black, line width=1mm] coordinates { (1997.41,8.03) (1998.33,8.03) (1998.33,8.13) (1999.66,8.13) (1999.66,8.13) (2000.00,8.13) (2000.00,8.28) (2000.58,8.28)(2000.58,8.44) (2001.00,8.44) (2001.00,8.64) (2001.66,8.64) (2001.66,8.64) (2002.33,8.64) (2002.33,8.90) (2003.33,8.90)};

\addplot[mark=none, solid, color=black, line width=1mm] coordinates { (2007.66,9.20) (2008.00,9.20) (2008.00,9.40) (2009.00,9.40) (2009.00,9.55)  (2010.00,9.55) (2010.00,9.60) (2011.00,9.60) (2011.00,9.70) (2012.00,9.70) (2012.00,9.80) (2013.00,9.80) (2013.00,9.90) (2014.00,9.90) (2014.00,10.00) (2015.00,10.00) (2015.00,10.10) (2016.00,10.10)  };

\addplot[mark=none, solid, color=black, line width=1mm] coordinates { (2016.58,10.35) (2017.00,10.35) (2017.00,10.65) (2018.00,10.65) };

\addplot[mark=none, solid, color=lightgray, line width=1mm] coordinates { (1997.41,6.41) (1998.33,6.41) (1998.33,6.65) (1999.66,6.65) (1999.66,6.65) (2000.00,6.65) (2000.00,6.80) (2000.58,6.80) (2000.58,6.95) (2001.00,6.95) (2001.00,7.16) (2001.66,7.16) (2001.66,7.16) (2002.33,7.16) (2002.33,7.40) (2003.33,7.40) };

\addplot[mark=none, solid, color=lightgray, line width=1mm] coordinates { (2007.66,7.70) (2008.00,7.70) (2008.00,7.90) (2009.00,7.90) (2009.00,8.05)  (2010.00,8.05) (2010.00,8.20) (2011.00,8.20) (2011.00,8.40) (2012.00,8.40) (2012.00,8.65) (2013.00,8.65) (2013.00,8.85) (2014.00,8.85) (2014.00,9.10) (2015.00,9.10) (2015.00,9.35) (2016.00,9.35)    };

\addplot[mark=none, solid, color=lightgray, line width=1mm] coordinates { (2016.58,9.85) (2017.00,9.85) (2017.00,10.40) (2018.00,10.40) };

\addplot[mark=none, solid, color=gray, line width=1mm] coordinates { (2016.00,8.50) (2016.58,8.50)   };

\legend{~West Germany,,,~East Germany,,,~Germany}
\end{axis}
\end{tikzpicture}
}
\end{subfigure}
\vskip \baselineskip
\begin{subfigure}{0.475\textwidth}
\centering
\caption{Roofing}
\scalebox{0.825}{
\begin{tikzpicture}

\begin{axis}[ytick={6,7,8,9,10,11,12,13,14}, height=7cm, width=14cm, grid=both, grid style = dotted, xtick={2000.45,2005.45,2010.45,2015.45},minor xtick={1999.45,2001.45,2002.45,2003.45,2004.45,2006.45,2007.45,2008.45,2009.45,2011.45,2012.45,2013.45,2014.45,2016.45,2017.45}, xlabel=Year, ylabel=Minimum Wage, ymin=5.75, ymax=14.25, xmin=1999.00, xmax=2018.00, legend pos = north west, y tick label style={/pgf/number format/.cd,fixed,fixed zerofill, precision=0,/tikz/.cd}, x tick label style={/pgf/number format/.cd,fixed,fixed zerofill, precision=0,/tikz/.cd}, legend style={font=\footnotesize}]

\addplot[mark=none, solid, color=black, line width=1mm] coordinates { (1997.75,8.18) (2000.66,8.18)  };
\addplot[mark=none, solid, color=black, line width=1mm] coordinates { (2001.66,8.95) (2002.66,8.95)  };

\addplot[mark=none, solid, color=lightgray, line width=1mm] coordinates { (1997.75,7.74) (2000.66,7.74)   };
\addplot[mark=none, solid, color=lightgray, line width=1mm] coordinates { (2001.66,8.44) (2002.6,8.44)   };

\addplot[mark=none, solid, color=gray, line width=1mm] coordinates { (2003.16,9.00) (2004.00,9.00)  };
\addplot[mark=none, solid, color=gray, line width=1mm] coordinates { (2004.41,9.30) (2004.41,9.30) (2005.00,9.30) (2005.00,9.65) (2006.00,9.65) (2006.00,10.00) (2007.00,10.00) (2007.00,10.00) (2008.00,10.00) (2008.00,10.20) (2009.00,10.20) (2009.00,10.40) (2010.00,10.40) (2010.00,10.60) (2011.00,10.60) (2011.00,10.80) (2012.00,10.80) (2012.00,11.00) (2013.00,11.00) (2013.00,11.20) (2014.00,11.20) (2014.00,11.55) (2015.00,11.55) (2015.00,11.85) (2016.00,11.85) (2016.00,12.05) (2017.00,12.05) (2017.00,12.25) (2018.00,12.25)   }; 

\legend{~West Germany \& Berlin,,~East Germany,,~Germany,,}
\end{axis}
\end{tikzpicture}
}
\end{subfigure}
\hfill
\begin{subfigure}{0.475\textwidth}
\centering
\caption{Painting \& Varnishing}
\scalebox{0.825}{
\begin{tikzpicture}

\begin{axis}[ytick={6,7,8,9,10,11,12,13,14}, height=7cm, width=14cm, grid=both, grid style = dotted, xtick={2000.45,2005.45,2010.45,2015.45},minor xtick={1999.45,2001.45,2002.45,2003.45,2004.45,2006.45,2007.45,2008.45,2009.45,2011.45,2012.45,2013.45,2014.45,2016.45,2017.45}, xlabel=Year, ylabel=Minimum Wage, ymin=5.75, ymax=14.25, xmin=1999.00, xmax=2018.00, legend pos = north west, y tick label style={/pgf/number format/.cd,fixed,fixed zerofill, precision=0,/tikz/.cd}, x tick label style={/pgf/number format/.cd,fixed,fixed zerofill, precision=0,/tikz/.cd}, legend style={font=\footnotesize}]

\addplot[mark=none, solid, color=black, line width=1mm] coordinates { (2003.91,7.69) (2004.24,7.69)  };
\addplot[mark=none, solid, color=black, line width=1mm] coordinates { (2004.41,7.69) (2005.24,7.69) (2005.24,7.85) (2008.16,7.85)  (2008.16,8.05) (2009.41,8.05)  };

\addplot[mark=none, solid, color=lightgray, line width=1mm] coordinates { (2003.91,7.00) (2004.24,7.00)  };
\addplot[mark=none, solid, color=lightgray, line width=1mm] coordinates { (2004.41,7.00) (2005.24,7.00) (2005.24,7.15) (2008.16,7.15)  (2008.16,7.50) (2009.41,7.50)  };

\addplot[mark=none, solid, color=gray, line width=1mm] coordinates { (2009.41,9.50) (2010.66,9.50) (2011.49,9.50) (2011.49,9.75) (2012.16,9.75)  };
\addplot[mark=none, solid, color=gray, line width=1mm] coordinates { (2012.41,9.75)  (2012.49,9.75) (2013.33,9.75) (2013.33,9.90) (2014.33,9.90) };
\addplot[mark=none, solid, color=gray, line width=1mm] coordinates { (2014.58,9.90) (2015.33,9.90) (2015.33,10.00) (2016.33,10.00)  (2016.33,10.10) (2017.33,10.10) (2017.33,10.35) (2018.33,10.35) (2018.33,10.60) (2019.33,10.60) (2019.33,10.85)  (2020.00,10.85)   };

\legend{~West Germany \& Berlin,,~East Germany,,~Germany,,}
\end{axis}
\end{tikzpicture}
}
\end{subfigure}

\floatfoot{\footnotesize\textsc{Note. ---} The figures show the level of sectoral minimum wages (in Euro per hour) between 1999 and 2017. The ticks on the x axes refer to 30 June of the respective year. Line interruptions point to periods in which no minimum wage was in force. In the electrical trade sector, the City of Berlin fell under the West German minimum wage regulation until 2003. Since its re-introduction in 2007, East German minimum wage levels apply in Berlin. AEntG = Arbeitnehmerentsendegesetz. AÜG = Arbeitnehmerüberlassungsgesetz. MiLoG = Mindestlohngesetz. TVG = Tarifvertragsgesetz. Sources: AEntG $\plus$ AÜG $\plus$ Bundesanzeiger $\plus$ MiLoG $\plus$ TVG.}
\end{figure}

\clearpage

\begin{figure}[!ht]
\addtocounter{figure}{-1}
\centering
\caption{Variation in Sectoral Minimum Wages (Cont.)}
\begin{subfigure}{0.475\textwidth}
\addtocounter{subfigure}{4}
\centering
\caption{Commercial Cleaning}
\scalebox{0.825}{
\begin{tikzpicture}

\begin{axis}[ytick={6,7,8,9,10,11,12,13,14}, height=7cm, width=14cm, grid=both, grid style = dotted, xtick={2000.45,2005.45,2010.45,2015.45},minor xtick={1999.45,2001.45,2002.45,2003.45,2004.45,2006.45,2007.45,2008.45,2009.45,2011.45,2012.45,2013.45,2014.45,2016.45,2017.45}, xlabel=Year, ylabel=Minimum Wage, ymin=5.75, ymax=14.25, xmin=1999.00, xmax=2018.00, legend pos = north west, y tick label style={/pgf/number format/.cd,fixed,fixed zerofill, precision=0,/tikz/.cd}, x tick label style={/pgf/number format/.cd,fixed,fixed zerofill, precision=0,/tikz/.cd}, legend style={font=\footnotesize}]

\addplot[mark=none, solid, color=black, line width=1mm] coordinates {  (2004.33,7.68) (2005.00,7.68) (2005.00,7.87) (2008.16,7.87) (2008.16,8.15) (2009.00,8.15) (2009.75,8.15) };
\addplot[mark=none, solid, color=black, line width=1mm] coordinates {  (2010.16,8.40) (2011.00,8.40) (2011.00,8.55) (2012.00,8.55) (2012.00,8.82) (2013.00,8.82) (2013.00,9.00) (2014.00,9.00) (2014.00,9.31) (2015.00,9.31) (2015.00,9.55) (2016.00,9.55)  };
\addplot[mark=none, solid, color=black, line width=1mm] coordinates {  (2016.16,9.80) (2017.00,9.80) (2017.00,10.00) (2018.00,10.00) };
\addplot[mark=none, solid, color=black, line width=1mm] coordinates {  (2018.16,10.30) (2019.00,10.30) (2019.00,10.30) (2020.00,10.56) };

\addplot[mark=none, solid, color=lightgray, line width=1mm] coordinates {  (2004.33,6.18) (2005.00,6.18) (2005.00,6.36) (2008.16,6.36) (2008.16,6.58) (2009.00,6.58) (2009.75,6.58) };
\addplot[mark=none, solid, color=lightgray, line width=1mm] coordinates {  (2010.16,6.83) (2011.00,6.83) (2011.00,7.00) (2012.00,7.00) (2012.00,7.33) (2013.00,7.33) (2013.00,7.56) (2014.00,7.56) (2014.00,7.96) (2015.00,7.96) (2015.00,8.50) (2016.00,8.50)  };
\addplot[mark=none, solid, color=lightgray, line width=1mm] coordinates {  (2016.16,8.70) (2017.00,8.70) (2017.00,9.05) (2018.00,9.05) };
\addplot[mark=none, solid, color=lightgray, line width=1mm] coordinates {  (2018.16,9.55) (2019.00,9.55) (2019.00,10.05) (2020.00,10.05) };

\addplot[mark=none, solid, color=gray, line width=1mm] coordinates {  (2016.00,8.50) (2016.16,8.50) };

\addplot[mark=none, solid, color=gray, line width=1mm] coordinates {  (2018.00,8.84) (2018.16,8.84) };

\legend{~West Germany \& Berlin,,,,~East Germany,,,,~Germany}
\end{axis}
\end{tikzpicture}
}
\end{subfigure}
\hfill
\begin{subfigure}{0.475\textwidth}
\centering
\caption{Waste Removal}
\scalebox{0.825}{
\begin{tikzpicture}

\begin{axis}[ytick={6,7,8,9,10,11,12,13,14}, height=7cm, width=14cm, grid=both, grid style = dotted, xtick={2000.45,2005.45,2010.45,2015.45},minor xtick={1999.45,2001.45,2002.45,2003.45,2004.45,2006.45,2007.45,2008.45,2009.45,2011.45,2012.45,2013.45,2014.45,2016.45,2017.45}, xlabel=Year, ylabel=Minimum Wage, ymin=5.75, ymax=14.25, xmin=1999.00, xmax=2018.00, legend pos = north west, y tick label style={/pgf/number format/.cd,fixed,fixed zerofill, precision=0,/tikz/.cd}, x tick label style={/pgf/number format/.cd,fixed,fixed zerofill, precision=0,/tikz/.cd}, legend style={font=\footnotesize}]

\addplot[mark=none, solid, color=gray, line width=1mm] coordinates { (2009.91,8.02) (2010.83,8.02) };
\addplot[mark=none, solid, color=gray, line width=1mm] coordinates { (2011.00,8.24) (2011.66,8.24) };
\addplot[mark=none, solid, color=gray, line width=1mm] coordinates { (2011.83,8.33) (2012.24,8.33) };
\addplot[mark=none, solid, color=gray, line width=1mm] coordinates { (2012.41,8.33) (2013.00,8.33) };
\addplot[mark=none, solid, color=gray, line width=1mm] coordinates { (2013.08,8.68) (2014.49,8.68) };
\addplot[mark=none, solid, color=gray, line width=1mm] coordinates { (2014.75,8.86) (2015.49,8.86) };
\addplot[mark=none, solid, color=gray, line width=1mm] coordinates { (2015.49,8.50) (2015.75,8.50) };
\addplot[mark=none, solid, color=gray, line width=1mm] coordinates { (2015.75,8.94) (2016.00,8.94) (2016.00,9.10) (2017.24,9.10) (2017.24,8.84) (2018.00,8.84) }; 

\legend{~Germany,,,,,,}

\end{axis}
\end{tikzpicture}
}
\end{subfigure}
\vskip \baselineskip
\begin{subfigure}{0.475\textwidth}
\centering
\caption{Nursing Care}
\scalebox{0.825}{
\begin{tikzpicture}

\begin{axis}[ytick={6,7,8,9,10,11,12,13,14}, height=7cm, width=14cm, grid=both, grid style = dotted, xtick={2000.45,2005.45,2010.45,2015.45},minor xtick={1999.45,2001.45,2002.45,2003.45,2004.45,2006.45,2007.45,2008.45,2009.45,2011.45,2012.45,2013.45,2014.45,2016.45,2017.45}, xlabel=Year, ylabel=Minimum Wage, ymin=5.75, ymax=14.25, xmin=1999.00, xmax=2018.00, legend pos = north west, y tick label style={/pgf/number format/.cd,fixed,fixed zerofill, precision=0,/tikz/.cd}, x tick label style={/pgf/number format/.cd,fixed,fixed zerofill, precision=0,/tikz/.cd}, legend style={font=\footnotesize}]

\addplot[mark=none, solid, color=black, line width=1mm] coordinates {  (2010.58,8.50) (2012.00,8.50) (2012.00,8.75)  (2013.49,8.75) (2013.49,9.00) (2015.00,9.00) (2015.00,9.40) (2016.00,9.40) (2016.00,9.75)  (2017.00,9.75) (2017.00,10.20)  (2018.00,10.20) }; 

\addplot[mark=none, solid, color=lightgray, line width=1mm] coordinates {  (2010.58,7.50) (2012.00,7.50) (2012.00,7.75)  (2013.49,7.75) (2013.49,8.00) (2015.00,8.00) (2015.00,8.65) (2016.00,8.65) (2016.00,9.00)  (2017.00,9.00) (2017.00,9.50)  (2018.00,9.50)  }; 

\legend{~West Germany \& Berlin,~East Germany}

\end{axis}
\end{tikzpicture}
}
\end{subfigure}
\hfill
\begin{subfigure}{0.475\textwidth}
\centering
\caption{Security}
\scalebox{0.825}{
\begin{tikzpicture}

\begin{axis}[ytick={6,7,8,9,10}, height=7cm, width=14cm, grid=both, grid style = dotted, xtick={2000.45,2005.45,2010.45,2015.45},minor xtick={1999.45,2001.45,2002.45,2003.45,2004.45,2006.45,2007.45,2008.45,2009.45,2011.45,2012.45,2013.45,2014.45,2016.45,2017.45}, xlabel=Year, ylabel=Minimum Wage\vphantom{{\LARGE I}}, ymin=6.35, ymax=9.25, xmin=1999.00, xmax=2018.00, legend pos = north west, y tick label style={/pgf/number format/.cd,fixed,fixed zerofill, precision=0,/tikz/.cd}, x tick label style={/pgf/number format/.cd,fixed,fixed zerofill, precision=0,/tikz/.cd}, legend style={font=\footnotesize}]

\addplot[name path = A, draw=none] coordinates {  (2011.41,8.60) (2012.16,8.60) (2012.16,8.75) (2013.00,8.75) (2013.00,8.90) (2014.00,8.90)  };

\addplot[name path = B, draw=none] coordinates {  (2011.41,6.53) (2012.16,6.53) (2012.16,7.00) (2013.00,7.00) (2013.00,7.76) (2014.00,7.76) };

\addplot[color=gray!40!white] fill between[of=A and B];










\addplot[mark=none, solid, color=gray, line width=1mm] coordinates {  (2015.00,8.50) (2017.00,8.50) (2017.00,8.84) (2018.00,8.84) (2019.00,8.84) (2019.00,9.19) (2020.00,9.19) };

\legend{,,~Federal States,~Germany}

\end{axis}
\end{tikzpicture}
}
\end{subfigure}
\vspace*{-0.3cm}
\floatfoot{\footnotesize\textsc{Note. ---} The figures show the level of sectoral minimum wages (in Euro per hour) between 1999 and 2017. The ticks on the x axes refer to 30 June of the respective year. Line interruptions point to periods in which no minimum wage was in force.  AEntG = Arbeitnehmerentsendegesetz. AÜG = Arbeitnehmerüberlassungsgesetz. MiLoG = Mindestlohngesetz. TVG = Tarifvertragsgesetz. Sources: AEntG $\plus$ AÜG $\plus$ Bundesanzeiger $\plus$ MiLoG $\plus$ TVG. }
\end{figure}

\clearpage

\begin{figure}[!ht]
\addtocounter{figure}{-1}
\centering
\caption{Variation in Sectoral Minimum Wages (Cont.)}
\begin{subfigure}{0.475\textwidth}
\addtocounter{subfigure}{8}
\centering
\caption{Temporary Work}
\scalebox{0.825}{
\begin{tikzpicture}

\begin{axis}[ytick={6,7,8,9,10,11,12,13,14}, height=7cm, width=14cm, grid=both, grid style = dotted, xtick={2000.45,2005.45,2010.45,2015.45},minor xtick={1999.45,2001.45,2002.45,2003.45,2004.45,2006.45,2007.45,2008.45,2009.45,2011.45,2012.45,2013.45,2014.45,2016.45,2017.45}, xlabel=Year, ylabel=Minimum Wage, ymin=5.75, ymax=14.25, xmin=1999.00, xmax=2018.00, legend pos = north west, y tick label style={/pgf/number format/.cd,fixed,fixed zerofill, precision=0,/tikz/.cd}, x tick label style={/pgf/number format/.cd,fixed,fixed zerofill, precision=0,/tikz/.cd}, legend style={font=\footnotesize}]

\addplot[mark=none, solid, color=black, line width=1mm] coordinates { (2012.00,7.89) (2012.83,7.89) (2012.83,8.19) (2013.83,8.19) };
\addplot[mark=none, solid, color=black, line width=1mm] coordinates { (2014.24,8.50) (2015.24,8.50) (2015.24,8.80) (2016.41,8.80) (2016.41,9.00) (2017.00,9.00)  };
\addplot[mark=none, solid, color=black, line width=1mm] coordinates { (2017.41,9.23) (2018.24,9.23) (2018.24,9.49) (2019.00,9.49) };
\addplot[mark=none, solid, color=black, line width=1mm] coordinates { (2019.24,9.79) (2019.75,9.79) (2019.75,9.96) (2020,9.96)  };

\addplot[mark=none, solid, color=lightgray, line width=1mm] coordinates { (2012.00,7.01) (2012.83,7.01) (2012.83,7.50) (2013.83,7.50) };
\addplot[mark=none, solid, color=lightgray, line width=1mm] coordinates { (2014.24,7.86) (2015.24,7.86) (2015.24,8.20) (2016.41,8.20) (2016.41,8.80) (2017.00,8.80)  };
\addplot[mark=none, solid, color=lightgray, line width=1mm] coordinates { (2017.41,8.91) (2018.24,8.91) (2018.24,9.27) (2019.00,9.27)   };
\addplot[mark=none, solid, color=lightgray, line width=1mm] coordinates { (2019.24,9.49) (2019.75,9.49) (2019.75,9.66) (2020,9.66)  };

\addplot[mark=none, solid, color=gray, line width=1mm] coordinates { (2017.00,8.50) (2017.41,8.50) };
\addplot[mark=none, solid, color=gray, line width=1mm] coordinates { (2019.00,9.49) (2019.24,9.49) };

\legend{~West Germany,,,,~East Germany \& Berlin,,,~Germany,}
\end{axis}
\end{tikzpicture}
}
\end{subfigure}
\hfill
\begin{subfigure}{0.475\textwidth}
\centering
\caption{Scaffolding}
\scalebox{0.825}{
\begin{tikzpicture}

\begin{axis}[ytick={6,7,8,9,10,11,12,13,14}, height=7cm, width=14cm, grid=both, grid style = dotted, xtick={2000.45,2005.45,2010.45,2015.45},minor xtick={1999.45,2001.45,2002.45,2003.45,2004.45,2006.45,2007.45,2008.45,2009.45,2011.45,2012.45,2013.45,2014.45,2016.45,2017.45}, xlabel=Year, ylabel=Minimum Wage, ymin=5.75, ymax=14.25, xmin=1999.00, xmax=2018.00, legend pos = north west, y tick label style={/pgf/number format/.cd,fixed,fixed zerofill, precision=0,/tikz/.cd}, x tick label style={/pgf/number format/.cd,fixed,fixed zerofill, precision=0,/tikz/.cd}, legend style={font=\footnotesize}]

\addplot[mark=none, solid, color=gray, line width=1mm] coordinates { (2013.58,10.00) (2014.16,10.00) };
\addplot[mark=none, solid, color=gray, line width=1mm] coordinates { (2014.66,10.25) (2015.33,10.25) (2015.33,10.50) (2016.24,10.50)};
\addplot[mark=none, solid, color=gray, line width=1mm] coordinates { (2016.24,8.50) (2016.33,8.50) };
\addplot[mark=none, solid, color=gray, line width=1mm] coordinates { (2016.33,10.70)  (2017.33,10.70) (2017.33,11.00) (2018.33,11.00) (2018.33,8.84) (2018.49,8.84) (2018.49,11.35) (2019.41,11.35) (2019.41,9.19) (2019.49,9.19) (2019.49,11.88) (2020.00,11.88) };

\legend{Germany,}

\end{axis}
\end{tikzpicture}
}
\end{subfigure}
\vskip \baselineskip
\begin{subfigure}{0.475\textwidth}
\centering
\caption{Stonemasonry}
\scalebox{0.825}{
\begin{tikzpicture}

\begin{axis}[ytick={6,7,8,9,10,11,12,13,14}, height=7cm, width=14cm, grid=both, grid style = dotted, xtick={2000.45,2005.45,2010.45,2015.45},minor xtick={1999.45,2001.45,2002.45,2003.45,2004.45,2006.45,2007.45,2008.45,2009.45,2011.45,2012.45,2013.45,2014.45,2016.45,2017.45}, xlabel=Year, ylabel=Minimum Wage, ymin=5.75, ymax=14.25, xmin=1999.00, xmax=2018.00, legend pos = north west, y tick label style={/pgf/number format/.cd,fixed,fixed zerofill, precision=0,/tikz/.cd}, x tick label style={/pgf/number format/.cd,fixed,fixed zerofill, precision=0,/tikz/.cd}, legend style={font=\footnotesize}]

\addplot[mark=none, solid, color=black, line width=1mm] coordinates { (2013.75,11.00) (2014.33,11.00) (2014.33,11.25)  (2015.33,11.25) };
\addplot[mark=none, solid, color=black, line width=1mm] coordinates { (2015.83,11.30) (2016.33,11.30) (2016.33,11.35) (2017.33,11.35) (2017.33,11.40)  (2018.33,11.40) };

\addplot[mark=none, solid, color=lightgray, line width=1mm] coordinates { (2013.75,10.13) (2014.33,10.13) (2014.33,10.66)  (2015.33,10.66) };
\addplot[mark=none, solid, color=lightgray, line width=1mm] coordinates { (2015.83,10.90) (2016.33,10.90) (2016.33,11.00) (2017.33,11.00) (2017.33,11.20)  (2018.33,11.20) };

\addplot[mark=none, solid, color=gray, line width=1mm] coordinates { (2015.33,8.50) (2015.83,8.50)  };
\addplot[mark=none, solid, color=gray, line width=1mm] coordinates { (2018.33,11.40) (2019.33,11.40) (2019.33,9.19) (2019.66,9.19) (2019.66,11.85) (2020.00,11.85) };

\legend{~West Germany \& Berlin,~East Germany,~Germany,}

\end{axis}
\end{tikzpicture}
}
\end{subfigure}
\hfill
\begin{subfigure}{0.475\textwidth}
\centering
\caption{Hairdressing}
\scalebox{0.825}{
\begin{tikzpicture}

\begin{axis}[ytick={6,7,8,9,10,11,12,13,14}, height=7cm, width=14cm, grid=both, grid style = dotted, xtick={2000.45,2005.45,2010.45,2015.45},minor xtick={1999.45,2001.45,2002.45,2003.45,2004.45,2006.45,2007.45,2008.45,2009.45,2011.45,2012.45,2013.45,2014.45,2016.45,2017.45}, xlabel=Year, ylabel=Minimum Wage, ymin=5.75, ymax=14.25, xmin=1999.00, xmax=2018.00, legend pos = north west, y tick label style={/pgf/number format/.cd,fixed,fixed zerofill, precision=0,/tikz/.cd}, x tick label style={/pgf/number format/.cd,fixed,fixed zerofill, precision=0,/tikz/.cd}, legend style={font=\footnotesize}]

\addplot[mark=none, solid, color=black, line width=1mm] coordinates {  (2013.83,7.50) (2014.58,7.50) (2014.58,8.00) (2015.58,8.00) };

\addplot[mark=none, solid, color=lightgray, line width=1mm] coordinates { (2013.83,6.50) (2014.58,6.50) (2014.58,7.50) (2015.58,7.50) };

\addplot[mark=none, solid, color=gray, line width=1mm] coordinates {  (2015.58,8.50) (2017.00,8.50) (2017.00,8.84) (2019.00,8.84) (2019.00,9.91) (2020.00,9.19) };

\legend{~West Germany,~East Germany \& Berlin,~Germany}
\end{axis}
\end{tikzpicture}
}
\end{subfigure}

\floatfoot{\footnotesize\textsc{Note. ---} The figures show the level of sectoral minimum wages (in Euro per hour) between 1999 and 2017. The ticks on the x axes refer to 30 June of the respective year. Line interruptions point to periods in which no minimum wage was in force. AEntG = Arbeitnehmerentsendegesetz. AÜG = Arbeitnehmerüberlassungsgesetz. MiLoG = Mindestlohngesetz. TVG = Tarifvertragsgesetz. Sources: AEntG $\plus$ AÜG $\plus$ Bundesanzeiger $\plus$ MiLoG $\plus$ TVG.}
\end{figure}

\clearpage

\begin{figure}[!ht]
\addtocounter{figure}{-1}
\centering
\caption{Variation in Sectoral Minimum Wages (Cont.)}
\begin{subfigure}{0.475\textwidth}
\addtocounter{subfigure}{12}
\centering
\caption{Chimney Sweeping}
\scalebox{0.825}{
\begin{tikzpicture}

\begin{axis}[ytick={6,7,8,9,10,11,12,13,14}, height=7cm, width=14cm, grid=both, grid style = dotted, xtick={2000.45,2005.45,2010.45,2015.45},minor xtick={1999.45,2001.45,2002.45,2003.45,2004.45,2006.45,2007.45,2008.45,2009.45,2011.45,2012.45,2013.45,2014.45,2016.45,2017.45}, xlabel=Year, ylabel=Minimum Wage, ymin=5.75, ymax=14.25, xmin=1999.00, xmax=2018.00, legend pos = north west, y tick label style={/pgf/number format/.cd,fixed,fixed zerofill, precision=0,/tikz/.cd}, x tick label style={/pgf/number format/.cd,fixed,fixed zerofill, precision=0,/tikz/.cd}, legend style={font=\footnotesize}]

\addplot[mark=none, solid, color=gray, line width=1mm] coordinates {  (2014.33,12.78) (2016.00,12.78) (2016.00,12.95) (2018.00,12.95)  }; 

\legend{~Germany}

\end{axis}
\end{tikzpicture}
}
\end{subfigure}
\hfill
\begin{subfigure}{0.475\textwidth}
\centering
\caption{Slaughtering \& Meat Processing}
\scalebox{0.825}{
\begin{tikzpicture}

\begin{axis}[ytick={6,7,8,9,10,11,12,13,14}, height=7cm, width=14cm, grid=both, grid style = dotted, xtick={2000.45,2005.45,2010.45,2015.45},minor xtick={1999.45,2001.45,2002.45,2003.45,2004.45,2006.45,2007.45,2008.45,2009.45,2011.45,2012.45,2013.45,2014.45,2016.45,2017.45}, xlabel=Year, ylabel=Minimum Wage, ymin=5.75, ymax=14.25, xmin=1999.00, xmax=2018.00, legend pos = north west, y tick label style={/pgf/number format/.cd,fixed,fixed zerofill, precision=0,/tikz/.cd}, x tick label style={/pgf/number format/.cd,fixed,fixed zerofill, precision=0,/tikz/.cd}, legend style={font=\footnotesize}]

\addplot[mark=none, solid, color=gray, line width=1mm] coordinates {  (2014.58,7.75) (2014.91,7.75)  (2014.91,8.00) (2015.75,8.00) (2015.75,8.60) (2016.91,8.60)  (2018.00,8.60) (2019.00,8.84) (2019.00,8.91) (2019.00,9.19) };

\legend{~Germany}

\end{axis}
\end{tikzpicture}
}
\end{subfigure}
\vskip \baselineskip
\begin{subfigure}{0.475\textwidth}
\centering
\caption{Textile \& Clothing}
\scalebox{0.825}{
\begin{tikzpicture}

\begin{axis}[ytick={6,7,8,9,10,11,12,13,14}, height=7cm, width=14cm, grid=both, grid style = dotted, xtick={2000.45,2005.45,2010.45,2015.45},minor xtick={1999.45,2001.45,2002.45,2003.45,2004.45,2006.45,2007.45,2008.45,2009.45,2011.45,2012.45,2013.45,2014.45,2016.45,2017.45}, xlabel=Year, ylabel=Minimum Wage, ymin=5.75, ymax=14.25, xmin=1999.00, xmax=2018.00, legend pos = north west, y tick label style={/pgf/number format/.cd,fixed,fixed zerofill, precision=0,/tikz/.cd}, x tick label style={/pgf/number format/.cd,fixed,fixed zerofill, precision=0,/tikz/.cd}, legend style={font=\footnotesize}]

\addplot[mark=none, solid, color=black, line width=1mm] coordinates { (2015.00,8.50) (2016.00,8.50) (2016.83,8.50) (2017.00,8.50) };

\addplot[mark=none, solid, color=lightgray, line width=1mm] coordinates { (2015.00,7.50) (2016.00,7.50) (2016.00,8.25) (2016.83,8.25) (2016.83,8.75) (2017.00,8.75) };

\addplot[mark=none, solid, color=gray, line width=1mm] coordinates { (2017.00,8.84) (2019.00,8.84) (2019.00,9.19) (2020.00,9.19)  };

\legend{~West Germany \& West Berlin,~East Germany \& East Berlin,~Germany}
\end{axis}
\end{tikzpicture}
}
\end{subfigure}
\hfill
\begin{subfigure}{0.475\textwidth}
\centering
\caption{Agriculture, Forestry \& Gardening}
\scalebox{0.825}{
\begin{tikzpicture}

\begin{axis}[ytick={6,7,8,9,10,11,12,13,14}, height=7cm, width=14cm, grid=both, grid style = dotted, xtick={2000.45,2005.45,2010.45,2015.45},minor xtick={1999.45,2001.45,2002.45,2003.45,2004.45,2006.45,2007.45,2008.45,2009.45,2011.45,2012.45,2013.45,2014.45,2016.45,2017.45}, xlabel=Year, ylabel=Minimum Wage, ymin=5.75, ymax=14.25, xmin=1999.00, xmax=2018.00, legend pos = north west, y tick label style={/pgf/number format/.cd,fixed,fixed zerofill, precision=0,/tikz/.cd}, x tick label style={/pgf/number format/.cd,fixed,fixed zerofill, precision=0,/tikz/.cd}, legend style={font=\footnotesize}]

\addplot[mark=none, solid, color=black, line width=1mm] coordinates {  (2015.00,7.40) (2016.00,7.40) (2016.00,8.00) (2017.00,8.00) };

\addplot[mark=none, solid, color=lightgray, line width=1mm] coordinates { (2015.00,7.20) (2016.00,7.20) (2016.00,7.90) (2017.00,7.90) };

\addplot[mark=none, solid, color=gray, line width=1mm] coordinates { (2017.00,8.60) (2017.83,8.60) (2017.83,9.10) (2018.00,9.10)  }; 

\legend{~West Germany,~East Germany \& Berlin,~Germany}
\end{axis}
\end{tikzpicture}
}
\end{subfigure}

\floatfoot{\footnotesize\textsc{Note. ---} The figures show the level of sectoral minimum wages (in Euro per hour) between 1999 and 2017. The ticks on the x axes refer to 30 June of the respective year. Line interruptions point to periods in which no minimum wage was in force. AEntG = Arbeitnehmerentsendegesetz. AÜG = Arbeitnehmerüberlassungsgesetz. MiLoG = Mindestlohngesetz. TVG = Tarifvertragsgesetz. Sources: AEntG $\plus$ AÜG $\plus$ Bundesanzeiger $\plus$ MiLoG $\plus$ TVG.}
\end{figure}

\end{landscape}

\clearpage

\section{Data: Further Evidence}
\label{sec:C}
\setcounter{table}{0} 
\setcounter{figure}{0} 

I assemble five datasets to empirically examine the monopsony argument for the German labor market. The three main sources for my baseline analyses comprise administrative information from the IEB and BHP data as well as survey information from the IAB Establishment Panel. To perform certain robustness checks, I additionally leverage two further data sources, namely the LIAB and AWFP data.

\paragraph{BHP.} The Establishment History Panel (BHP) differentiates between regular full-time workers, regular part-time workers, marginal part-time workers, and apprentices. Marginal part-time jobs involve reduced social security contributions for work contracts below a legally defined income threshold. Since April 1, 1999, the monthly salary must not exceed 325 Euro. On April 1, 2003 and January 1, 2013, the threshold was lifted to 400 and 450 Euro per month, respectively. Throughout the study, I disregard apprentices who are, in most sectors, exempt from the minimum wage. Unfortunately, the BHP data do not allow for identifying white-collar workers who are in some sectors not subject to the minimum wage regulation. In the underlying IEB data, wages are right-censored at the upper-earnings limit of social security contributions. \citet{GanzerEtAl2020} provide a detailed description of the BHP implementation of the two-step wage imputation in \citet{CardEtAl2013}. In principle, however, the imputation is unlikely to have any relevance for the minimum wage effects since the censoring limit lies far above the minimum wage levels in the sectors under study.

\paragraph{Descriptive Statistics.} Table \ref{tab:C1} shows descriptive statistics for the BHP variables before and after the minimum wage introduction in the sectors for June 30 of the years 1999-2017. The full sample comprises 6,865,711 firm-year observations, of which 2,981,029 observations (43.4 percent) refer to years before the minimum wage was introduced for the first time in the sectors and 3,884,682 observations (56.6 percent) refer to the years thereafter.

\afterpage{
\begin{landscape}

\begin{table}[!ht]
\caption{Descriptive Statistics}
\label{tab:C1}
\begin{threeparttable}
\begin{subtable}{0.475\textwidth}
\caption{\normalsize Before Minimum Wage Introduction in the Sectors}
\scalebox{0.8}{
\begin{tabular}{L{6.5cm}C{1cm}C{0.8cm}C{0.8cm}C{0.8cm}C{0.8cm}C{1.5cm}} \hline
\multirow{3.4}{*}{}  & \multirow{3.4}{*}{Mean} & \multirow{3.4}{*}{P25} & \multirow{3.4}{*}{P50} & \multirow{3.4}{*}{\shortstack{P75}} & \multirow{3.4}{*}{\shortstack{SD}} & \multirow{3.4}{*}{\shortstack{Obser- \\ vations}} \\
          &   &   &   &   &  &  \\
          &   &   &   &   &  &  \\ \hline
          &   &   &   &   &  &  \\[-0.4cm]
\textbf{Mean Daily Wage (in Euro)}      &        &           &           &           &           &               \\
~~Regular Full-Time                     & 52.04  & 35.4 & 48.0 & 65.6 & 24.4 & 2,205,172       \\
\textbf{Employment (in Heads)}          &        &           &           &           &           &               \\
~~Regular Full-Time                     & 7.106  & 0.00 & 1.00 & 4.00 & 31.6 & 2,981,029    \\
~~Overall                               & 11.41  & 1.00 & 3.00 & 7.00 & 45.7 & 2,981,029       \\
~~Regular Part-Time                     & 1.885  & 0.00 & 0.00 & 1.00 & 13.3 & 2,981,029        \\
~~Marginal Part-Time                    & 2.421  & 0.00 & 1.00 & 2.00 & 17.1 & 2,981,029        \\
\textbf{Hourly Minimum Wage (in Euro)}  &        &           &           &           &           &               \\
~~Overall                               & n/a    & n/a       & n/a       & n/a       & n/a       & n/a    \\
\textbf{Territory (0/1)}                &        &           &           &           &           &               \\
~~West Germany                          & 0.811  &           &           &           &           & 2,981,029       \\
~~East Germany                          & 0.168  &           &           &           &           & 2,981,029       \\
~~City of Berlin                        & 0.021  &           &           &           &           & 2,981,029       \\
\textbf{Minimum Wage Sector (0/1)}      &        &           &           &           &           &               \\
~~Main Construction                     & n/a  &           &           &           &           & n/a      \\
~~Electrical Trade                      & n/a  &           &           &           &           & n/a      \\
~~Roofing                               & n/a  &           &           &           &           & n/a      \\
~~Painting \& Varnishing                & 0.044  &           &           &           &           & 2,981,029        \\
~~Commercial Cleaning                   & 0.031  &           &           &           &           & 2,981,029        \\
~~Waste Removal                         & 0.026  &           &           &           &           & 2,981,029        \\
~~Nursing Care                          & 0.058  &           &           &           &           & 2,981,029        \\
~~Security                              & 0.016  &           &           &           &          & 2,981,029       \\
~~Temporary Work                        & 0.036  &           &           &           &          & 2,981,029       \\
~~Scaffolding                           & 0.015  &           &           &           &          & 2,981,029       \\
~~Stonemasonry                          & 0.023  &           &           &           &          & 2,981,029       \\
~~Hairdressing                          & 0.241  &           &           &           &          & 2,981,029       \\
~~Chimney Sweeping                      & 0.039  &           &           &           &          & 2,981,029       \\
~~Slaughtering \& Meat Processing       & 0.071  &           &           &           &          & 2,981,029       \\
~~Textile \& Clothing                   & 0.042  &           &           &           &          & 2,981,029       \\
~~Agriculture, Forestry \& Gardening    & 0.360  &           &           &           &          & 2,981,029       \\ \hline
\end{tabular}
}
\end{subtable}\hfill%
\begin{subtable}{0.475\textwidth}
\caption{\normalsize After Minimum Wage Introduction in the Sectors}
\scalebox{0.8}{
\begin{tabular}{L{6.5cm}C{1cm}C{0.8cm}C{0.8cm}C{0.8cm}C{0.8cm}C{1.5cm}} \hline
\multirow{3.4}{*}{}  & \multirow{3.4}{*}{Mean} & \multirow{3.4}{*}{P25} & \multirow{3.4}{*}{P50} & \multirow{3.4}{*}{\shortstack{P75}} & \multirow{3.4}{*}{\shortstack{SD}} & \multirow{3.4}{*}{\shortstack{Obser- \\ vations}} \\
          &   &   &   &   &  &  \\
          &   &   &   &   &  &  \\ \hline
         &   &   &   &   &  &  \\[-0.4cm]
\textbf{Mean Daily Wage (in Euro)}      &        &           &           &           &           &               \\
~~Regular Full-Time                     & 69.51  & 54.1 & 68.9 & 82.9 & 25.4 & 2,986,985       \\
\textbf{Employment (in Heads)}          &        &           &           &           &           &               \\
~~Regular Full-Time                     & 7.606  & 1.00 & 2.00 & 6.00 & 29.7 & 3,884,682    \\
~~Overall                               & 12.93  & 2.00 & 4.00 & 9.00 & 50.4 & 3,884,682       \\
~~Regular Part-Time                     & 2.350  & 0.00 & 0.00 & 1.00 & 17.0 & 3,884,682        \\
~~Marginal Part-Time                    & 2.975  & 0.00 & 1.00 & 2.00 & 22.1 & 3,884,682        \\
\textbf{Hourly Minimum Wage (in Euro)}  &        &           &           &           &           &               \\
~~Overall                               & 8.956     & 8.18        & 9.40        & 10.4        & 2.41        & 3,884,682      \\
\textbf{Territory (0/1)}                &        &           &           &           &           &               \\
~~West Germany                          & 0.773  &           &           &           &           & 3,884,682       \\
~~East Germany                          & 0.191  &           &           &           &           & 3,884,682       \\
~~City of Berlin                        & 0.036  &           &           &           &           & 3,884,682       \\
\textbf{Minimum Wage Sector (0/1)}      &        &           &           &           &           &               \\
~~Main Construction                     & 0.391  &           &           &           &           & 3,884,682        \\
~~Electrical Trade                      & 0.149  &           &           &           &           & 3,884,682        \\
~~Roofing                               & 0.061  &           &           &           &           & 3,884,682        \\
~~Painting \& Varnishing                & 0.087  &           &           &           &           & 3,884,682        \\
~~Commercial Cleaning                   & 0.094  &           &           &           &           & 3,884,682        \\
~~Waste Removal                         & 0.014  &           &           &           &           & 3,884,682        \\
~~Nursing Care                          & 0.038  &           &           &           &           & 3,884,682        \\
~~Security                              & 0.009  &           &           &           &          & 3,884,682       \\
~~Temporary Work                        & 0.019  &           &           &           &          & 3,884,682       \\
~~Scaffolding                           & 0.003  &           &           &           &          & 3,884,682       \\
~~Stonemasonry                          & 0.004  &           &           &           &          & 3,884,682       \\
~~Hairdressing                          & 0.050  &           &           &           &          & 3,884,682       \\
~~Chimney Sweeping                      & 0.008  &           &           &           &          & 3,884,682       \\
~~Slaughtering \& Meat Processing       & 0.007  &           &           &           &          & 3,884,682       \\
~~Textile \& Clothing                   & 0.005  &           &           &           &          & 3,884,682       \\
~~Agriculture, Forestry \& Gardening    & 0.061  &           &           &           &          & 3,884,682       \\ \hline
\end{tabular}
}
\end{subtable}
\end{threeparttable}
\vspace*{-0.35cm}
\floatfoot{\footnotesize\textsc{Note. ---} The tables shows descriptive statistics of the main variables before and after the minimum wage came into effect for the first time in the respective sector. All statistics reflect firm-year observations. In the sectors of main construction, electrical trade, and roofing, sectoral minimum wages were already into force before the period of analysis. PX = Xth Percentile. SD = Standard Deviation. Source: BHP, 1999-2017.}
\end{table}

\end{landscape}
}

Before the minimum wage introduction in the sectors, the average firm employs 11.4 workers, namely 7.1 regular full-time workers, 1.9 regular part-time workers, and 2.4 marginal part-time workers. For lack of information on individual hours worked, average daily wages can only be meaningfully interpreted for regular full-time workers (who are supposed to work a similar number of hours). On average, the mean daily wage of regular full-time workers is 52.04 Euro (per calendar day) which, by assuming a 40-hours week, implies an hourly wage of 9.11 Euro. The 25th percentile, the median, and the 75th percentile of mean daily wages of regular full-time workers are 35.4, 48.0, and 65.6 Euro per calendar day, respectively. The IEB data feature a major structural break in on April 1, 1999. From this date onward, the IEB further include marginal part-time workers. To avoid spurious changes in the labor market concentration over time (which is calculated on overall employment), I begin the analysis in the year 1999. Since minimum wages were introduced in the sectors of main construction, electrical trade, and roofing already before June 30, 1999, these sector do not enter the multivariate analysis in Section \ref{sec:6} and the event-study analysis in Section \ref{sec:7}.

After the minimum wage introduction in the sectors, the average firm employs 12.9 workers, namely 7.6 regular full-time workers, 2.4 regular part-time workers, and 2.0 marginal part-time workers. On average, the mean daily wage of regular full-time workers is 69.51 Euro (per calendar day) which, by assuming a 40-hours week, implies an hourly wage of 12.16 Euro. The 25th percentile, the median, and the 75th percentile of average daily wages of regular full-time workers are 54.1, 68.9, and 82.9 Euro per calendar day, respectively. The average hourly minimum wage is 8.96 Euro per hour.

\paragraph{LIAB.} In general, about 0.6 percent of firms in the BHP data can be linked with survey information from the IAB Establishment Panel. The linkage of both datasets is known as the ``Linked Employer-Employee Dataset of the IAB'' (LIAB), which allows me to perform a set of important robustness checks. First, I build on the production-function approach and recover firm-level markdowns and markups for firms in the LIAB subsample (see Appendix \ref{sec:F}) because the IAB Establishment Panel inquires firms about their annual revenues, annual expenditure on intermediate goods, and investment behavior (which allows for an approximation of the capital stock). The availability of markdowns allows to inspect the performance of concentration indices as a measure of monopsony power (see Section \ref{sec:5}) and represents a more comprehensive moderating variable on the minimum wage effects on employment (see Section \ref{sec:7}). Second, dividing the information on annual revenues in the IAB Establishment Panel by the number of workers yields a proxy for firm productivity, which represents a potentially confounding moderating variable of the minimum wage effects on employment (see Section \ref{sec:7}). The LIAB sample for firms in the minimum wage sector contains 41,005 observations but, due to item non-response, a smaller number of observations enters the above-mentioned robustness checks.

\paragraph{AWFP.} In the complementary event-study analysis of the first-time introduction of minimum wages in the sectors in Section \ref{sec:7}, I make use of the ``Administrative Wage and Labor Market Flow Panel'' (AWFP) data \citep{StueberSeth2019}. The AWFP data consolidate IEB notifications on individual workers at the firm level but, unlike the BHP data, is not constructed at yearly but quarterly frequency. In each, the quarterly reference dates refer to March 31, June 30, September 30, and December 31. The lower temporal aggregation of the AWFP data is a key advantage over the BHP data when analyzing the effects right before and after the minimum wage introduction. After filtering on information within 10 quarters before and 12 quarters after the event, the final event-study AWFP dataset comprises 5,325,320 firm-quarter observations.

\clearpage

\section{Labor Market Concentration: Further Evidence}
\label{sec:D}
\setcounter{table}{0} 
\setcounter{figure}{0} 

In this section, I provide information on the statistical properties of concentration indices, the delineation of the underlying labor markets, and additional descriptive evidence on labor market concentration in the Germany.

\paragraph{Herfindahl-Hirschman Index.} \citet{Marfels1971} defines absolute concentration (of labor markets) as the accumulation of a given number of objects (workers) on subjects (employers). I follow standard practice and construct measures of absolute labor market concentration on the basis of the Herfindahl-Hirschman Index (HHI), which equals the sum of squared market shares \citep{Hirschman1945,Herfindahl1950}. Specifically, I calculate labor market concentration as
\begin{equation}
\label{eq:D1}
H\!H\!I_{mt} = \sum_{j=1}^{J} s_{jmt}^{2}
\end{equation}
where $ s_{jmt} = \frac{L_{jmt}}{\sum_{j=1}^{J} L_{jmt}} $ represent the share of firm $j$ in employment of labor market $m$ in year $t$. When calculating these employment shares, I use the sum of regular full-time, regular part-time, and marginal part-time in the firm or market, respectively. In an alternative formulation, \citet{Adelman1969} shows that HHI is a function of the number of employers $J$ (i.e., the fewness dimension) and the variance of market shares $\sigma^{2}$ (i.e., the unevenness dimension): $ H\!H\!I = J \, \sigma^{2} + \frac{1}{J} $. If employers have equal size (i.e., $\sigma^{2}=0$), the HHI collapses to $\frac{1}{J}$.

HHI values range from 0 to 1, higher values signal greater concentration. In many countries, the HHI offers a guideline for antitrust policy. The E.U.\ Commission \citeyearpar{EC2004} scrutinizes the intensity of product market competition by means of three intervals for HHI values: low (0.0-0.1), medium (0.1-0.2) and high levels of concentration (0.2-1.0). These intervals were defined in an ad-hoc manner for the evaluation of product markets in the European Union. In the U.S., similar thresholds have been legislated and, since recently, additionally refer to labor markets \citep{FTC2023}.

Unlike estimates of simple population means, absolute concentration indices are usually biased when they are derived from random samples. In terms of HHI, \citet{AbelEtAl2020} simulate that random sampling of workers results in an upward bias. The bias stems from two sources: On the one hand, HHI is a decreasing function in the absolute number of employers in the market. However, sampling workers will underestimate the true number of employers in the market, thus artificially increasing the expected value of estimated employment shares. On the other hand, even when unbiased estimates for employment shares are available, $E[\hat{s}_{j}]=s_{j}$, Jensen's inequality implies that the squaring of employment shares results in: $ E[\hat{s}^{2}_{j}]>s_{j}^{2} $. In addition, the sampling bias is exacerbated under random sampling of employers instead of workers. In this study, I am minimizing such a bias by leveraging the near-universe of workers that appear in the IEB data. In fact, all workers in the German labor market enter the calculation expect civil servants, self-employed, and family workers who are not liable to social security contributions.

\paragraph{Delineation of Labor Markets.} The use of concentration indices necessitates an appropriate definition of labor markets. Following the literature, I operationalize labor markets on the basis of observable firm characteristics, namely industry and workplace (e.g., \citealp{Rinz2022}). On the one hand, \citet{Neal1995} shows that displaced workers who switch industry suffer a greater earnings loss than those workers that take up a new job in their pre-displacement industry, thus lending support to industry-specific human capital. On the other hand, labor markets also feature a local dimension as the attractiveness of jobs to applicants sharply decays with distance \citep{ManningPetrongolo2017,MarinescuRathelot2018}.

As baseline specification, I define a labor market $m$ as a combination of a 4-digit NACE industry class $i$ and a commuting zone $z$. In terms of industry, the NACE industry classification is designed to group together lines of commerce with related operative tasks and, therefore, similar industry-specific human capital. I use time-consistent information on the 2008 version of the German ``WZ'' Classification of Economic Activities \citep{Destatis2008}, which derives its four leading digits from the Statistical Classification of Economic Activities in the European Community (NACE). The 4-digit classification entails $I=615$ industry classes, and 91 of these entries relate to (subsets of the) minimum wage sectors. In robustness checks, I employ the broader 3-digit and the narrower 5-digit classification, which differentiate between 272 industry groups and 839 industry subclasses, respectively.

In terms of regions, I employ the graph-theoretical method from \citet{KroppSchwengler2016} to merge 401 administrative districts (3-digit NUTS regions) to more adequate commuting zones with strong interactions within but few connections between zones. The method seeks to maximize modularity which is a normalized measure for the quality of a division of a network into modules. Specifically, the method uses the concept of dominant flows to combine pairs of administrative districts between which commuting shares exceed a certain threshold. Based on commuting patterns from the German Federal Employment Agency for the years 1999-2017, the modularity-maximizing threshold value for the identification of dominant flows amounts to 7.0 percent. This threshold delivers a delineation of $R=51$ commuting zones (see Figure \ref{fig:C1}), raising modularity from a value of 0.63 to 0.85. By means of this functional delineation, the share of commuters between regions shrinks from 38.9 to 9.8 percent.

\afterpage{
\begin{landscape}

\begin{figure}[!ht]
\centering
\caption{Commuting Zones}
\label{fig:C1}

\scalebox{0.85}{
\begin{minipage}[t][5cm][c]{0.225\textwidth}
\begin{flushleft}
\vspace*{10.5cm}
{\scriptsize
1 Hamburg, City \vphantom{/} \\[-0.05cm]
2 Braunschweig, City \vphantom{/} \\[-0.05cm]
3 Goettingen, District \vphantom{/} \\[-0.05cm]
4 Region Hanover \vphantom{/} \\[-0.05cm]
5 Aurich, District \vphantom{/} \\[-0.05cm]
6 Emsland, District \vphantom{/} \\[-0.05cm]
7 Bremen, City \vphantom{/} \\[-0.05cm]
8 Duisburg, City \vphantom{/} \\[-0.05cm]
9 Cologne, City \vphantom{/} \\[-0.05cm]
10 Region Aachen \vphantom{/} \\[-0.05cm]
11 Guetersloh, District \vphantom{/} \\[-0.05cm]
12 Siegen-Wittgenstein, District \vphantom{/} \\[-0.05cm]
13 Frankfurt am Main, City \vphantom{/} \\[-0.05cm]
14 Kassel, City \vphantom{/} \\[-0.05cm]
15 Mayen-Koblenz, District \vphantom{/} \\[-0.05cm]
16 Trier, City \vphantom{/} \\[-0.05cm]
17 Stuttgart, City \vphantom{/} \\[-0.05cm]
18 Karlsruhe, District \vphantom{/} \\[-0.05cm]
19 Rhein-Neckar-Kreis\vphantom{/} \\[-0.05cm]
20 Freiburg im Breisgau, City \vphantom{/} \\[-0.05cm]
21 Ortenaukreis \vphantom{/} \\[-0.05cm]
22 Konstanz, District \vphantom{/} \\[-0.05cm]
23 Loerrach, District \vphantom{/} \\[-0.05cm]
24 Waldshut, District \vphantom{/} \\[-0.05cm]
25 Ulm, City \vphantom{/} \\[-0.05cm]
26 Ravensburg, District \vphantom{/} \\[-0.05cm]
27 Munich, City \vphantom{/} \\[-0.05cm]
28 Regensburg, City \vphantom{/} \\[-0.05cm]
29 Bayreuth, City \vphantom{/} \\[-0.05cm]
30 Coburg, District \vphantom{/} \\[-0.05cm]
31 Hof, District \vphantom{/} \\[-0.05cm]
32 Nuremberg, City \vphantom{/} \\[-0.05cm]
33 Schweinfurt, City \vphantom{/} \\[-0.05cm]
34 Wuerzburg, City \vphantom{/} \\[-0.05cm]
35 Region Saarbruecken \vphantom{/} \\[-0.05cm]
36 Berlin, City \vphantom{/} \\[-0.05cm]
37 Spree-Neisse, District \vphantom{/} \\[-0.05cm]
38 Rostock, City \vphantom{/} \\[-0.05cm]
39 Mecklenburgische Seenplatte, District \vphantom{/} \\[-0.05cm]
40 Vorpommern-Ruegen, District \vphantom{/} \\[-0.05cm]
41 Vorpommern-Greifswald, District \vphantom{/} \\[-0.05cm]
42 Mittelsachsen, District \vphantom{/} \\[-0.05cm]
43 Vogtlandkreis \vphantom{/} \\[-0.05cm]
44 Dresden, City \vphantom{/} \\[-0.05cm]
45 Leipzig, City \vphantom{/} \\[-0.05cm]
46 Halle (Saale), City \vphantom{/} \\[-0.05cm]
47 Magdeburg, City \vphantom{/} \\[-0.05cm]
48 Anhalt-Bitterfeld, District \vphantom{/} \\[-0.05cm]
49 Harz, District \vphantom{/} \\[-0.05cm]
50 Stendal, District \vphantom{/} \\[-0.05cm]
51 Erfurt, City \vphantom{/} \\[-0.05cm]
}

\end{flushleft}
\end{minipage}
}
\begin{minipage}[t][5cm][c]{0.48\textwidth}
\begin{center}
\vspace*{7.9cm}
\includegraphics[page=1,width=0.8\textwidth]{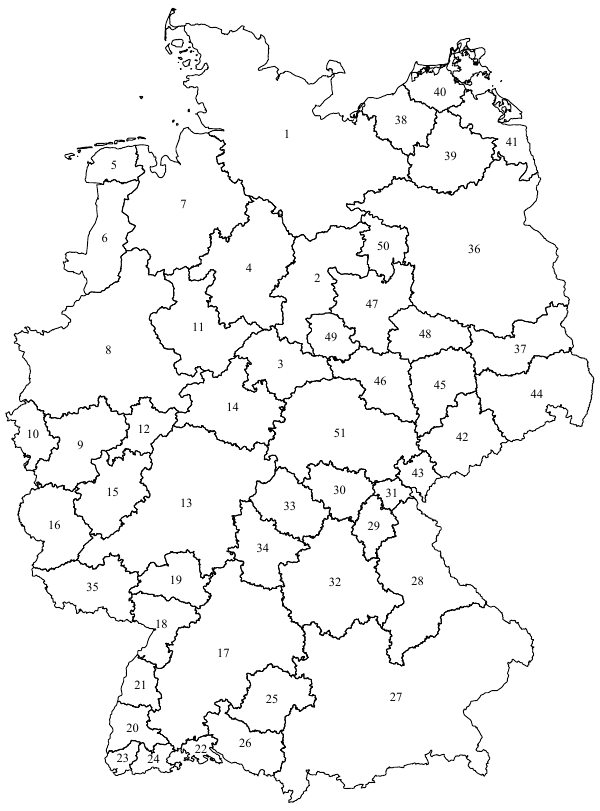}
\end{center}
\end{minipage}

\vspace*{8cm}
\vspace*{-0.3cm}
\floatfoot{\footnotesize\textsc{Note. ---} The figure illustrates the delineation of commuting zones based on 401 German districts (3-digit NUTS regions). On the left-hand side, labels indicate each commuting zone's central district to which peripheral districts were assigned based on the concept of dominant flows. The map on the right-hand side illustrates the delineation of 51 commuting zones using the graph-theoretical method from \citet{KroppSchwengler2016} and register data on German commuting patterns between 1999 and 2017. NUTS = Statistical Nomenclature of Territorial Units in the European Community. Source: Official Statistics of the German Federal Employment Agency, 1999-2017.}
\end{figure}

\end{landscape}
}

\paragraph{Baseline Concentration.} Table \ref{tab:D1} displays descriptive statistics for labor market concentration in Germany. Given its baseline version for employment in pairs of 4-digit NACE industries and commuting zones, the average HHI equals 0.342 which, by taking the reciprocal, is equivalent to 2.9 equally-sized firms in the labor market. At the 25th percentile, the median, and the 75th percentile, the equivalent number of competitors is 14.5 (HHI=0.069), 4.6 (HHI=0.217), and 1.9 (HHI=0.526). A large fraction of labor markets features near-zero HHI values, resembling the notion of atomistic labor markets. For higher HHI values, the density becomes increasingly smaller. However, a spike appears at the upper end of the distribution, indicating a considerable portion of labor markets with a single employer (11.7 percent). Overall, 15.8 of the labor markets feature medium levels of concentration, operationalized by HHI values within the thresholds of 0.10 and 0.20 from E.U.\ antitrust policy. Likewise, 51.8 percent of the labor markets exceed the value of 0.20, mirroring high levels of concentration.

\afterpage{
\begin{landscape}
\begin{table}[ht]
\centering
\begin{threeparttable}
\caption{Descriptive Statistics for Labor Market Concentration}
\label{tab:D1}
\begin{tabular}{L{7cm}C{1.3cm}C{1.3cm}C{1.3cm}C{1.3cm}C{1.3cm}C{1.3cm}C{1.5cm}C{1.5cm}C{2.2cm}} \hline
\multirow{3.4}{*}{}  & \multirow{3.4}{*}{Mean} & \multirow{3.4}{*}{P25} & \multirow{3.4}{*}{P50} & \multirow{3.4}{*}{P75} &  \multirow{3.4}{*}{\shortstack{Mini-\vphantom{/} \\ mum\vphantom{/} }} &  \multirow{3.4}{*}{\shortstack{Maxi-\vphantom{/} \\ mum\vphantom{/} }} &  \multirow{3.4}{*}{\shortstack{Share\vphantom{/} \\ (0.1-0.2)\vphantom{/}}} & \multirow{3.4}{*}{\shortstack{Share\vphantom{/} \\ (0.2-1.0)\vphantom{/} }} & \multirow{3.4}{*}{\shortstack{\!\!Obser- \\ \!\!vations}} \\
&   &   &   &   &   &    &   &   & \\
&   &   &   &   &   &    &   &   & \\ \hline
&   &   &   &   &   &    &   &   & \\[-0.4cm]

\textbf{Baseline:}  &   &   &   &   &   &    &   &   & \\

\multirow{1}{*}{\shortstack{HHI, NACE-4 $\times$ CZ, Employment}}   & \multirow{1}{*}{0.342}   & \multirow{1}{*}{0.069} & \multirow{1}{*}{0.217}  & \multirow{1}{*}{0.526}  & \multirow{1}{*}{0.000}  & \multirow{1}{*}{1.000} & \multirow{1}{*}{0.158}  & \multirow{1}{*}{0.518}  &  \multirow{1}{*}{~476,768}   \\[0.2cm]

\textbf{With Weights:}  &   &   &   &   &   &    &   &   & \\

\multirow{1}{*}{Worker-Weighted}   & \multirow{1}{*}{0.090}   & \multirow{1}{*}{0.006} & \multirow{1}{*}{0.022}  & \multirow{1}{*}{0.081}  & \multirow{1}{*}{0.000}  & \multirow{1}{*}{1.000} & \multirow{1}{*}{0.087}  & \multirow{1}{*}{0.129}  &  \multirow{1}{*}{638,048,138}    \\

\multirow{1}{*}{Firm-Weighted}   & \multirow{1}{*}{0.033}   & \multirow{1}{*}{0.002} & \multirow{1}{*}{0.008}  & \multirow{1}{*}{0.027}  & \multirow{1}{*}{0.000}  & \multirow{1}{*}{1.000} & \multirow{1}{*}{0.044}  & \multirow{1}{*}{0.034}  &  \multirow{1}{*}{~52,911,664}    \\[0.2cm]

\textbf{Alternative Industrial Definition:}  &  &   &   &   &   &    &   &   & \\

\multirow{1}{*}{NACE-3 Industries}   & \multirow{1}{*}{0.264}   & \multirow{1}{*}{0.039} & \multirow{1}{*}{0.133}  & \multirow{1}{*}{0.382}  & \multirow{1}{*}{0.000}  & \multirow{1}{*}{1.000} & \multirow{1}{*}{0.160}  & \multirow{1}{*}{0.402}  &  \multirow{1}{*}{228,885}    \\

\multirow{1}{*}{NACE-5 Industries}  & \multirow{1}{*}{0.357}   & \multirow{1}{*}{0.081} & \multirow{1}{*}{0.236}  & \multirow{1}{*}{0.551}  & \multirow{1}{*}{0.000}  & \multirow{1}{*}{1.000} & \multirow{1}{*}{0.162}  & \multirow{1}{*}{0.544}  &  \multirow{1}{*}{647,974}    \\

\multirow{1}{*}{Flow-Adjusted NACE-4 Industries}  & \multirow{1}{*}{0.168}   & \multirow{1}{*}{0.019} & \multirow{1}{*}{0.075}  & \multirow{1}{*}{0.226}  & \multirow{1}{*}{0.000}  & \multirow{1}{*}{1.000} & \multirow{1}{*}{0.161}  & \multirow{1}{*}{0.279}  &  \multirow{1}{*}{~593,042}    \\[0.2cm]

\textbf{Alternative Spatial Definition:}  &  &   &   &   &   &    &   &   & \\

\multirow{1}{*}{NUTS-3 Regions}  & \multirow{1}{*}{0.491}   & \multirow{1}{*}{0.169} & \multirow{1}{*}{0.414}  & \multirow{1}{*}{0.901}  & \multirow{1}{*}{0.000}  & \multirow{1}{*}{1.000} & \multirow{1}{*}{0.134}  & \multirow{1}{*}{0.706}  &  \multirow{1}{*}{~2,774,477}    \\[0.2cm]

\textbf{Alternative Object:}  &  &   &   &   &   &    &   &   & \\

\multirow{1}{*}{Hires}  & \multirow{1}{*}{0.351}   & \multirow{1}{*}{0.075} & \multirow{1}{*}{0.225}  & \multirow{1}{*}{0.538}  & \multirow{1}{*}{0.000}  & \multirow{1}{*}{1.000} & \multirow{1}{*}{0.160}  & \multirow{1}{*}{0.530}  &  \multirow{1}{*}{426,400}    \\

\multirow{1}{*}{Wage Bill}  & \multirow{1}{*}{0.379}   & \multirow{1}{*}{0.093} & \multirow{1}{*}{0.265}  & \multirow{1}{*}{0.597}  & \multirow{1}{*}{0.000}  & \multirow{1}{*}{1.000} & \multirow{1}{*}{0.157}  & \multirow{1}{*}{0.573}  &  \multirow{1}{*}{~476,767}    \\[0.2cm]

\textbf{Alternative Concentration Index:}  &  &   &   &   &   &    &   &   & \\

\multirow{1}{*}{Rosenbluth Index}  & \multirow{1}{*}{0.325}   & \multirow{1}{*}{0.048} & \multirow{1}{*}{0.187}  & \multirow{1}{*}{0.517}  & \multirow{1}{*}{0.000}  & \multirow{1}{*}{1.000} & \multirow{1}{*}{0.141}  & \multirow{1}{*}{0.483}  &  \multirow{1}{*}{476,768}    \\

\multirow{1}{*}{1-Firm Concentration Ratio}   & \multirow{1}{*}{0.437}   & \multirow{1}{*}{0.164} & \multirow{1}{*}{0.359}  & \multirow{1}{*}{0.667}  & \multirow{1}{*}{0.000}  & \multirow{1}{*}{1.000} & \multirow{1}{*}{0.156}  & \multirow{1}{*}{0.693}  &  \multirow{1}{*}{476,768}    \\

\multirow{1}{*}{Inverse Number of Firms}  & \multirow{1}{*}{0.225}   & \multirow{1}{*}{0.018} & \multirow{1}{*}{0.077}  & \multirow{1}{*}{0.250}  & \multirow{1}{*}{0.000}  & \multirow{1}{*}{1.000} & \multirow{1}{*}{0.123}  & \multirow{1}{*}{0.332}  &  \multirow{1}{*}{476,768}   \\

\multirow{1}{*}{Exponential Index}  & \multirow{1}{*}{0.295}   & \multirow{1}{*}{0.040} & \multirow{1}{*}{0.153}  & \multirow{1}{*}{0.452}  & \multirow{1}{*}{0.000}  & \multirow{1}{*}{1.000} & \multirow{1}{*}{0.150}  & \multirow{1}{*}{0.436}  &  \multirow{1}{*}{476,768}    \\

\multirow{1}{*}{Market Share}  & \multirow{1}{*}{0.009}   & \multirow{1}{*}{0.000} & \multirow{1}{*}{0.001}  & \multirow{1}{*}{0.002}  & \multirow{1}{*}{0.000}  & \multirow{1}{*}{1.000} & \multirow{1}{*}{0.008}  & \multirow{1}{*}{0.009}  &  \multirow{1}{*}{~52,911,664}    \\[0.2cm]

\textbf{Subsample:}  &  &   &   &   &   &    &   &   & \\

\multirow{1}{*}{Only Minimum Wage Sectors}  & \multirow{1}{*}{0.314}   & \multirow{1}{*}{0.055} & \multirow{1}{*}{0.181}  & \multirow{1}{*}{0.498}  & \multirow{1}{*}{0.000}  & \multirow{1}{*}{1.000} & \multirow{1}{*}{0.161}  & \multirow{1}{*}{0.474}  &  \multirow{1}{*}{ 65,772}  \\ \hline

\end{tabular}
\begin{tablenotes}[para]
\footnotesize\textsc{Note. ---} The table displays descriptive statistics for labor market concentration in Germany. Labor markets refer to pair-wise combinations of industries and commuting zones, and are tracked with annual frequency. CZ = Commuting Zone. HHI = Herfindahl-Hirschman Index. NACE-X = X-Digit Statistical Nomenclature of Economic Activities in the European Community. NUTS-X = X-Digit Statistical Nomenclature of Territorial Units. PX = Xth Percentile. Source: IEB, 1999-2017.
\end{tablenotes}
\end{threeparttable}
\end{table}
\end{landscape}
}

\paragraph{Unit of Observation.} Figure \ref{fig:D2} depicts the cumulative distribution of labor market concentration by unit of observation. Unweighted HHI values obscure the share of workers and firms that engage in labor markets with high concentration. When weighting the labor market HHI values by the number of workers, the average HHI reduces to 0.090, implying that the average worker engages in a labor market with 11.1 equally-sized employers. In contrast, the average firm encounters an equivalent of 30.3 competitors (HHI=0.033). Overall, 12.9 percent of workers face high levels of labor market concentration, indicating that labor markets with fewer workers tend to be the more concentrated markets. 3.4 percent of firms operate in highly concentrated labor markets. This relatively small share stems from the fact that, by construction, low-HHI markets feature a large number of firms.

\paragraph{Alternative Labor Market Definitions.} In a next step, I examine the sensitivity of the baseline concentration measure to different labor market definitions. When adopting the broader 3-digit NACE classification, HHI values become somewhat smaller (HHI=0.264). Given this definition, 40.2 percent of all markets feature a high level of concentration. For the narrower 5-digit NACE classification, the HHI values become slightly larger (HHI=0.357), and the fraction of highly concentrated markets equals 54.4 percent. In general, the flow-adjusted HHI (based on 4-digit industries), which treats jobs in related industries as additional outside options, delivers lower levels of concentration than the standard HHI (see Appendix \ref{sec:E}). Specifically, the average flow-adjusted HHI turns out only half as large (HHI=0.168) as the average standard HHI. Despite the flow adjustment, 27.9 percent of labor markets show up high levels of concentration. I also construct HHI values for a spatial division into 401 administrative districts (3-digit NUTS regions), which the delineation of commuting zones is based on. District-level HHI values turn out considerably lower (HHI=0.491), and 70.6 percent of markets are highly concentrated markets.

\paragraph{Alternative Objects and Subjects.} The use of measures of employment concentration can be justified by the Cournot oligopsony model, in which the average employment-weighted markdown is a function of the employment-based HHI in the market (e.g., \citealp{Arnold2021}; \citealp{BenmelechEtAl2022}; \citealp{DodiniEtAl2024}). Nevertheless, parts of the literature rely on concentration of new hires (e.g., \citealp{MarinescuEtAl2021}; \citealp{BassaniniEtAl2024}) or vacancies (e.g., \citealp{AzarEtAl2022}; \citealp{SchubertEtAl2022}), arguing that these measures more accurately capture the availability of job opportunities than the mere stock of employment. While micro-level data on the (near-)universe of vacancies is not available for the German labor market, I can calculate the concentration of new hires using the IEB data. Specifically, I define hires on the basis of inflows, namely the number of workers who work in a firm on June 30 of the respective year but were not employed in the same firm in the year before. As marginal employment is recorded in the IEB only from 1999 onwards, the year 1999 does not permit differentiating between newly hired and incumbent marginal workers. Therefore, I can only meaningfully calculate the concentration of new hires for the years 2000-2017. Specifically, new hires turn out only slightly more concentrated (HHI=0.351) than employment.

In a representative agent framework with differentiated jobs, \citet{BergerEtAl2022} demonstrate that markdowns positively depend on the wage bill HHI. Using the IEB data, I also construct measures of wage bill concentration (HHI=0.379).

The prevalence of companies with more than a single firm leads to an underestimation of labor market concentration to the extent that firms within the same company do not compete for workers \citep{MarinescuEtAl2021}. According to the IAB Establishment Panel, the bias from multi-firm companies is seemingly small because 72.7 percent of surveyed firms constitute single-firm companies.

\paragraph{Alternative Concentration Measures.} The Herfindahl-Hirschman Index derives popularity from the fact that the measure reflects the two determinants of absolute concentration -- the fewness and unevenness dimension of competitors -- in a simple fashion. Specifically, the HHI is the arithmetic mean of market shares, whereby these shares are weighted with themselves. The squaring of market shares assigns relatively high weights to large firms in the market \citep{CurryGeorge1983}. To test the sensitivity, I calculate five alternative measures that emphasize different aspects of the absolute concentration of workers on employers: the Rosenbluth Index (RBI), the K-Firm Concentration Ratio (CRK), the Inverse Number of Firms (INF), the Exponential Index (EXP), and the underlying market shares. In line with the HHI, their values lie in the unit interval. The formulas for the alternative concentration indices, where $j$ denotes the rank of firms in descending order of market shares, look as follows: 
\begin{itemize}
\item Rosenbluth Index (RBI):\, $ RB\!I_{mt} =  1 \, / \, (  2\,\sum_{j=1}^{J} s_{jmt} \, j - 1 ) $
\item K-Firm Concentration Ratio (CRK):\, $ C\!RK_{mt} =  \sum_{j=1}^{K} s_{jmt} $
\item Inverse Number of Firms (INF):\, $ I\!N\!F_{mt} =  1/J $
\item Exponential Index (EXP):\, $ E\!X\!P_{mt} = \prod_{j=1}^{J} s_{jmt}^{s_{jmt}}$
\item Market Share:\, $s_{jmt}$
\end{itemize}
The Rosenbluth Index employs ranks of firms (in descending order of market shares) as weights, thus attributing less weight to larger firms \citep{Rosenbluth1955,HallTideman1967}. The average Rosenbluth Index falls slightly short of the corresponding HHI value (RBI=0.325). The K-Firm Concentration Ratio sums up the shares of the $K$ largest competitors in the market, applying unit weights to a fixed set of market shares. On average, the largest firm accounts for 43.7 percent of employment in the labor market (CR1=0.437). The Inverse Number of Firms measures is the reciprocal the number of competitors in the market, and 4.4 firms operate in the average labor market (INF=0.225). However, the INF distribution is highly right-skewed with a median of only 0.077, which is equivalent to 13 firms. The Exponential Index is the geometric mean of market shares weighted with themselves and turns out somewhat lower in general (EXP=0.295). Finally, at the level of firms, the market shares average 0.9 percent. The average turns out relatively low because, by construction, slightly concentrated labor markets with many competitors receive larger weights.

\paragraph{Within- and Between-Variation.} Figure \ref{fig:D3} visualizes the development of labor market concentration in Germany over time. The average HHI remained relatively stable during the period of analysis. Between 1999 and 2017, there was a slight decrease in the average market-level HHI from 0.347 to 0.338. The alternative indices experience a similar trend.

In contrast, concentration indices exhibit a markedly stronger variation between than within labor markets. Figure \ref{fig:D4} uses boxplots to visualize the distribution of labor market concentration by 4-digit NACE industries in the German labor market. Plausibly, widespread lines of business (such as restaurants, medical practices, retail trade or legal activities) constitute the least concentrated industries. In particular, I report relatively low values of labor market concentration for industries that relate to the following minimum wage sectors: hairdressing, roofing, painting and varnishing, and electrical trade. Vice versa, highly specialized industries tend to show a greater degree of concentration, such as industries from the minimum wage sectors of waste removal, textile and clothing, and agriculture, forestry and gardening. By and large, labor market concentration in the 91 industries that belong to the minimum wage sectors (HHI=0.314) does not markedly differ from the universe of 615 4-digit industries.

\begin{figure}[!ht]
\centering
\caption{Labor Market Concentration by Commuting Zone}
\label{fig:D5}
\scalebox{0.80}{
\includegraphics{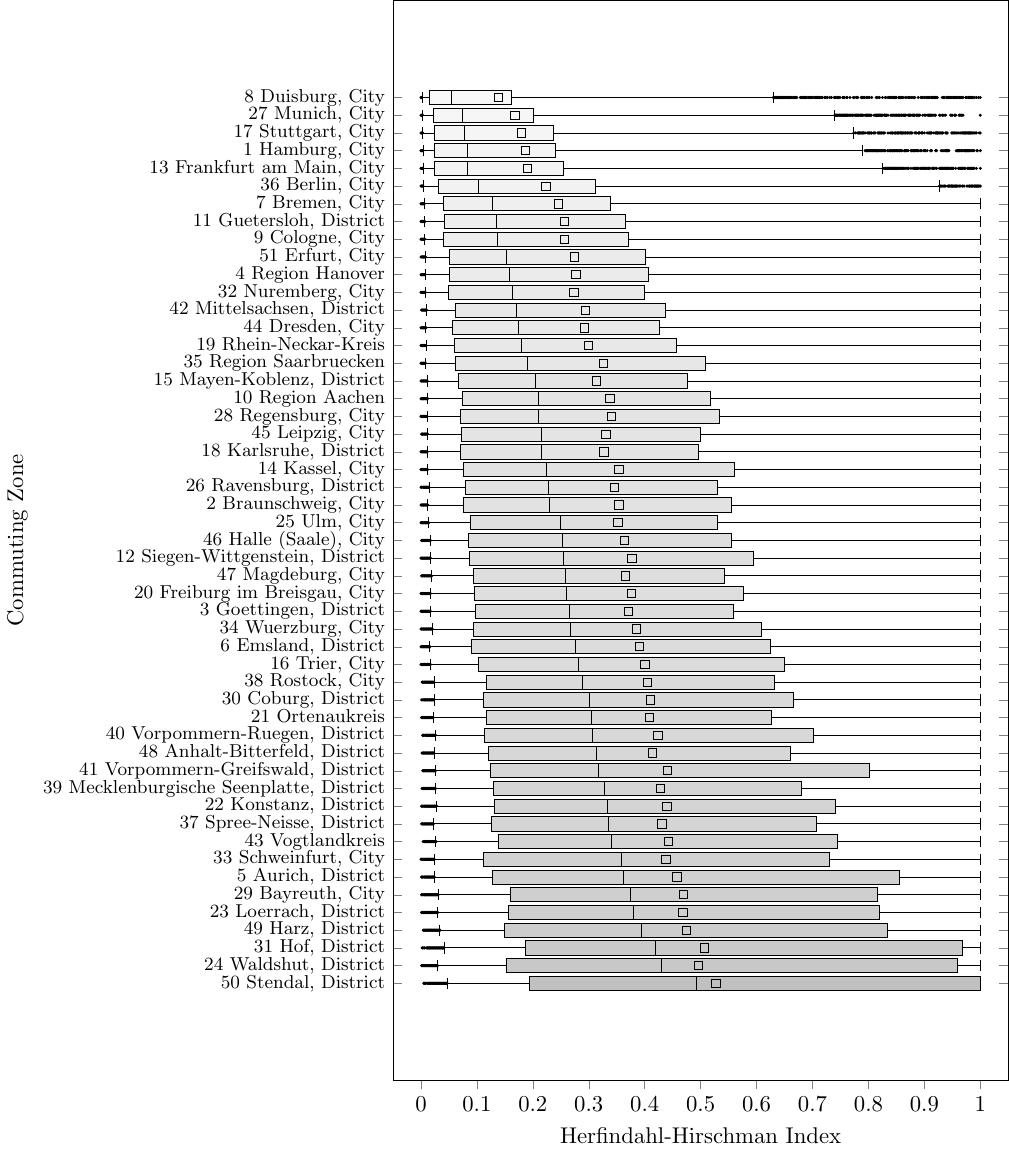}
}
\floatfoot{ \scriptsize \textsc{Note. ---} The figure visualizes the distribution of labor market concentration by commuting zones in Germany. Labor market concentration refers to employment-based HHIs for pair-wise combinations of 4-digit NACE industries and commuting zones, and is tracked with annual frequency. In each of the boxplots, the center marks the median of the HHI distribution whereas left and right margins represent the 25th and 75th percentile. Lower and upper whiskers indicate the 5th and 95th percentile, respectively. Hollow squares illustrate the underlying means. Dots represent outliers (i.e., values below the 5th or above the 95th percentile). HHI = Herfindahl-Hirschman Index. NACE = Statistical Nomenclature of Economic Activities in the European Community. Source: IEB, 1999-2017.}
\end{figure}

\begin{figure}[!ht]
\centering
\caption{Labor Market Concentration by 3-Digit NUTS Regions}
\label{fig:D6}
\scalebox{1}{

\includegraphics[page=1,width=.95\textwidth]{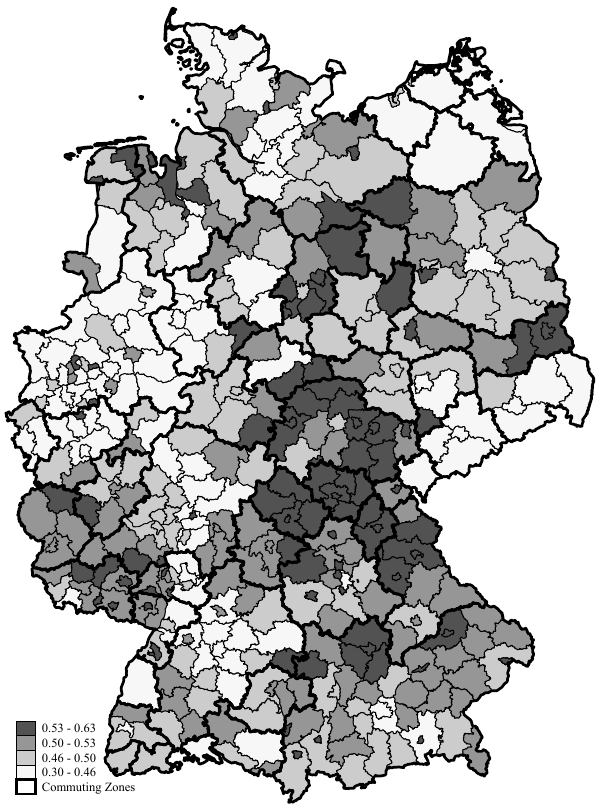}
}
\floatfoot{\footnotesize\textsc{Note. ---} The map displays average labor market concentration by 3-digit NUTS regions in Germany. Labor market concentration refers to employment-based HHIs for pair-wise combinations of 4-digit NACE industries and commuting zones, and is tracked with annual frequency. HHI = Herfindahl-Hirschman Index. NACE = Statistical Nomenclature of Economic Activities in the European Community. NUTS-3 = 3-Digit Statistical Nomenclature of Territorial Units in the European Community. Source: IEB, 1999-2017.}
\end{figure}

Figure \ref{fig:D5} uses boxplots to visualize the distribution of labor market concentration by commuting zones. The map in Figure \ref{fig:D6} displays average labor market concentration by the underlying 3-digit NUTS regions. Both figures highlight that there is also a strong heterogeneity of HHI values across commuting zones, with labor markets generally being more concentrated in rural than in urban regions.

\clearpage

\vspace*{\fill}

\begin{figure}[!ht]
\centering
\caption{Cumulative Distribution of Labor Market Concentration by Unit}
\label{fig:D2}
\scalebox{0.80}{
\begin{tikzpicture}
\begin{axis}[ytick={0,0.1,0.2,0.3,0.4,0.5,0.6,0.7,0.8,0.9,1}, height=14cm, width=14cm, grid=major, grid style = dotted, xtick={0,0.2,0.4,0.6,0.8,1}, xlabel=Herfindahl-Hirschman Index,ylabel=Cumulative Distribution, xmin=-0.05, xmax=1.05, ymin = -0.05, ymax=1.05, legend pos = south east, y tick label style={/pgf/number format/.cd,fixed,fixed zerofill, precision=1,/tikz/.cd},x tick label style={/pgf/number format/.cd,fixed,fixed zerofill, precision=1,/tikz/.cd}]

\addplot[mark=none, solid, color=black, line width=1mm] coordinates { (0,0)  (.0041,.37) (.0061,.453) (.0063,.461) (.0089,.532) (.012,.592) (.014,.627) (.015,.642) (.017,.672) (.019,.693) (.021,.704) (.024,.728) (.026,.744) (.029,.765) (.029,.765) (.031,.779) (.032,.782) (.032,.785) (.034,.791) (.034,.794) (.035,.797) (.035,.8) (.04,.82) (.042,.825) (.043,.828) (.059,.869) (.059,.87) (.066,.882) (.071,.89) (.071,.89) (.073,.892) (.083,.905) (.086,.909) (.086,.909) (.097,.919) (.099,.921) (.104,.925) (.108,.928) (.11,.929) (.111,.93) (.116,.933) (.123,.937) (.125,.938) (.127,.94) (.128,.94) (.14,.946) (.141,.946) (.141,.946) (.142,.947) (.156,.953) (.158,.953) (.161,.954) (.186,.962) (.19,.963) (.203,.966) (.203,.966) (.203,.966) (.223,.971) (.225,.971) (.235,.973) (.239,.973) (.284,.979) (.303,.981) (.311,.982) (.316,.983) (.356,.986) (.374,.987) (.414,.989) (.438,.99) (.44,.99) (.496,.992) (.497,.992) (.51,.993) (.537,.994) (.603,.995) (.643,.996) (.645,.996) (.682,.997) (.686,.997) (.733,.997) (.753,.997) (.802,.998) (.811,.998) (.945,.999) (.974,.999) (1,.999) (1,1) (1,.999) (1,.999) (1,.999) (1,.999) (1,1) (1,1) (1,.999) (1,.999) (1,.999) (1,1)};

\addplot[mark=none, dash pattern = on 7.6pt off 2pt, color=lightgray, line width=1mm] coordinates { (0,0)  (.0041,.173) (.0061,.245) (.0063,.254) (.0089,.314) (.012,.368) (.014,.4) (.015,.416) (.017,.447) (.019,.473) (.021,.485) (.024,.515) (.026,.531) (.029,.557) (.029,.557) (.031,.575) (.032,.58) (.032,.583) (.034,.591) (.034,.594) (.035,.599) (.035,.603) (.04,.629) (.042,.636) (.043,.64) (.059,.697) (.059,.698) (.066,.717) (.071,.729) (.071,.729) (.073,.733) (.083,.754) (.086,.76) (.086,.76) (.097,.778) (.099,.781) (.104,.789) (.108,.795) (.11,.797) (.111,.798) (.116,.804) (.123,.811) (.125,.813) (.127,.816) (.128,.816) (.14,.828) (.141,.829) (.141,.829) (.142,.83) (.156,.842) (.158,.843) (.161,.846) (.186,.863) (.19,.865) (.203,.872) (.203,.872) (.203,.872) (.223,.883) (.225,.884) (.235,.888) (.239,.89) (.284,.908) (.303,.916) (.311,.919) (.316,.92) (.356,.93) (.374,.934) (.414,.942) (.438,.946) (.44,.946) (.496,.955) (.497,.955) (.51,.957) (.537,.961) (.603,.968) (.643,.973) (.645,.973) (.682,.976) (.686,.977) (.733,.979) (.753,.981) (.802,.984) (.811,.984) (.945,.992) (.974,.994) (1,.999) (1,.998) (1,1) (1,.997) (1,.997) (1,.999) (1,.997) (1,.998) (1,.997) (1,.998) (1,.999) (1,1)};

\addplot[mark=none, dash pattern = on 11.4pt off 2pt, color=darkgray, line width=1mm] coordinates { (0,0)  (.0041,.018) (.0061,.028) (.0063,.029) (.0089,.042) (.012,.057) (.014,.067) (.015,.072) (.017,.082) (.019,.09) (.021,.095) (.024,.107) (.026,.115) (.029,.128) (.029,.129) (.031,.138) (.032,.14) (.032,.142) (.034,.146) (.034,.148) (.035,.151) (.035,.153) (.04,.17) (.042,.174) (.043,.177) (.059,.224) (.059,.225) (.066,.243) (.071,.255) (.071,.256) (.073,.26) (.083,.284) (.086,.291) (.086,.291) (.097,.313) (.099,.318) (.104,.327) (.108,.337) (.11,.34) (.111,.342) (.116,.351) (.123,.363) (.125,.366) (.127,.371) (.128,.371) (.14,.392) (.141,.394) (.141,.394) (.142,.396) (.156,.418) (.158,.42) (.161,.426) (.186,.461) (.19,.466) (.203,.483) (.203,.483) (.203,.483) (.223,.508) (.225,.51) (.235,.521) (.239,.525) (.284,.573) (.303,.59) (.311,.597) (.316,.601) (.356,.635) (.374,.649) (.414,.678) (.438,.691) (.44,.693) (.496,.721) (.497,.721) (.51,.739) (.537,.756) (.603,.788) (.643,.802) (.645,.803) (.682,.815) (.686,.816) (.733,.828) (.753,.832) (.802,.844) (.811,.846) (.945,.873) (.974,.879) (1,.894) (1,.896) (1,.912) (1,.936) (1,.94) (1,.956) (1,.966) (1,.967) (1,.983) (1,.988) (1,.989) (1,1)};

\legend{~Firms,~Workers,~Labor Markets}
\end{axis}
\end{tikzpicture}
}
\floatfoot{\footnotesize\textsc{Note. ---} The figure illustrates cumulative distribution functions of labor market concentration in Germany for three different units of observation: firms, workers, and labor markets. Labor market concentration refers to employment-based HHIs for pair-wise combinations of 4-digit NACE industries and commuting zones, and is tracked with annual frequency. The cumulative distribution functions of firms and workers are generated on the basis of respective frequency weights. HHI = Herfindahl-Hirschman Index. NACE = Statistical Nomenclature of Economic Activities in the European Community. Source: IEB, 1999-2017.}
\end{figure}

\vspace*{\fill}
\clearpage
\vspace*{\fill}

\begin{figure}[!ht]
\centering
\caption{Trend in Labor Market Concentration by Concentration Measure}
\label{fig:D3}
\scalebox{0.80}{
\begin{tikzpicture}
\begin{axis}[ytick={0.2,0.3,0.4,0.5,0.6}, height=14cm, width=14cm, grid=major, grid style = dotted, xtick={2000,2005,2010,2015}, xlabel=Year, ylabel=Average Labor Market Concentration, ymin=0.175, ymax=0.625, xmin=1998, xmax=2018, legend pos = north east, y tick label style={/pgf/number format/.cd,fixed,fixed zerofill, precision=1,/tikz/.cd}]

\addplot[mark=none, solid, color=black, line width=1mm] coordinates { (1999,.44199985) (2000,.44107428) (2001,.44248751) (2002,.44290555) (2003,.44248438) (2004,.4393515) (2005,.43923345) (2006,.43783769) (2007,.43961072) (2008,.43818289) (2009,.43567145) (2010,.43556038) (2011,.4343293) (2012,.43393832) (2013,.43369806) (2014,.43286532) (2015,.43322325) (2016,.43345851) (2017,.43419126) };

\addplot[mark=none, dash pattern = on 7.6pt off 2pt, color=lightgray, line width=1mm] coordinates { (1999,.34692895) (2000,.34619635) (2001,.34754792) (2002,.34764645) (2003,.34689844) (2004,.34425455) (2005,.34407568) (2006,.34256384) (2007,.34418085) (2008,.34237429) (2009,.3392415) (2010,.33927205) (2011,.3381308) (2012,.33807713) (2013,.33773673) (2014,.33728316) (2015,.33761665) (2016,.33752686) (2017,.33815241) };

\addplot[mark=none, dash pattern = on 11.4pt off 2pt, color=darkgray, line width=1mm] coordinates { (1999,.32981464) (2000,.32935843) (2001,.33077019) (2002,.33072925) (2003,.33005968) (2004,.32742405) (2005,.32720327) (2006,.32584572) (2007,.32718578) (2008,.32502899) (2009,.32172766) (2010,.32193482) (2011,.32106307) (2012,.32145905) (2013,.32101721) (2014,.3210429) (2015,.32121494) (2016,.32095286) (2017,.32179528) };

\addplot[mark=none, dash pattern = on 4.4pt off 2pt, color=gray, line width=1mm] coordinates { (1999,.29962295) (2000,.2990967) (2001,.30031231) (2002,.30015108) (2003,.2993955) (2004,.2970908) (2005,.29696983) (2006,.29552904) (2007,.29661399) (2008,.29464909) (2009,.29124704) (2010,.2915822) (2011,.29072464) (2012,.29092741) (2013,.29039735) (2014,.29044524) (2015,.29058278) (2016,.29020461) (2017,.29092625) };

\addplot[mark=none, solid, color=black!25!white, line width=1mm] coordinates { (1999,.23043729) (2000,.22991556) (2001,.23043101) (2002,.2304945) (2003,.23013274) (2004,.22862332) (2005,.22873777) (2006,.22740556) (2007,.22710556) (2008,.22530168) (2009,.22174202) (2010,.22222735) (2011,.22157301) (2012,.22194794) (2013,.22110565) (2014,.22167072) (2015,.22184791) (2016,.22136095) (2017,.22219706) };

\legend{~1-Firm Concentration Ratio, Herfindahl-Hirschman Index, Rosenbluth Index, Exponential Index, Inverse Number of Firms}

\end{axis}
\end{tikzpicture}
}
\floatfoot{\footnotesize\textsc{Note. ---} The figure reports means of selected concentration indices over time to visualize the development of labor market concentration in Germany. Labor market concentration refers to employment-based concentration indices for pair-wise combinations of 4-digit NACE industries and commuting zones, and is tracked with annual frequency. NACE = Statistical Nomenclature of Economic Activities in the European Community. Source: IEB, 1999-2017.}
\end{figure}

\vspace*{\fill}

\clearpage


\begin{figure}[!ht]
\centering
\caption{Labor Market Concentration by 4-Digit NACE Industry}
\label{fig:D4}
\scalebox{0.80}{
\includegraphics{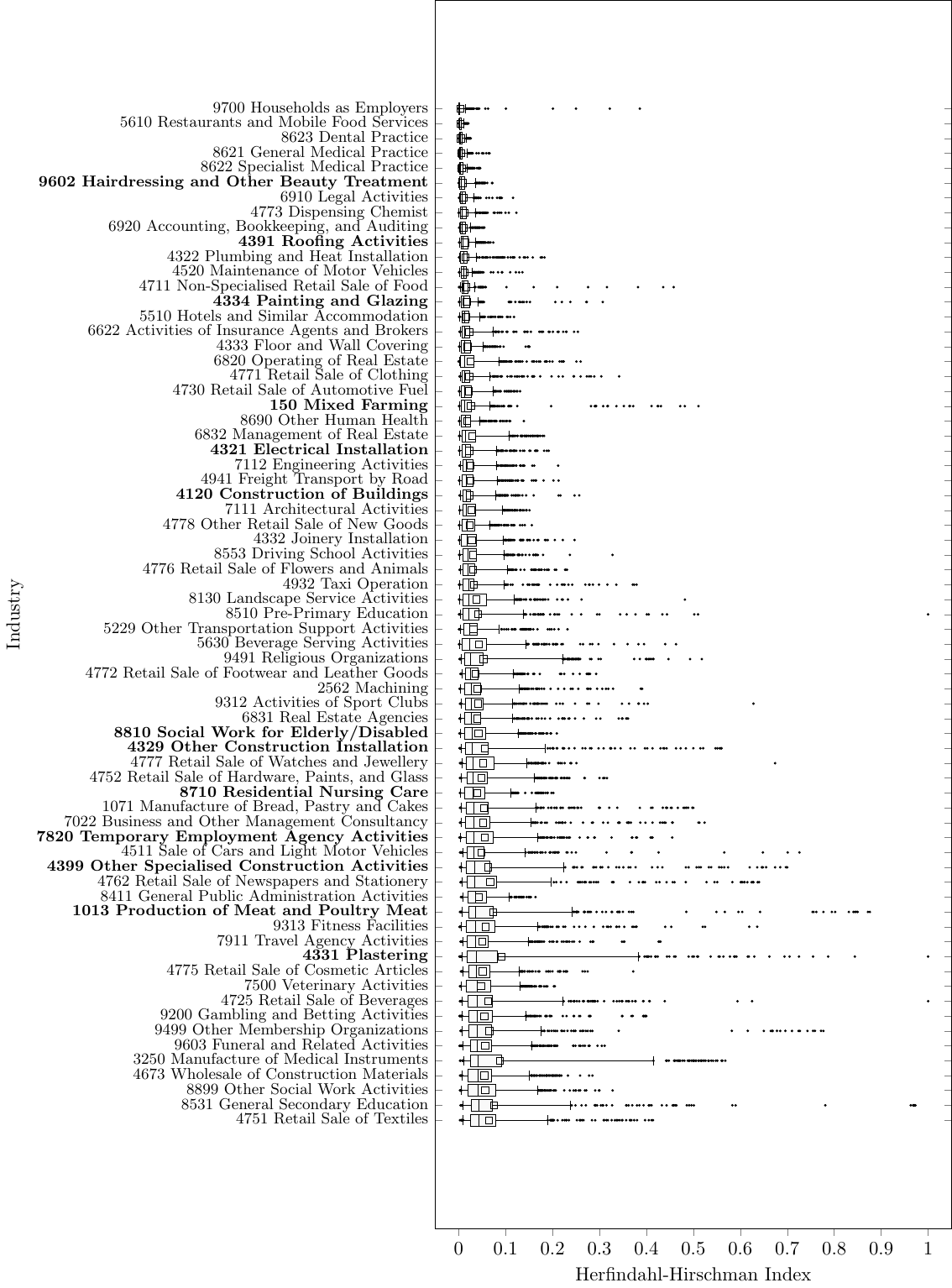}
}
\floatfoot{ \scriptsize \textsc{Note. ---} The figure visualizes the distribution of labor market concentration by 4-digit NACE industries in the Germany. Labor market concentration refers to employment-based HHIs for pair-wise combinations of 4-digit NACE industries and commuting zones, and is tracked with annual frequency. In each of the boxplots, the center marks the median whereas left and right margins represent the 25th and 75th percentile. Lower and upper whiskers indicate the 5th and 95th percentile. Hollow squares illustrate the underlying means. Dots represent outliers (i.e., values below the 5th or above the 95th percentile). Bold industries refer to minimum wage sectors. HHI = Herfindahl-Hirschman Index. NACE = Statistical Nomenclature of Economic Activities in the European Community. Source: IEB, 1999-2017.}
\end{figure}
\begin{figure}[!ht]
\addtocounter{figure}{-1}
\centering
\caption{Labor Market Concentration by 4-Digit NACE Industry (Cont.)}
\scalebox{0.80}{
\includegraphics{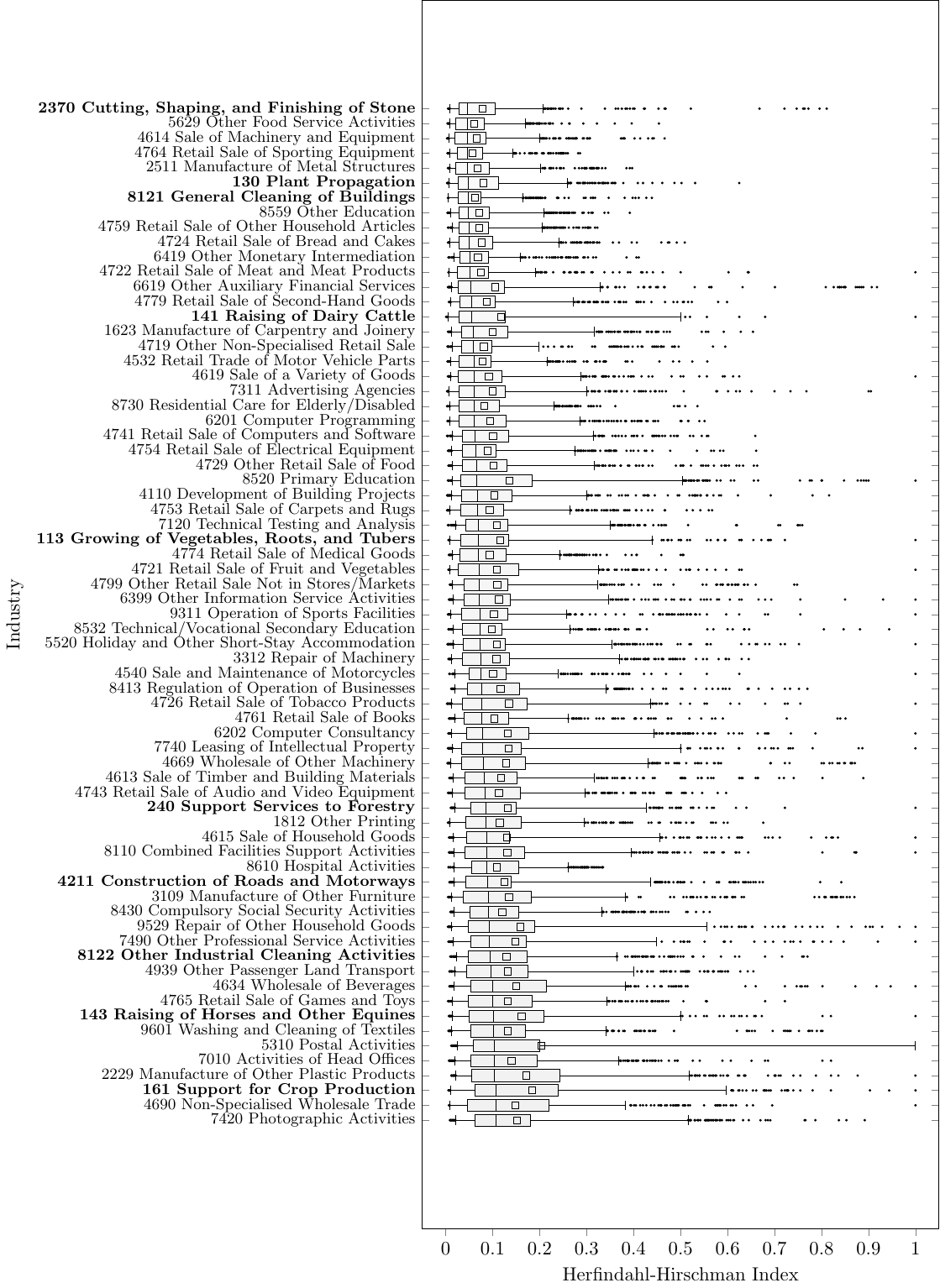}
}
\floatfoot{ \scriptsize \textsc{Note. ---} The figure visualizes the distribution of labor market concentration by 4-digit NACE industries in the Germany. Labor market concentration refers to employment-based HHIs for pair-wise combinations of 4-digit NACE industries and commuting zones, and is tracked with annual frequency. In each of the boxplots, the center marks the median whereas left and right margins represent the 25th and 75th percentile. Lower and upper whiskers indicate the 5th and 95th percentile. Hollow squares illustrate the underlying means. Dots represent outliers (i.e., values below the 5th or above the 95th percentile). Bold industries refer to minimum wage sectors. HHI = Herfindahl-Hirschman Index. NACE = Statistical Nomenclature of Economic Activities in the European Community. Source: IEB, 1999-2017.}
\end{figure}
\begin{figure}[!ht]
\addtocounter{figure}{-1}
\centering
\caption{Labor Market Concentration by 4-Digit NACE Industry (Cont.)}
\scalebox{0.80}{
\includegraphics{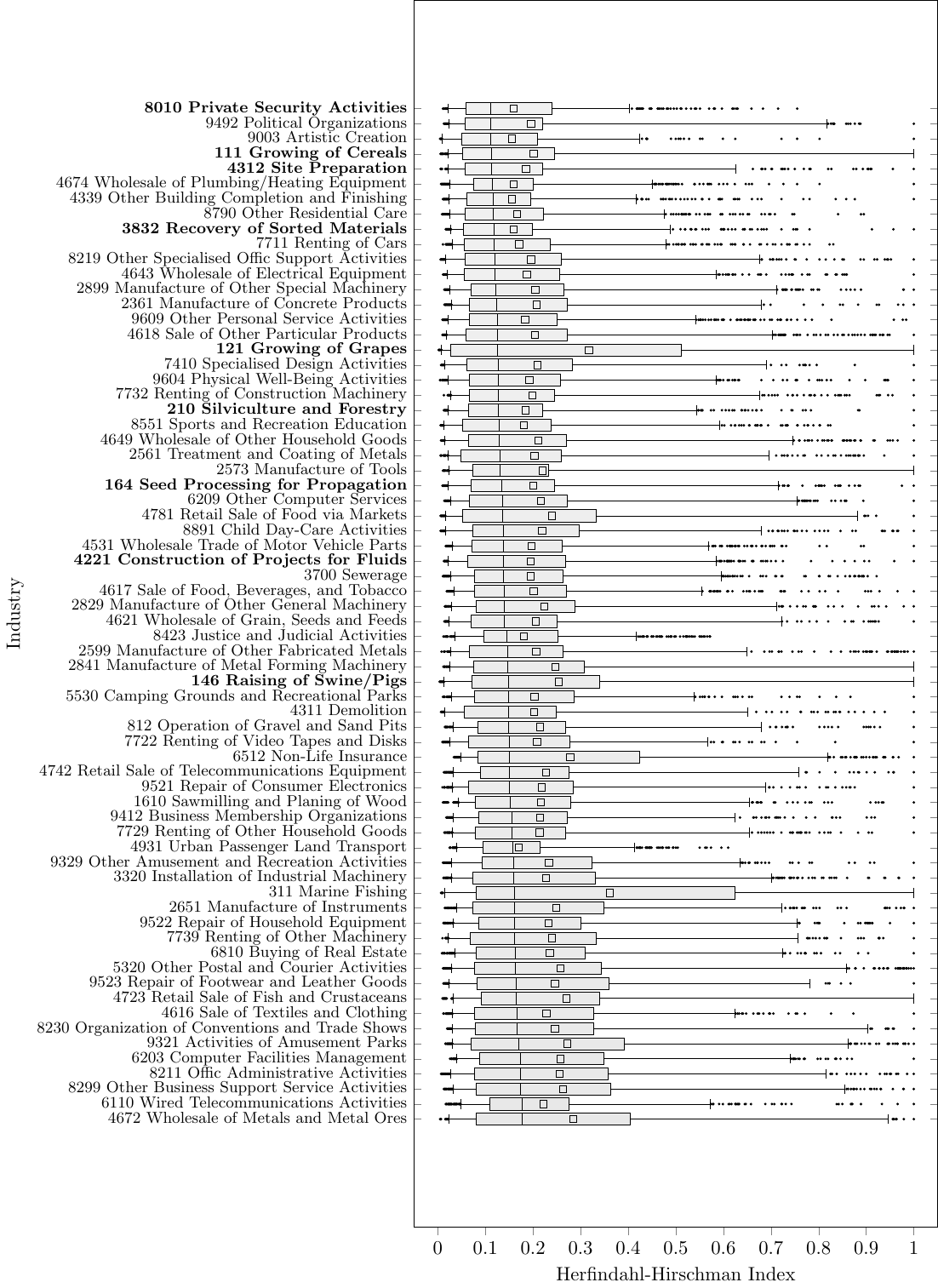}
}
\floatfoot{ \scriptsize \textsc{Note. ---} The figure visualizes the distribution of labor market concentration by 4-digit NACE industries in the Germany. Labor market concentration refers to employment-based HHIs for pair-wise combinations of 4-digit NACE industries and commuting zones, and is tracked with annual frequency. In each of the boxplots, the center marks the median whereas left and right margins represent the 25th and 75th percentile. Lower and upper whiskers indicate the 5th and 95th percentile. Hollow squares illustrate the underlying means. Dots represent outliers (i.e., values below the 5th or above the 95th percentile). Bold industries refer to minimum wage sectors. HHI = Herfindahl-Hirschman Index. NACE = Statistical Nomenclature of Economic Activities in the European Community. Source: IEB, 1999-2017.}
\end{figure}
\begin{figure}[!ht]
\addtocounter{figure}{-1}
\centering
\caption{Labor Market Concentration by 4-Digit NACE Industry (Cont.)}
\scalebox{0.80}{
\includegraphics{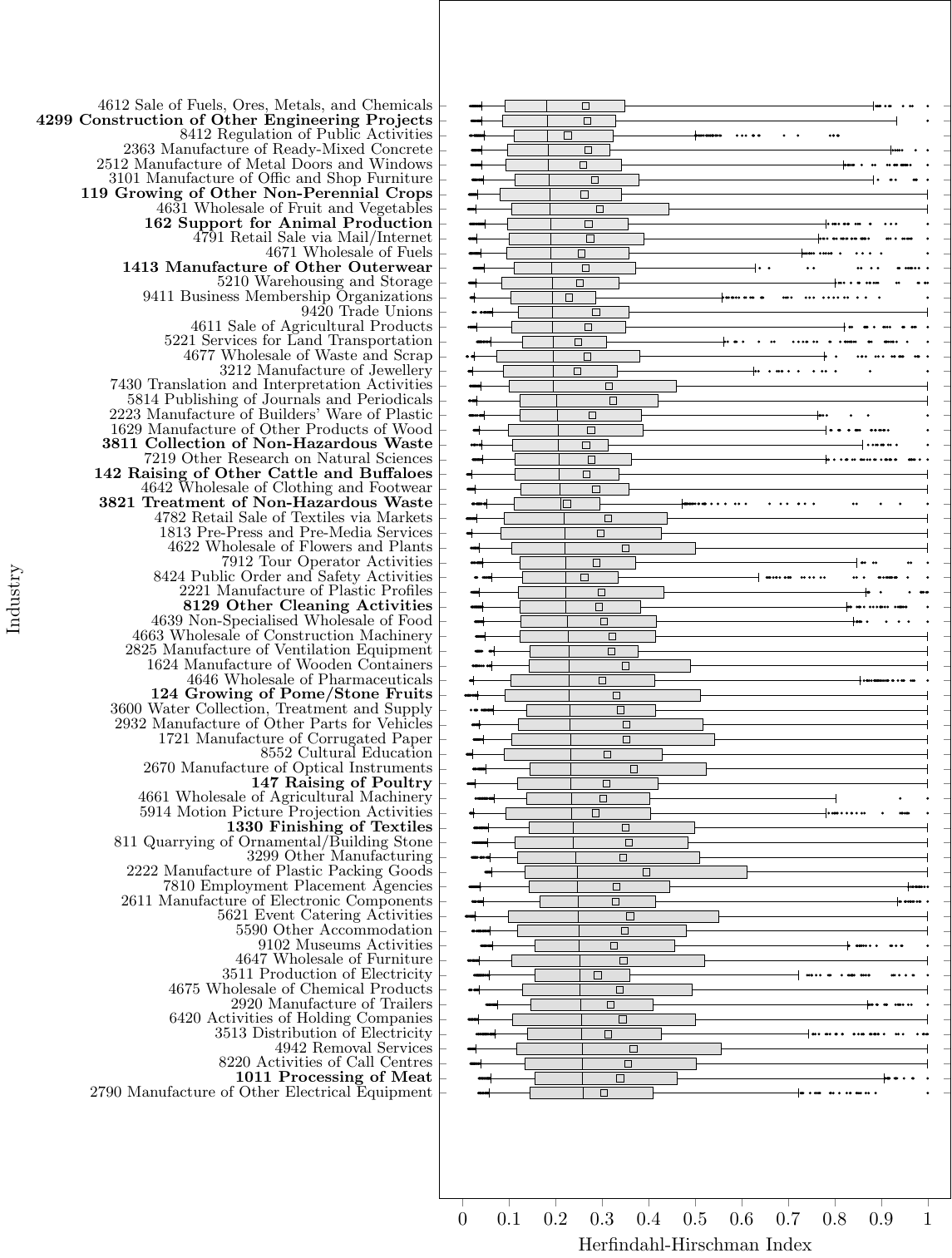}
}
\floatfoot{ \scriptsize \textsc{Note. ---} The figure visualizes the distribution of labor market concentration by 4-digit NACE industries in the Germany. Labor market concentration refers to employment-based HHIs for pair-wise combinations of 4-digit NACE industries and commuting zones, and is tracked with annual frequency. In each of the boxplots, the center marks the median whereas left and right margins represent the 25th and 75th percentile. Lower and upper whiskers indicate the 5th and 95th percentile. Hollow squares illustrate the underlying means. Dots represent outliers (i.e., values below the 5th or above the 95th percentile). Bold industries refer to minimum wage sectors. HHI = Herfindahl-Hirschman Index. NACE = Statistical Nomenclature of Economic Activities in the European Community. Source: IEB, 1999-2017.}
\end{figure}
\begin{figure}[!ht]
\addtocounter{figure}{-1}
\centering
\caption{Labor Market Concentration by 4-Digit NACE Industry (Cont.)}
\scalebox{0.80}{
\includegraphics{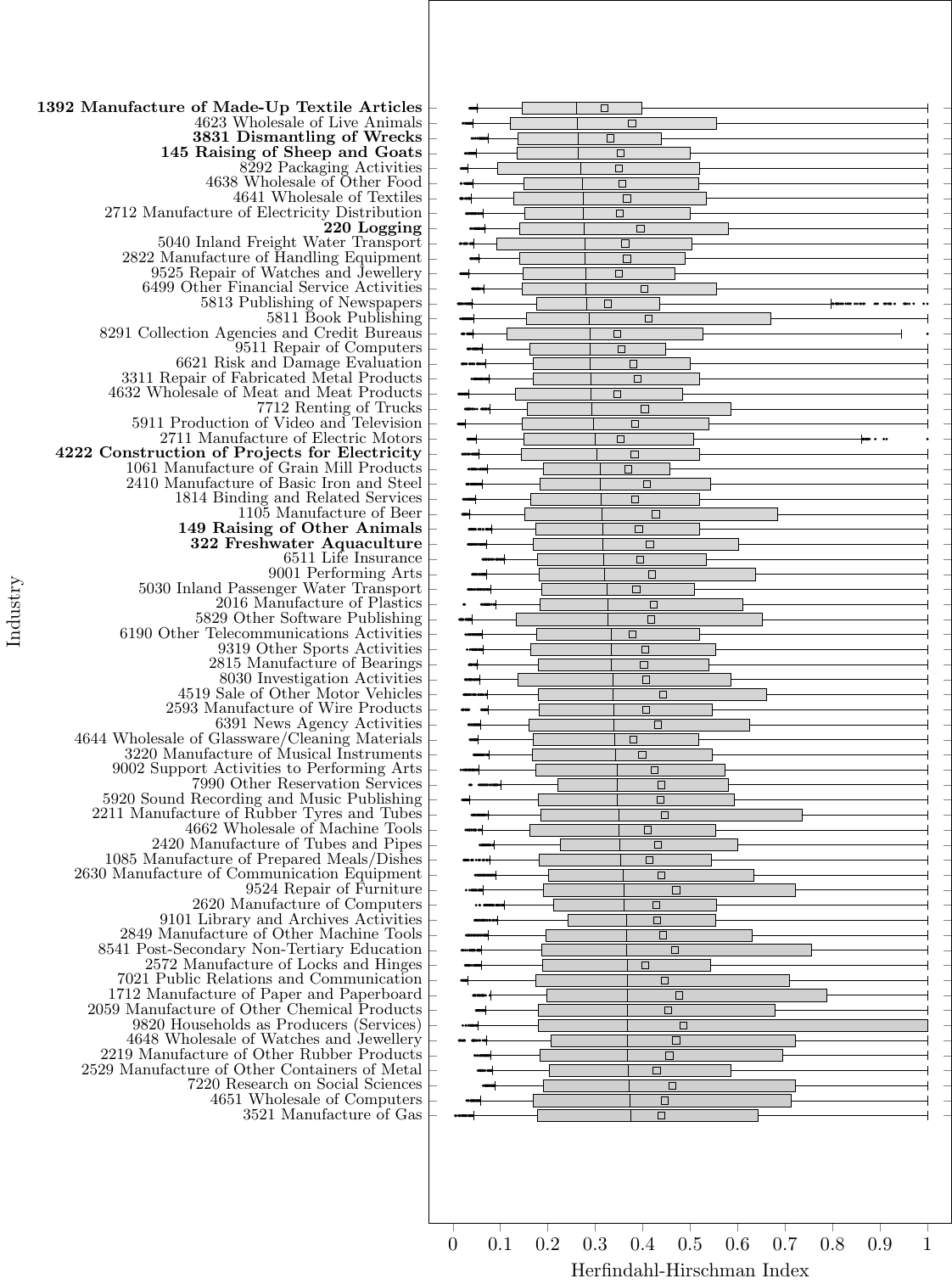}
}
\floatfoot{ \scriptsize \textsc{Note. ---} The figure visualizes the distribution of labor market concentration by 4-digit NACE industries in the Germany. Labor market concentration refers to employment-based HHIs for pair-wise combinations of 4-digit NACE industries and commuting zones, and is tracked with annual frequency. In each of the boxplots, the center marks the median whereas left and right margins represent the 25th and 75th percentile. Lower and upper whiskers indicate the 5th and 95th percentile. Hollow squares illustrate the underlying means. Dots represent outliers (i.e., values below the 5th or above the 95th percentile). Bold industries refer to minimum wage sectors. HHI = Herfindahl-Hirschman Index. NACE = Statistical Nomenclature of Economic Activities in the European Community. Source: IEB, 1999-2017.}
\end{figure}
\begin{figure}[!ht]
\addtocounter{figure}{-1}
\centering
\caption{Labor Market Concentration by 4-Digit NACE Industry (Cont.)}
\scalebox{0.80}{
\includegraphics{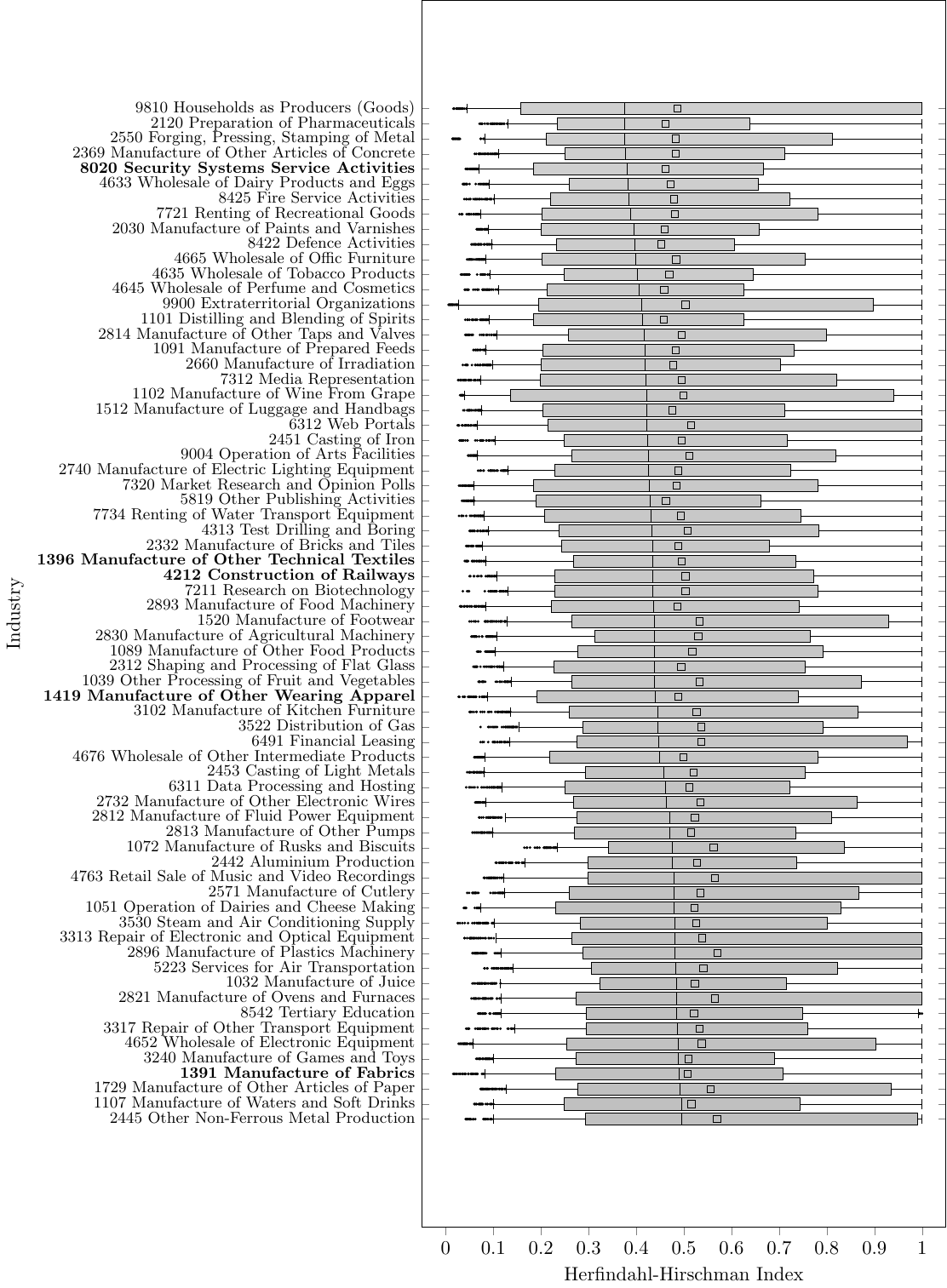}
}
\floatfoot{ \scriptsize \textsc{Note. ---} The figure visualizes the distribution of labor market concentration by 4-digit NACE industries in the Germany. Labor market concentration refers to employment-based HHIs for pair-wise combinations of 4-digit NACE industries and commuting zones, and is tracked with annual frequency. In each of the boxplots, the center marks the median whereas left and right margins represent the 25th and 75th percentile. Lower and upper whiskers indicate the 5th and 95th percentile. Hollow squares illustrate the underlying means. Dots represent outliers (i.e., values below the 5th or above the 95th percentile). Bold industries refer to minimum wage sectors. HHI = Herfindahl-Hirschman Index. NACE = Statistical Nomenclature of Economic Activities in the European Community. Source: IEB, 1999-2017.}
\end{figure}
\begin{figure}[!ht]
\addtocounter{figure}{-1}
\centering
\caption{Labor Market Concentration by 4-Digit NACE Industry (Cont.)}
\scalebox{0.80}{
\includegraphics{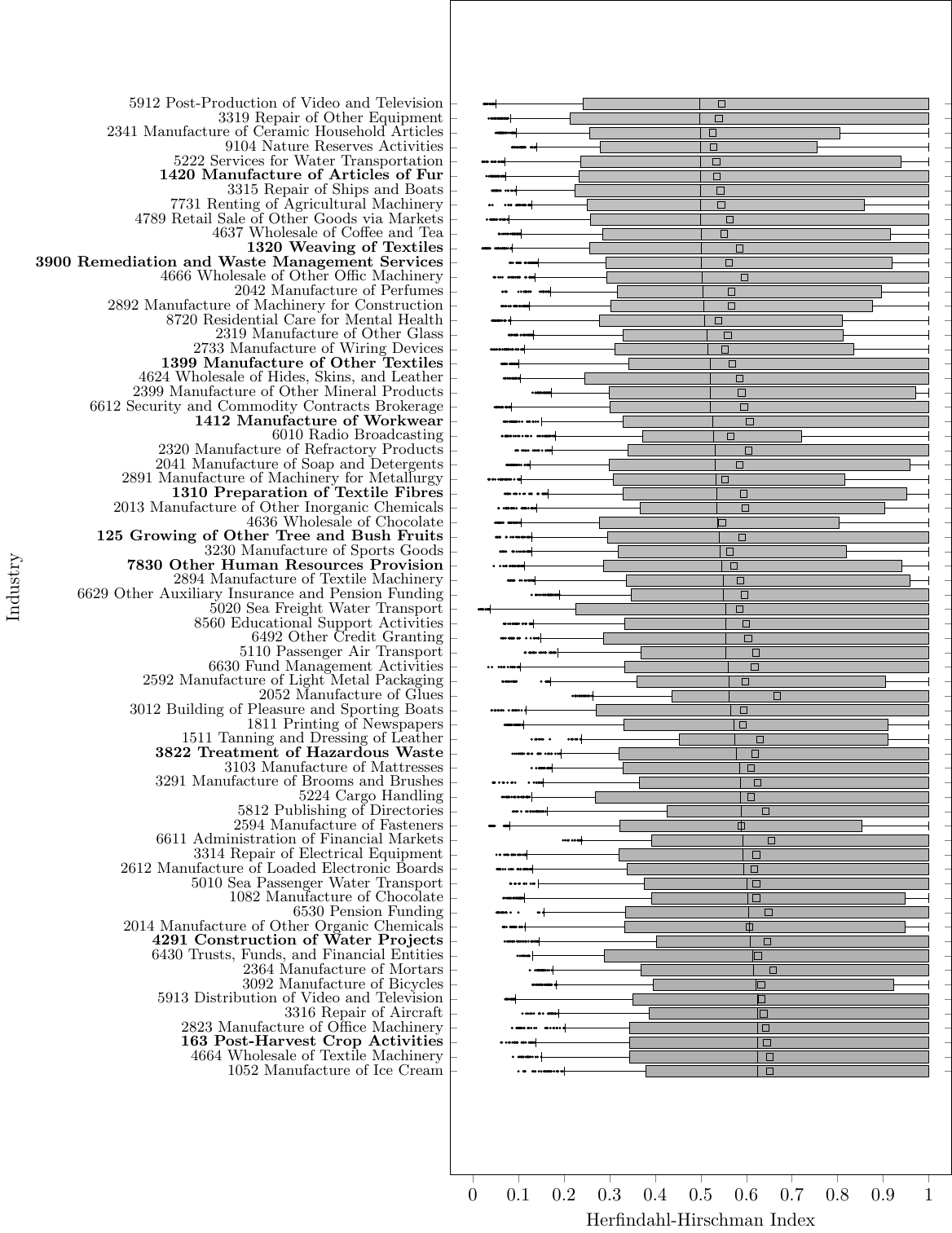}
}
\floatfoot{ \scriptsize \textsc{Note. ---} The figure visualizes the distribution of labor market concentration by 4-digit NACE industries in the Germany. Labor market concentration refers to employment-based HHIs for pair-wise combinations of 4-digit NACE industries and commuting zones, and is tracked with annual frequency. In each of the boxplots, the center marks the median whereas left and right margins represent the 25th and 75th percentile. Lower and upper whiskers indicate the 5th and 95th percentile. Hollow squares illustrate the underlying means. Dots represent outliers (i.e., values below the 5th or above the 95th percentile). Bold industries refer to minimum wage sectors. HHI = Herfindahl-Hirschman Index. NACE = Statistical Nomenclature of Economic Activities in the European Community. Source: IEB, 1999-2017.}
\end{figure}
\begin{figure}[!ht]
\addtocounter{figure}{-1}
\centering
\caption{Labor Market Concentration by 4-Digit NACE Industry (Cont.)}
\scalebox{0.80}{
\includegraphics{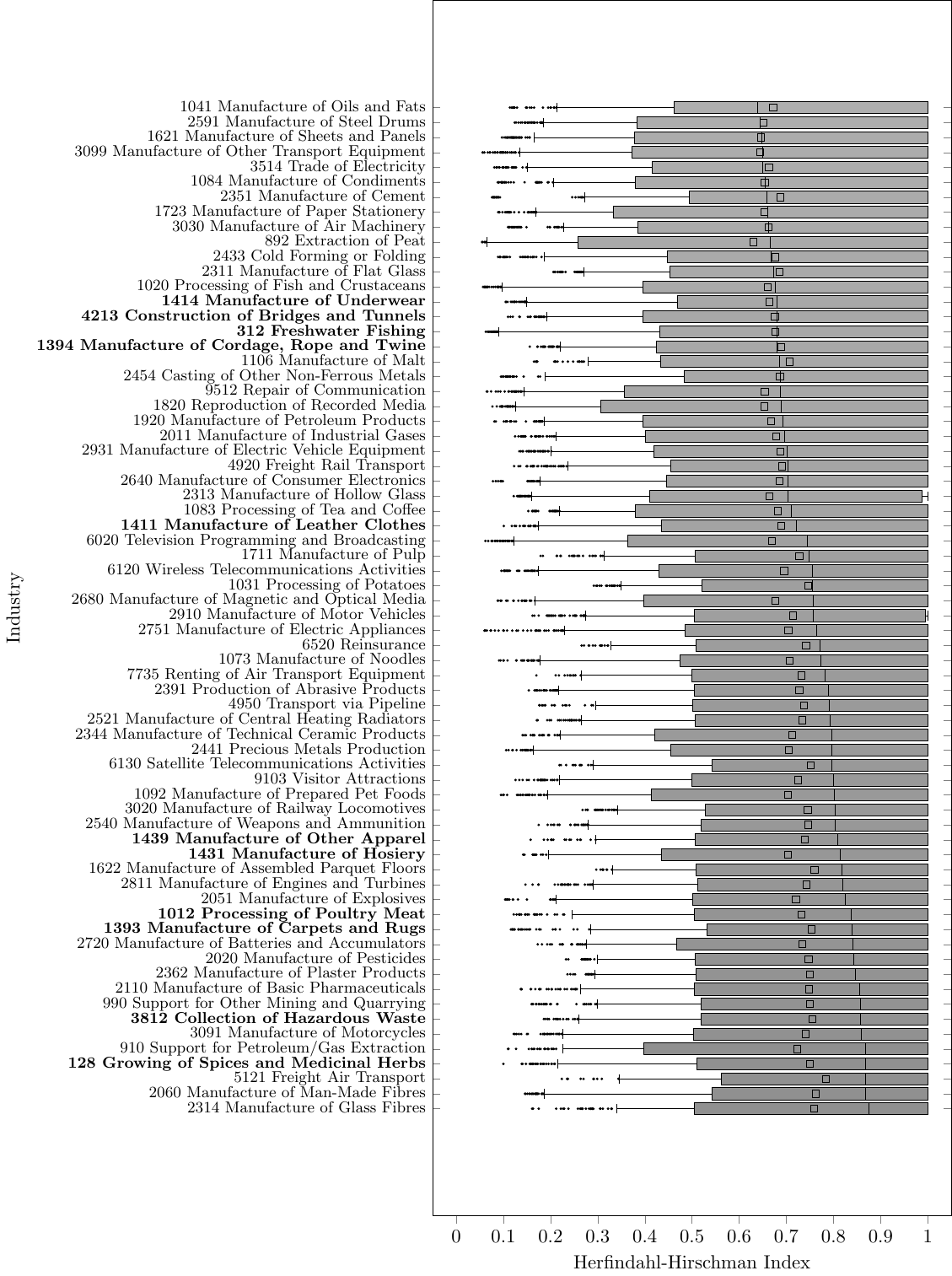}
}
\floatfoot{ \scriptsize \textsc{Note. ---} The figure visualizes the distribution of labor market concentration by 4-digit NACE industries in the Germany. Labor market concentration refers to employment-based HHIs for pair-wise combinations of 4-digit NACE industries and commuting zones, and is tracked with annual frequency. In each of the boxplots, the center marks the median whereas left and right margins represent the 25th and 75th percentile. Lower and upper whiskers indicate the 5th and 95th percentile. Hollow squares illustrate the underlying means. Dots represent outliers (i.e., values below the 5th or above the 95th percentile). Bold industries refer to minimum wage sectors. HHI = Herfindahl-Hirschman Index. NACE = Statistical Nomenclature of Economic Activities in the European Community. Source: IEB, 1999-2017.}
\end{figure}
\begin{figure}[!ht]
\addtocounter{figure}{-1}
\centering
\caption{Labor Market Concentration by 4-Digit NACE Industry (Cont.)}
\scalebox{0.80}{
\includegraphics{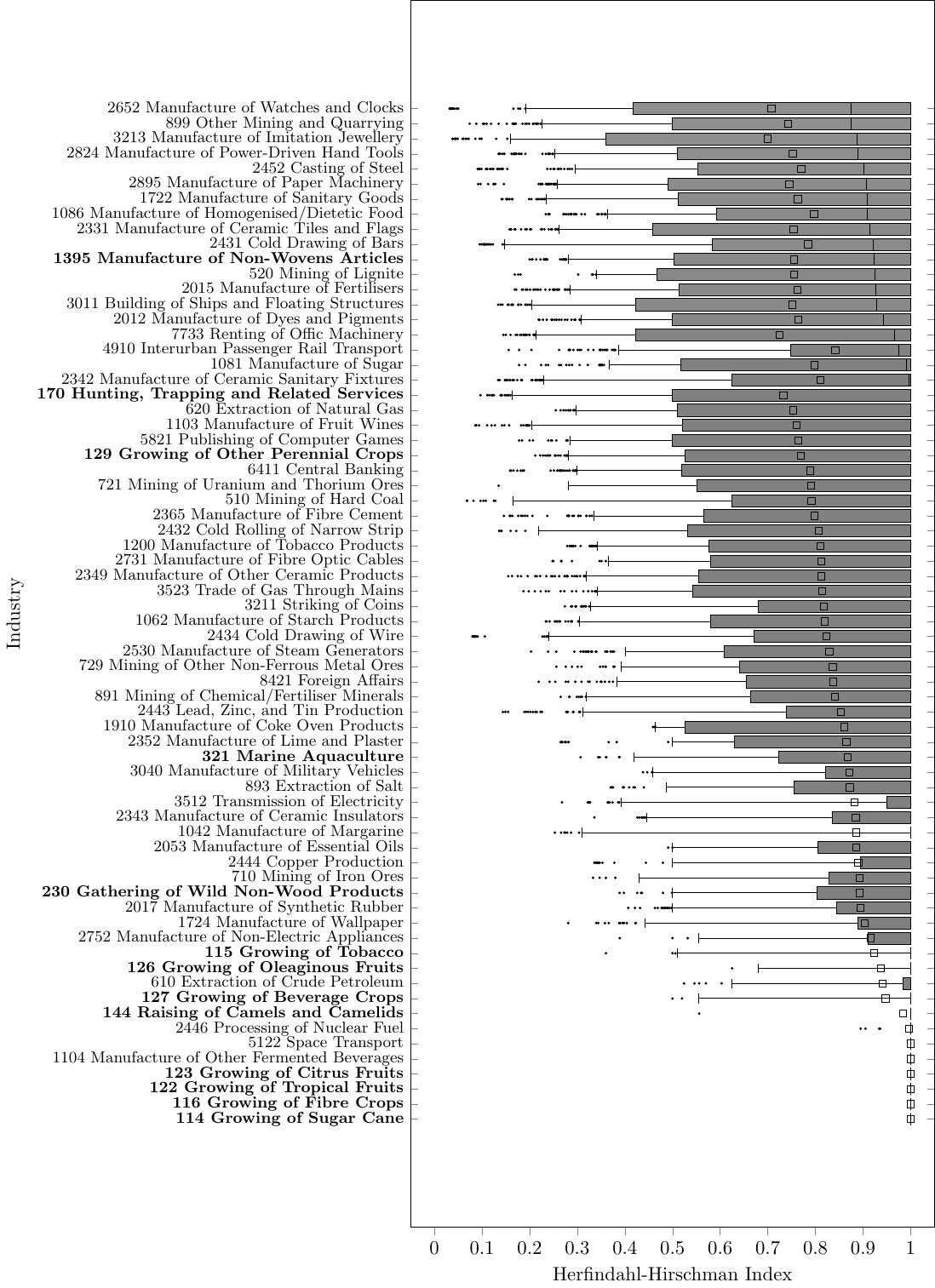}
}
\floatfoot{ \scriptsize \textsc{Note. ---} The figure visualizes the distribution of labor market concentration by 4-digit NACE industries in the Germany. Labor market concentration refers to employment-based HHIs for pair-wise combinations of 4-digit NACE industries and commuting zones, and is tracked with annual frequency. In each of the boxplots, the center marks the median whereas left and right margins represent the 25th and 75th percentile. Lower and upper whiskers indicate the 5th and 95th percentile. Hollow squares illustrate the underlying means. Dots represent outliers (i.e., values below the 5th or above the 95th percentile). Bold industries refer to minimum wage sectors. HHI = Herfindahl-Hirschman Index. NACE = Statistical Nomenclature of Economic Activities in the European Community. Source: IEB, 1999-2017.}
\end{figure}

\clearpage
\vspace*{\fill}

\vspace*{\fill}
\clearpage
\vspace*{\fill}

\vspace*{\fill}
\clearpage

\section{Flow-Adjusted Herfindahl-Hirschman Index}
\label{sec:E}
\setcounter{table}{0} 
\setcounter{figure}{0} 

The calculation of indices of labor market concentration necessitates a definition of what constitutes the relevant labor market. In practice, researchers use combinations of industries (or occupations) and regions to delineate labor markets. In doing so, however, labor markets are implicitly divided into discrete segments between which workers are not supposed to move \citep{Manning2021}. Thus, if the labor market is defined overly narrow (broad), labor market concentration will be overestimated (underestimated).

In terms of the spatial division, administrative regions (such as federal states or districts) do not necessarily overlap with the true regional scope of the underlying labor market. To address this problem, researchers follow standard practice and construct functional labor market regions to capture commuting flows more adequately than administrative regions. In terms of the industrial (or occupational) division of the labor market, researchers normally rely on available classifications of industries (or occupations). However, in contrast to regions, it is not standard practice to group together related sub-industries (or sub-occupations) to improve the delineation of these classifications.

\paragraph{Flow Adjustment.} To overcome this shortcoming, I closely follow \citet{Arnold2021} and implement a data-driven approach to consider jobs in differently classified industries as additional outside options. Specifically, I calculate a flow-adjusted version of the Herfindahl-Hirschman Index that builds on mobility patterns across labor markets to attribute weights to job opportunities in all other industries. The underlying idea is that the relative value of job opportunities in different industries can be inferred from labor market flows within and between these industries. Let $P(h|i)$ denote the probability that a worker in industry $i$ in year $t$ is employed in industry $h$ in year $t+1$. When working in industry $i$, the relative value of a job opportunity in industry $h$ (compared to industry $i$) then is:
\begin{equation}
\label{eq:E1}
\omega_{ih} = \frac{P(h|i)}{P(i|i)} \cdot \frac{L_{i}}{L_{h}}
\end{equation}
To infer weights from transitions, it is necessary to take into account that the number of inflows in a labor market positively depends on the number of job opportunities in the market. For this reason, the relative transition probabilities are normalized by employment in the respective industries. Note that, by construction, the industry under study always receives unit weight: $\omega_{ii}=1$. When calculating the transition probabilities with the IEB data, I pool mobility patterns over commuting zones and time (1999-2017) in order to obtain a meaningful and stable weight matrix.

Given the weight matrix, the flow-adjusted Herfindahl-Hirschman Index
\begin{equation}
\label{eq:E2}
Adjusted \,\ H\!H\!I_{izt} = \sum_{j=1}^{J} \tilde{s}_{jizt}^{2}
\end{equation}
is the squared sum of flow-adjusted employment shares for industry $i$ in commuting zone $z$ at time $t$. In particular, the flow-adjusted employment share $\tilde{s}$ is calculated as follows:
\begin{equation}
\label{eq:E3}
\tilde{s}_{jizt} = \frac{\sum_{h=1}^{I} \omega_{ih} \cdot L_{jhzt}}{\sum_{h=1}^{I} \omega_{ih} \cdot \sum_{j=1}^{J} L_{jhzt}}
\end{equation}
When calculating these flow-adjusted shares, jobs in the same industry and commuting zone receive unit weight, jobs in all other industries but the same commuting zone receive the weight $\omega_{ih}$, and jobs in all other commuting zone receive zero weight. With no flows between industries, the neighboring industries receive zero weight and the flow-based HHI collapses to the standard HHI with simple employment shares in the industry-by-commuting-zone pair: $H\!H\!I_{izt} = \sum_{j=1}^{J} s_{jizt}^{2}$. With random flows between industries, all industries receive the same weight and the flow-based HHI collapses to the standard HHI for commuting zones only.

\paragraph{Descriptive Statistics.} The histogram in Figure \ref{fig:E1} shows illustrates the distribution of the flow-based Herfindahl-Hirschman Index. For employment in pairs of 4-digit NACE industries and commuting zones, the average flow-adjusted HHI turns out only half as large (0.168)  as the average standard HHI. At the 25th percentile, the median, and the 75th percentile, the flow-adjusted HHI is 0.019, 0.075, and 0.226, respectively. Despite the flow adjustment, 16.1 and 27.9 percent of labor markets show up medium or high levels of concentration.

\clearpage
\vspace*{\fill}

\begin{figure}[!ht]
\centering
\caption{Distribution of Flow-Adjusted HHI}
\label{fig:E1}
\scalebox{0.80}{
\begin{tikzpicture}
\pgfplotsset{set layers}
\begin{axis}[area style, xlabel=Herfindahl-Hirschman Index, ylabel=Density, xtick align=outside, xtick pos = left, height=14cm, width=14cm, xmin=-0.05, xmax=1.05, ymin = -0.01, ymax=0.33, ytick={0,0.06,0.12,0.18,0.24,0.30},xtick={0,0.2,0.4,0.6,0.8,1}, grid=major, set layers, major grid style = {dotted, /pgfplots/on layer=axis background}, legend pos = north east, y tick label style={/pgf/number format/.cd,fixed,fixed zerofill, precision=2,/tikz/.cd}, x tick label style={/pgf/number format/.cd,fixed,fixed zerofill, precision=1,/tikz/.cd}]
\draw[dashed, color=blue] (0.019,0) -- (0.019,0.31) node[above right, color=blue, xshift=-0.5cm]{P25=0.019};
\draw[dashed, color=blue] (0.075,0) -- (0.075,0.29) node[above right, color=blue, xshift=-0.5cm]{P50=0.075};
\draw[dashed, color=blue] (0.168,0) -- (0.168,0.27) node[above right, color=blue, xshift=-0.5cm]{Mean=0.168};
\draw[dashed, color=blue] (0.226,0) -- (0.226,0.25) node[above right, color=blue, xshift=-0.5cm]{P75=0.226};
\addplot+[ybar interval, mark=no, color=black, fill=gray!5!white] plot coordinates { (0,.29506341) (.025,.12039282)};
\addplot+[ybar interval, mark=no, color=black, fill=gray!5!white] plot coordinates { (.025,.12039282) (.05,.08348818)};
\addplot+[ybar interval, mark=no, color=black, fill=gray!5!white] plot coordinates { (.05,.08348818) (.075,.06376783)};
\addplot+[ybar interval, mark=no, color=black, fill=gray!5!white] plot coordinates { (.075,.06376783) (.1,.05088341)};
\addplot+[ybar interval, mark=no, color=black, fill=gray!20!white] plot coordinates { (.1,.05088341) (.125,.04303743)};
\addplot+[ybar interval, mark=no, color=black, fill=gray!20!white] plot coordinates { (.125,.04303743) (.15000001,.03586761)};
\addplot+[ybar interval, mark=no, color=black, fill=gray!20!white] plot coordinates { (.15000001,.03586761) (.175,.03008219)};
\addplot+[ybar interval, mark=no, color=black, fill=gray!20!white] plot coordinates { (.175,.03008219) (.2,.0261887)};
\addplot+[ybar interval, mark=no, color=black, fill=gray!60!white] plot coordinates { (.2,.0261887) (.22499999,.023484)};
\addplot+[ybar interval, mark=no, color=black, fill=gray!60!white] plot coordinates { (.22499999,.023484) (.25,.02034089)};
\addplot+[ybar interval, mark=no, color=black, fill=gray!60!white] plot coordinates { (.25,.02034089) (.27500001,.01829719)};
\addplot+[ybar interval, mark=no, color=black, fill=gray!60!white] plot coordinates { (.27500001,.01829719) (.30000001,.01573244)};
\addplot+[ybar interval, mark=no, color=black, fill=gray!60!white] plot coordinates { (.30000001,.01573244) (.32499999,.01438009)};
\addplot+[ybar interval, mark=no, color=black, fill=gray!60!white] plot coordinates { (.32499999,.01438009) (.34999999,.01294006)};
\addplot+[ybar interval, mark=no, color=black, fill=gray!60!white] plot coordinates { (.34999999,.01294006) (.375,.01181872)};
\addplot+[ybar interval, mark=no, color=black, fill=gray!60!white] plot coordinates { (.375,.01181872) (.40000001,.01097393)};
\addplot+[ybar interval, mark=no, color=black, fill=gray!60!white] plot coordinates { (.40000001,.01097393) (.42500001,.00998917)};
\addplot+[ybar interval, mark=no, color=black, fill=gray!60!white] plot coordinates { (.42500001,.00998917) (.44999999,.0092641)};
\addplot+[ybar interval, mark=no, color=black, fill=gray!60!white] plot coordinates { (.44999999,.0092641) (.47499999,.00865031)};
\addplot+[ybar interval, mark=no, color=black, fill=gray!60!white] plot coordinates { (.47499999,.00865031) (.5,.00769929)};
\addplot+[ybar interval, mark=no, color=black, fill=gray!60!white] plot coordinates { (.5,.00769929) (.52499998,.00661505)};
\addplot+[ybar interval, mark=no, color=black, fill=gray!60!white] plot coordinates { (.52499998,.00661505) (.55000001,.00615133)};
\addplot+[ybar interval, mark=no, color=black, fill=gray!60!white] plot coordinates { (.55000001,.00615133) (.57499999,.00596754)};
\addplot+[ybar interval, mark=no, color=black, fill=gray!60!white] plot coordinates { (.57499999,.00596754) (.60000002,.00531497)};
\addplot+[ybar interval, mark=no, color=black, fill=gray!60!white] plot coordinates { (.60000002,.00531497) (.625,.00500302)};
\addplot+[ybar interval, mark=no, color=black, fill=gray!60!white] plot coordinates { (.625,.00500302) (.64999998,.00496592)};
\addplot+[ybar interval, mark=no, color=black, fill=gray!60!white] plot coordinates { (.64999998,.00496592) (.67500001,.00473154)};
\addplot+[ybar interval, mark=no, color=black, fill=gray!60!white] plot coordinates { (.67500001,.00473154) (.69999999,.00452413)};
\addplot+[ybar interval, mark=no, color=black, fill=gray!60!white] plot coordinates { (.69999999,.00452413) (.72500002,.00457978)};
\addplot+[ybar interval, mark=no, color=black, fill=gray!60!white] plot coordinates { (.72500002,.00457978) (.75,.00428975)};
\addplot+[ybar interval, mark=no, color=black, fill=gray!60!white] plot coordinates { (.75,.00428975) (.77499998,.00406717)};
\addplot+[ybar interval, mark=no, color=black, fill=gray!60!white] plot coordinates { (.77499998,.00406717) (.80000001,.00419363)};
\addplot+[ybar interval, mark=no, color=black, fill=gray!60!white] plot coordinates { (.80000001,.00419363) (.82499999,.00419532)};
\addplot+[ybar interval, mark=no, color=black, fill=gray!60!white] plot coordinates { (.82499999,.00419532) (.85000002,.00413461)};
\addplot+[ybar interval, mark=no, color=black, fill=gray!60!white] plot coordinates { (.85000002,.00413461) (.875,.00382266)};
\addplot+[ybar interval, mark=no, color=black, fill=gray!60!white] plot coordinates { (.875,.00382266) (.89999998,.00390192)};
\addplot+[ybar interval, mark=no, color=black, fill=gray!60!white] plot coordinates { (.89999998,.00390192) (.92500001,.00387156)};
\addplot+[ybar interval, mark=no, color=black, fill=gray!60!white] plot coordinates { (.92500001,.00387156) (.94999999,.00416665)};
\addplot+[ybar interval, mark=no, color=black, fill=gray!60!white] plot coordinates { (.94999999,.00416665) (.97500002,.00316166)};
\addplot+[ybar interval, mark=no, color=black, fill=gray!60!white] plot coordinates { (.97500002,.00316166) (1,.0001)};
\legend{{~Low (56.0 Percent)},,,,{~Medium (16.1 Percent)},,,,{~High (27.9 Percent)}}
\end{axis}
\end{tikzpicture}
}
\floatfoot{\footnotesize\textsc{Note. ---} The figure illustrates the distribution of labor market concentration in Germany. Labor market concentration refers to flow-adjusted HHIs for pair-wise combinations of 4-digit NACE industries and commuting zones, and is tracked with annual frequency. HHI = Herfindahl-Hirschman Index. NACE = Statistical Nomenclature of Economic Activities in the European Community. Source: IEB, 1999-2017.}
\end{figure}

\vspace*{\fill}
\clearpage

\section{Firm-Level Markdowns and Markups}
\label{sec:F}
\setcounter{table}{0} 
\setcounter{figure}{0} 

In his seminal paper, \citet{Hall1988} developed a tractable method to recover average price-cost markups by relying on easily available production data, meanwhile known as the ``Production Function Approach''. The method is based on the insight that markups in the product market manifest in a wedge between an input's cost share and revenue share. \citet{DeLoecker2011} and \citet{DeLoeckerWarzynski2012} generalized Hall's framework to allow for different markups across producers and a more flexible production technology. In their framework, the markup is reflected by the difference between the output elasticity of a variable input and the input's revenue share. In addition to product market imperfections, \citet{DobbelaereMairesse2013} refined the methodology to simultaneously determine imperfections in the factor market. To correct for price-cost markups, the method relies on a variable input $M$ which, unlike labor $L$, is not subject to factor market imperfections. Allowing for a finitely elastic supply of workers to the firm, the authors show that markdowns in the labor market can then be recovered as
\begin{equation}
\label{eq:F1}
\frac{MRPL}{W} = \frac{\,\,\, \text{elasticity of output w.r.t. labor}\,\,\, }{\,\,\, \text{labor share} \,\,\cdot\,\, \text{price-cost markup}\,\,\,} = \frac{ \,\,\varepsilon^{Q}_{L} \,\,}{\,\, \alpha_{L}\,\,  } \cdot \frac{\,\,\alpha_{M}\,\,}{\,\,\varepsilon^{Q}_{M}\,\, }
\end{equation}
where $\alpha$ denotes the share of expenditures on a certain input in total revenues and $\varepsilon$ denotes the elasticity of output $Q$ with respect to a certain input. Thus, the markdown mirrors the ratio of the elasticity of output with respect to labor to the respective share of labor in total revenues, corrected by the markup in the product market (which is itself the ratio of the elasticity of output with respect to input $M$ to the respective share of input $M$ in total revenues). While input shares are readily available from standard production data, the output elasticities of the input factors must be estimated using regression techniques.

\paragraph{Econometric Framework.} To estimate output elasticities, I follow \citet{DobbelaereEtAl2024} and specify the parameters of a three-factor Translog production function, which is a logarithmic second-order Taylor approximation to an arbitrary twice-differentiable production function. Specifically, I estimate the following output equation, separately for different 1-digit industries:
\begin{equation}
\label{eq:F2}
\begin{gathered}
q_{jit} \,=\, \beta_{0} \,+\, \beta_{l} \, l_{jit} \,+\, \beta_{m} \, m_{jit} \,+\, \beta_{k} \, k_{jit} \,+\, \beta_{ll} \, l^{2}_{jit} \,+\, \beta_{mm} \, m^{2}_{jit} \,+\, \beta_{kk} \, k^{2}_{jit} \\
\,+\, \beta_{lm} \, l_{jit} \, m_{jit} \,+\, \beta_{lk} \, l_{jit} \, k_{jit} \,+\, \beta_{mk} \, m_{jit} \, k_{jit} \,+\,  \delta_{j} \,+\, \zeta_{t} \,+\, \epsilon_{jit}
\end{gathered}
\end{equation}
where $q$, $l$, $m$, and $k$ reflect the quantities of output, labor, intermediate goods, and capital (all in logs). The equation incorporates firm fixed effects $\delta_{j}$ to account for time-invariant productivity differences between producers as well as year fixed effects $\zeta_{t}$ to address advances from technological change. Assuming that the markets for intermediate goods are perfectly competitive (to correct for price-cost markups), I retrieve the relevant output elasticities
\begin{equation}
\label{eq:F3}
\hat{\varepsilon}^{Q}_{L} \,=\, \hat{\beta}_{l} \,+\, 2 \, \hat{\beta}_{ll} \, l^{2}_{jit}  \,+\, \hat{\beta}_{lm} \, m_{jit} \,+\, \hat{\beta}_{lk}  k_{jit}
\end{equation}
\begin{equation}
\label{eq:F4}
\hat{\varepsilon}^{Q}_{M} \,=\, \hat{\beta}_{m} \,+\, 2 \, \hat{\beta}_{mm} \, m^{2}_{jit} \,+\, \hat{\beta}_{lm} \, l_{jit} \,+\, \hat{\beta}_{mk}  k_{jit}
\end{equation}
which vary over firms and time. To finally recover markdowns, I follow Equation (\ref{eq:F1}) and combine the estimated output elasticities (\ref{eq:F3}) and (\ref{eq:F4}) with survey information on factor shares of labor and intermediate goods.

\paragraph{Implementation for the German Labor Market.} To estimate markdowns for the German labor market, I assemble data from three different sources for the years 1999-2017:
\begin{itemize}
\item administrative information on overall headcount employment and the time-consistent 1-digit industry affiliation from the BHP data,
\item survey information on annual revenues (in Euro), the monthly wage bill (in Euro, as of June of the respective year), annual expenditure on intermediate goods (in Euro), expenditure for investment (in Euro), and type of investments from the IAB Establishment Panel, and
\item national consumer price levels from the German Federal Statistical Office (Destatis).
\end{itemize}
As information on revenues, intermediate goods, and investment is asked retrospectively in the IAB Establishment, I additionally make use of the 2018 wave and move this information one year into the past. Building on the modified perpentual-inventory approach \citep{Mueller2008,Mueller2010,Mueller2017}, I approximate firms' capital stock from information on investment in the IAB Establishment Panel. To arrive at real values, I deflate all nominal variables using the national consumer price levels. I eliminate information from the finance and insurance sector, for which the revenue measure refers to balance sheet total (and is thus not comparable with the other sectors). Moreover, I discard 1-digit industries that enter the regression with less than 250 observations. Thus, the regression sample consists of 91,083 observations and 13,850 firms that belong to 30 1-digit sectors. As output is not observed directly, I normalize revenues by $exp(\hat{\epsilon}_{jit})$ to adjust for idiosyncratic components that are independent of input use and productivity.
To correct for survey-driven measurement error, I trim the resulting markdowns at the 5th and 95th percentile of their yearly distribution. When reporting the results, I use cross-sectional survey weights to ensure that the markdowns are representative for the German labor market.

\paragraph{Descriptive Statistics.} Table \ref{tab:F1} delivers descriptives statistics for the firm-level markdowns. Across firms from all sectors between 1999 and 2017, the firm-level markdowns averaged 1.456 which, by taking the reciprocal, implies that workers are paid about 69 percent of their marginal revenue product. When filtering for firms that belong to the subset of minimum wage sectors, the average markdown turns out lower, implying that workers obtain 81 percent of their MRPL, mirroring that minimum wage are supposed to counteract markdowns. My finding that markdowns point towards substantial monopsony power is well in line with studies that adopt a complementary approach and semi-structurally estimate wage elasticities of labor supply to the firm on related data from Germany \citep{HirschEtAl2010,BachmannFrings2017,HirschEtAl2018,BachmannEtAl2022,HirschEtAl2022}. These studies typically arrive at elasticities in the range between 1.5 and 4, implying that workers earn between 65 and 80 percent of their marginal revenue product.

\paragraph{Relationship between Markdowns and Labor Market Concentration.} My result that the average labor market is highly concentrated (HHI=0.342) is consistent with both aforementioned markdowns and available labor supply elasticities in a sense that all measures point to considerable monopsony power in the German labor market. In a next step, I directly relate firms' markdowns to their labor market's  HHI value (for the baseline labor market definition). On average, workers' share in their marginal revenue product turns out larger in labor markets with low HHI values (69.2 percent) than in markets with medium (67.1 percent) or high HHI values (60.8 percent). The finding that markdowns show up even in labor markets with low HHI values suggests that alternative sources of monopsony power, such as job differentiation or search frictions, give rise to monopsony power even in ``thick'' labor markets with many employers \citep{Manning2003b}.

Binned scatter plots in Figure \ref{fig:F1} underline that there is a positive relationship between firms' markdowns and HHI values (in logs) -- both for all sectors and the subset of minimum wage sectors. Thus, markdowns clearly increase with higher labor market concentration. In quantitative terms, Table \ref{tab:F2} regresses firm-level markdowns on different measures of labor market concentration. Using the baseline HHI, a 100 percent HHI increase (e.g., a reduction in the number of equally-sized employers from 10 to 5 firms) raises firms' markdowns on average by 0.037. For the subset of firms in minimum wage sectors, this effect turns out even more pronounced. The positive relationship also holds for HHI values based on 3-digit NACE industries, 5-digit NACE industries, 3-digit NUTS regions, and when using firms' market share as explanatory variable. Additional regressions show that, on average, the prevalence of a sectoral minimum wage is associated with lower markdowns. Specifically, an increase of a sectoral minimum wage by 1 Euro significantly reduces the average markdown by 0.025. Thus, higher minimum wages move wages closer to marginal productivity, providing tentative evidence that the minimum wage policies limit employers' wage-setting power. By and large, the presented evidence underlines that higher labor market concentration confers more monopsony power upon employers in the German labor market -- as put forward by the ``structure-conduct-performance'' literature.

\paragraph{Relationship between Markups and Labor Market Concentration.} To ease the later interpretation of the multivariate results through the lens of the monopsony model, it is important to rule out any interplay between labor and product markets. In general, industry-based measures of local labor market concentration might also reflect product market concentration, notably when most of the products are sold locally \citep{Manning2021}. However, certain implications of firms' wage-setting power in the labor market could, in principle, also arise from price-setting power in the product market. For instance, monopolists reduce their output and, like monopsonists, employ fewer workers than in the absence of market power.

For lack of concentration indices in German product markets, I examine the relationship between labor market power and product market power using estimated price-cost markups. These firm-level markups are recovered in course of the markdown estimation. Unlike markdowns, the plausibility of these markups can be validated using survey information on firms. Since 2008, the IAB Establishment Panel asks firms about the intensity of product market competition using a four-point scale \citep{HirschEtAl2014}: none (13.6 percent), low (14.2 percent), medium (36.8 percent), or high product market competition (35.4 percent). If the estimated markups indeed mirror product market power, higher values of self-rated product market competition should feature lower price-cost markups. To examine this conjecture, Table \ref{tab:F3} features regressions of the firm-level markups on binary variables for firms' perceived strength of product market competition. In line, the results show that higher product market competition involves increasingly smaller markups, lending credence to the markup (and the associated markdown) estimation.

Table \ref{tab:F4} shows OLS regressions of firm-level markups on different measures of labor market concentration. Indeed, there is a positive, albeit weak, relationship between labor market concentration and markups. Reassuringly, however, the relationship collapses when looking at minimum wage sectors only, mitigating the concern that labor market concentration could reflect product market power in the later multivariate analyses. Furthermore, the relationship becomes even negative when excluding service sectors whose product markets are more local by nature. Thus, in the later analysis, I will corroborate the robustness of the multivariate results under exclusion of the service sectors. When filtering out service sectors, I roughly follow the definition from \citet{MianSufi2014} and discard those sectors that are neither classified as ``tradable'' or ``construction'' industries: namely commercial cleaning, waste removal, nursing care, security, temporary work, hairdressing, and chimney sweeping.

\begin{landscape}

\vspace*{\fill}

\begin{table}[!ht]
\centering
\begin{threeparttable}
\caption{Descriptive Statistics for Markdowns}
\label{tab:F1}
\begin{tabular}{L{7.5cm}C{2.5cm}C{2.5cm}C{2.5cm}C{2.5cm}C{2.5cm}} \hline
\multirow{5.4}{*}{}  & \multirow{5.4}{*}{Mean} & \multirow{5.4}{*}{\shortstack{Standard \vphantom{/} \\ Deviation}} & \multirow{5.4}{*}{\shortstack{Implied \vphantom{/} \\ Share in \vphantom{/} \\ MRPL }} & \multirow{5.4}{*}{\shortstack{Implied \vphantom{/} \\ Average \vphantom{/} \\ LS Elasticity \vphantom{/} \\ to the Firm }} & \multirow{5.4}{*}{Observations}  \\
&   &   &   &  &  \\
&   &   &   &  &  \\
&   &   &   &  &  \\
&   &   &   &  &  \\ \hline
&   &   &   &  &  \\[-0.4cm]
\multirow{1}{*}{All Sectors}   & \multirow{1}{*}{1.456}   & \multirow{1}{*}{1.003} & \multirow{1}{*}{0.687} & \multirow{1}{*}{2.194}  & \multirow{1}{*}{77,359}     \\[0.2cm]
\multirow{1}{*}{All Sectors: Low HHI (0-0.1)}   & \multirow{1}{*}{1.445}   & \multirow{1}{*}{0.999} & \multirow{1}{*}{0.692} & \multirow{1}{*}{2.248} & \multirow{1}{*}{57,341}     \\[0.2cm]
\multirow{1}{*}{All Sectors: Medium HHI (0.1-0.2)}   & \multirow{1}{*}{1.490}   & \multirow{1}{*}{1.025} & \multirow{1}{*}{0.671}  & \multirow{1}{*}{2.039} & \multirow{1}{*}{8,545}     \\[0.2cm]
\multirow{1}{*}{All Sectors: High HHI (0.2-1)}   & \multirow{1}{*}{1.644}   & \multirow{1}{*}{1.038} & \multirow{1}{*}{0.608} & \multirow{1}{*}{1.552} & \multirow{1}{*}{11,473}     \\[0.2cm]
\multirow{1}{*}{Only Minimum Wage Sectors}   & \multirow{1}{*}{1.238}   & \multirow{1}{*}{0.818} & \multirow{1}{*}{0.808} & \multirow{1}{*}{4.203} & \multirow{1}{*}{14,812}     \\ \hline
\end{tabular}
\begin{tablenotes}[para]
\footnotesize\textsc{Note. ---} The table displays descriptive statistics for firm-level markdowns in the German labor market. Building on estimated Translog production functions, the markdowns were calculated using administrative information on firms' employment, survey information on firms' annual revenues and monthly wage bill (as of June), and yearly information on price indices from Germany's national accounts. Markdowns were trimmed at the 5th and 95th percentile of their distribution. The results are weighted using cross-sectional survey weights. The share of workers' wages in the marginal product of labor is calculated as the inverse of the average markdown. The average labor supply elasticity to the firm is calculated by inserting the average markdown into Equation (\ref{eq:A5}). LS = Labor Supply. MRPL = Marginal Revenue Product of Labor. Sources: BHP $\plus$ IAB Establishment Panel $\plus$ Destatis, 1999-2017.
\end{tablenotes}
\end{threeparttable}
\end{table}

\vspace*{\fill}
\clearpage
\vspace*{\fill}

\begin{figure}[!ht]
\centering
\caption{Markdowns and Labor Market Concentration}
\label{fig:F1}
\vspace*{0.5cm}
\begin{subfigure}{0.475\textwidth}
\caption{\normalsize{All Sectors}}
\centering
\scalebox{0.675}{
\begin{tikzpicture}
\begin{axis}[xlabel=Log Herfindahl-Hirschman Index, ylabel=Markdown, xmin=-8.25,xmax=0.25, ymin=0.7, ymax=2.1, height=14cm, width=14cm, grid=major, legend pos = south east, xtick={-8,-6,-4,-2,0}, ytick={0.8,1.0,1.2,1.4,1.6,1.8,2}, grid style = dotted, y tick label style={/pgf/number format/.cd,fixed,fixed zerofill, precision=1,/tikz/.cd}]
\addplot[only marks, mark=o, mark options={solid,scale=1}, color=black] coordinates {   (-8.3810158,1.3681197) (-8.0126333,1.3781189) (-7.7101464,1.5534039) (-7.5213175,1.3810197) (-7.3198504,1.3426489) (-7.1039352,1.3993123) (-6.9562769,1.4389935) (-6.8716488,1.3624583) (-6.7678695,1.5126907) (-6.6878977,1.6645806) (-6.6141281,1.4294553) (-6.5359912,1.4402201) (-6.4829979,1.3309084) (-6.4108477,1.3624796) (-6.3328066,1.3576493) (-6.2769899,1.3741533) (-6.227355,1.3919284) (-6.1739941,1.26025) (-6.1192608,1.2913454) (-6.0783963,1.3729709) (-6.036737,1.3959441) (-5.9849744,1.2882189) (-5.9384251,1.4239107) (-5.9015064,1.2592554) (-5.8651996,1.1750072) (-5.8255811,1.2514119) (-5.7848725,1.3096747) (-5.740346,1.2998675) (-5.6927433,1.2788428) (-5.6418486,1.4148871) (-5.5911679,1.5481954) (-5.5472021,1.3305027) (-5.5042591,1.4986511) (-5.4590492,1.2947283) (-5.4096365,1.3263122) (-5.3669477,1.3134407) (-5.3309774,1.3700956) (-5.2940907,1.4872591) (-5.2560391,1.4592739) (-5.2153854,1.4079138) (-5.1760607,1.4157619) (-5.1307368,1.3734349) (-5.0813894,1.4147193) (-5.0302639,1.3658535) (-4.9838147,1.3210058) (-4.9338484,1.5158079) (-4.883986,1.4850529) (-4.8364062,1.5306619) (-4.7881474,1.351264) (-4.7375278,1.5019089) (-4.6910057,1.5118146) (-4.6470714,1.497631) (-4.598568,1.4990004) (-4.5536571,1.5373282) (-4.5045848,1.5825619) (-4.4579072,1.5398258) (-4.4068947,1.5175711) (-4.3617496,1.5287668) (-4.3208313,1.5785311) (-4.2755241,1.5171012) (-4.2325277,1.5176735) (-4.188406,1.3361088) (-4.1426272,1.4751107) (-4.0992618,1.3949744) (-4.0536461,1.5334154) (-3.9986453,1.4030195) (-3.9481802,1.4494951) (-3.895906,1.5006984) (-3.8432293,1.569711) (-3.7908387,1.465904) (-3.7343278,1.4844596) (-3.6821866,1.4710985) (-3.6343637,1.7707514) (-3.5861561,1.6440489) (-3.5317874,1.6345822) (-3.4763408,1.625558) (-3.412642,1.4439633) (-3.3497286,1.4946423) (-3.2876344,1.5307797) (-3.2275467,1.5288254) (-3.1637237,1.4345319) (-3.0894063,1.480062) (-3.0155101,1.6157546) (-2.941258,1.5901703) (-2.8712609,1.5257988) (-2.7903976,1.4634656) (-2.6995194,1.4705441) (-2.6033566,1.4353304) (-2.4996049,1.4765879) (-2.3969927,1.5095426) (-2.2976322,1.5149651) (-2.1942594,1.4601902) (-2.0755315,1.4858356) (-1.9523979,1.4790052) (-1.8150043,1.4595171) (-1.6613925,1.5851184) (-1.4616264,1.5211838) (-1.2270637,1.6932878) (-.90345788,1.6879805) (-.35255817,1.6776911) };
\addplot[domain=-8:0, color=black, solid, line width=0.75mm] {.0373665  *x+  1.628547   };
\legend{~Observations,~Linear Fit}
\end{axis}
\end{tikzpicture}
}
\end{subfigure}
\hfill
\begin{subfigure}{0.475\textwidth}
\addtocounter{subfigure}{0}
\caption{\normalsize{Only Minimum Wage Sectors}}
\centering
\scalebox{0.675}{
\begin{tikzpicture}
\begin{axis}[xlabel=Log Herfindahl-Hirschman Index, ylabel=Markdown, xmin=-8.25,xmax=0.25, ymin=0.7, ymax=2.1, height=14cm, width=14cm, grid=major, legend pos = south east, xtick={-8,-6,-4,-2,0}, ytick={0.8,1.0,1.2,1.4,1.6,1.8,2.0}, grid style = dotted, y tick label style={/pgf/number format/.cd,fixed,fixed zerofill, precision=1,/tikz/.cd}]
\addplot[only marks, mark=o, mark options={solid,scale=1}, color=black] coordinates {  (-8.1917677,.68537325) (-7.5988631,.8583917) (-7.3371673,.63205159) (-7.1247654,.91980934) (-7.0491848,.88361675) (-6.9668069,.94590372) (-6.8374376,1.5226945) (-6.711134,1.2671306) (-6.5765939,1.3839582) (-6.5219135,1.1760623) (-6.4797478,1.2222004) (-6.4193168,1.1412838) (-6.3186026,1.2594832) (-6.2756782,.95675874) (-6.2292814,1.0062056) (-6.1782875,1.0360522) (-6.127418,1.1449897) (-6.0950499,1.0107259) (-6.0565386,1.2908535) (-5.9896927,1.0984885) (-5.9168463,1.0080265) (-5.8593707,.98070234) (-5.8071527,1.1490519) (-5.75108,1.2065347) (-5.6939983,1.3513919) (-5.644587,1.2987227) (-5.6019511,1.4011929) (-5.5680218,1.306668) (-5.5314827,1.4871035) (-5.5034056,1.198155) (-5.4784331,1.2081714) (-5.4441371,1.0842962) (-5.4062533,1.1744847) (-5.3787494,1.3195049) (-5.3529506,1.3372391) (-5.3357,1.3558685) (-5.3034339,1.2340502) (-5.2628131,1.1130153) (-5.2170739,1.1682065) (-5.1749177,1.2605106) (-5.123693,1.2205338) (-5.0802169,1.1534737) (-5.0439782,1.1003865) (-5.0015483,1.2076035) (-4.9569783,1.1138459) (-4.9197135,1.2466091) (-4.8681345,1.2982408) (-4.8157196,1.3634478) (-4.7655339,1.2561328) (-4.7284236,1.2401252) (-4.692709,1.326133) (-4.6578999,1.3119915) (-4.6241455,1.6497947) (-4.5895405,1.3326145) (-4.5436091,1.3664016) (-4.49476,1.4704237) (-4.4393663,1.2675432) (-4.387764,1.2994437) (-4.3530278,1.0872389) (-4.3208489,1.2834131) (-4.2876582,1.3720945) (-4.2503781,1.3101661) (-4.2070313,1.2071942) (-4.1575079,1.3015566) (-4.1137214,1.241815) (-4.0696568,1.289959) (-4.0182076,1.28432) (-3.9687357,1.4548881) (-3.9230833,1.4763354) (-3.8813162,1.2819638) (-3.83951,1.1919394) (-3.8031518,1.1796494) (-3.7566321,1.2627953) (-3.7060225,1.2306617) (-3.6598446,1.3708563) (-3.6138992,1.4704788) (-3.5560262,1.3516237) (-3.5094876,1.4457195) (-3.4556985,1.2924061) (-3.4007971,.98193139) (-3.3457355,1.5035526) (-3.2601199,1.3396981) (-3.1893497,1.4908383) (-3.0911846,1.3447787) (-3.010812,1.2677956) (-2.9251273,1.2419677) (-2.8525014,1.2686533) (-2.7357662,1.4383082) (-2.6176379,1.4190735) (-2.4987113,1.102927) (-2.400763,1.1122001) (-2.2988307,.96932459) (-2.2035735,1.1181467) (-2.088706,1.0214338) (-1.9823685,.95848733) (-1.8528621,1.4184411) (-1.6747675,1.3666129) (-1.4355257,1.4691561) (-1.1854224,1.6408569) (-.58139652,1.745092) };
\addplot[domain=-8:0, color=black, solid, line width=0.75mm] {.0592996*x+ 1.512051 };
\legend{~Observations,~Linear Fit}
\end{axis}
\end{tikzpicture}
}
\end{subfigure}
\floatfoot{\footnotesize\textsc{Note: ---} The figures display binned scatter plots with 100 hundred markers to depict the relationship between firm-level markdowns and log labor market concentration (in terms of employment-based HHI). Building on estimated Translog production functions, the markdowns were approximated using administrative information on firms' employment, survey information on firms' annual revenue and monthly wage bill (as of June), and yearly information on price levels from Germany's national accounts. Labor markets are pair-wise combinations of 4-digit NACE industries and commuting zones. The results are weighted using cross-sectional survey weights. HHI = Herfindahl-Hirschman Index. NACE = Statistical Nomenclature of Economic Activities in the European Community. Sources: IEB $\plus$ BHP $\plus$ IAB Establishment Panel $\plus$ Destatis, 1999-2017.}
\end{figure}

\vspace*{\fill}
\clearpage
\vspace*{\fill}

\begin{table}[!ht]
\centering
\scalebox{0.90}{
\begin{threeparttable}
\caption{Relationship between Markdowns, Labor Market Concentration, and Minimum Wages}
\label{tab:F2}
\begin{tabular}{L{4cm}C{2.5cm}C{2.5cm}C{2.5cm}C{2.5cm}C{2.5cm}C{2.5cm}C{2.5cm}C{2.5cm}} \hline
&&&&&&&& \\[-0.3cm]
\multirow{6.4}{*}{} & \multirow{6.4}{*}{\shortstack{(1) \\ \textbf{All Sectors\vphantom{/}} \\ Markdown }} & \multirow{6.4}{*}{\shortstack{(2) \\ \textbf{Baseline\vphantom{/}} \\ Markdown }} & \multirow{6.4}{*}{\shortstack{(3) \\ \textbf{Min Wage\vphantom{/}} \\ \textbf{Prevalence\vphantom{/}} \\ Markdown }} & \multirow{6.4}{*}{\shortstack{(4) \\ \textbf{Min Wage\vphantom{/}} \\ \textbf{Level\vphantom{/}} \\ Markdown }} & \multirow{6.4}{*}{\shortstack{(5) \\ \textbf{NACE-3\vphantom{/}} \\ \textbf{Industries\vphantom{/}} \\ Markdown }} & \multirow{6.4}{*}{\shortstack{(6) \\ \textbf{NACE-5\vphantom{/}} \\ \textbf{Industries\vphantom{/}} \\ Markdown }} & \multirow{6.4}{*}{\shortstack{(7) \\ \textbf{NUTS-3\vphantom{/}} \\ \textbf{Regions\vphantom{/}} \\ Markdown }} & \multirow{6.4}{*}{\shortstack{(8) \\ \textbf{Market\vphantom{/}} \\ \textbf{Share\vphantom{/}} \\ Markdown }} \\
&&&&&&&& \\
&&&&&&&& \\
&&&&&&&& \\
&&&&&&&& \\
&&&&&&&& \\[0.3cm] \hline
&&&&&&&& \\[-0.3cm]
\multirow{2.4}{*}{Log HHI} &   \multirow{2.4}{*}{\shortstack{\hphantom{***}0.037***  \\ (0.010)}}   &  \multirow{2.4}{*}{\shortstack{\hphantom{***}0.059***  \\ (0.019)}}  &   &  &   \multirow{2.4}{*}{\shortstack{\hphantom{***}0.059***  \\ (0.017)}}   &  \multirow{2.4}{*}{\shortstack{\hphantom{***}0.103***  \\ (0.017)}} &   \multirow{2.4}{*}{\shortstack{\hphantom{***}0.063***  \\ (0.023)}}   &      \\
&&&&&&&& \\[0.2cm]
\multirow{2.4}{*}{\shortstack[l]{Minimum Wage \\ Prevalence (0=No, 1=Yes)}} &   &   &  \multirow{2.4}{*}{\shortstack{\hphantom{***}-0.251***\hphantom{-}  \\ (0.032)}}  &  &  &  &  & \\
&&&&&&&& \\[0.2cm]
\multirow{2.4}{*}{\shortstack[l]{Minimum Wage \\ Level}} &   &   &   &  \multirow{2.4}{*}{\shortstack{\hphantom{***}-0.025***\hphantom{-} \\ (0.003)}}    &  &  &  &     \\
&&&&&&&& \\[0.2cm]
\multirow{2.4}{*}{Log Market Share} &   &   &   &    &  &  &  &  \multirow{2.4}{*}{\shortstack{\hphantom{***}0.147***  \\ (0.013)}}   \\
&&&&&&&& \\[0.2cm]  \hline
&&&&&&&& \\[-0.4cm]
\multirow{3.4}{*}{\shortstack[l]{Labor Market \\ Definition \\ (Object)}}  & \multirow{3.4}{*}{\shortstack{NACE-4 \\ $\times$ CZ \\ (Employment)}} & \multirow{3.4}{*}{\shortstack{NACE-4 \\ $\times$ CZ \\ (Employment)}} & \multirow{3.4}{*}{\shortstack{NACE-4 \\ $\times$ CZ \\ (Employment)}} & \multirow{3.4}{*}{\shortstack{NACE-4 \\ $\times$ CZ \\ (Employment)}} & \multirow{3.4}{*}{\shortstack{NACE-3 \\ $\times$ CZ \\ (Employment)}} & \multirow{3.4}{*}{\shortstack{NACE-5 \\ $\times$ CZ \\ (Employment)}} & \multirow{3.4}{*}{\shortstack{NACE-4 \\ $\times$ NUTS-3 \\ (Employment)}} & \multirow{3.4}{*}{\shortstack{NACE-4 \\ $\times$ CZ \\ (Employment)}} \\
&&&&&&&&  \\
&&&&&&&& \\
\multirow{3.4}{*}{\shortstack[l]{Sample}}   & \multirow{3.4}{*}{All Sectors} & \multirow{3.4}{*}{\shortstack{Only Minimum \vphantom{/} \\ Wage Sectors }}  & \multirow{3.4}{*}{All Sectors}  & \multirow{3.4}{*}{All Sectors} & \multirow{3.4}{*}{\shortstack{Only Minimum \vphantom{/} \\ Wage Sectors }}  & \multirow{3.4}{*}{\shortstack{Only Minimum \vphantom{/} \\ Wage Sectors }}   & \multirow{3.4}{*}{\shortstack{Only Minimum \vphantom{/} \\ Wage Sectors }} & \multirow{3.4}{*}{\shortstack{Only Minimum \vphantom{/} \\ Wage Sectors }} \\
&&&&&&&& \\
&&&&&&&& \\[0.2cm]
Observations &  77,359          & 14,812          & 77,359          & 77,359       &  14,812          & 14,812          & 14,812          & ~14,812          \\[0.2cm] \hline
\end{tabular}
\begin{tablenotes}[para]
\footnotesize\textsc{Note. ---} The table displays OLS regressions of firm-level markdowns on labor market concentration (in terms of HHI) and the prevalence/level of sectoral minimum wages. Building on estimated Translog production function, the markdowns were calculated using administrative information on firms' employment, survey information on firms' annual revenues and monthly wage bill (as of June), and yearly information on the price levels from Germany's national accounts. Markdowns were trimmed at the 5th and 95th percentile of their distribution. The results are estimated using cross-sectional survey weights. Standard errors (in parentheses) are clustered at the labor market level. CZ = Commuting Zone. HHI = Herfindahl-Hirschman Index. MW = Minimum Wage. NACE-X = X-Digit Statistical Nomenclature of Economic Activities in the European Community. NUTS-3 = 3-Digit Statistical Nomenclature of Territorial Units.  * = p$<$0.10. ** = p$<$0.05. *** = p$<$0.01. Sources: IEB $\plus$ BHP $\plus$ IAB Establishment Panel $\plus$ Destatis, 1999-2017.
\end{tablenotes}
\end{threeparttable}
}
\end{table}

\vspace*{\fill}
\clearpage
\vspace*{\fill}

\begin{table}[!ht]
\centering
\scalebox{0.90}{
\begin{threeparttable}
\caption{Relationship between Markups and Product Market Competition}
\label{tab:F3}
\begin{tabular}{L{4cm}C{5.8cm}C{5.8cm}} \hline
&& \\[-0.3cm]
\multirow{5.4}{*}{} & \multirow{5.4}{*}{\shortstack{(1) \\ \textbf{All Sectors\vphantom{/}} \\ Markup }} & \multirow{5.4}{*}{\shortstack{(2) \\ \textbf{Baseline\vphantom{/}} \\ Markup }}  \\
&& \\
&& \\
&& \\
&&\\[0.3cm] \hline
&& \\[-0.3cm]
\multicolumn{3}{l}{\textbf{Product Market Competition (Reference Category: None)}} \\
\multirow{2.4}{*}{Low} &   \multirow{2.4}{*}{\shortstack{-0.005\hphantom{-}  \\ (0.014)}}   &  \multirow{2.4}{*}{\shortstack{-0.012\hphantom{-}  \\ (0.028)}}    \\
&&  \\
\multirow{2.4}{*}{Medium} &   \multirow{2.4}{*}{\shortstack{\hphantom{*}-0.027*\hphantom{-}  \\ (0.014)}}   &  \multirow{2.4}{*}{\shortstack{-0.032\hphantom{-}  \\ (0.030)}} \\
&&  \\
\multirow{2.4}{*}{High} &   \multirow{2.4}{*}{\shortstack{\hphantom{***}-0.069***\hphantom{-}  \\ (0.015)}}   &  \multirow{2.4}{*}{\shortstack{\hphantom{**}-0.072**\hphantom{-}  \\ (0.030)}}   \\
&& \\[0.2cm] \hline
\multirow{3.4}{*}{\shortstack[l]{Sample}}   & \multirow{3.4}{*}{All Sectors} & \multirow{3.4}{*}{\shortstack{Only Minimum \vphantom{/} \\ Wage Sectors \vphantom{/} }}   \\
&& \\
&& \\[0.2cm]
Years &  2008-2017    & 2008-2017    \\[0.2cm]
Observations &  42,599          & 7,871            \\[0.2cm] \hline
\end{tabular}
\begin{tablenotes}[para]
\footnotesize\textsc{Note. ---} The table displays OLS regressions of estimated markups on binary variables for the intensity of product market competition. Building on estimated Translog production function, the markups were calculated using administrative information on firms' employment, survey information on firms' annual revenues and monthly wage bill (as of June), and yearly information on the price levels from Germany's national accounts. Markups were trimmed at the 5th and 95th percentile of their distribution. From 2008 onwards, the IAB Establishment Panel asks firms to self-assess the strength of product market competition based on a four-dimensional scale: no, low, medium, or high product market competition. The results are weighted using cross-sectional survey weights. Standard errors (in parentheses) are clustered at the firm level. * = p$<$0.10. ** = p$<$0.05. *** = p$<$0.01. Sources: IEB $\plus$ BHP $\plus$ IAB Establishment Panel, 2008-2017.
\end{tablenotes}
\end{threeparttable}
}
\end{table}

\vspace*{\fill}
\clearpage
\vspace*{\fill}

\begin{table}[!ht]
\centering
\scalebox{0.90}{
\begin{threeparttable}
\caption{Relationship between Markups, Labor Market Concentration, and Services}
\label{tab:F4}
\begin{tabular}{L{4cm}C{2.4cm}C{2.4cm}C{2.4cm}C{2.4cm}C{2.4cm}C{2.4cm}C{2.4cm}C{2.4cm}} \hline
&&&&&&&& \\[-0.3cm]
\multirow{6.4}{*}{} & \multirow{6.4}{*}{\shortstack{(1) \\ \textbf{All Sectors\vphantom{/}} \\ Markup }} & \multirow{6.4}{*}{\shortstack{(2) \\ \textbf{Baseline\vphantom{/}} \\ Markup }} & \multirow{6.4}{*}{\shortstack{(3) \\ \textbf{Without\vphantom{/}} \\ \textbf{Services\vphantom{/}} \\ Markup }} & \multirow{6.4}{*}{\shortstack{(4) \\ \textbf{Without\vphantom{/}} \\ \textbf{Services\vphantom{/}} \\ Markup }} & \multirow{6.4}{*}{\shortstack{(5) \\ \textbf{NACE-3\vphantom{/}} \\ \textbf{Industries\vphantom{/}} \\ Markup }} & \multirow{6.4}{*}{\shortstack{(6) \\ \textbf{NACE-5\vphantom{/}} \\ \textbf{Industries\vphantom{/}} \\ Markup }} & \multirow{6.4}{*}{\shortstack{(7) \\ \textbf{NUTS-3\vphantom{/}} \\ \textbf{Regions\vphantom{/}} \\ Markup }} & \multirow{6.4}{*}{\shortstack{(8) \\ \textbf{Market\vphantom{/}} \\ \textbf{Share\vphantom{/}} \\ Markup }}  \\
&&&&&&&& \\
&&&&&&&& \\
&&&&&&&& \\
&&&&&&&& \\
&&&&&&&& \\[0.3cm] \hline
&&&&&&&& \\[-0.3cm]
\multirow{2.4}{*}{Log HHI} &   \multirow{2.4}{*}{\shortstack{\hphantom{*}0.005*  \\ (0.003)}}   &  \multirow{2.4}{*}{\shortstack{-0.006\hphantom{-}  \\ (0.007)}}  & \multirow{2.4}{*}{\shortstack{\hphantom{***}-0.011***\hphantom{-}  \\ (0.004)}}  & \multirow{2.4}{*}{\shortstack{\hphantom{***}-0.021***\hphantom{-}  \\ (0.007)}} &   \multirow{2.4}{*}{\shortstack{-0.010\hphantom{-}  \\ (0.007)}}   &  \multirow{2.4}{*}{\shortstack{\hphantom{**}-0.014**\hphantom{-}  \\ (0.006)}} &   \multirow{2.4}{*}{\shortstack{-0.005\hphantom{-}  \\ (0.007)}}   &     \\
&&&&&&&& \\[0.2cm]
\multirow{2.4}{*}{Log Market Share} &     &    &   &  &    &   &   &  \multirow{2.4}{*}{\shortstack{\hphantom{***}-0.028***\hphantom{-}  \\ (0.005)}}    \\
&&&&&&&& \\[0.2cm]  \hline
&&&&&&&& \\[-0.4cm]
\multirow{3.4}{*}{\shortstack[l]{Labor Market \\ Definition \\ (Object)}}  & \multirow{3.4}{*}{\shortstack{NACE-4 \\ $\times$ CZ \\ (Employment)}} & \multirow{3.4}{*}{\shortstack{NACE-4 \\ $\times$ CZ \\ (Employment)}} & \multirow{3.4}{*}{\shortstack{NACE-4 \\ $\times$ CZ \\ (Employment)}} & \multirow{3.4}{*}{\shortstack{NACE-4 \\ $\times$ CZ \\ (Employment)}} & \multirow{3.4}{*}{\shortstack{NACE-3 \\ $\times$ CZ \\ (Employment)}} & \multirow{3.4}{*}{\shortstack{NACE-5 \\ $\times$ CZ \\ (Employment)}} & \multirow{3.4}{*}{\shortstack{NACE-4 \\ $\times$ NUTS-3 \\ (Employment)}} & \multirow{3.4}{*}{\shortstack{NACE-4 \\ $\times$ CZ \\ (Employment)}} \\
&&&&&&&&  \\
&&&&&&&& \\
\multirow{3.4}{*}{\shortstack[l]{Sample}}   & \multirow{3.4}{*}{All Sectors} & \multirow{3.4}{*}{\shortstack{Only Minimum \vphantom{/} \\ Wage Sectors }}  & \multirow{3.4}{*}{All Sectors}   & \multirow{3.4}{*}{\shortstack{Only Minimum \vphantom{/} \\ Wage Sectors }} & \multirow{3.4}{*}{\shortstack{Only Minimum \vphantom{/} \\ Wage Sectors }}  & \multirow{3.4}{*}{\shortstack{Only Minimum \vphantom{/} \\ Wage Sectors }}   & \multirow{3.4}{*}{\shortstack{Only Minimum \vphantom{/} \\ Wage Sectors }} & \multirow{3.4}{*}{\shortstack{Only Minimum \vphantom{/} \\ Wage Sectors }} \\
&&&&&&&& \\
&&&&&&&& \\[0.2cm]
Without Services &  No    & No    & Yes    & Yes &  No    & No    & No    & No    \\[0.2cm]
Observations &  81,000          & 14,802          & 44,141          & 11,285       &  14,802          & 14,802          & 14,802          & ~14,802          \\[0.2cm] \hline
\end{tabular}
\begin{tablenotes}[para]
\footnotesize\textsc{Note. ---} The table displays OLS regressions of firm-level markups on labor market concentration (in terms of HHI). Building on estimated Translog production function, the markups were calculated using administrative information on firms' employment, survey information on firms' annual revenues and monthly wage bill (as of June), and yearly information on the price levels from Germany's national accounts. Markups were trimmed at the 5th and 95th percentile of their distribution. In Columns (3) and (4), I exclude the service sectors for which product and labor market power is supposed to overlap stronly. The results are estimated using cross-sectional survey weights. Standard errors (in parentheses) are clustered at the labor market level. CZ = Commuting Zone. HHI = Herfindahl-Hirschman Index. MW = Minimum Wage. NACE-X = X-Digit Statistical Nomenclature of Economic Activities in the European Community. NUTS-3 = 3-Digit Statistical Nomenclature of Territorial Units.  * = p$<$0.10. ** = p$<$0.05. *** = p$<$0.01. Sources: IEB $\plus$ BHP $\plus$ IAB Establishment Panel $\plus$ Destatis, 1999-2017.
\end{tablenotes}
\end{threeparttable}
}
\end{table}

\vspace*{\fill}
\clearpage

\end{landscape}

\section{Effects of Labor Market Concentration: Further Evidence}
\label{sec:G}
\setcounter{table}{0} 
\setcounter{figure}{0} 

In this section, I provide a set of robustness checks and heterogeneity analyses to further inspect the effect of labor market concentration on wages and employment. Moreover, I scrutinize the potentially confounding role of common labor supply or demand shocks as well as product market power to facilitate the causal interpretation of the baseline regression results.

\paragraph{Sensitivity and Heterogeneity.} To test the sensitivity of the baseline wage effect, namely -0.020 in Column (4) of Table \ref{tab:1}, I perform additional IV regressions with alternative specifications in Table \ref{tab:G1}, labor market definitions in Table \ref{tab:G2}, and concentration indices in Table \ref{tab:G3}. In these checks, most of the effects remain in close vicinity to the baseline, while some turn out even more pronounced. When omitting industry-by-year fixed effects, the wage effect becomes more negative by an order of magnitude. By contrast, using 3-digit-industry-by-year fixed effects (rather than 2-digit-industry-by-year fixed effects) leaves the effect unaltered. Albeit desirable, including 4-digit-industry-by-year fixed effects would absorb overly much variation because the instrumental variable is essentially defined at this level. In terms of the instrumental variable, the use of leave-one-out HHI (as an alternative to the leave-one-out inverse number of firms) does not change the results. In terms of the labor market definition, 3-digit NACE industries and 3-digit NUTS regions deliver more negative effects whereas the results hardly change for 5-digit NACE industries or the flow-based HHI formulation for 4-digit NACE industries. HHI values, that refer to new hires or the wage bill (rather than employment) do not produce markedly different results. The same holds true for specifications that employ the Rosenbluth Index, the Inverse Number of Firms, the Exponential Index, or the underlying market shares. The 1-Firm Concentration Ratio also delivers a negative but markedly stronger negative effect on wages.

Next, I carry out separate regressions for West German and East German firms in Table \ref{tab:G4}. Importantly, the negative effect of labor market on wages manifests in both parts of the country. In Table \ref{tab:G5}, I explore whether the effect of labor market concentration varies at different percentiles of the wage distribution in a firm. The results indicate that a 100 percent increase in HHI reduces the 25th percentile by 0.8 percent, the median by 1.8 percent and the 75th percentile by 2.8 percent. Hence, labor market concentration enables firms to push down large portions of the wage distribution, with greater monopsonistic conduct at the top than at the bottom end of the distribution. This heterogeneity mirrors related evidence that non-routine cognitive workers in Germany are subject to a higher degree of monopsony power than non-routine manual or routine workers \citep{BachmannEtAl2022}.

I examine the robustness of the baseline employment effect, namely -0.034 in Column (4) of Table \ref{tab:2}, with respect to alternative specifications in Table \ref{tab:G6}, labor market definitions in Table \ref{tab:G7}, and concentration indices in Table \ref{tab:G8}. Again, the effects turn out to be mostly similar to the baseline, occasionally even larger.

In Table \ref{tab:G9}, regressions show that the negative effect of labor market concentration on employment is evident both for West and East German firms. As my baseline effects always refer to regular full-time workers, Table \ref{tab:G10} provides employment effects for other groups of workers for whom no interpretaple wage measure is available in the data. Importantly, I also arrive at a similarly negative effect of labor market concentration on overall employment, which is the sum of regular full-time, regular part-time, and marginal part-time workers. Moreover, the regressions point out that more concentrated labor markets come along with lower employment of marginal part-time workers. By contrast, I report a zero HHI effect on employment of regular part-time workers. A leading explanation for this result is that, when monopsonists curtail their demanded volume of work along the intensive margin, some regular full-time workers work fewer hours and, thus, become regular part-time workers, seemingly counterbalancing a genuinely negative HHI effect on regular part-time employment.

\paragraph{Causal Interpretation.} The leave-one-out instrument, as given by Equation (\ref{eq:4}), is not fully exogenous when labor supply or labor demand shocks are correlated across commuting zones (e.g., due to common technology shocks). Specifically, the exogeneity assumption is violated when the instrument exerts an additional direct effect, $\Phi\neq0$, on the outcome variable in the second-stage regression (i.e., other than through labor market concentration). When allowing for instrument endogeneity, the second-stage regression takes the following form:
\begin{equation}
ln\,Y_{jizt} \,=\,  \theta \cdot \widehat{ln\,H\!H\!I}_{izt} \,+\, \Phi \cdot \overline{ln\,I\!N\!F}^{\,-z}_{it}  \,+\, \delta_{j}  \,+\, \zeta_{zt} \,+\, \zeta_{it} \,+\, \varepsilon_{jizt}
\end{equation}
To assess the magnitude of the potential bias, I formulate a range of values for this direct effect between zero (exogeneity) and the estimated coefficient $\Phi^{min}$ of the following reduced-form regression of the outcome variable on the instrument itself:
\begin{equation}
ln\,Y_{jizt} \,=\,  \Phi^{min} \cdot \overline{ln\,I\!N\!F}^{\,-z}_{it}  \,+\, \delta_{j}  \,+\, \zeta_{zt}  \,+\, \zeta_{it} \,+\, \varepsilon_{jizt}
\end{equation}
Given this range, I apply the plausibly exogenous regression method \citep{ConleyEtAl2012} to derive bounds for the causal effect of labor market concentration on the outcomes of interest.

Table \ref{tab:G11} systematically delivers the results when applying the plausibly exogenous regression method to the baseline wage and employment regression, respectively. In terms of wages, the reduced-form regression yields a coefficient of -0.0167. This coefficient reflects the maximum of instrument endogeneity that can enter the second-stage regression of the wage variable on the first-stage variation in labor market concentration. Under the assumption that the instrument's direct effect on wages in the second stage ranges between zero (full exogeneity) and -0.0167 (reduced-form effect), the bounds for the second-stage effect of labor market concentration on wages range between -0.027 and 0.007. Thus, the causal effect of labor market concentration on wages is negative provided that the direct effect of the instrument on wages in the second stage is greater than -0.0115, or smaller than 68.9 (=0.0115/0.0167) percent of the reduced-form effect. Hence, the negative effect of labor market concentration on wages remains even for a large degree of potential instrument endogeneity.

In terms of employment, the reduced-form regression for employment delivers a coefficient of -0.0278. The plausible interval for the second-stage effect of HHI on employment ranges between -0.051 and 0.017. Accordingly, the true effect of labor market concentration on employment is negative provided that the size of the instrument's direct effect on employment in the second stage is smaller than 50.7 (=0.0141/0.0278) percent of the reduced-form effect. Again, the negative effect of labor market concentration on employment holds for large potential violations of exogeneity of the instrument.

As discussed in Section \ref{sec:5}, industry-based measures of local labor market concentration might pick up product market concentration unless the latter has a national rather than a local dimension. Under imperfect competition in product markets, firms with monopoly power will lower output and employment \citep{MarinescuEtAl2021} and, consequently, market wages might decline even in the absence of monopsony power. These lower wages are set competitively and remain tied to marginal productivity along an infinitely elastic labor supply curve to the firm. In the presence of rent sharing, however, monopolists will pass on parts of their profits to their employees, thus paying higher wages \citep{QiuSojourner2023}. Taken together, the effect of product market power on wages is theoretically ambiguous whereas the effect on employment is clearly negative. Reassuringly, the analysis in Appendix \ref{sec:F} has shown that labor market concentration positively correlates with markdowns and that such a positive relationship with price-cost markups is not evident, ruling out a confounding impact of product market power. As an additional check, Table \ref{tab:G12} re-runs the regressions without the service sectors where product markets tend to have more of a local nature and thus might overlap with labor markets. In line with the absence of a positive relationship between HHI values and markups in these sectors, the effects of labor market concentration on wages and employment change hardly change.

By and large, I document robust evidence that firms employ fewer workers at lower wage levels when operating in more concentrated labor markets. The finding that higher labor market concentration reduces both wages and employment corroborates the prediction of monopsony theory in a sense that employers with labor market power suppress wages along a positively-sloped labor supply curve to the firm. Division of the baseline employment elasticity by the respective wage elasticity yields a wage elasticity of labor supply of 1.7 (=0.034/0.020). Albeit conceptually different, this value appears plausible in a sense that it lies in the ballpark of estimated wage elasticities of labor supply to the single firm from dynamic monopsony models on the German labor market.

\begin{landscape}

\vspace*{\fill}

\begin{table}[!ht]
\centering
\scalebox{0.90}{
\begin{threeparttable}
\caption{Effects of Labor Market Concentration on Wages by Specification}
\label{tab:G1}
\begin{tabular}{L{4cm}C{3.75cm}C{3.75cm}C{3.75cm}} \hline
&&& \\[-0.3cm]
\multirow{7.4}{*}{}  & \multirow{7.4}{*}{\shortstack{(1) \\ \textbf{Without \vphantom{/}} \\  \textbf{NACE-2-by-Year \vphantom{/}} \\ \textbf{Fixed Effects \vphantom{/}} \\ Log \\ Mean Daily\vphantom{/} \\ Wages }} & \multirow{7.4}{*}{\shortstack{(2) \\ \textbf{With \vphantom{/}} \\ \textbf{NACE-3-by-Year \vphantom{/}} \\ \textbf{Fixed Effects \vphantom{/}}  \\ Log \\ Mean Daily\vphantom{/} \\ Wages }} & \multirow{7.4}{*}{\shortstack{(3) \\ \textbf{Other Instrument \vphantom{/}} \\ Log \\ Mean Daily\vphantom{/} \\ Wages }} \\
&&& \\&&& \\&&& \\&&& \\&&& \\
&&& \\[0.3cm] \hline
&&& \\[-0.3cm]
\multirow{2.4}{*}{Log HHI} &  \multirow{2.4}{*}{\shortstack{\hphantom{***}-0.046***\hphantom{-}  \\ (0.006)}}  & \multirow{2.4}{*}{\shortstack{\hphantom{**}-0.019**\hphantom{-}  \\ (0.008)}} & \multirow{2.4}{*}{\shortstack{\hphantom{***}-0.023***\hphantom{-}  \\ (0.005)}}   \\
&&& \\[0.2cm] 
\multirow{2.4}{*}{\shortstack{Fixed Effects}}  & \multirow{2.4}{*}{\shortstack{Firm \\ CZ $\times$ Year}} &  \multirow{2.4}{*}{\shortstack{Firm \\ CZ $\times$ Year \\ NACE-3 $\times$ Year}} & \multirow{2.4}{*}{\shortstack{Firm \\ CZ $\times$ Year \\ NACE-2 $\times$ Year}}  \\
&&& \\[0.4cm] \hline
&&& \\[-0.2cm]
Instrument  & $ \overline{\text{Log}\,\,\text{INF}} \,\,\backslash\, \{c\} $ & $ \overline{\text{Log}\,\,\text{INF}} \,\,\backslash\, \{c\} $ & $ \overline{\text{Log}\,\,\text{HHI}} \,\,\backslash\, \{c\} $   \\
\multirow{3.4}{*}{\shortstack[l]{Labor Market \\ Definition \\ (Object)}}  & \multirow{3.4}{*}{\shortstack{NACE-4 \\ $\times$ CZ \\ (Employment)}} & \multirow{3.4}{*}{\shortstack{NACE-4 \\ $\times$ CZ \\ (Employment)}} & \multirow{3.4}{*}{\shortstack{NACE-4 \\ $\times$ CZ \\ (Employment)}}  \\
&&& \\
&&& \\[0.2cm] 
Observations &  2,139,426       & 2,139,157       & 2,139,400             \\
\multirow{2.4}{*}{\shortstack{First-Stage Effect}} &  \multirow{2.4}{*}{\shortstack{\hphantom{***}0.820***  \\ (0.039)}}  & \multirow{2.4}{*}{\shortstack{\hphantom{***}0.707***  \\ (0.044)}}  &  \multirow{2.4}{*}{\shortstack{\hphantom{***}0.928***  \\ (0.054)}}   \\
&&& \\[0.2cm] 
F: Instrument & 430.8 & 261.7 & ~295.8 \\[0.2cm]  \hline
\end{tabular}
\begin{tablenotes}[para]
\footnotesize\textsc{Note. ---} The table displays fixed effects regressions of log daily wages (in terms of firm-level averages of regular full-time workers) on log labor market concentration (in terms of HHI). The instrumental variable refers to the leave-one-out average of the log inverse number of firms (or log HHI) in all other commuting zones but for the same industry and time period. Standard errors (in parentheses) are clustered at the labor market level. CZ = Commuting Zone. HHI = Herfindahl-Hirschman Index. INF = Inverse Number of Firms. NACE-X = X-Digit Statistical Nomenclature of Economic Activities in the European Community. * = p$<$0.10. ** = p$<$0.05. *** = p$<$0.01. Sources: IEB $\plus$ BHP, 1999-2014.
\end{tablenotes}
\end{threeparttable}
}
\end{table}

\vspace*{\fill}
\clearpage

\begin{table}[!ht]
\centering
\scalebox{0.90}{
\begin{threeparttable}
\caption{Effects of Labor Market Concentration on Wages by Labor Market Definition}
\label{tab:G2}
\begin{tabular}{L{4cm}C{2.7cm}C{2.7cm}C{2.7cm}C{2.7cm}C{2.7cm}C{2.7cm}} \hline
&&&&&& \\[-0.3cm]
\multirow{7.4}{*}{}  & \multirow{7.4}{*}{\shortstack{(1) \\ \textbf{NACE-3 \vphantom{/}} \\ \textbf{Industries \vphantom{/}} \\ Log \\ Mean Daily\vphantom{/} \\ Wages }} & \multirow{7.4}{*}{\shortstack{(2) \\ \textbf{NACE-5 \vphantom{/}} \\ \textbf{Industries \vphantom{/}} \\ Log \\ Mean Daily\vphantom{/} \\ Wages }} & \multirow{7.4}{*}{\shortstack{(3) \\ \textbf{Flow-Adjusted \vphantom{/}} \\ \textbf{NACE-4 \vphantom{/}} \\ \textbf{Industries \vphantom{/}} \\ Log \\ Mean Daily\vphantom{/} \\ Wages }} & \multirow{7.4}{*}{\shortstack{(4) \\ \textbf{NUTS-3 \vphantom{/}} \\ \textbf{Regions \vphantom{/}} \\ Log \\ Mean Daily\vphantom{/} \\ Wages }} & \multirow{7.4}{*}{\shortstack{(5) \\ \textbf{HHI \vphantom{/}} \\ \textbf{Based on \vphantom{/}} \\ \textbf{Hires \vphantom{/}}  \\ Log \\ Mean Daily\vphantom{/} \\ Wages }} & \multirow{7.4}{*}{\shortstack{(6) \\ \textbf{HHI \vphantom{/}} \\ \textbf{Based on \vphantom{/}} \\ \textbf{Wage Bill \vphantom{/}} \\ Log \\ Mean Daily\vphantom{/} \\ Wages }} \\
&&&&&& \\
&&&&&& \\
&&&&&& \\
&&&&&& \\
&&&&&& \\
&&&&&& \\[0.3cm] \hline
&&&&&& \\[-0.3cm]
\multirow{2.4}{*}{Log HHI} &  \multirow{2.4}{*}{\shortstack{\hphantom{***}-0.038***\hphantom{-}  \\ (0.007)}}  & \multirow{2.4}{*}{\shortstack{\hphantom{***}-0.019***\hphantom{-}  \\ (0.004)}} & & \multirow{2.4}{*}{\shortstack{\hphantom{***}-0.034***\hphantom{-}  \\ (0.004)}} & \multirow{2.4}{*}{\shortstack{\hphantom{***}-0.027***\hphantom{-}  \\ (0.004)}} & \multirow{2.4}{*}{\shortstack{\hphantom{***}-0.023***\hphantom{-}  \\ (0.005)}}  \\
&&&&&& \\[0.2cm]
\multirow{2.4}{*}{Log Adjusted HHI} &    &    &   \multirow{2.4}{*}{\shortstack{\hphantom{***}-0.023***\hphantom{-}  \\ (0.005)}}  &  &  &   \\
&&&&&& \\[0.2cm]
\multirow{2.4}{*}{\shortstack{Fixed Effects}}  & \multirow{2.4}{*}{\shortstack{Firm \\ CZ $\times$ Year \\ NACE-2 $\times$ Year}} &  \multirow{2.4}{*}{\shortstack{Firm \\ CZ $\times$ Year \\ NACE-2 $\times$ Year}} &  \multirow{2.4}{*}{\shortstack{Firm \\ CZ $\times$ Year \\ NACE-2 $\times$ Year}} & \multirow{2.4}{*}{\shortstack{Firm \\ NUTS-3 $\times$ Year \\ NACE-2 $\times$ Year}} &  \multirow{2.4}{*}{\shortstack{Firm \\ CZ $\times$ Year \\ NACE-2 $\times$ Year}} &  \multirow{2.4}{*}{\shortstack{Firm \\ CZ $\times$ Year \\ NACE-2 $\times$ Year}} \\
&&&&&& \\[0.4cm] \hline
&&&&&& \\[-0.2cm]
Instrument  & $ \overline{\text{Log}\,\,\text{INF}} \,\,\backslash\, \{c\} $ & $ \overline{\text{Log}\,\,\text{INF}} \,\,\backslash\, \{c\} $  & $ \overline{\text{Log}\,\,\text{INF}} \,\,\backslash\, \{c\} $ & $ \overline{\text{Log}\,\,\text{INF}} \,\,\backslash\, \{c\} $ & $ \overline{\text{Log}\,\,\text{INF}} \,\,\backslash\, \{c\} $  & $ \overline{\text{Log}\,\,\text{INF}} \,\,\backslash\, \{c\} $ \\
\multirow{3.4}{*}{\shortstack[l]{Labor Market \\ Definition \\ (Object)}}  & \multirow{3.4}{*}{\shortstack{NACE-3 \\ $\times$ CZ \\ (Employment)}} & \multirow{3.4}{*}{\shortstack{NACE-5 \\ $\times$ CZ \\ (Employment)}} & \multirow{3.4}{*}{\shortstack{NACE-4 \\ $\times$ CZ \\ (Employment)}} & \multirow{3.4}{*}{\shortstack{NACE-4 \\ $\times$ NUTS-3 \\ (Employment)}} & \multirow{3.4}{*}{\shortstack{NACE-4 \\ $\times$ CZ \\ (Hires)}}  & \multirow{3.4}{*}{\shortstack{NACE-4 \\ $\times$ CZ \\ (Wage Bill)}} \\
&&&&&&  \\
&&&&&&  \\
\multirow{2.4}{*}{Sample}  & \multirow{2.4}{*}{1999-2014} & \multirow{2.4}{*}{1999-2014} & \multirow{2.4}{*}{1999-2014} & \multirow{2.4}{*}{1999-2014}  & \multirow{2.4}{*}{2000-2014} & \multirow{2.4}{*}{1999-2014}  \\
&&&&&& \\[0.2cm] 
Observations &  2,139,400       & 2,139,400       & 2,139,400       & 2,139,400    & 1,958,594    & ~2,139,400        \\
\multirow{2.4}{*}{\shortstack{First-Stage Effect}} &  \multirow{2.4}{*}{\shortstack{\hphantom{***}0.966***  \\ (0.076)}}  &  \multirow{2.4}{*}{\shortstack{\hphantom{***}0.814***  \\ (0.035)}}  & \multirow{2.4}{*}{\shortstack{\hphantom{***}0.720***  \\ (0.042)}}  & \multirow{2.4}{*}{\shortstack{\hphantom{***}0.753***  \\ (0.015)}}    & \multirow{2.4}{*}{\shortstack{\hphantom{***}0.795***  \\ (0.034)}}  & \multirow{2.4}{*}{\shortstack{\hphantom{***}0.722***  \\ (0.042)}}   \\
&&&&&& \\[0.2cm] 
F Statistic & 159.9 & 543.6 & 287.6 & 2474.8 & 563.1 & ~291.8 \\[0.2cm]  \hline
\end{tabular}
\begin{tablenotes}[para]
\footnotesize\textsc{Note. ---} The table displays fixed effects regressions of log daily wages (in terms of firm-level averages of regular full-time workers) on log labor market concentration (in terms of HHI). The instrumental variable refers to the leave-one-out average of the log inverse number of firms in all other regions but for the same industry and time period. Standard errors (in parentheses) are clustered at the labor market level. CZ = Commuting Zone. HHI = Herfindahl-Hirschman Index. INF = Inverse Number of Firms. NACE-X = X-Digit Statistical Nomenclature of Economic Activities in the European Community. NUTS-X = X-Digit Statistical Nomenclature of Territorial Units. * = p$<$0.10. ** = p$<$0.05. *** = p$<$0.01. Sources: IEB $\plus$ BHP, 1999-2014.
\end{tablenotes}
\end{threeparttable}
}
\end{table}

\clearpage

\begin{table}[!ht]
\centering
\scalebox{0.90}{
\begin{threeparttable}
\caption{Effects of Labor Market Concentration on Wages by Concentration Measure}
\label{tab:G3}
\begin{tabular}{L{4cm}C{2.7cm}C{2.7cm}C{2.7cm}C{2.7cm}C{2.7cm}} \hline
&&&&& \\[-0.3cm]
\multirow{7.4}{*}{} & \multirow{7.4}{*}{\shortstack{(1) \\ \textbf{Rosenbluth\vphantom{/}} \\ \textbf{Index\vphantom{/}} \\ Log \\ Mean Daily\vphantom{/} \\ Wages }} & \multirow{7.4}{*}{\shortstack{(2) \\ \textbf{1-Firm\vphantom{/}} \\ \textbf{Concentration\vphantom{/}} \\ \textbf{Ratio\vphantom{/}} \\ Log \\ Mean Daily\vphantom{/} \\ Wages }} & \multirow{7.4}{*}{\shortstack{(3) \\ \textbf{Inverse\vphantom{/}} \\ \textbf{Number\vphantom{/}} \\ \textbf{of Firms\vphantom{/}} \\ Log \\ Mean Daily\vphantom{/} \\ Wages }} & \multirow{7.4}{*}{\shortstack{(4) \\ \textbf{Exponential\vphantom{/}} \\ \textbf{Index\vphantom{/}} \\ Log \\ Mean Daily\vphantom{/} \\ Wages }} & \multirow{7.4}{*}{\shortstack{(5) \\ \textbf{Market\vphantom{/}} \\ \textbf{Share\vphantom{/}} \\ Log \\ Mean Daily\vphantom{/} \\ Wages }}  \\
&&&&& \\
&&&&& \\
&&&&& \\
&&&&& \\
&&&&& \\
&&&&& \\[0.3cm] \hline
&&&&& \\[-0.3cm]
\multirow{2.4}{*}{Log RBI}  & \multirow{2.4}{*}{\shortstack{\hphantom{***}-0.019***\hphantom{-} \\ (0.004)}}   &   &    &    &     \\
&&&&&  \\
\multirow{2.4}{*}{Log CR1}  &       & \multirow{2.4}{*}{\shortstack{\hphantom{***}-0.030***\hphantom{-} \\ (0.006)}}       &    &    &     \\
&&&&&  \\
\multirow{2.4}{*}{Log INF}  &       &        & \multirow{2.4}{*}{\shortstack{\hphantom{***}-0.018***\hphantom{-} \\ (0.004)}}    &   &     \\
&&&&&  \\
\multirow{2.4}{*}{Log EXP}  &       &        &       &  \multirow{2.4}{*}{\shortstack{\hphantom{***}-0.019***\hphantom{-} \\ (0.004)}}   &     \\
&&&&&  \\
\multirow{2.4}{*}{Log Market Share}  &       &       &       &   &  \multirow{2.4}{*}{\shortstack{\hphantom{***}-0.021***\hphantom{-} \\ (0.004)}} \\
&&&&&  \\[0.2cm] 
\multirow{2.4}{*}{\shortstack{Fixed Effects}}  & \multirow{2.4}{*}{\shortstack{Firm \\ CZ $\times$ Year \\ NACE-2 $\times$ Year}} &  \multirow{2.4}{*}{\shortstack{Firm \\ CZ $\times$ Year \\ NACE-2 $\times$ Year}} & \multirow{2.4}{*}{\shortstack{Firm \\ CZ $\times$ Year \\ NACE-2 $\times$ Year}} &  \multirow{2.4}{*}{\shortstack{Firm \\ CZ $\times$ Year\\ NACE-2 $\times$ Year}} &  \multirow{2.4}{*}{\shortstack{Firm \\ CZ $\times$ Year\\ NACE-2 $\times$ Year}}  \\
&&&&& \\[0.4cm] \hline
&&&&& \\[-0.2cm]
Instrument  & $ \overline{\text{Log}\,\,\text{INF}} \,\,\backslash\, \{c\} $ & $ \overline{\text{Log}\,\,\text{INF}} \,\,\backslash\, \{c\} $ & $ \overline{\text{Log}\,\,\text{INF}} \,\,\backslash\, \{c\} $ & $ \overline{\text{Log}\,\,\text{INF}} \,\,\backslash\, \{c\} $  & $ \overline{\text{Log}\,\,\text{INF}} \,\,\backslash\, \{c\} $    \\[0.2cm]
Observations &  2,139,400       & 2,139,400       & 2,139,400       & 2,139,400       & ~2,139,400       \\
\multirow{2.4}{*}{\shortstack{First-Stage Effect}} &  \multirow{2.4}{*}{\shortstack{\hphantom{***}0.889***  \\ (0.017)}} & \multirow{2.4}{*}{\shortstack{\hphantom{***}0.564***  \\ (0.050)}} & \multirow{2.4}{*}{\shortstack{\hphantom{***}0.920***  \\ (0.016)}} & \multirow{2.4}{*}{\shortstack{\hphantom{***}0.889***  \\ (0.019)}} & \multirow{2.4}{*}{\shortstack{\hphantom{***}0.801***  \\ (0.017)}}  \\
&&&&& \\[0.2cm] 
F Statistic & 2793.0 & 126.5 & 3425.7 & 2185.7  & ~2233.2 \\[0.2cm]  \hline
\end{tabular}
\begin{tablenotes}[para]
\footnotesize\textsc{Note. ---} The table displays fixed effects regressions of log daily wages (in terms of firm-level averages of regular full-time workers) on log labor market concentration. The instrumental variable refers to the leave-one-out average of the log inverse number of firms in all other commuting zones but for the same industry and time period. Labor markets are combinations of 4-digit NACE industries and commuting zone. Standard errors (in parentheses) are clustered at the labor market level. CR1 = 1-Firm Concentration Ratio. CZ = Commuting Zone. EXP = Exponential Index. INF = Inverse Number of Firms. NACE-X = X-Digit Statistical Nomenclature of Economic Activities in the European Community. RBI = Rosenbluth Index. * = p$<$0.10. ** = p$<$0.05. *** = p$<$0.01. Sources: IEB $\plus$ BHP, 1999-2014.
\end{tablenotes}
\end{threeparttable}
}
\end{table}

\clearpage
\vspace*{\fill}

\begin{table}[!ht]
\centering
\scalebox{0.90}{
\begin{threeparttable}
\caption{Effects of Labor Market Concentration on Wages by Territory}
\label{tab:G4}
\begin{tabular}{L{4cm}C{5.8cm}C{5.8cm}} \hline
&& \\[-0.3cm]
\multirow{6.4}{*}{} & \multirow{6.4}{*}{\shortstack{(1) \\ \textbf{West\vphantom{/}} \\ \textbf{Germany\vphantom{/}} \\ Log \\ Mean Daily\vphantom{/} \\ Wages }} & \multirow{6.4}{*}{\shortstack{(2) \\ \textbf{East\vphantom{/}} \\ \textbf{Germany\vphantom{/}} \\ Log \\ Mean Daily\vphantom{/} \\ Wages }}  \\
&& \\
&& \\
&& \\
&& \\
&& \\[0.2cm] \hline
&& \\[-0.3cm]
\multirow{2.4}{*}{Log HHI} &  \multirow{2.4}{*}{\shortstack{\hphantom{***}-0.024***\hphantom{-}  \\ (0.005)}}  & \multirow{2.4}{*}{\shortstack{\hphantom{**}-0.015**\hphantom{-}  \\ (0.007)}}   \\
&& \\[0.2cm] 
\multirow{2.4}{*}{\shortstack{Fixed Effects}}  & \multirow{2.4}{*}{\shortstack{Firm \\ CZ $\times$ Year \\ NACE-2 $\times$ Year}} &  \multirow{2.4}{*}{\shortstack{Firm \\ CZ $\times$ Year \\ NACE-2 $\times$ Year}}  \\
&& \\[0.4cm] \hline
&& \\[-0.2cm]
Instrument  & $ \overline{\text{Log}\,\,\text{INF}} \,\,\backslash\, \{c\} $ & $ \overline{\text{Log}\,\,\text{INF}} \,\,\backslash\, \{c\} $   \\
\multirow{3.4}{*}{\shortstack[l]{Labor Market \\ Definition \\ (Object)}}  & \multirow{3.4}{*}{\shortstack{NACE-4 \\ $\times$ CZ \\ (Employment)}} & \multirow{3.4}{*}{\shortstack{NACE-4 \\ $\times$ CZ \\ (Employment)}}  \\
&& \\
&& \\
\multirow{2.4}{*}{Sample}  & \multirow{2.4}{*}{West Germany} & \multirow{2.4}{*}{\shortstack{East Germany \\[-0.1cm] Berlin}}  \\
&& \\[0.2cm] 
Observations &  1,676,435       & ~462,850            \\
\multirow{2.4}{*}{\shortstack{First-Stage Effect}} &  \multirow{2.4}{*}{\shortstack{\hphantom{***}0.876***  \\ (0.053)}}  & \multirow{2.4}{*}{\shortstack{\hphantom{***}0.676***  \\ (0.033)}}   \\
&& \\[0.2cm] 
F Statistic & 274.1 & ~412.7   \\[0.2cm]  \hline
\end{tabular}
\begin{tablenotes}[para]
\footnotesize\textsc{Note. ---} The table displays fixed effects regressions of log daily wages (in terms of firm-level averages of regular full-time workers) on log labor market concentration (in terms of HHI). The instrumental variable refers to the leave-one-out average of the log inverse number of firms in all other commuting zones but for the same industry and time period. Standard errors (in parentheses) are clustered at the labor market level. CZ = Commuting Zone. HHI = Herfindahl-Hirschman Index. INF = Inverse Number of Firms. NACE-X = X-Digit Statistical Nomenclature of Economic Activities in the European Community. * = p$<$0.10. ** = p$<$0.05. *** = p$<$0.01. Sources: IEB $\plus$ BHP, 1999-2014.
\end{tablenotes}
\end{threeparttable}
}
\end{table}

\vspace*{\fill}
\clearpage
\vspace*{\fill}

\begin{table}[!ht]
\centering
\scalebox{0.90}{
\begin{threeparttable}
\caption{Effects of Labor Market Concentration on Wages by Percentile}
\label{tab:G5}
\begin{tabular}{L{4cm}C{3.75cm}C{3.75cm}C{3.75cm}} \hline
&&& \\[-0.3cm]
\multirow{6.4}{*}{} & \multirow{6.4}{*}{\shortstack{(1) \\ \textbf{25th\vphantom{/}} \\ \textbf{Percentile\vphantom{/}} \\ Log \\ P25 Daily\vphantom{/} \\ Wages }} & \multirow{6.4}{*}{\shortstack{(2) \\ \textbf{Median\vphantom{/}} \\ Log \\ P50 Daily\vphantom{/} \\ Wages }} & \multirow{6.4}{*}{\shortstack{(3) \\ \textbf{75th\vphantom{/}} \\ \textbf{Percentile\vphantom{/}} \\ Log \\ P75 Daily\vphantom{/} \\ Wages }}  \\
&&& \\
&&& \\
&&& \\
&&& \\
&&& \\[0.3cm] \hline
&&& \\[-0.3cm]
\multirow{2.4}{*}{Log HHI} &  \multirow{2.4}{*}{\shortstack{\hphantom{**}-0.008**\hphantom{-}  \\ (0.004)}}  & \multirow{2.4}{*}{\shortstack{\hphantom{***}-0.018***\hphantom{-}  \\ (0.004)}} & \multirow{2.4}{*}{\shortstack{\hphantom{***}-0.028***\hphantom{-}  \\ (0.005)}}   \\
&&& \\[0.2cm] 
\multirow{2.4}{*}{\shortstack{Fixed Effects}}  & \multirow{2.4}{*}{\shortstack{Firm \\ CZ $\times$ Year \\ NACE-2 $\times$ Year}} &  \multirow{2.4}{*}{\shortstack{Firm \\ CZ $\times$ Year \\ NACE-2 $\times$ Year}} & \multirow{2.4}{*}{\shortstack{Firm \\ CZ $\times$ Year \\ NACE-2 $\times$ Year}}  \\
&&& \\[0.4cm] \hline
&&& \\[-0.2cm]
Instrument  & $ \overline{\text{Log}\,\,\text{INF}} \,\,\backslash\, \{c\} $ & $ \overline{\text{Log}\,\,\text{INF}} \,\,\backslash\, \{c\} $ & $ \overline{\text{Log}\,\,\text{INF}} \,\,\backslash\, \{c\} $   \\
\multirow{3.4}{*}{\shortstack[l]{Labor Market \\ Definition \\ (Object)}}  & \multirow{3.4}{*}{\shortstack{NACE-4 \\ $\times$ CZ \\ (Employment)}} & \multirow{3.4}{*}{\shortstack{NACE-4 \\ $\times$ CZ \\ (Employment)}} & \multirow{3.4}{*}{\shortstack{NACE-4 \\ $\times$ CZ \\ (Employment)}}  \\
&&& \\
&&& \\[0.2cm] 
Observations &  2,139,400       & 2,139,400       & ~2,139,400           \\
\multirow{2.4}{*}{\shortstack{First-Stage Effect}} & \multirow{2.4}{*}{\shortstack{\hphantom{***}0.820***  \\ (0.042)}}  &  \multirow{2.4}{*}{\shortstack{\hphantom{***}0.820***  \\ (0.042)}} & \multirow{2.4}{*}{\shortstack{\hphantom{***}0.820***  \\ (0.042)}}   \\
&&& \\[0.2cm] 
F Statistic & 387.0 & 387.0 & ~387.0  \\[0.2cm]  \hline
\end{tabular}
\begin{tablenotes}[para]
\footnotesize\textsc{Note. ---} The table displays fixed effects regressions of log daily wages (in terms of firm-level percentiles of regular full-time workers) on log labor market concentration (in terms of HHI). The instrumental variable refers to the leave-one-out average of the log inverse number of firms in all other commuting zones but for the same industry and time period. Standard errors (in parentheses) are clustered at the labor market level. CZ = Commuting Zone. HHI = Herfindahl-Hirschman Index. INF = Inverse Number of Firms. NACE-X = X-Digit Statistical Nomenclature of Economic Activities in the European Community. PX = Xth Percentile. * = p$<$0.10. ** = p$<$0.05. *** = p$<$0.01. Sources: IEB $\plus$ BHP, 1999-2014.
\end{tablenotes}
\end{threeparttable}
}
\end{table}

\vspace*{\fill}
\clearpage
\vspace*{\fill}

\begin{table}[!ht]
\centering
\scalebox{0.90}{
\begin{threeparttable}
\caption{Effects of Labor Market Concentration on Employment by Specification}
\label{tab:G6}
\begin{tabular}{L{4cm}C{3.75cm}C{3.75cm}C{3.75cm}} \hline
&&& \\[-0.3cm]
\multirow{7.4}{*}{}  & \multirow{7.4}{*}{\shortstack{(1) \\ \textbf{Without \vphantom{/}} \\  \textbf{NACE-2-by-Year \vphantom{/}} \\ \textbf{Fixed Effects \vphantom{/}} \\ Log \\ Regular FT\vphantom{/} \\ Employment }} & \multirow{7.4}{*}{\shortstack{(2) \\ \textbf{With \vphantom{/}} \\ \textbf{NACE-3-by-Year \vphantom{/}} \\ \textbf{Fixed Effects \vphantom{/}}  \\ Log \\ Regular FT\vphantom{/} \\ Employment }} & \multirow{7.4}{*}{\shortstack{(3) \\ \textbf{Other Instrument \vphantom{/}} \\ Log \\ Regular FT\vphantom{/} \\ Employment }} \\
&&& \\&&& \\&&& \\&&& \\&&& \\&&& \\[0.3cm] \hline
&&& \\[-0.3cm]
\multirow{2.4}{*}{Log HHI} &  \multirow{2.4}{*}{\shortstack{\hphantom{***}-0.156***\hphantom{-}  \\ (0.018)}}  & \multirow{2.4}{*}{\shortstack{\hphantom{*}-0.034*\hphantom{-}  \\ (0.018)}} & \multirow{2.4}{*}{\shortstack{\hphantom{***}-0.035***\hphantom{-}  \\ (0.013)}}   \\
&&& \\[0.2cm] 
\multirow{2.4}{*}{\shortstack{Fixed Effects}}  & \multirow{2.4}{*}{\shortstack{Firm \\ CZ $\times$ Year}} &  \multirow{2.4}{*}{\shortstack{Firm \\ CZ $\times$ Year \\ NACE-3 $\times$ Year}} & \multirow{2.4}{*}{\shortstack{Firm \\ CZ $\times$ Year \\ NACE-2 $\times$ Year}}  \\
&&& \\[0.4cm] \hline
&&& \\[-0.2cm]
Instrument  & $ \overline{\text{Log}\,\,\text{INF}} \,\,\backslash\, \{c\} $ & $ \overline{\text{Log}\,\,\text{INF}} \,\,\backslash\, \{c\} $ & $ \overline{\text{Log}\,\,\text{HHI}} \,\,\backslash\, \{c\} $   \\
\multirow{3.4}{*}{\shortstack[l]{Labor Market \\ Definition \\ (Object)}}  & \multirow{3.4}{*}{\shortstack{NACE-4 \\ $\times$ CZ \\ (Employment)}} & \multirow{3.4}{*}{\shortstack{NACE-4 \\ $\times$ CZ \\ (Employment)}} & \multirow{3.4}{*}{\shortstack{NACE-4 \\ $\times$ CZ \\ (Employment)}}  \\
&&& \\
&&& \\[0.2cm] 
Observations &  2,139,426       & 2,139,157       & 2,139,400             \\
\multirow{2.4}{*}{\shortstack{First-Stage Effect}} &  \multirow{2.4}{*}{\shortstack{\hphantom{***}0.820***  \\ (0.039)}}  & \multirow{2.4}{*}{\shortstack{\hphantom{***}0.707***  \\ (0.044)}}  &  \multirow{2.4}{*}{\shortstack{\hphantom{***}0.928***  \\ (0.054)}}   \\
&&& \\[0.2cm] 
F: Instrument & 430.8 & 261.7 & ~295.8 \\[0.2cm]  \hline
\end{tabular}
\begin{tablenotes}[para]
\footnotesize\textsc{Note. ---} The table displays fixed effects regressions of log employment (in terms of regular full-time workers per firm) on log labor market concentration (in terms of HHI). The instrumental variable refers to the leave-one-out average of the log inverse number of firms (or log HHI) in all other commuting zones but for the same industry and time period. Standard errors (in parentheses) are clustered at the labor market level. CZ = Commuting Zone. FT = Full-Time. HHI = Herfindahl-Hirschman Index. INF = Inverse Number of Firms. NACE-X = X-Digit Statistical Nomenclature of Economic Activities in the European Community. PT = Part-Time. * = p$<$0.10. ** = p$<$0.05. *** = p$<$0.01. Sources: IEB $\plus$ BHP, 1999-2014.
\end{tablenotes}
\end{threeparttable}
}
\end{table}

\vspace*{\fill}
\clearpage

\begin{table}[!ht]
\centering
\scalebox{0.90}{
\begin{threeparttable}
\caption{Effects of Labor Market Concentration on Employment by Labor Market Definition}
\label{tab:G7}
\begin{tabular}{L{4cm}C{2.7cm}C{2.7cm}C{2.7cm}C{2.7cm}C{2.7cm}C{2.7cm}} \hline
&&&&&& \\[-0.3cm]
\multirow{7.4}{*}{}  & \multirow{7.4}{*}{\shortstack{(1) \\ \textbf{NACE-3 \vphantom{/}} \\ \textbf{Industries \vphantom{/}} \\ Log \\ Regular FT\vphantom{/} \\ Employment }} & \multirow{7.4}{*}{\shortstack{(2) \\ \textbf{NACE-5 \vphantom{/}} \\ \textbf{Industries \vphantom{/}} \\ Log \\ Regular FT\vphantom{/} \\ Employment }} & \multirow{7.4}{*}{\shortstack{(3) \\ \textbf{Flow-Adjusted \vphantom{/}} \\ \textbf{NACE-4 \vphantom{/}} \\ \textbf{Industries \vphantom{/}} \\ Log \\ Regular FT\vphantom{/} \\ Employment }} & \multirow{7.4}{*}{\shortstack{(4) \\ \textbf{NUTS-3 \vphantom{/}} \\ \textbf{Regions \vphantom{/}} \\ Log \\ Regular FT\vphantom{/} \\ Employment }} & \multirow{7.4}{*}{\shortstack{(5) \\ \textbf{HHI \vphantom{/}} \\ \textbf{Based on \vphantom{/}} \\ \textbf{Hires \vphantom{/}}  \\ Log \\ Regular FT\vphantom{/} \\ Employment }} & \multirow{7.4}{*}{\shortstack{(6) \\ \textbf{HHI \vphantom{/}} \\ \textbf{Based on \vphantom{/}} \\ \textbf{Wage Bill \vphantom{/}} \\ Log \\ Regular FT\vphantom{/} \\ Employment }} \\
&&&&&& \\
&&&&&& \\
&&&&&& \\
&&&&&& \\
&&&&&& \\
&&&&&& \\[0.3cm] \hline
&&&&&& \\[-0.3cm]
\multirow{2.4}{*}{Log HHI} &  \multirow{2.4}{*}{\shortstack{\hphantom{***}-0.077***\hphantom{-}  \\ (0.020)}}  & \multirow{2.4}{*}{\shortstack{\hphantom{***}-0.032***\hphantom{-}  \\ (0.011)}} & & \multirow{2.4}{*}{\shortstack{\hphantom{***}-0.064***\hphantom{-}  \\ (0.009)}} & \multirow{2.4}{*}{\shortstack{\hphantom{***}-0.037***\hphantom{-}  \\ (0.012)}} & \multirow{2.4}{*}{\shortstack{\hphantom{***}-0.038***\hphantom{-}  \\ (0.013)}}   \\
&&&&&& \\
\multirow{2.4}{*}{Log Adjusted HHI} &    &   & \multirow{2.4}{*}{\shortstack{\hphantom{***}-0.039***\hphantom{-}  \\ (0.013)}}  &  &  &  \\
&&&&&& \\[0.2cm] 
\multirow{2.4}{*}{\shortstack{Fixed Effects}}  & \multirow{2.4}{*}{\shortstack{Firm \\ CZ $\times$ Year \\ NACE-2 $\times$ Year}} & \multirow{2.4}{*}{\shortstack{Firm \\ CZ $\times$ Year \\ NACE-2 $\times$ Year}} &  \multirow{2.4}{*}{\shortstack{Firm \\ CZ $\times$ Year \\ NACE-2 $\times$ Year}} & \multirow{2.4}{*}{\shortstack{Firm \\ NUTS-3 $\times$ Year \\ NACE-2 $\times$ Year}} &  \multirow{2.4}{*}{\shortstack{Firm \\ CZ $\times$ Year \\ NACE-2 $\times$ Year}} &  \multirow{2.4}{*}{\shortstack{Firm \\ CZ $\times$ Year \\ NACE-2 $\times$ Year}} \\
&&&&&& \\[0.4cm] \hline
&&&&&& \\[-0.2cm]
Instrument  & $ \overline{\text{Log}\,\,\text{INF}} \,\,\backslash\, \{c\} $ & $ \overline{\text{Log}\,\,\text{INF}} \,\,\backslash\, \{c\} $ & $ \overline{\text{Log}\,\,\text{INF}} \,\,\backslash\, \{c\} $ & $ \overline{\text{Log}\,\,\text{INF}} \,\,\backslash\, \{c\} $    & $ \overline{\text{Log}\,\,\text{INF}} \,\,\backslash\, \{c\} $  &   $ \overline{\text{Log}\,\,\text{INF}} \,\,\backslash\, \{c\} $  \\
\multirow{3.4}{*}{\shortstack[l]{Labor Market \\ Definition \\ (Object)}}  & \multirow{3.4}{*}{\shortstack{NACE-3 \\ $\times$ CZ \\ (Employment)}} & \multirow{3.4}{*}{\shortstack{NACE-5 \\ $\times$ CZ \\ (Employment)}} & \multirow{3.4}{*}{\shortstack{NACE-4 \\ $\times$ CZ \\ (Employment)}} & \multirow{3.4}{*}{\shortstack{NACE-4 \\ $\times$ NUTS-3 \\ (Employment)}} & \multirow{3.4}{*}{\shortstack{NACE-4 \\ $\times$ CZ \\ (Hires)}}  & \multirow{3.4}{*}{\shortstack{NACE-4 \\ $\times$ CZ \\ (Wage Bill)}}  \\
&&&&&&  \\
&&&&&&  \\
\multirow{2.4}{*}{Sample}  & \multirow{2.4}{*}{1999-2014} & \multirow{2.4}{*}{1999-2014} & \multirow{2.4}{*}{1999-2014}  & \multirow{2.4}{*}{1999-2014}  & \multirow{2.4}{*}{2000-2014}  & \multirow{2.4}{*}{1999-2014} \\
&&&&&& \\[0.2cm] 
Observations &  2,139,400       & 2,139,400       & 2,139,400       & 2,139,400    & 1,958,594    & ~2,139,400        \\
\multirow{2.4}{*}{\shortstack{First-Stage Effect}} &  \multirow{2.4}{*}{\shortstack{\hphantom{***}0.966***  \\ (0.076)}}  &  \multirow{2.4}{*}{\shortstack{\hphantom{***}0.814***  \\ (0.035)}}  & \multirow{2.4}{*}{\shortstack{\hphantom{***}0.720***  \\ (0.042)}}  & \multirow{2.4}{*}{\shortstack{\hphantom{***}0.753***  \\ (0.015)}}    & \multirow{2.4}{*}{\shortstack{\hphantom{***}0.795***  \\ (0.034)}}  & \multirow{2.4}{*}{\shortstack{\hphantom{***}0.722***  \\ (0.042)}}   \\
&&&&&& \\[0.2cm] 
F Statistic & 159.9 & 543.6 & 287.6 & 2474.8 & 563.1 & ~291.8 \\[0.2cm]  \hline
\end{tabular}
\begin{tablenotes}[para]
\footnotesize\textsc{Note. ---} The table displays fixed effects regressions of log employment (in terms of regular full-time workers per firm) on log labor market concentration (in terms of HHI). The instrumental variable refers to the leave-one-out average of the log inverse number of firms in all other regions but for the same industry and time period. Standard errors (in parentheses) are clustered at the labor market level. CZ = Commuting Zone. FT = Full-Time. HHI = Herfindahl-Hirschman Index. INF = Inverse Number of Firms. NACE-X = X-Digit Statistical Nomenclature of Economic Activities in the European Community. NUTS-X = X-Digit Statistical Nomenclature of Territorial Units. * = p$<$0.10. ** = p$<$0.05. *** = p$<$0.01. Sources: IEB $\plus$ BHP, 1999-2014.
\end{tablenotes}
\end{threeparttable}
}
\end{table}

\clearpage

\begin{table}[!ht]
\centering
\scalebox{0.90}{
\begin{threeparttable}
\caption{Effects of Labor Market Concentration on Employment by Concentration Measure}
\label{tab:G8}
\begin{tabular}{L{4cm}C{2.7cm}C{2.7cm}C{2.7cm}C{2.7cm}C{2.7cm}} \hline
&&&&& \\[-0.3cm]
\multirow{7.4}{*}{} & \multirow{7.4}{*}{\shortstack{(1) \\ \textbf{Rosenbluth\vphantom{/}} \\ \textbf{Index\vphantom{/}} \\ Log \\ Regular FT\vphantom{/} \\ Employment }} & \multirow{7.4}{*}{\shortstack{(2) \\ \textbf{1-Firm\vphantom{/}} \\ \textbf{Concentration\vphantom{/}} \\ \textbf{Ratio\vphantom{/}} \\ Log \\ Regular FT\vphantom{/} \\ Employment }} & \multirow{7.4}{*}{\shortstack{(3) \\ \textbf{Inverse\vphantom{/}} \\ \textbf{Number\vphantom{/}} \\ \textbf{of Firms\vphantom{/}} \\ Log \\ Regular FT\vphantom{/} \\ Employment }} & \multirow{7.4}{*}{\shortstack{(4) \\ \textbf{Exponential\vphantom{/}} \\ \textbf{Index\vphantom{/}} \\ Log \\ Regular FT\vphantom{/} \\ Employment }} & \multirow{7.4}{*}{\shortstack{(5) \\ \textbf{Market\vphantom{/}} \\ \textbf{Share\vphantom{/}} \\ Log \\ Regular FT\vphantom{/} \\ Employment }}  \\
&&&&& \\
&&&&& \\
&&&&& \\
&&&&& \\
&&&&& \\
&&&&& \\[0.3cm] \hline
&&&&& \\[-0.3cm]
\multirow{2.4}{*}{Log RBI}  & \multirow{2.4}{*}{\shortstack{\hphantom{***}-0.031***\hphantom{-} \\ (0.010)}}   &   &   &   & \\
&&&&&  \\
\multirow{2.4}{*}{Log CR1}  &      & \multirow{2.4}{*}{\shortstack{\hphantom{***}-0.049***\hphantom{-} \\ (0.017)}}       &    &   & \\
&&&&&  \\
\multirow{2.4}{*}{Log INF}  &      &       & \multirow{2.4}{*}{\shortstack{\hphantom{***}-0.030***\hphantom{-} \\ (0.010)}}   &   & \\
&&&&&  \\
\multirow{2.4}{*}{Log EXP}  &      &       &       &  \multirow{2.4}{*}{\shortstack{\hphantom{***}-0.031***\hphantom{-} \\ (0.010)}} & \\
&&&&&  \\
\multirow{2.4}{*}{Log Market Share}  &      &      &       &  & \multirow{2.4}{*}{\shortstack{\hphantom{***}-0.035***\hphantom{-} \\ (0.011)}}  \\
&&&&&  \\[0.2cm] 
\multirow{2.4}{*}{\shortstack{Fixed Effects}}  & \multirow{2.4}{*}{\shortstack{Firm \\ CZ $\times$ Year \\ NACE-2 $\times$ Year}} &  \multirow{2.4}{*}{\shortstack{Firm \\ CZ $\times$ Year \\ NACE-2 $\times$ Year}} & \multirow{2.4}{*}{\shortstack{Firm \\ CZ $\times$ Year \\ NACE-2 $\times$ Year}} &  \multirow{2.4}{*}{\shortstack{Firm \\ CZ $\times$ Year \\ NACE-2 $\times$ Year}} &  \multirow{2.4}{*}{\shortstack{Firm \\ CZ $\times$ Year \\ NACE-2 $\times$ Year}}  \\
&&&&& \\[0.4cm] \hline
&&&&& \\[-0.2cm]
Instrument  & $ \overline{\text{Log}\,\,\text{INF}} \,\,\backslash\, \{c\} $ & $ \overline{\text{Log}\,\,\text{INF}} \,\,\backslash\, \{c\} $ & $ \overline{\text{Log}\,\,\text{INF}} \,\,\backslash\, \{c\} $ & $ \overline{\text{Log}\,\,\text{INF}} \,\,\backslash\, \{c\} $   & $ \overline{\text{Log}\,\,\text{INF}} \,\,\backslash\, \{c\} $  \\[0.2cm]
Observations &  2,139,400       & 2,139,400       & 2,139,400       & 2,139,400       & ~2,139,400       \\
\multirow{2.4}{*}{\shortstack{First-Stage Effect}} &  \multirow{2.4}{*}{\shortstack{\hphantom{***}0.889***  \\ (0.017)}} & \multirow{2.4}{*}{\shortstack{\hphantom{***}0.564***  \\ (0.050)}} & \multirow{2.4}{*}{\shortstack{\hphantom{***}0.920***  \\ (0.016)}} & \multirow{2.4}{*}{\shortstack{\hphantom{***}0.889***  \\ (0.019)}} & \multirow{2.4}{*}{\shortstack{\hphantom{***}0.801***  \\ (0.017)}}  \\
&&&&& \\[0.2cm] 
F Statistic & 2793.0 & 126.5 & 3425.7 & 2185.7  & ~2233.2 \\[0.2cm]  \hline
\end{tabular}
\begin{tablenotes}[para]
\footnotesize\textsc{Note. ---} The table displays fixed effects regressions of log employment (in terms of regular full-time workers per firm) on log labor market concentration. The instrumental variable refers to the leave-one-out average of the log inverse number of firms in all other commuting zones but for the same industry and time period. Labor markets are combinations of 4-digit NACE industries and commuting zone. Standard errors (in parentheses) are clustered at the labor market level. CR1 = 1-Firm Concentration Ratio. CZ = Commuting Zone. EXP = Exponential Index. FT = Full-Time. INF = Inverse Number of Firms. NACE-X = X-Digit Statistical Nomenclature of Economic Activities in the European Community. RBI = Rosenbluth Index. * = p$<$0.10. ** = p$<$0.05. *** = p$<$0.01. Sources: IEB $\plus$ BHP, 1999-2014.
\end{tablenotes}
\end{threeparttable}
}
\end{table}

\clearpage
\vspace*{\fill}

\begin{table}[!ht]
\centering
\scalebox{0.90}{
\begin{threeparttable}
\caption{Effects of Labor Market Concentration on Employment by Territory}
\label{tab:G9}
\begin{tabular}{L{4cm}C{5.8cm}C{5.8cm}} \hline
&& \\[-0.3cm]
\multirow{6.4}{*}{} & \multirow{6.4}{*}{\shortstack{(1) \\ \textbf{West\vphantom{/}} \\ \textbf{Germany\vphantom{/}} \\ Log \\ Regular FT\vphantom{/} \\ Employment }} & \multirow{6.4}{*}{\shortstack{(2) \\ \textbf{East\vphantom{/}} \\ \textbf{Germany\vphantom{/}} \\ Log \\ Regular FT\vphantom{/} \\ Employment }}  \\
&& \\
&& \\
&& \\
&& \\
&& \\[0.3cm] \hline
&& \\[-0.3cm]
\multirow{2.4}{*}{Log HHI} &  \multirow{2.4}{*}{\shortstack{\hphantom{***}-0.044***\hphantom{-}  \\ (0.013)}}  & \multirow{2.4}{*}{\shortstack{\hphantom{*}-0.031*\hphantom{-}  \\ (0.017)}}   \\
&& \\[0.2cm] 
\multirow{2.4}{*}{\shortstack{Fixed Effects}}  & \multirow{2.4}{*}{\shortstack{Firm \\ CZ $\times$ Year \\ NACE-2 $\times$ Year}} &  \multirow{2.4}{*}{\shortstack{Firm \\ CZ $\times$ Year \\ NACE-2 $\times$ Year}}  \\
&& \\[0.4cm] \hline
&& \\[-0.2cm]
Instrument  & $ \overline{\text{Log}\,\,\text{INF}} \,\,\backslash\, \{c\} $ & $ \overline{\text{Log}\,\,\text{INF}} \,\,\backslash\, \{c\} $   \\
\multirow{3.4}{*}{\shortstack[l]{Labor Market \\ Definition \\ (Object)}}  & \multirow{3.4}{*}{\shortstack{NACE-4 \\ $\times$ CZ \\ (Employment)}} & \multirow{3.4}{*}{\shortstack{NACE-4 \\ $\times$ CZ \\ (Employment)}}  \\
&& \\
&& \\
\multirow{2.4}{*}{Sample}  & \multirow{2.4}{*}{West Germany} & \multirow{2.4}{*}{\shortstack{East Germany \\[-0.1cm] Berlin}}  \\
&& \\[0.2cm] 
Observations &  1,676,435       & ~462,850            \\
\multirow{2.4}{*}{\shortstack{First-Stage Effect}} &  \multirow{2.4}{*}{\shortstack{\hphantom{***}0.876***  \\ (0.053)}}  & \multirow{2.4}{*}{\shortstack{\hphantom{***}0.676***  \\ (0.033)}}   \\
&& \\[0.2cm] 
F Statistic & 274.1 & ~412.7   \\[0.2cm]  \hline
\end{tabular}
\begin{tablenotes}[para]
\footnotesize\textsc{Note. ---} The table displays fixed effects regressions of log employment (in terms of regular full-time workers per firm) on log labor market concentration (in terms of HHI). The instrumental variable refers to the leave-one-out average of the log inverse number of firms in all other commuting zones but for the same industry and time period. Standard errors (in parentheses) are clustered at the labor market level. CZ = Commuting Zone. FT = Full-Time. HHI = Herfindahl-Hirschman Index. INF = Inverse Number of Firms. NACE-X = X-Digit Statistical Nomenclature of Economic Activities in the European Community. * = p$<$0.10. ** = p$<$0.05. *** = p$<$0.01. Sources: IEB $\plus$ BHP, 1999-2014.
\end{tablenotes}
\end{threeparttable}
}
\end{table}

\vspace*{\fill}
\clearpage
\vspace*{\fill}

\begin{table}[!ht]
\centering
\scalebox{0.90}{
\begin{threeparttable}
\caption{Effects of Labor Market Concentration on Employment by Labor Outcome}
\label{tab:G10}
\begin{tabular}{L{4cm}C{3.75cm}C{3.75cm}C{3.75cm}} \hline
&&& \\[-0.3cm]
\multirow{6.4}{*}{} & \multirow{6.4}{*}{\shortstack{(1) \\ \textbf{All\vphantom{/}} \\  \textbf{Workers\vphantom{/}} \\ Log \\ Overall\vphantom{/} \\ Employment }} & \multirow{6.4}{*}{\shortstack{(2) \\ \textbf{Regular PT\vphantom{/}} \\  \textbf{Workers\vphantom{/}} \\ Log \\ Regular PT\vphantom{/} \\ Employment }} & \multirow{6.4}{*}{\shortstack{(3) \\ \textbf{Marginal PT\vphantom{/}} \\  \textbf{Workers\vphantom{/}} \\ Log \\ Marginal\vphantom{/} \\ Employment }}   \\
&&& \\
&&& \\
&&& \\
&&& \\
&&& \\[0.3cm] \hline
&&& \\[-0.3cm]
\multirow{2.4}{*}{Log HHI} &  \multirow{2.4}{*}{\shortstack{\hphantom{***}-0.040***\hphantom{-}  \\ (0.010)}}  & \multirow{2.4}{*}{\shortstack{0.004  \\ (0.011)}} & \multirow{2.4}{*}{\shortstack{\hphantom{***}-0.030***\hphantom{-}  \\ (0.011)}}   \\
&&& \\[0.2cm] 
\multirow{2.4}{*}{\shortstack{Fixed Effects}}  & \multirow{2.4}{*}{\shortstack{Firm \\ CZ $\times$ Year \\ NACE-2 $\times$ Year}} &  \multirow{2.4}{*}{\shortstack{Firm \\ CZ $\times$ Year \\ NACE-2 $\times$ Year}} & \multirow{2.4}{*}{\shortstack{Firm \\ CZ $\times$ Year \\ NACE-2 $\times$ Year}}  \\
&&& \\[0.4cm] \hline
&&& \\[-0.2cm]
Instrument  & $ \overline{\text{Log}\,\,\text{INF}} \,\,\backslash\, \{c\} $ & $ \overline{\text{Log}\,\,\text{INF}} \,\,\backslash\, \{c\} $ & $ \overline{\text{Log}\,\,\text{INF}} \,\,\backslash\, \{c\} $   \\
\multirow{3.4}{*}{\shortstack[l]{Labor Market \\ Definition \\ (Object)}}  & \multirow{3.4}{*}{\shortstack{NACE-4 \\ $\times$ CZ \\ (Employment)}} & \multirow{3.4}{*}{\shortstack{NACE-4 \\ $\times$ CZ \\ (Employment)}} & \multirow{3.4}{*}{\shortstack{NACE-4 \\ $\times$ CZ \\ (Employment)}}  \\
&&& \\
&&& \\[0.2cm] 
Observations &  2,825,341       & 823,125         & ~1,654,825           \\
\multirow{2.4}{*}{\shortstack{First-Stage Effect}} & \multirow{2.4}{*}{\shortstack{\hphantom{***}0.830***  \\ (0.042)}}  &  \multirow{2.4}{*}{\shortstack{\hphantom{***}0.813***  \\ (0.031)}} & \multirow{2.4}{*}{\shortstack{\hphantom{***}0.834***  \\ (0.042)}}   \\
&&& \\[0.2cm] 
F Statistic & 390.1 & 674.8 & ~402.8  \\[0.2cm]  \hline
\end{tabular}
\begin{tablenotes}[para]
\footnotesize\textsc{Note. ---} The table displays fixed effects regressions of log employment on log labor market concentration (in terms of HHI). The instrumental variable refers to the leave-one-out average of the log inverse number of firms in all other commuting zones but for the same industry and time period. Standard errors (in parentheses) are clustered at the labor market level. CZ = Commuting Zone. FT = Full-Time. HHI = Herfindahl-Hirschman Index. INF = Inverse Number of Firms. NACE-X = X-Digit Statistical Nomenclature of Economic Activities in the European Community. PT = Part-Time. * = p$<$0.10. ** = p$<$0.05. *** = p$<$0.01. Sources: IEB $\plus$ BHP, 1999-2014.
\end{tablenotes}
\end{threeparttable}
}
\end{table}

\vspace*{\fill}
\clearpage
\vspace*{\fill}

\begin{table}[!ht]
\centering
\scalebox{0.85}{
\begin{threeparttable}
\caption{Plausibly Exogenous Regressions}
\label{tab:G11}
\begin{tabular}{L{4cm}C{5.8cm}C{5.8cm}} \hline
&& \\[-0.3cm]
\multirow{5.4}{*}{} & \multirow{5.4}{*}{\shortstack{(1) \\ \textbf{Wages\vphantom{/}} \\ Log \\ Mean Daily\vphantom{/} \\ Wages }} & \multirow{5.4}{*}{\shortstack{(2) \\ \textbf{Employment\vphantom{/}} \\ Log \\ Regular FT\vphantom{/} \\ Employment }}  \\
&& \\
&& \\
&& \\
&& \\[0.3cm] \hline
&& \\[-0.3cm]
\multirow{2.4}{*}{$ \overline{\text{Log}\,\,\text{INF}} \,\,\backslash\, \{c\} $} &  \multirow{2.4}{*}{\shortstack{\hphantom{***}$\Phi^{\text{min}}=\,\,$-0.0167***\hphantom{-}  \\ (0.0035)}}  & \multirow{2.4}{*}{\shortstack{\hphantom{***}$\Phi^{\text{min}}=\,\,$-0.0278***\hphantom{-}  \\ (0.0089)}}   \\
&& \\[0.2cm] 
\multirow{2.4}{*}{\shortstack{Fixed Effects}}  & \multirow{2.4}{*}{\shortstack{Firm \\ CZ $\times$ Year \\ NACE-2 $\times$ Year}} &  \multirow{2.4}{*}{\shortstack{Firm \\ CZ $\times$ Year \\ NACE-2 $\times$ Year}}  \\
&& \\[0.4cm] \hline
&& \\[-0.4cm]

\multirow{3.4}{*}{\shortstack[l]{Labor Market \\ Definition \\ (Object)}}  & \multirow{3.4}{*}{\shortstack{NACE-4 \\ $\times$ CZ \\ (Employment)}} & \multirow{3.4}{*}{\shortstack{NACE-4 \\ $\times$ CZ \\ (Employment)}}  \\
&& \\
&& \\[0.2cm] 
Observations &  2,139,400        & 2,139,400      \\[0.2cm] 
$\big[\Phi^{\text{min}}\,;\,\Phi^{\text{max}}\,\big] $& ~[\,-0.0167\,;\,0.0000\,] & ~[\,-0.0278\,;\,0.0000\,] \\[0.2cm]
$\big[\,\theta^{0.05}\,;\,\theta^{0.95} \,\big] $& ~[\,-0.0268\,;\,0.0065\,] & ~[\,-0.0508\,;\,0.0165\,] \\[0.2cm]
$\theta^{0.95}<0$, when ... & \,\,~$\Phi\,>\,\,$-0.0115\hphantom{-} & \,\,~$\Phi\,>\,\,$-0.0141\hphantom{-}  \\[0.2cm]  \hline
\end{tabular}
\end{threeparttable}
}
\floatfoot{\footnotesize\textsc{Note. ---} The table displays results from reduced-form and plausibly exogenous regressions to examine the potential bias that may arise from endogeneity of the instrument. The upper part of the table shows reduced-form regressions of log daily wages and employment on the instrument. The instrumental variable refers to the leave-one-out industry average of the log inverse number of firms in all other commuting zones but for the same industry and time period. Standard errors (in parentheses) are clustered at the labor market level. In the lower part of the table, I apply the results from the reduced-form regressions to derive lower and upper bounds for the causal effect of labor market concentration on the outcomes of interest. In the second-stage regression, the exogeneity assumption is violated when the instrumental variable exerts an additional direct effect, $\Phi\neq0$, on the outcome variable (i.e., other than through labor market concentration). To assess the potential magnitude of the bias, I formulate a range of values for this direct effect between zero (exogeneity) and the estimated reduced-form coefficient $\Phi^{\text{min}}$. The reduced-form coefficient reflects the maximum of instrument endogeneity that can enter the second-stage regression of the outcome variable on the first-stage variation in HHI. Given the formulated range, I apply the plausibly exogenous regression method \citep{ConleyEtAl2012} to derive the 90 percent confidence interval for the causal effect of labor market concentration on wages and employment. Furthermore, I specify the threshold for the direct effect of the instrumental variable in the second-stage regression above which the interval of the causal effect $\theta$ of labor market concentration on the outcome variable would be fully to the left of zero (i.e., the effect would be significantly negative). CZ = Commuting Zone. INF = Inverse Number of Firms. NACE-4 = 4-Digit Statistical Nomenclature of Economic Activities in the European Community. * = p$<$0.10. ** = p$<$0.05. *** = p$<$0.01. Sources: IEB $\plus$ BHP, 1999-2014.}
\end{table}

\clearpage
\vspace*{\fill}

\begin{table}[!ht]
\centering
\scalebox{0.90}{
\begin{threeparttable}
\caption{Effects of Labor Market Concentration on Wages and Employment Without Services}
\label{tab:G12}
\begin{tabular}{L{4cm}C{5.8cm}C{5.8cm}} \hline
&& \\[-0.3cm]
\multirow{6.4}{*}{} & \multirow{6.4}{*}{\shortstack{(1) \\ \textbf{Without\vphantom{/}} \\ \textbf{Services\vphantom{/}} \\ Log \\ Mean Daily\vphantom{/}  \\ Wages }} & \multirow{6.4}{*}{\shortstack{(2) \\ \textbf{Without\vphantom{/}} \\ \textbf{Services\vphantom{/}} \\ Log \\ Regular FT\vphantom{/} \\ Employment }}  \\
&& \\
&& \\
&& \\
&& \\
&& \\[0.3cm] \hline
&& \\[-0.3cm]
\multirow{2.4}{*}{Log HHI} &  \multirow{2.4}{*}{\shortstack{\hphantom{***}-0.024***\hphantom{-}  \\ (0.005)}}  & \multirow{2.4}{*}{\shortstack{\hphantom{***}-0.035***\hphantom{-}  \\ (0.011)}}   \\
&& \\[0.2cm] 
\multirow{2.4}{*}{\shortstack{Fixed Effects}}  & \multirow{2.4}{*}{\shortstack{Firm \\ CZ $\times$ Year \\ NACE-2 $\times$ Year}} &  \multirow{2.4}{*}{\shortstack{Firm \\ CZ $\times$ Year \\ NACE-2 $\times$ Year}}  \\
&& \\[0.4cm] \hline
&& \\[-0.2cm]
Instrument  & $ \overline{\text{Log}\,\,\text{INF}} \,\,\backslash\, \{c\} $ & $ \overline{\text{Log}\,\,\text{INF}} \,\,\backslash\, \{c\} $   \\
\multirow{3.4}{*}{\shortstack[l]{Labor Market \\ Definition \\ (Object)}}  & \multirow{3.4}{*}{\shortstack{NACE-4 \\ $\times$ CZ \\ (Employment)}} & \multirow{3.4}{*}{\shortstack{NACE-4 \\ $\times$ CZ \\ (Employment)}}  \\
&& \\
&& \\
\multirow{2.4}{*}{Sample}  & \multirow{2.4}{*}{\shortstack{Without \\ Services}} & \multirow{2.4}{*}{\shortstack{Without \\ Services}}  \\
&& \\[0.2cm] 
Observations &  1,123,481       & ~1,123,481          \\
\multirow{2.4}{*}{\shortstack{First-Stage Effect}} & \multirow{2.4}{*}{\shortstack{\hphantom{***}0.800***  \\ (0.039)}}  &  \multirow{2.4}{*}{\shortstack{\hphantom{***}0.800***  \\ (0.039)}}   \\
&& \\[0.2cm] 
F Statistic & 410.9 & ~410.9   \\[0.2cm]  \hline
\end{tabular}
\begin{tablenotes}[para]
\footnotesize\textsc{Note. ---} The table displays fixed effects regressions of log daily wages (in terms of firm-level averages of regular full-time workers) and log employment (of regular full-time workers per firm) on log labor market concentration (in terms of HHI). The instrumental variable refers to the leave-one-out average of the log inverse number of firms in all other commuting zones but for the same industry and time period. Standard errors (in parentheses) are clustered at the labor market level. CZ = Commuting Zone. FT = Full-Time. HHI = Herfindahl-Hirschman Index. INF = Inverse Number of Firms. NACE-X = X-Digit Statistical Nomenclature of Economic Activities in the European Community. * = p$<$0.10. ** = p$<$0.05. *** = p$<$0.01. Sources: IEB $\plus$ BHP, 1999-2014.
\end{tablenotes}
\end{threeparttable}
}
\end{table}

\vspace*{\fill}

\end{landscape}

\clearpage

\section{Minimum Wage Effects: Further Evidence}
\label{sec:H}
\setcounter{table}{0} 
\setcounter{figure}{0} 

This section assembles additional evidence on the minimum wage effects on wages and employment. First, I inspect the relationship between minimum wage levels and the baseline HHI values. Second, I provide visual evidence for the baseline minimum wage effects on employment. Third, I provide full details on the sensitivity and heterogeneity analyses for the minimum wage effects.

\paragraph{Relationship between Minimum Wages and HHI.} In terms of the moderating variable, the use of average HHI values will give rise to endogeneity in case minimum wage changes correlate with HHI levels. To gauge the magnitude of this potential bias, Table \ref{tab:H1} regresses time-varying values of the baseline HHI on the level of sectoral minimum wages to shed light on the correlation between both variables. Conditional on the covariates from the baseline specification, the variables do not exhibit any correlation at all (p=0.90), invalidating the concern that averaging over HHI might induce endogeneity.

\paragraph{Heterogeneity of Wage Effects.} In Table \ref{tab:H2}, I estimate the minimum wage effects of wages separately for the West and East German labor market. In either case, the main effect is significantly positive but turns out more pronounced in East Germany where general wage levels tend to be lower. Both interaction effects turn out to be positive but only the coefficient for West Germany remains significant.

First and foremost, minimum wages push up wages at the bottom of the wage distribution. Hence, higher percentiles of the wage distribution are supposed to show accordingly smaller wage effects. To test this conjecture, Table \ref{tab:H3} displays the effects of sectoral minimum wages on the 25th percentile, the median, and the 75th percentile of the firm-specific wage distribution of regular full-time workers. In zero-HHI markets, a minimum wage increase by 10 percent significantly raises the 25th percentile of the wage distribution by 1.4 percent. In, this effect weakens for higher percentiles of the distribution where wage rates increasingly exceed the wage floors: an analogous minimum wage increase moves the median wage by 0.9 percent and the 75th percentile of the wage distribution by just 0.5 percent. Yet, the fact that the middle and upper part of the distribution shifts mirrors the large Kaitz Indices found in Table \ref{tab:B2}. For each of the three percentiles, increases in sectoral minimum wages bite harder in more concentrated labor markets, although only the interaction effects at the median and 75th percentile are significantly positive. Interestingly, the interaction effect for the 75th percentile is still substantive, reflecting evidence that, in highly concentrated minimum wage sectors, employers can suppress wages not only at the bottom but also at the top of their firm-specific wage distribution (from Table \ref{tab:G5}).

\paragraph{Minimum Wage Effects on Employment.} Figure \ref{fig:H1} visualizes the minimum wage elasticities of employment by HHI for the baseline specification in Column (4) of Table \ref{tab:4}.

\paragraph{Non-Linearities.} I perform multiple checks to examine the sensitivity and heterogeneity of the estimated minimum wage effects on employment. As my analysis takes place at the firm level and the firm-weighted HHI distribution is not uniform, the identification of the linear interaction effect may not reflect all parts of the HHI range equally. To ascertain that the positive HHI gradient is not just a result of the linearity assumption, I construct interaction effects between minimum wage levels and indicator variables that separate low- from high-HHI labor markets. Following E.U.\ antitrust thresholds, Table \ref{tab:H4}) displays the effects for divisions of the HHI range into two, three, four and five domains. Across these specifications, the estimates exhibit the same pattern as before: negative employment effects in slightly concentrated markets that weaken for higher HHI domains and, ultimately, become positive in highly concentrated labor markets.

\paragraph{Sensitivity and Heterogeneity of Employment Effects.} I perform sensitivity checks for alternative specifications in Table \ref{tab:H5}, labor market definitions in Table \ref{tab:H6}, and concentration measures in Table \ref{tab:H7}. In general, the results are robust to various modifications of the baseline equation, thus lending further support to the monopsony argument.

In line with the insignificant correlation between HHI values and sectoral minimum wage levels (see Table \ref{tab:H1}), the use of time-varying or predetermined HHI values as moderating variable does not lead to marked changes in the elasticity estimates. In particular, as a more rigorous alternative to HHI averages, predetermined HHI values eliminate any reverse causality between minimum wage effects and HHI values but, as a drawback, provide a less representative picture of the underlying labor market concentration over the period of study. When constructing predetermined HHI values, I pick the HHI value in the year before the firm was first subject to minimum wage legislation in the period under study. For lack of time-consistent information on 1998, I resort to the 1999 HHI value if needed. In a further robustness check, I demonstrate that omitting sector-specific linear time trends does not alter the pattern of minimum wage effects. Furthermore note that, given the log-linear formulation, the identification of the baseline effects solely rests on adjustments of minimum wage levels. Hence, variation from the introduction of a sectoral minimum wage does not enter the baseline regression. Even in the absence of a minimum wage regulation, the availability of unemployment benefits or quasi-fixed cost of labor supply define a lower wage ceiling, a so-called ``implicit minimum wage''. To capture full variation in minimum wage legislation, I assign operating firms an implicit minimum wage in the year preceding the minimum wage introduction, proxied by the fifth percentile of the hourly wage distribution. For lack of IEB information on individual hours worked, I construct the fifth percentile of hourly wage rates by dividing weekly earnings of regular full-time workers by 40 working hours per week. In the augmented regression, the main effect remains negative while the HHI interaction effect turns out to be positive yet again.

By and large, the observed pattern of elasticities is not sensitive to broader or narrower definitions of labor markets. The interaction effect remains positive both for 3-digit or 5-digit industries but turns insignificant for the broader definition. However, the most convincing way to test the sensitivity of the industry dimension is to take additionally into account employment opportunities outside the discrete segments of these classifications. Crucially, when performing such a flow adjustment (see Appendix \ref{sec:E}), the positive effect of labor market concentration on the employment effects of minimum wages strengthens by an order of magnitude. In terms of the regional dimension, the use of 3-digit NUTS regions does not result in marked changes. As an alternative to the stock of employment, the monopsony argument also holds true when constructing the HHI based on new hires or the wage bill. Moreover, the results remain unchanged when employing alternative concentration indices -- namely the Rosenbluth Index, the 1-Firm Concentration Ratio, the Inverse Number of Firms, and the Exponential Index -- or the underlying market shares as moderating variables.

In a next step, I perform separate regressions by West Germany and East Germany in Table \ref{tab:H8}. In line with a stronger bite, the negative employment effects in low-HHI markets turn out more pronounced in East than in West Germany. Importantly, Table \ref{tab:H9} shows that my findings in favor of the monopsony argument my findings in favor of the monopsony argument carry over to overall employment, which comprises regular full-time, regular part-time and marginal part-time workers. Interestingly, the subgroup of regular part-time workers exhibits positive employment effects not only in highly concentrated but, to a lesser degree, also in slightly concentrated labor markets. Since the subgroup of regular full-time workers shows a fairly strong decline in slightly concentrated labor markets, an obvious explanation for the increase in regular-part time employment in low-HHI markets is an employer-driven reduction in individual working hours that transforms some regular full-time workers into regular part-time workers. In addition, with higher minimum wage levels, the monthly income of marginal part-time workers increasingly exceeds the threshold above which marginal part-time jobs automatically turn into regular part-time jobs. This mechanism is supposed to be more pronounced in highly concentrated markets where wage increases tend to be stronger. Accordingly, the effect on marginal employment in high-HHI markets is the least positive among the different groups of workers and turns out not to be statistically different from the effect in low-HHI markets.

\paragraph{Non-Monotonous Relationship.} Table \ref{tab:H10} displays the full results when regressing log employment on a full set of interaction terms between log sectoral minimum wages, labor market concentration (measured as average HHI), and indicator variables for five quintiles of bindingness of the sectoral minimum wage in a firm (measured as average Kaitz Index).

\paragraph{Minimum Wage Introduction in the Sectors.} Building on the timing-based event-study specification, Figure \ref{fig:H2} visualizes the estimated dynamic treatment effects for firms' average daily wages of regular full-time workers before and after the event. It is important to note that, unlike for employment, the quarterly values for average daily wages are subject to some temporal aggregation. The reason is that, by default, workers' total earnings per job are notified only on a calendar-year basis in the IEB data. Thus, average daily wages at quarterly reference dates in the AWFP data become increasingly smoothed to the extent that workers experience wage changes over the course of the calendar year. For instance, suppose that there is a minimum wage hike on October 1, 2014. Due to the practice of calender-year-wise reporting in the IEB, the increase in earnings is averaged over the entire calendar year rather than attributed only to the fourth quarter. Thus, average daily wages of the affected workers in the first, second, and third quarter would be over-reported in the AWFP data whereas daily wages in the fourth quarter of 2014 would be under-reported.

\begin{landscape}

\vspace*{\fill}

\begin{table}[!ht]
\centering
\scalebox{0.90}{
\begin{threeparttable}
\caption{Minimum Wage Effects on HHI}
\label{tab:H1}
\begin{tabular}{L{4cm}C{12cm}} \hline
\multirow{6}{*}{} & \multirow{6.4}{*}{\shortstack{(1) \\ \textbf{HHI\vphantom{/}} \\ Herfindahl-\vphantom{/} \\ Hirschman\vphantom{/} \\ Index }}  \\
&  \\
&  \\
&  \\
&  \\
&  \\[0.2cm] \hline
\multirow{3.4}{*}{Log Minimum Wage} &  \multirow{3.4}{*}{\shortstack{0.000288 \\ (0.001820) \\ p = 0.874 }}  \\
& \\
& \\[0.2cm] 
Control Variables  & Yes   \\
\multirow{2.4}{*}{\shortstack{Fixed Effects}}  & \multirow{2.4}{*}{\shortstack{Firm \\ CZ $\times$ Year}} \\
& \\[0.2cm] \hline
& \\[-0.4cm]
\multirow{3.4}{*}{\shortstack[l]{Labor Market \\ Definition \\ (Object)}}  & \multirow{3.4}{*}{\shortstack{NACE-4 \\ $\times$ CZ \\ (Employment)}} \\
& \\
& \\[0.2cm] 
Observations &  2,700,155     \\[0.2cm]
Adjusted R$^2$ &  0.927  \\[0.2cm] \hline
\end{tabular}
\begin{tablenotes}[para]
\footnotesize\textsc{Note. ---} The table displays a fixed effects regression of labor market concentration (measured as average HHI) on log sectoral minimum wage levels. The set of control variables includes log sectoral employment, the sectoral share of firms subject to a collective bargaining agreement, and sector-specific linear time trends. Standard errors (in parentheses) are clustered at the sector-by-federal-state level. CZ = Commuting Zone. HHI = Herfindahl-Hirschman Index. NACE-4 = 4-Digit Statistical Nomenclature of Economic Activities in the European Community. * = p$<$0.10. ** = p$<$0.05. *** = p$<$0.01. Sources: IEB $\plus$ BHP $\plus$ IAB Establishment Panel, 1999-2017.
\end{tablenotes}
\end{threeparttable}
}
\end{table}

\vspace*{\fill}
\clearpage
\vspace*{\fill}

\begin{table}[!ht]
\centering
\scalebox{0.90}{
\begin{threeparttable}
\caption{Minimum Wage Effects on Wages by Territory}
\label{tab:H2}
\begin{tabular}{L{4cm}C{5.8cm}C{5.8cm}} \hline
&& \\[-0.3cm]
\multirow{6}{*}{} & \multirow{6.4}{*}{\shortstack{(1) \\ \textbf{West\vphantom{/}} \\ \textbf{Germany\vphantom{/}} \\ Log \\ Mean Daily\vphantom{/} \\ Wages }} & \multirow{6.4}{*}{\shortstack{(2) \\ \textbf{East\vphantom{/}} \\ \textbf{Germany\vphantom{/}}  \\ Log \\ Mean Daily\vphantom{/} \\ Wages }}   \\
&& \\
&& \\
&& \\
&& \\
&& \\[0.3cm] \hline
&& \\[-0.3cm]
\multirow{2.4}{*}{Log Minimum Wage} &  \multirow{2.4}{*}{\shortstack{\hphantom{***}0.060***  \\ (0.021)}}  & \multirow{2.4}{*}{\shortstack{\hphantom{***}0.111***  \\ (0.036)}}    \\
&& \\
\multirow{2.4}{*}{\shortstack[l]{Log Minimum Wage \\ $\times$ $\overline{\text{HHI}}$}} &  \multirow{2.4}{*}{\shortstack{\hphantom{***}0.427*** \\ (0.121)}}  & \multirow{2.4}{*}{\shortstack{0.052 \\ (0.085)}}    \\
&& \\[0.2cm] 
Control Variables  & Yes & Yes   \\
\multirow{2.4}{*}{\shortstack{Fixed Effects}}  & \multirow{2.4}{*}{\shortstack{Firm \\ CZ $\times$ Year}} & \multirow{2.4}{*}{\shortstack{Firm \\ CZ $\times$ Year}} \\
&& \\[0.2cm] \hline
&& \\[-0.4cm]
\multirow{3.4}{*}{\shortstack[l]{Labor Market \\ Definition \\ (Object)}}  & \multirow{3.4}{*}{\shortstack{NACE-4 \\ $\times$ CZ \\ (Employment)}} & \multirow{3.4}{*}{\shortstack{NACE-4 \\ $\times$ CZ \\ (Employment)}}  \\
&& \\
&& \\
\multirow{2.4}{*}{Sample}  & \multirow{2.4}{*}{West Germany} & \multirow{2.4}{*}{\shortstack{East Germany \\[-0.1cm] Berlin}}  \\
&& \\[0.2cm] 
Observations &  2,063,033    & 637,122         \\[0.2cm]
Adjusted R$^2$ &  0.805    & 0.790         \\[0.2cm] \hline
\end{tabular}
\begin{tablenotes}[para]
\footnotesize\textsc{Note. ---} The table displays fixed effects regressions of log daily wages (in terms of firm-level averages of regular full-time workers) on log sectoral minimum wage levels as well as their interaction effect with labor market concentration (measured as average HHI over time). The set of control variables includes log sectoral employment, the sectoral share of firms subject to a collective bargaining agreement, and sector-specific linear time trends. Standard errors (in parentheses) are clustered at the sector-by-federal-state level. CZ = Commuting Zone. HHI = Herfindahl-Hirschman Index. NACE-4 = 4-Digit Statistical Nomenclature of Economic Activities in the European Community. * = p$<$0.10. ** = p$<$0.05. *** = p$<$0.01. Sources: IEB $\plus$ BHP $\plus$ IAB Establishment Panel, 1999-2017.
\end{tablenotes}
\end{threeparttable}
}
\end{table}

\vspace*{\fill}
\clearpage
\vspace*{\fill}

\begin{table}[!ht]
\centering
\scalebox{0.90}{
\begin{threeparttable}
\caption{Minimum Wage Effects on Wages by Percentile}
\label{tab:H3}
\begin{tabular}{L{4cm}C{3.75cm}C{3.75cm}C{3.75cm}} \hline
&&& \\[-0.3cm]
\multirow{6.4}{*}{} & \multirow{6.4}{*}{\shortstack{(1) \\ \textbf{25th\vphantom{/}} \\ \textbf{Percentile\vphantom{/}} \\ Log \\ P25 Daily\vphantom{/} \\ Wages }} & \multirow{6.4}{*}{\shortstack{(2) \\ \textbf{Median\vphantom{/}} \\ Log \\ P50 Daily\vphantom{/} \\ Wages }} & \multirow{6.4}{*}{\shortstack{(3) \\ \textbf{75th\vphantom{/}} \\ \textbf{Percentile\vphantom{/}} \\ Log \\ P75 Daily\vphantom{/} \\ Wages }}  \\
&&&  \\
&&&  \\
&&&  \\
&&&  \\
&&&  \\[0.2cm] \hline
&&& \\[-0.3cm]
\multirow{2.4}{*}{Log Minimum Wage} &  \multirow{2.4}{*}{\shortstack{\hphantom{***}0.141***  \\ (0.023)}}  & \multirow{2.4}{*}{\shortstack{\hphantom{***}0.093***  \\ (0.021)}} & \multirow{2.4}{*}{\shortstack{\hphantom{***}0.051***  \\ (0.019)}}    \\
&&&  \\
\multirow{2.4}{*}{\shortstack[l]{Log Minimum Wage \\ $\times$ $\overline{\text{HHI}}$}} &  \multirow{2.4}{*}{\shortstack{0.157 \\ (0.098)}}  & \multirow{2.4}{*}{\shortstack{\hphantom{***}0.268*** \\ (0.084)}} & \multirow{2.4}{*}{\shortstack{\hphantom{***}0.281*** \\ (0.084)}}    \\
&&&  \\[0.2cm] 
&&&  \\[-0.2cm]
Control Variables  & Yes & Yes  & Yes  \\
\multirow{2.4}{*}{\shortstack{Fixed Effects}}  & \multirow{2.4}{*}{\shortstack{Firm \\ CZ $\times$ Year}} & \multirow{2.4}{*}{\shortstack{Firm \\ CZ $\times$ Year}} & \multirow{2.4}{*}{\shortstack{Firm \\ CZ $\times$ Year}} \\
&&& \\[0.2cm] \hline
&&& \\[-0.4cm]
\multirow{3.4}{*}{\shortstack[l]{Labor Market \\ Definition \\ (Object)}}  & \multirow{3.4}{*}{\shortstack{NACE-4 \\ $\times$ CZ \\ (Employment)}} & \multirow{3.4}{*}{\shortstack{NACE-4 \\ $\times$ CZ \\ (Employment)}} & \multirow{3.4}{*}{\shortstack{NACE-4 \\ $\times$ CZ \\ (Employment)}} \\
&&& \\
&&& \\[0.2cm] 
Observations &  2,700,155    & 2,700,155    & 2,700,155      \\[0.2cm]
Adjusted R$^2$ &  0.695    & 0.797    &   0.802         \\[0.2cm] \hline
\end{tabular}
\begin{tablenotes}[para]
\footnotesize\textsc{Note. ---} The table displays fixed effects regressions of log daily wages (in terms of firm-level percentiles of regular full-time workers) on log sectoral minimum wage levels as well as their interaction effect with labor market concentration (measured as average HHI over time). The set of control variables includes log sectoral employment, the sectoral share of firms subject to a collective bargaining agreement, and sector-specific linear time trends. Standard errors (in parentheses) are clustered at the sector-by-federal-state level. CZ = Commuting Zone. HHI = Herfindahl-Hirschman Index. NACE-4 = 4-Digit Statistical Nomenclature of Economic Activities in the European Community. PX = Xth Percentile. * = p$<$0.10. ** = p$<$0.05. *** = p$<$0.01. Sources: IEB $\plus$ BHP $\plus$ IAB Establishment Panel, 1999-2017.
\end{tablenotes}
\end{threeparttable}
}
\end{table}

\vspace*{\fill}
\end{landscape}

\clearpage
\vspace*{\fill}

\begin{figure}[!ht]
\centering
\caption{Minimum Wage Elasticity of Employment}
\label{fig:H1}
\scalebox{0.80}{
\begin{tikzpicture}
\begin{axis}[xlabel=Herfindahl-Hirschman Index, ylabel=Minimum Wage Elasticity of Employment, xmin=-0.05,xmax=1.05, ymax=1.60, ymin=-0.60, height=14cm, width=14cm, grid=major, grid style=dotted, ytick={-0.50,-0.25,0,0.25,0.50,0.75,1.00,1.25,1.50}, extra y tick style={grid=none}, extra y ticks={0}, xtick={0,0.2,0.4,0.6,0.8,1.0}, legend pos = south east, y tick label style={/pgf/number format/.cd,fixed,fixed zerofill, precision=2,/tikz/.cd}, x tick label style={/pgf/number format/.cd,fixed,fixed zerofill, precision=1,/tikz/.cd}]
\addplot[domain=0:1, color=black, solid, line width=1.6pt] { -.2316022712913724 + 1.129695467818866 * x };
\addplot[draw=none, name path = A] coordinates {   (0,-.17589681) (.00401606,-.17176521) (.00803213,-.16756558) (.01204819,-.16329664) (.01606426,-.15895739) (.02008032,-.15454705) (.02409638,-.15006511) (.02811245,-.14551139) (.03212851,-.14088592) (.03614458,-.13618906) (.04016064,-.13142145) (.04417671,-.12658395) (.04819277,-.12167773) (.05220884,-.11670411) (.0562249,-.1116647) (.06024097,-.10656121) (.06425703,-.10139555) (.06827309,-.09616971) (.07228915,-.09088581) (.07630522,-.08554602) (.08032128,-.08015256) (.08433735,-.07470766) (.08835341,-.06921357) (.09236947,-.06367251) (.09638554,-.05808665) (.10040161,-.05245815) (.10441767,-.04678911) (.10843374,-.04108154) (.1124498,-.0353374) (.11646587,-.02955857) (.12048193,-.02374685) (.12449799,-.01790396) (.12851405,-.01203154) (.13253012,-.00613115) (.13654618,-.00020429) (.14056225,.00574767) (.14457831,.01172336) (.14859438,.01772155) (.15261044,.02374102) (.15662651,.0297807) (.16064256,.03583947) (.16465864,.04191638) (.16867469,.04801046) (.17269076,.05412085) (.17670682,.06024668) (.18072289,.06638722) (.18473895,.07254167) (.18875502,.07870938) (.19277108,.08488965) (.19678715,.09108192) (.20080322,.09728556) (.20481928,.10350002) (.20883535,.10972482) (.21285141,.11595942) (.21686748,.12220342) (.22088353,.12845634) (.2248996,.13471779) (.22891566,.14098737) (.23293173,.14726475) (.23694779,.15354952) (.24096386,.15984142) (.24497992,.16614009) (.24899599,.17244528) (.25301206,.17875671) (.2570281,.18507403) (.26104417,.19139713) (.26506025,.1977257) (.26907632,.20405951) (.27309236,.21039833) (.27710843,.21674202) (.2811245,.22309038) (.28514057,.22944319) (.28915662,.23580025) (.29317269,.24216148) (.29718876,.24852669) (.30120483,.25489572) (.30522087,.26126841) (.30923694,.26764464) (.31325302,.27402434) (.31726909,.28040731) (.32128513,.28679341) (.3253012,.29318261) (.32931727,.29957479) (.33333334,.3059698) (.33734939,.31236756) (.34136546,.31876799) (.34538153,.32517102) (.3493976,.33157656) (.35341364,.33798441) (.35742971,.34439465) (.36144578,.35080716) (.36546186,.35722187) (.3694779,.36363861) (.37349397,.37005746) (.37751004,.37647828) (.38152611,.38290104) (.38554215,.38932559) (.38955823,.39575201) (.3935743,.40218019) (.39759037,.40861008) (.40160644,.4150416) (.40562248,.42147467) (.40963855,.42790934) (.41365463,.43434554) (.4176707,.4407832) (.42168674,.44722223) (.42570281,.45366272) (.42971888,.46010455) (.43373495,.46654767) (.437751,.47299203) (.44176707,.47943771) (.44578314,.48588458) (.44979921,.49233261) (.45381525,.49878174) (.45783132,.50523204) (.46184739,.5116834) (.46586347,.51813585) (.46987951,.5245893) (.47389558,.53104377) (.47791165,.53749925) (.48192772,.54395568) (.48594376,.55041295) (.48995984,.55687124) (.49397591,.56333035) (.49799198,.56979036) (.50200802,.57625121) (.50602412,.58271295) (.51004016,.5891754) (.51405621,.59563857) (.51807231,.6021027) (.52208835,.60856742) (.52610439,.61503291) (.53012049,.62149924) (.53413653,.62796611) (.53815264,.63443381) (.54216868,.64090204) (.54618472,.64737093) (.55020082,.6538406) (.55421686,.66031075) (.5582329,.66678154) (.562249,.67325306) (.56626505,.67972499) (.57028115,.68619764) (.57429719,.69267076) (.57831323,.69914442) (.58232933,.70561874) (.58634537,.71209341) (.59036142,.71856868) (.59437752,.72504455) (.59839356,.73152077) (.60240966,.73799759) (.6064257,.74447483) (.61044174,.75095248) (.61445785,.75743073) (.61847389,.76390928) (.62248999,.77038842) (.62650603,.77686787) (.63052207,.78334773) (.63453817,.78982812) (.63855422,.79630882) (.64257026,.80278993) (.64658636,.80927151) (.6506024,.8157534) (.6546185,.82223576) (.65863454,.82871836) (.66265059,.83520138) (.66666669,.84168482) (.67068273,.84816849) (.67469877,.85465258) (.67871487,.86113703) (.68273091,.86762172) (.68674701,.87410688) (.69076306,.88059223) (.6947791,.88707793) (.6987952,.89356399) (.70281124,.90005028) (.70682728,.90653688) (.71084338,.91302383) (.71485943,.91951102) (.71887553,.92599857) (.72289157,.9324863) (.72690761,.93897432) (.73092371,.94546264) (.73493975,.95195121) (.7389558,.95844001) (.7429719,.96492916) (.74698794,.9714185) (.75100404,.97790813) (.75502008,.98439795) (.75903612,.990888) (.76305223,.99737835) (.76706827,1.0038688) (.77108431,1.0103596) (.77510041,1.0168507) (.77911645,1.0233419) (.78313255,1.0298334) (.78714859,1.036325) (.79116464,1.0428169) (.79518074,1.049309) (.79919678,1.0558013) (.80321288,1.0622938) (.80722892,1.0687865) (.81124496,1.0752794) (.81526107,1.0817724) (.81927711,1.0882657) (.82329315,1.0947591) (.82730925,1.1012528) (.83132529,1.1077466) (.83534139,1.1142406) (.83935744,1.1207348) (.84337348,1.1272291) (.84738958,1.1337237) (.85140562,1.1402184) (.85542166,1.1467131) (.85943776,1.1532083) (.86345381,1.1597034) (.86746991,1.1661988) (.87148595,1.1726943) (.87550199,1.1791899) (.87951809,1.1856859) (.88353413,1.1921817) (.88755018,1.1986779) (.89156628,1.2051742) (.89558232,1.2116705) (.89959842,1.2181672) (.90361446,1.2246639) (.9076305,1.2311606) (.9116466,1.2376577) (.91566265,1.2441547) (.91967869,1.250652) (.92369479,1.2571493) (.92771083,1.2636468) (.93172693,1.2701445) (.93574297,1.2766422) (.93975902,1.2831401) (.94377512,1.2896382) (.94779116,1.2961363) (.9518072,1.3026345) (.9558233,1.3091329) (.95983934,1.3156314) (.96385545,1.32213) (.96787149,1.3286287) (.97188753,1.3351275) (.97590363,1.3416265) (.97991967,1.3481255) (.98393571,1.3546246) (.98795182,1.3611239) (.99196786,1.3676233) (.99598396,1.3741229) (1,1.3806224) };
\addplot[draw=none, name path = B] coordinates {   (0,-.28730774) (.00401606,-.28236547) (.00803213,-.27749124) (.01204819,-.27268633) (.01606426,-.26795173) (.02008032,-.2632882) (.02409638,-.25869629) (.02811245,-.25417614) (.03212851,-.24972776) (.03614458,-.24535075) (.04016064,-.24104451) (.04417671,-.23680814) (.04819277,-.2326405) (.05220884,-.22854026) (.0562249,-.22450581) (.06024097,-.22053544) (.06425703,-.21662726) (.06827309,-.21277922) (.07228915,-.20898928) (.07630522,-.20525521) (.08032128,-.2015748) (.08433735,-.19794585) (.08835341,-.19436608) (.09236947,-.19083329) (.09638554,-.18734528) (.10040161,-.18389989) (.10441767,-.18049508) (.10843374,-.17712879) (.1124498,-.17379907) (.11646587,-.17050405) (.12048193,-.16724192) (.12449799,-.16401094) (.12851405,-.16080952) (.13253012,-.15763603) (.13654618,-.15448906) (.14056225,-.15136714) (.14457831,-.14826898) (.14859438,-.14519329) (.15261044,-.14213893) (.15662651,-.13910472) (.16064256,-.13608965) (.16465864,-.1330927) (.16867469,-.13011293) (.17269076,-.12714945) (.17670682,-.12420144) (.18072289,-.12126809) (.18473895,-.1183487) (.18875502,-.11544254) (.19277108,-.11254897) (.19678715,-.10966735) (.20080322,-.10679713) (.20481928,-.10393775) (.20883535,-.10108867) (.21285141,-.09824943) (.21686748,-.09541956) (.22088353,-.09259862) (.2248996,-.08978621) (.22891566,-.08698194) (.23293173,-.08418544) (.23694779,-.08139638) (.24096386,-.07861441) (.24497992,-.07583923) (.24899599,-.07307055) (.25301206,-.07030809) (.2570281,-.06755161) (.26104417,-.06480083) (.26506025,-.06205552) (.26907632,-.05931547) (.27309236,-.05658047) (.27710843,-.05385029) (.2811245,-.05112476) (.28514057,-.0484037) (.28915662,-.04568696) (.29317269,-.04297432) (.29718876,-.04026565) (.30120483,-.03756081) (.30522087,-.03485966) (.30923694,-.03216205) (.31325302,-.02946785) (.31726909,-.02677695) (.32128513,-.02408924) (.3253012,-.02140458) (.32931727,-.01872287) (.33333334,-.01604402) (.33734939,-.01336795) (.34136546,-.01069453) (.34538153,-.00802368) (.3493976,-.00535532) (.35341364,-.00268938) (.35742971,-.00002576) (.36144578,.00263562) (.36546186,.0052948) (.3694779,.00795186) (.37349397,.01060689) (.37751004,.01325994) (.38152611,.01591107) (.38554215,.01856031) (.38955823,.02120776) (.3935743,.02385347) (.39759037,.02649747) (.40160644,.02913982) (.40562248,.03178055) (.40963855,.03441975) (.41365463,.03705743) (.4176707,.03969364) (.42168674,.04232841) (.42570281,.04496181) (.42971888,.04759386) (.43373495,.0502246) (.437751,.05285404) (.44176707,.05548226) (.44578314,.05810928) (.44979921,.06073511) (.45381525,.06335978) (.45783132,.06598336) (.46184739,.06860585) (.46586347,.07122727) (.46987951,.07384766) (.47389558,.07646705) (.47791165,.07908546) (.48192772,.08170292) (.48594376,.08431943) (.48995984,.08693504) (.49397591,.08954976) (.49799198,.09216363) (.50200802,.09477663) (.50602412,.09738884) (.51004016,.1000002) (.51405621,.10261079) (.51807231,.10522064) (.52208835,.1078297) (.52610439,.11043803) (.53012049,.11304568) (.53413653,.11565259) (.53815264,.11825886) (.54216868,.12086441) (.54618472,.12346931) (.55020082,.12607361) (.55421686,.12867725) (.5582329,.13128027) (.562249,.13388273) (.56626505,.13648456) (.57028115,.13908586) (.57429719,.14168656) (.57831323,.14428671) (.58232933,.14688636) (.58634537,.14948544) (.59036142,.15208401) (.59437752,.1546821) (.59839356,.15727967) (.60240966,.15987678) (.6064257,.16247338) (.61044174,.16506952) (.61445785,.16766523) (.61847389,.17026044) (.62248999,.17285527) (.62650603,.17544962) (.63052207,.17804356) (.63453817,.18063711) (.63855422,.18323022) (.64257026,.18582293) (.64658636,.18841529) (.6506024,.19100721) (.6546185,.19359881) (.65863454,.19618998) (.66265059,.1987808) (.66666669,.20137131) (.67068273,.20396142) (.67469877,.20655119) (.67871487,.20914066) (.68273091,.21172975) (.68674701,.21431856) (.69076306,.21690701) (.6947791,.21949515) (.6987952,.222083) (.70281124,.22467053) (.70682728,.22725774) (.71084338,.22984472) (.71485943,.23243135) (.71887553,.23501775) (.72289157,.23760383) (.72690761,.24018963) (.73092371,.2427752) (.73493975,.24536046) (.7389558,.24794547) (.7429719,.25053027) (.74698794,.25311476) (.75100404,.25569904) (.75502008,.25828305) (.75903612,.26086682) (.76305223,.26345038) (.76706827,.26603368) (.77108431,.26861674) (.77510041,.27119961) (.77911645,.27378222) (.78313255,.27636465) (.78714859,.27894685) (.79116464,.28152883) (.79518074,.28411061) (.79919678,.28669217) (.80321288,.28927356) (.80722892,.29185471) (.81124496,.29443568) (.81526107,.29701647) (.81927711,.29959705) (.82329315,.30217746) (.82730925,.30475768) (.83132529,.3073377) (.83534139,.30991757) (.83935744,.31249726) (.84337348,.31507674) (.84738958,.3176561) (.85140562,.32023525) (.85542166,.32281423) (.85943776,.32539308) (.86345381,.32797173) (.86746991,.33055025) (.87148595,.3331286) (.87550199,.33570677) (.87951809,.33828485) (.88353413,.34086272) (.88755018,.34344044) (.89156628,.34601805) (.89558232,.3485955) (.89959842,.3511728) (.90361446,.35374996) (.9076305,.35632697) (.9116466,.35890388) (.91566265,.36148062) (.91967869,.36405724) (.92369479,.36663374) (.92771083,.36921006) (.93172693,.37178633) (.93574297,.37436241) (.93975902,.37693837) (.94377512,.37951425) (.94779116,.38208997) (.9518072,.38466555) (.9558233,.38724107) (.95983934,.38981643) (.96385545,.39239171) (.96787149,.39496681) (.97188753,.39754182) (.97590363,.40011677) (.97991967,.40269157) (.98393571,.40526626) (.98795182,.40784085) (.99196786,.41041532) (.99598396,.41298974) (1,.415564) };
\addplot[color=gray!40!white] fill between[of=A and B];
\addplot[loosely dashed] coordinates {(-0.05,0) (1.05,0)};
\legend{~Employment Elasticity,,,~90\% Confidence Interval,}
\end{axis}
\end{tikzpicture}
}
\floatfoot{\footnotesize\textsc{Note. ---} The figure illustrates estimated minimum wage elasticities of employment. Estimates stem from fixed effects regressions of log employment (in terms of regular full-time workers per firm) on log sectoral minimum wage levels as well as their interaction effect with labor market concentration (measured as average HHI over time). The thick line reports point estimates for different levels of concentration. The grey shade represents 90 percent confidence intervals. Sources: IEB $\plus$ BHP $\plus$ IAB Establishment Panel, 1999-2017.}
\end{figure}

\vspace*{\fill}
\clearpage
\vspace*{\fill}

\begin{table}[!ht]
\centering
\scalebox{0.90}{
\begin{threeparttable}
\caption{Minimum Wage Effects on Employment by Categorial Interaction Effects}
\label{tab:H4}
\begin{tabular}{L{4cm}C{2.7cm}C{2.7cm}C{2.7cm}C{2.7cm}} \hline
&&&& \\[-0.3cm]
\multirow{6}{*}{} & \multirow{6.4}{*}{\shortstack{(1) \\ \textbf{Two\vphantom{/}} \\ \textbf{Categories\vphantom{/}} \\ Log \\ Regular FT\vphantom{/} \\ Employment }} & \multirow{6.4}{*}{\shortstack{(2) \\ \textbf{Three\vphantom{/}} \\ \textbf{Categories\vphantom{/}} \\ Log \\ Regular FT\vphantom{/} \\ Employment }} & \multirow{6.4}{*}{\shortstack{(3) \\ \textbf{Four\vphantom{/}} \\ \textbf{Categories\vphantom{/}} \\ Log \\ Regular FT\vphantom{/} \\ Employment }} & \multirow{6.4}{*}{\shortstack{(4) \\ \textbf{Five\vphantom{/}} \\ \textbf{Categories\vphantom{/}} \\ Log \\ Regular FT\vphantom{/} \\ Employment }} \\
&&&&  \\
&&&&  \\
&&&&  \\
&&&&  \\
&&&&  \\[0.3cm] \hline
&&&& \\[-0.3cm]
\multirow{2.4}{*}{Log Minimum Wage} &  \multirow{2.4}{*}{\shortstack{\hphantom{***}-0.217***\hphantom{-} \\ (0.033)}}  & \multirow{2.4}{*}{\shortstack{\hphantom{***}-0.222***\hphantom{-} \\ (0.033)}} & \multirow{2.4}{*}{\shortstack{\hphantom{***}-0.226***\hphantom{-} \\ (0.033)}}  & \multirow{2.4}{*}{\shortstack{\hphantom{***}-0.225***\hphantom{-} \\ (0.033)}}  \\
&&&&  \\
\multirow{2.4}{*}{\shortstack[l]{Log Minimum Wage \\ $\times$ $\overline{\text{HHI}}$ (0.00-0.05)}} &   &  & \multirow{2.4}{*}{\shortstack{Reference \\ Group}} & \multirow{2.4}{*}{\shortstack{Reference \\ Group}}   \\
&&&&  \\
\multirow{2.4}{*}{\shortstack[l]{Log Minimum Wage \\ $\times$ $\overline{\text{HHI}}$ (0.00-0.10)}} &   &  \multirow{2.4}{*}{\shortstack{Reference \\ Group}}  &  &      \\
&&&&  \\
\multirow{2.4}{*}{\shortstack[l]{Log Minimum Wage \\ $\times$ $\overline{\text{HHI}}$ (0.00-0.20)}} &  \multirow{2.4}{*}{\shortstack{Reference \\ Group}} &    &  &      \\
&&&&  \\
\multirow{2.4}{*}{\shortstack[l]{Log Minimum Wage \\ $\times$ $\overline{\text{HHI}}$ (0.05-0.10)}} &   &   &  \multirow{2.4}{*}{\shortstack{0.115 \\ (0.099)}} & \multirow{2.4}{*}{\shortstack{0.116 \\ (0.099)}}      \\
&&&&  \\
\multirow{2.4}{*}{\shortstack[l]{Log Minimum Wage \\ $\times$ $\overline{\text{HHI}}$ (0.10-0.20)}} &   &  \multirow{2.4}{*}{\shortstack{\hphantom{***}0.288*** \\ (0.088)}} &  \multirow{2.4}{*}{\shortstack{\hphantom{***}0.298*** \\ (0.092)}} &  \multirow{2.4}{*}{\shortstack{\hphantom{***}0.299*** \\ (0.092)}}     \\
&&&&  \\
\multirow{2.4}{*}{\shortstack[l]{Log Minimum Wage \\ $\times$ $\overline{\text{HHI}}$ (0.20-0.40)}} &   &   &   &  \multirow{2.4}{*}{\shortstack{\hphantom{**}0.204** \\ (0.095)}}    \\
&&&&  \\
\multirow{2.4}{*}{\shortstack[l]{Log Minimum Wage \\ $\times$ $\overline{\text{HHI}}$ (0.20-1.00)}} &  \multirow{2.4}{*}{\shortstack{\hphantom{***}0.260*** \\ (0.088)}} & \multirow{2.4}{*}{\shortstack{\hphantom{***}0.275*** \\ (0.090)}}  &  \multirow{2.4}{*}{\shortstack{\hphantom{***}0.286*** \\ (0.092)}} &       \\
&&&&  \\
\multirow{2.4}{*}{\shortstack[l]{Log Minimum Wage \\ $\times$ $\overline{\text{HHI}}$ (0.40-1.00)}} &   &   &   & \multirow{2.4}{*}{\shortstack{\hphantom{***}0.555*** \\ (0.188)}}      \\
&&&&  \\[0.2cm] 
Control Variables  & Yes & Yes  & Yes & Yes  \\
\multirow{2.4}{*}{\shortstack{Fixed Effects}}  & \multirow{2.4}{*}{\shortstack{Firm \\ CZ $\times$ Year}} & \multirow{2.4}{*}{\shortstack{Firm \\ CZ $\times$ Year}} & \multirow{2.4}{*}{\shortstack{Firm \\ CZ $\times$ Year}}  & \multirow{2.4}{*}{\shortstack{Firm \\ CZ $\times$ Year}} \\
&&&& \\[0.2cm] \hline
&&&& \\[-0.4cm]
\multirow{3.4}{*}{\shortstack[l]{Labor Market \\ Definition \\ (Object)}}  & \multirow{3.4}{*}{\shortstack{NACE-4 \\ $\times$ CZ \\ (Employment)}} & \multirow{3.4}{*}{\shortstack{NACE-4 \\ $\times$ CZ \\ (Employment)}} & \multirow{3.4}{*}{\shortstack{NACE-4 \\ $\times$ CZ \\ (Employment)}}  & \multirow{3.4}{*}{\shortstack{NACE-4 \\ $\times$ CZ \\ (Employment)}} \\
&&&& \\
&&&& \\[0.2cm] 
Observations &  2,700,155    & 2,700,155    & 2,700,155    & 2,700,155      \\[0.2cm]
Adjusted R$^2$ &  0.877    & 0.877    &   0.877    & 0.877      \\[0.2cm] \hline
\end{tabular}
\begin{tablenotes}[para]
\footnotesize\textsc{Note. ---} The table displays fixed effects regressions of log employment (in terms of regular full-time workers per firm) on log sectoral minimum wage levels as well as their categorial interaction effect with labor market concentration (measured as average HHI over time). The set of control variables includes log sectoral employment, the sectoral share of firms subject to a collective bargaining agreement, and sector-specific linear time trends. Standard errors (in parentheses) are clustered at the sector-by-federal-state level. CZ = Commuting Zone. FT = Full-Time. HHI = Herfindahl-Hirschman Index. NACE-4 = 4-Digit Statistical Nomenclature of Economic Activities in the European Community. * = p$<$0.10. ** = p$<$0.05. *** = p$<$0.01. Sources: IEB $\plus$ BHP $\plus$ IAB Establishment Panel, 1999-2017.
\end{tablenotes}
\end{threeparttable}
}
\end{table}

\vspace*{2.5cm}
\vspace*{\fill}

\begin{landscape}

\clearpage
\vspace*{\fill}

\begin{table}[!ht]
\centering
\scalebox{0.90}{
\begin{threeparttable}
\caption{Minimum Wage Effects on Employment by Specification}
\label{tab:H5}
\begin{tabular}{L{4cm}C{2.7cm}C{2.7cm}C{2.7cm}C{2.7cm}} \hline
&&&& \\[-0.3cm]
\multirow{6}{*}{} & \multirow{6.4}{*}{\shortstack{(1) \\ \textbf{Time-Varying\vphantom{/}} \\ \textbf{HHI\vphantom{/}} \\ Log \\ Regular FT\vphantom{/} \\ Employment }} & \multirow{6.4}{*}{\shortstack{(2) \\ \textbf{Predetermined\vphantom{/}} \\ \textbf{HHI\vphantom{/}} \\ Log \\ Regular FT\vphantom{/} \\ Employment }} & \multirow{6.4}{*}{\shortstack{(3) \\ \textbf{Without\vphantom{/}} \\ \textbf{Time Trends\vphantom{/}} \\ Log \\ Regular FT\vphantom{/} \\ Employment }}  & \multirow{6.4}{*}{\shortstack{(4) \\ \textbf{With Implicit\vphantom{/}} \\ \textbf{Min Wage\vphantom{/}} \\ Log \\ Regular FT\vphantom{/} \\ Employment }} \\
&&&&   \\
&&&&   \\
&&&&   \\
&&&&   \\
&&&&   \\[0.2cm] \hline
&&&& \\[-0.3cm]
\multirow{2.4}{*}{Log Minimum Wage} & \multirow{2.4}{*}{\shortstack{\hphantom{***}-0.227***\hphantom{-}  \\ (0.033)}}  & \multirow{2.4}{*}{\shortstack{\hphantom{***}-0.232***\hphantom{-}  \\ (0.034)}} & \multirow{2.4}{*}{\shortstack{\hphantom{***}-0.309***\hphantom{-}  \\ (0.029)}} & \multirow{2.4}{*}{\shortstack{\hphantom{***}-0.247***\hphantom{-}  \\ (0.028)}}   \\
&&&&  \\
\multirow{2.4}{*}{\shortstack[l]{Log Minimum Wage \\ $\times$ $\overline{\text{HHI}}$}}   & \multirow{2.4}{*}{\shortstack{\hphantom{***}0.715*** \\ (0.241)}}   &  \multirow{2.4}{*}{\shortstack{\hphantom{***}1.074*** \\ (0.268)}}  &   \multirow{2.4}{*}{\shortstack{\hphantom{***}1.155*** \\ (0.288)}} &  \multirow{2.4}{*}{\shortstack{\hphantom{***}0.820*** \\ (0.168)}} \\
&&&&  \\[0.2cm] 
Control Variables  & Yes & Yes  & Yes & Yes  \\
\multirow{2.4}{*}{\shortstack{Fixed Effects}}  & \multirow{2.4}{*}{\shortstack{Firm \\ CZ $\times$ Year}} & \multirow{2.4}{*}{\shortstack{Firm \\ CZ $\times$ Year}} & \multirow{2.4}{*}{\shortstack{Firm \\ CZ $\times$ Year}} & \multirow{2.4}{*}{\shortstack{Firm \\ CZ $\times$ Year}} \\
&&&& \\[0.2cm] \hline
&&&& \\[-0.4cm]
\multirow{3.4}{*}{\shortstack[l]{Labor Market \\ Definition \\ (Object)}}  & \multirow{3.4}{*}{\shortstack{NACE-4 \\ $\times$ CZ \\ (Employment)}}  & \multirow{3.4}{*}{\shortstack{NACE-4 \\ $\times$ CZ \\ (Employment)}} & \multirow{3.4}{*}{\shortstack{NACE-4 \\ $\times$ CZ \\ (Employment)}}  & \multirow{3.4}{*}{\shortstack{NACE-4 \\ $\times$ CZ \\ (Employment)}} \\
&&&&  \\
&&&&  \\
\multirow{2.4}{*}{Specification}  & \multirow{2.4}{*}{\shortstack{Time-Varying\vphantom{/} \\[-0.1cm]HHI}} & \multirow{2.4}{*}{\shortstack{Predetermined\vphantom{/} \\[-0.1cm]HHI}}  & \multirow{2.4}{*}{\shortstack{Without\vphantom{/} \\[-0.1cm] Time Trends}}  & \multirow{2.4}{*}{\shortstack{With Implicit\vphantom{/} \\[-0.1cm] Minimum Wage}}  \\
&&&& \\[0.2cm] 
Observations &  2,700,155    & 2,699,813    & 2,700,155    & 2,867,705      \\[0.2cm]
Adjusted R$^2$ &  0.877    & 0.877    &   0.877    & 0.878      \\[0.2cm] \hline
\end{tabular}
\begin{tablenotes}[para]
\footnotesize\textsc{Note. ---} The table displays fixed effects regressions of log employment (in terms of regular full-time workers per firm) on log sectoral minimum wage levels as well as their interaction effect with labor market concentration. The set of control variables includes log sectoral employment, the sectoral share of firms subject to a collective bargaining agreement, and sector-specific linear time trends. Standard errors (in parentheses) are clustered at the sector-by-federal-state level. CZ = Commuting Zone. FT = Full-Time. HHI = Herfindahl-Hirschman Index. NACE-4 = 4-Digit Statistical Nomenclature of Economic Activities in the European Community. * = p$<$0.10. ** = p$<$0.05. *** = p$<$0.01. Sources: IEB $\plus$ BHP $\plus$ IAB Establishment Panel, 1999-2017.
\end{tablenotes}
\end{threeparttable}
}
\end{table}

\vspace*{\fill}
\clearpage

\vspace*{0.1cm}

\begin{table}[!ht]
\centering
\scalebox{0.90}{
\begin{threeparttable}
\caption{Minimum Wage Effects on Employment by Labor Market Definition}
\label{tab:H6}
\begin{tabular}{L{4cm}C{2.7cm}C{2.7cm}C{2.7cm}C{2.7cm}C{2.7cm}C{2.7cm}} \hline
&&&&&& \\[-0.3cm]
\multirow{7.4}{*}{}  & \multirow{7.4}{*}{\shortstack{(1) \\ \textbf{NACE-3 \vphantom{/}} \\ \textbf{Industries \vphantom{/}} \\ Log \\ Regular FT\vphantom{/} \\ Employment }} & \multirow{7.4}{*}{\shortstack{(2) \\ \textbf{NACE-5 \vphantom{/}} \\ \textbf{Industries \vphantom{/}} \\ Log \\ Regular FT\vphantom{/} \\ Employment }} & \multirow{7.4}{*}{\shortstack{(3) \\ \textbf{Flow-Adjusted \vphantom{/}} \\ \textbf{NACE-4 \vphantom{/}} \\ \textbf{Industries \vphantom{/}} \\ Log \\ Regular FT\vphantom{/} \\ Employment }} & \multirow{7.4}{*}{\shortstack{(4) \\ \textbf{NUTS-3 \vphantom{/}} \\ \textbf{Regions \vphantom{/}} \\ Log \\ Regular FT\vphantom{/} \\ Employment }} & \multirow{7.4}{*}{\shortstack{(5) \\ \textbf{HHI \vphantom{/}} \\ \textbf{Based on \vphantom{/}} \\ \textbf{Hires \vphantom{/}}  \\ Log \\ Regular FT\vphantom{/} \\ Employment }} & \multirow{7.4}{*}{\shortstack{(6) \\ \textbf{HHI \vphantom{/}} \\ \textbf{Based on \vphantom{/}} \\ \textbf{Wage Bill \vphantom{/}} \\ Log \\ Regular FT\vphantom{/} \\ Employment }} \\
&&&&&&   \\
&&&&&&   \\
&&&&&&   \\
&&&&&&   \\
&&&&&&   \\
&&&&&&   \\[0.3cm] \hline
&&&&&& \\[-0.3cm]
\multirow{2.4}{*}{Log Minimum Wage} & \multirow{2.4}{*}{\shortstack{\hphantom{***}-0.221***\hphantom{-}  \\ (0.033)}}  & \multirow{2.4}{*}{\shortstack{\hphantom{***}-0.230***\hphantom{-}  \\ (0.034)}} & \multirow{2.4}{*}{\shortstack{\hphantom{***}-0.233***\hphantom{-}  \\ (0.034)}} & \multirow{2.4}{*}{\shortstack{\hphantom{***}-0.249***\hphantom{-}  \\ (0.036)}} & \multirow{2.4}{*}{\shortstack{\hphantom{***}-0.208***\hphantom{-}  \\ (0.033)}} & \multirow{2.4}{*}{\shortstack{\hphantom{***}-0.237***\hphantom{-}  \\ (0.034)}}  \\
&&&&&&  \\
\multirow{2.4}{*}{\shortstack[l]{Log Minimum Wage \\ $\times$ $\overline{\text{HHI}}$}}   & \multirow{2.4}{*}{\shortstack{0.457 \\ (0.448)}}   &  \multirow{2.4}{*}{\shortstack{\hphantom{***}0.861*** \\ (0.234)}}  &   & \multirow{2.4}{*}{\shortstack{\hphantom{***}0.624*** \\ (0.137)}} &  \multirow{2.4}{*}{\shortstack{\hphantom{*}0.425* \\ (0.254)}} &  \multirow{2.4}{*}{\shortstack{\hphantom{***}1.147*** \\ (0.272)}} \\
&&&&&&  \\
\multirow{2.4}{*}{\shortstack[l]{Log Minimum Wage \\ $\times$ $\overline{\text{Adjusted HHI}}$}}   &   &  & \multirow{2.4}{*}{\shortstack{\hphantom{***}1.592*** \\ (0.420)}}  &   &  & \\
&&&&&&  \\[0.2cm] 
Control Variables  & Yes & Yes  & Yes & Yes & Yes & Yes \\
\multirow{2.4}{*}{\shortstack{Fixed Effects}}  & \multirow{2.4}{*}{\shortstack{Firm \\ CZ $\times$ Year}} & \multirow{2.4}{*}{\shortstack{Firm \\ CZ $\times$ Year}} & \multirow{2.4}{*}{\shortstack{Firm \\ CZ $\times$ Year}} & \multirow{2.4}{*}{\shortstack{Firm \\ NUTS-3 $\times$ Year}} & \multirow{2.4}{*}{\shortstack{Firm \\ CZ $\times$ Year}} & \multirow{2.4}{*}{\shortstack{Firm \\ CZ $\times$ Year}} \\
&&&&&& \\[0.2cm] \hline
&&&&&& \\[-0.4cm]
\multirow{3.4}{*}{\shortstack[l]{Labor Market \\ Definition \\ (Object)}}  & \multirow{3.4}{*}{\shortstack{NACE-3 \\ $\times$ CZ \\ (Employment)}} & \multirow{3.4}{*}{\shortstack{NACE-5 \\ $\times$ CZ \\ (Employment)}} & \multirow{3.4}{*}{\shortstack{NACE-4 \\ $\times$ CZ \\ (Employment)}} & \multirow{3.4}{*}{\shortstack{NACE-4 \\ $\times$ NUTS-3 \\ (Employment)}}  & \multirow{3.4}{*}{\shortstack{NACE-4 \\ $\times$ CZ \\ (Hires)}}  & \multirow{3.4}{*}{\shortstack{NACE-4 \\ $\times$ CZ \\ (Wage Bill)}} \\
&&&&&&  \\
&&&&&&  \\
\multirow{2.4}{*}{Sample}  & \multirow{2.4}{*}{1999-2017} & \multirow{2.4}{*}{1999-2017} & \multirow{2.4}{*}{1999-2017} & \multirow{2.4}{*}{1999-2017}  & \multirow{2.4}{*}{2000-2017} & \multirow{2.4}{*}{1999-2017} \\
&&&&&& \\[0.2cm] 
Observations &  2,700,155    & 2,700,155    & 2,700,155    & 2,700,155    & 2,582,560    & ~2,700,155      \\[0.2cm]
Adjusted R$^2$ &  0.877    & 0.877    &   0.877    & 0.877   & 0.880  & ~0.877  \\[0.2cm] \hline
\end{tabular}
\begin{tablenotes}[para]
\footnotesize\textsc{Note. ---} The table displays fixed effects regressions of log employment (in terms of regular full-time workers per firm) on log sectoral minimum wage levels as well as their interaction effect with labor market concentration (measured as average HHI over time). The set of control variables includes log sectoral employment, the sectoral share of firms subject to a collective bargaining agreement, and sector-specific linear time trends. Standard errors (in parentheses) are clustered at the sector-by-federal-state level. CZ = Commuting Zone. FT = Full-Time. HHI = Herfindahl-Hirschman Index. NACE-X = X-Digit Statistical Nomenclature of Economic Activities in the European Community. NUTS-X = X-Digit Statistical Nomenclature of Territorial Units. * = p$<$0.10. ** = p$<$0.05. *** = p$<$0.01. Sources: IEB $\plus$ BHP $\plus$ IAB Establishment Panel, 1999-2017.
\end{tablenotes}
\end{threeparttable}
}
\end{table}

\clearpage

\begin{table}[!ht]
\centering
\scalebox{0.90}{
\begin{threeparttable}
\caption{Minimum Wage Effects on Employment by Concentration Measure}
\label{tab:H7}
\begin{tabular}{L{4cm}C{2.7cm}C{2.7cm}C{2.7cm}C{2.7cm}C{2.7cm}} \hline
&&&&& \\[-0.3cm]
\multirow{7.4}{*}{} & \multirow{7.4}{*}{\shortstack{(1) \\ \textbf{Rosenbluth\vphantom{/}} \\ \textbf{Index\vphantom{/}} \\ Log \\ Regular FT\vphantom{/} \\ Employment }} & \multirow{7.4}{*}{\shortstack{(2) \\ \textbf{1-Firm\vphantom{/}} \\ \textbf{Concentration\vphantom{/}} \\ \textbf{Ratio\vphantom{/}} \\ Log \\ Regular FT\vphantom{/} \\ Employment }} & \multirow{7.4}{*}{\shortstack{(3) \\ \textbf{Inverse\vphantom{/}} \\ \textbf{Number\vphantom{/}} \\ \textbf{of Firms\vphantom{/}} \\ Log \\ Regular FT\vphantom{/} \\ Employment }} & \multirow{7.4}{*}{\shortstack{(4) \\ \textbf{Exponential\vphantom{/}} \\ \textbf{Index\vphantom{/}} \\ Log \\ Regular FT\vphantom{/} \\ Employment }} & \multirow{7.4}{*}{\shortstack{(5) \\ \textbf{Market\vphantom{/}} \\ \textbf{Share\vphantom{/}} \\ Log \\ Regular FT\vphantom{/} \\ Employment }}  \\
&&&&&   \\
&&&&&   \\
&&&&&   \\
&&&&&   \\
&&&&&   \\
&&&&&   \\[0.3cm] \hline
&&&&& \\[-0.3cm]
\multirow{2.4}{*}{Log Minimum Wage} & \multirow{2.4}{*}{\shortstack{\hphantom{***}-0.227***\hphantom{-}  \\ (0.033)}}  & \multirow{2.4}{*}{\shortstack{\hphantom{***}-0.256***\hphantom{-}  \\ (0.037)}} & \multirow{2.4}{*}{\shortstack{\hphantom{***}-0.222***\hphantom{-}  \\ (0.033)}} & \multirow{2.4}{*}{\shortstack{\hphantom{***}-0.226***\hphantom{-}  \\ (0.033)}} & \multirow{2.4}{*}{\shortstack{\hphantom{***}-0.225***\hphantom{-}  \\ (0.033)}}  \\
&&&&&  \\
\multirow{2.4}{*}{\shortstack[l]{Log Minimum Wage \\ $\times$ $\overline{\text{RBI}}$}}  & \multirow{2.4}{*}{\shortstack{\hphantom{***}1.196*** \\ (0.302)}}   &   &   & & \\
&&&&&  \\
\multirow{2.4}{*}{\shortstack[l]{Log Minimum Wage \\ $\times$ $\overline{\text{CR1}}$}}  & & \multirow{2.4}{*}{\shortstack{\hphantom{***}0.856*** \\ (0.246)}}       &    &  & \\
&&&&&  \\
\multirow{2.4}{*}{\shortstack[l]{Log Minimum Wage \\ $\times$ $\overline{\text{INF}}$}}  & &       & \multirow{2.4}{*}{\shortstack{\hphantom{***}1.449*** \\ (0.331)}}   &  & \\
&&&&&  \\
\multirow{2.4}{*}{\shortstack[l]{Log Minimum Wage \\ $\times$ $\overline{\text{EXP}}$}}  & &       &       &  \multirow{2.4}{*}{\shortstack{\hphantom{***}1.297*** \\ (0.329)}} & \\
&&&&&  \\[0.2cm] 
\multirow{2.4}{*}{\shortstack[l]{Log Minimum Wage \\ $\times$ $\overline{\text{Market Share}}$}}  &        &       &       &  & \multirow{2.4}{*}{\shortstack{\hphantom{***}1.739*** \\ (0.313)}}  \\
&&&&&  \\[0.2cm] 
Control Variables  & Yes & Yes  & Yes & Yes & Yes \\
\multirow{2.4}{*}{\shortstack{Fixed Effects}}  & \multirow{2.4}{*}{\shortstack{Firm \\ CZ $\times$ Year}} & \multirow{2.4}{*}{\shortstack{Firm \\ CZ $\times$ Year}} & \multirow{2.4}{*}{\shortstack{Firm \\ CZ $\times$ Year}} & \multirow{2.4}{*}{\shortstack{Firm \\ CZ $\times$ Year}} & \multirow{2.4}{*}{\shortstack{Firm \\ CZ $\times$ Year}} \\
&&&&& \\[0.2cm] \hline
&&&&& \\[-0.2cm]
Observations &  2,700,155    & 2,700,155    & 2,700,155    & 2,700,155    & 2,700,155     \\[0.2cm]
Adjusted R$^2$ &  0.877    & 0.877    &   0.877   &   0.877 &   0.877     \\[0.2cm] \hline
\end{tabular}
\begin{tablenotes}[para]
\footnotesize\textsc{Note. ---} The table displays fixed effects regressions of log employment (in terms of regular full-time workers per firm) on log sectoral minimum wage levels as well as their interaction effect with labor market concentration (measured as average over time). The set of control variables includes log sectoral employment, the sectoral share of firms subject to a collective bargaining agreement, and sector-specific linear time trends. Labor markets are combinations of 4-digit NACE industries and commuting zone. Standard errors (in parentheses) are clustered at the sector-by-federal-state level. CR1 = 1-Firm Concentration Ratio. CZ = Commuting Zone. FT = Full-Time. EXP = Exponential Index. INF = Inverse Number of Firms. NACE-4 = 4-Digit Statistical Nomenclature of Economic Activities in the European Community. RBI = Rosenbluth Index. * = p$<$0.10. ** = p$<$0.05. *** = p$<$0.01. Sources: IEB $\plus$ BHP $\plus$ IAB Establishment Panel, 1999-2017.
\end{tablenotes}
\end{threeparttable}
}
\end{table}

\clearpage
\vspace*{\fill}

\begin{table}[!ht]
\centering
\scalebox{0.90}{
\begin{threeparttable}
\caption{Minimum Wage Effects on Employment by Territory}
\label{tab:H8}
\begin{tabular}{L{4cm}C{5.8cm}C{5.8cm}} \hline
&& \\[-0.3cm]
\multirow{6}{*}{} & \multirow{6.4}{*}{\shortstack{(1) \\ \textbf{West\vphantom{/}} \\ \textbf{Germany\vphantom{/}} \\ Log \\ Regular FT\vphantom{/} \\ Employment }} & \multirow{6.4}{*}{\shortstack{(2) \\ \textbf{West\vphantom{/}} \\ \textbf{Germany\vphantom{/}} \\ Log \\ Regular FT\vphantom{/} \\ Employment }}   \\
&&      \\
&&      \\
&&      \\
&&      \\
&&      \\[0.3cm] \hline
&& \\[-0.3cm]
\multirow{2.4}{*}{Log Minimum Wage} &  \multirow{2.4}{*}{\shortstack{\hphantom{***}-0.181***\hphantom{-} \\ (0.038)}}  & \multirow{2.4}{*}{\shortstack{\hphantom{***}-0.338***\hphantom{-} \\ (0.067)}}    \\
&&  \\
\multirow{2.4}{*}{\shortstack[l]{Log Minimum Wage \\ $\times$ $\overline{\text{HHI}}$}} &  \multirow{2.4}{*}{\shortstack{\hphantom{***}1.037*** \\ (0.377)}}  & \multirow{2.4}{*}{\shortstack{\hphantom{**}1.249** \\ (0.485)}}    \\
&&  \\[0.2cm] 
Control Variables  & Yes & Yes    \\
\multirow{2.4}{*}{\shortstack{Fixed Effects}}  & \multirow{2.4}{*}{\shortstack{Firm \\ CZ $\times$ Year}} & \multirow{2.4}{*}{\shortstack{Firm \\ CZ $\times$ Year}}  \\
&& \\[0.2cm] \hline
&& \\[-0.4cm]
\multirow{3.4}{*}{\shortstack[l]{Labor Market \\ Definition \\ (Object)}}  & \multirow{3.4}{*}{\shortstack{NACE-4 \\ $\times$ CZ \\ (Employment)}} & \multirow{3.4}{*}{\shortstack{NACE-4 \\ $\times$ CZ \\ (Employment)}}   \\
&& \\
&& \\
\multirow{2.4}{*}{Sample}  & \multirow{2.4}{*}{West Germany} & \multirow{2.4}{*}{\shortstack{East Germany \\[-0.1cm] Berlin}}  \\
&& \\[0.2cm] 
Observations &  2,063,033    & 637,122         \\[0.2cm]
Adjusted R$^2$ &  0.881    & 0.861     \\[0.2cm] \hline
\end{tabular}
\begin{tablenotes}[para]
\footnotesize\textsc{Note. ---} The table displays fixed effects regressions of log employment (in terms of regular full-time workers per firm) on log sectoral minimum wage levels as well as their interaction effect with labor market concentration (measured as average HHI over time). The set of control variables includes log sectoral employment, the sectoral share of firms subject to a collective bargaining agreement, and sector-specific linear time trends. Standard errors (in parentheses) are clustered at the sector-by-federal-state level. CZ = Commuting Zone. FT = Full-Time. HHI = Herfindahl-Hirschman Index. NACE-4 = 4-Digit Statistical Nomenclature of Economic Activities in the European Community. * = p$<$0.10. ** = p$<$0.05. *** = p$<$0.01. Sources: IEB $\plus$ BHP $\plus$ IAB Establishment Panel, 1999-2017.
\end{tablenotes}
\end{threeparttable}
}
\end{table}

\vspace*{\fill}
\clearpage

\vspace*{\fill}

\begin{table}[!ht]
\centering
\scalebox{0.90}{
\begin{threeparttable}
\caption{Minimum Wage Effects on Employment by Labor Outcome}
\label{tab:H9}
\begin{tabular}{L{4cm}C{3.75cm}C{3.75cm}C{3.75cm}} \hline
&&& \\[-0.3cm]
\multirow{6.4}{*}{} & \multirow{6.4}{*}{\shortstack{(1) \\ \textbf{All\vphantom{/}} \\  \textbf{Workers\vphantom{/}} \\ Log \\ Overall\vphantom{/} \\ Employment }} & \multirow{6.4}{*}{\shortstack{(2) \\ \textbf{Regular PT\vphantom{/}} \\  \textbf{Workers\vphantom{/}} \\ Log \\ Regular PT\vphantom{/} \\ Employment }} & \multirow{6.4}{*}{\shortstack{(3) \\ \textbf{Marginal PT\vphantom{/}} \\  \textbf{Workers\vphantom{/}} \\ Log \\ Marginal\vphantom{/} \\ Employment }}  \\
&&&      \\
&&&      \\
&&&      \\
&&&      \\
&&&      \\[0.3cm] \hline
&&& \\[-0.3cm]
\multirow{2.4}{*}{Log Minimum Wage} &  \multirow{2.4}{*}{\shortstack{\hphantom{***}-0.068***\hphantom{-} \\ (0.025)}}  & \multirow{2.4}{*}{\shortstack{\hphantom{***}0.357*** \\ (0.063)}} & \multirow{2.4}{*}{\shortstack{\hphantom{***}-0.069***\hphantom{-} \\ (0.025)}}     \\
&&&  \\
\multirow{2.4}{*}{\shortstack[l]{Log Minimum Wage \\ $\times$ $\overline{\text{HHI}}$}} &  \multirow{2.4}{*}{\shortstack{\hphantom{***}1.005*** \\ (0.275)}}  & \multirow{2.4}{*}{\shortstack{\hphantom{***}1.385*** \\ (0.373)}} & \multirow{2.4}{*}{\shortstack{0.341 \\ (0.245)}}     \\
&&&  \\[0.2cm] 
Control Variables  & Yes & Yes  & Yes   \\
\multirow{2.4}{*}{\shortstack{Fixed Effects}}  & \multirow{2.4}{*}{\shortstack{Firm \\ CZ $\times$ Year}} & \multirow{2.4}{*}{\shortstack{Firm \\ CZ $\times$ Year}} & \multirow{2.4}{*}{\shortstack{Firm \\ CZ $\times$ Year}}  \\
&&& \\[0.2cm] \hline
&&& \\[-0.4cm]
\multirow{3.4}{*}{\shortstack[l]{Labor Market \\ Definition \\ (Object)}}  & \multirow{3.4}{*}{\shortstack{NACE-4 \\ $\times$ CZ \\ (Employment)}} & \multirow{3.4}{*}{\shortstack{NACE-4 \\ $\times$ CZ \\ (Employment)}} & \multirow{3.4}{*}{\shortstack{NACE-4 \\ $\times$ CZ \\ (Employment)}} \\
&&& \\
&&& \\[0.2cm] 
Observations &  3,507,692    & 1,240,992    & 1,957,415       \\[0.2cm]
Adjusted R$^2$ &  0.886    & 0.890    &   0.824         \\[0.2cm] \hline
\end{tabular}
\begin{tablenotes}[para]
\footnotesize\textsc{Note. ---} The table displays fixed effects regressions of log employment on log sectoral minimum wage levels as well as their interaction effect with labor market concentration (measured as average HHI over time). The set of control variables includes log sectoral employment, the sectoral share of firms subject to a collective bargaining agreement, and sector-specific linear time trends. Overall employment is defined as the sum of regular full-time workers, regular part-time workers and workers in marginal employment. Standard errors (in parentheses) are clustered at the sector-by-federal-state level. CZ = Commuting Zone. HHI = Herfindahl-Hirschman Index. NACE-4 = 4-Digit Statistical Nomenclature of Economic Activities in the European Community. PT = Part-Time. * = p$<$0.10. ** = p$<$0.05. *** = p$<$0.01. Sources: IEB $\plus$ BHP $\plus$ IAB Establishment Panel, 1999-2017.
\end{tablenotes}
\end{threeparttable}
}
\end{table}

\vspace*{\fill}

\end{landscape}

\clearpage
\vspace*{\fill}

\begin{table}[!ht]
\centering
\scalebox{0.90}{
\begin{threeparttable}
\caption{Minimum Wage Effects on Employment by Bindingness}
\label{tab:H10}
\begin{tabular}{L{10cm}C{6cm}} \hline
\multirow{6}{*}{} & \multirow{6.4}{*}{\shortstack{(1) \\ \textbf{By Kaitz\vphantom{/}} \\ \textbf{Index\vphantom{/}} \\ Log \\ Regular FT\vphantom{/} \\ Employment }}  \\
&  \\
&  \\
&  \\
&  \\
&  \\[0.2cm] \hline
\multirow{2.4}{*}{Log Minimum Wage} & \multirow{2.4}{*}{\shortstack{\hphantom{***}-0.230***\hphantom{-} \\ (0.047)}}    \\
&  \\
\multirow{2.4}{*}{Log Minimum Wage $\times$ $\overline{\text{2nd Kaitz Quintile}}$ } &  \multirow{2.4}{*}{\shortstack{\hphantom{**}0.088** \\ (0.035)}}    \\
&  \\
\multirow{2.4}{*}{Log Minimum Wage $\times$ $\overline{\text{3rd Kaitz Quintile}}$ } &  \multirow{2.4}{*}{\shortstack{\hphantom{*}0.099* \\ (0.056)}}      \\
&  \\
\multirow{2.4}{*}{Log Minimum Wage $\times$ $\overline{\text{4th Kaitz Quintile}}$ } &  \multirow{2.4}{*}{\shortstack{-0.094\hphantom{-} \\ (0.071)}}      \\
&  \\
\multirow{2.4}{*}{Log Minimum Wage $\times$ $\overline{\text{5th Kaitz Quintile}}$ } &  \multirow{2.4}{*}{\shortstack{\hphantom{***}-0.302***\hphantom{-} \\ (0.106)}}      \\
&  \\
\multirow{2.4}{*}{Log Minimum Wage $\times$ $\overline{\text{HHI}}$ } &    \multirow{2.4}{*}{\shortstack{\hphantom{***}0.993*** \\ (0.362)}}    \\
&  \\
\multirow{2.4}{*}{Log Minimum Wage $\times$ $\overline{\text{HHI}}$ $\times$ $\overline{\text{2nd Kaitz Quintile}}$ } &    \multirow{2.4}{*}{\shortstack{0.553 \\ (0.341)}}    \\
&  \\
\multirow{2.4}{*}{Log Minimum Wage $\times$ $\overline{\text{HHI}}$ $\times$ $\overline{\text{3rd Kaitz Quintile}}$ } &    \multirow{2.4}{*}{\shortstack{0.435 \\ (0.508)}}     \\
&  \\
\multirow{2.4}{*}{Log Minimum Wage $\times$ $\overline{\text{HHI}}$ $\times$ $\overline{\text{4th Kaitz Quintile}}$ } &   \multirow{2.4}{*}{\shortstack{0.073 \\ (0.701)}}      \\
&  \\
\multirow{2.4}{*}{Log Minimum Wage $\times$ $\overline{\text{HHI}}$ $\times$ $\overline{\text{5th Kaitz Quintile}}$ } &   \multirow{2.4}{*}{\shortstack{\hphantom{**}-1.485**\hphantom{-} \\ (0.685)}}      \\
& \\[0.2cm] 
Control Variables  & Yes   \\
\multirow{2.4}{*}{\shortstack{Fixed Effects}}  & \multirow{2.4}{*}{\shortstack{Firm \\ CZ $\times$ Year}}  \\
& \\[0.2cm] \hline
 \\[-0.4cm]
\multirow{3.4}{*}{\shortstack[l]{Labor Market \\ Definition \\ (Object)}}  & \multirow{3.4}{*}{\shortstack{NACE-4 \\ $\times$ CZ \\ (Employment)}}  \\
& \\
& \\[0.2cm] 
Observations &  2,700,155           \\[0.2cm]
Adjusted R$^2$ &  0.877         \\[0.2cm] \hline
\end{tabular}
\begin{tablenotes}[para]
\footnotesize\textsc{Note. ---} The table displays fixed effects regressions of log employment (of regular full-time workers per firm) on a full set of interaction terms between log sectoral minimum wage levels, labor market concentration (measured as average HHI over time), and five quintile groups of minimum wage bindingness (measured as average Kaitz Index). The set of control variables includes log sectoral employment, the sectoral share of firms subject to a collective bargaining agreement, and sector-specific linear time trends. The Kaitz Index is calculated as the ratio of the sectoral minimum wage to the median hourly wage rate of regular full-time workers within a firm. Standard errors (in parentheses) are clustered at the sector-by-federal-state level. CZ = Commuting Zone. FT = Full-Time. HHI = Herfindahl-Hirschman Index. NACE-4 = 4-Digit Statistical Nomenclature of Economic Activities in the European Community. * = p$<$0.10. ** = p$<$0.05. *** = p$<$0.01. Sources: IEB $\plus$ BHP $\plus$ IAB Establishment Panel, 1999-2017.
\end{tablenotes}
\end{threeparttable}
}
\end{table}

\vspace*{\fill}
\clearpage

\begin{landscape}


\begin{figure}[!t]
\centering
\caption{Wage Effects of Minimum Wage Introduction}
\label{fig:H2}
\centering
\scalebox{0.70}{
\begin{tikzpicture}
\begin{axis}[xlabel={Quarters since Minimum Wage Introduction in the Sector}, ylabel={De-Trended Semi-Elasticity}, xmin=-11, xmax=11, ymin=-0.05, ymax=0.11, height=13.5cm, width=20cm,ymajorgrids, xtick={-10,-9,-8,-7,-6,-5,-4,-3,-2,-1,0,1,2,3,4,5,6,7,8,9,10}, xticklabels={-10,-9,-8,-7,-6,-5,-4,-3,-2,-1,0,1,2,3,4,5,6,7,8,9,{10-12}}, ytick={-0.04,-0.02,0,0.02,0.04,0.06,0.08,0.10}, grid style = dotted,  tick label style={/pgf/number format/.cd,fixed,fixed zerofill, precision=2,/tikz/.cd}, x tick label style={/pgf/number format/.cd,fixed,fixed zerofill, precision=0,/tikz/.cd},legend pos= north west, scaled ticks=false,label style= {font=\Large}]

\addplot[mark=none,solid, color=gray, line width=0.25mm] coordinates {   (-10.15,.00044159) (-9.1499996,.00088697) (-8.1499996,.0005833) (-7.1500001,-.00274531) (-6.1500001,.00064634) (-5.1500001,-.0000854) (-4.1500001,-.00109152) (-3.1500001,.00094801) (-2.1500001,-.00019289) (-1,.0006089) (-.15000001,-.00065497) (.85000002,.00474034) (1.85,.00807319) (2.8499999,.01028548) (3.8499999,.01136965) (4.8499999,.01252422) (5.8499999,.01820134) (6.8499999,.01904065) (7.8499999,.02290926) (8.8500004,.02275922) (9.8500004,.02124334) };
\addplot[only marks,mark=square*,mark options={fill=white}, color=gray, error bar legend, error bars/.cd, y dir=both, y explicit] coordinates {  (-10.15,.00044159)+-(0,.0108199) (-9.1499996,.00088697)+-(0,.00969148) (-8.1499996,.0005833)+-(0,.00848455) (-7.1500001,-.00274531)+-(0,.00706065) (-6.1500001,.00064634)+-(0,.00580344) (-5.1500001,-.0000854)+-(0,.00430685) (-4.1500001,-.00109152)+-(0,.00433074) (-3.1500001,.00094801)+-(0,.00250214) (-2.1500001,-.00019289)+-(0,.00142403) (-1,.0006089)+-(0,0) (-.15000001,-.00065497)+-(0,.0018947) (.85000002,.00474034)+-(0,.0026952) (1.85,.00807319)+-(0,.00373021) (2.8499999,.01028548)+-(0,.00477088) (3.8499999,.01136965)+-(0,.00492628) (4.8499999,.01252422)+-(0,.00596071) (5.8499999,.01820134)+-(0,.00652123) (6.8499999,.01904065)+-(0,.00739276) (7.8499999,.02290926)+-(0,.00759087) (8.8500004,.02275922)+-(0,.00837903) (9.8500004,.02124334)+-(0,.00993747)  };

\addplot[mark=none,solid, color=black, line width=0.25mm] coordinates {   (-9.8500004,.00692222) (-8.8500004,.00593855) (-7.8499999,-.00756208) (-6.8499999,-.0014671) (-5.8499999,.00704409) (-4.8499999,.00271663) (-3.8499999,-.02268423) (-2.8499999,-.01698226) (-1.85,.00785907) (-1,.01821512) (.15000001,.01687064) (1.15,.02406852) (2.1500001,.0340685) (3.1500001,.04170976) (4.1500001,.04165706) (5.1500001,.04951986) (6.1500001,.04931643) (7.1500001,.06137929) (8.1499996,.06169878) (9.1499996,.0631543) (10.15,.06428698) };
\addplot[only marks,mark=square*,mark options={fill=black}, color=black, error bar legend, error bars/.cd, y dir=both, y explicit] coordinates {  (-9.8500004,.00692222)+-(0,.02776966) (-8.8500004,.00593855)+-(0,.02589502) (-7.8499999,-.00756208)+-(0,.02192273) (-6.8499999,-.0014671)+-(0,.0201296) (-5.8499999,.00704409)+-(0,.01886354) (-4.8499999,.00271663)+-(0,.0169587) (-3.8499999,-.02268423)+-(0,.01140144) (-2.8499999,-.01698226)+-(0,.01028949) (-1.85,.00785907)+-(0,.00738886) (-1,.01821512)+-(0,0) (.15000001,.01687064)+-(0,.00985906) (1.15,.02406852)+-(0,.01224134) (2.1500001,.0340685)+-(0,.01347148) (3.1500001,.04170976)+-(0,.01534357) (4.1500001,.04165706)+-(0,.01903283) (5.1500001,.04951986)+-(0,.02060154) (6.1500001,.04931643)+-(0,.02066853) (7.1500001,.06137929)+-(0,.02161265) (8.1499996,.06169878)+-(0,.02465376) (9.1499996,.0631543)+-(0,.02516704) (10.15,.06428698)+-(0,.02682657)  };

\addplot[loosely dashed, color=blue] coordinates {(-0.5,1) (-0.5,-1)};

\legend{,~HHI=0~,,~HHI=1~,}

\end{axis}
\end{tikzpicture}
}
\vspace{-0.25cm}
\floatfoot{\footnotesize\textsc{Note. ---} The figure illustrates estimated semi-elasticities from the sector-wise introduction of minimum wages. The timing-based event-study model regresses log daily wages (in terms of averages of regular full-time workers) on interactions between event-time dummy variables and labor market concentration (measured as average HHI over time), firm fixed effects, and quarter-by-commuting-zone fixed effects. The coefficients were corrected for linear pre-trends. The markers illustrate point estimates for polar HHI levels. Each point estimate features a 90 percent confidence interval. Standard errors (in parentheses) are clustered at the sector-by-federal-state level.  HHI = Herfindahl- Hirschman Index. Sources: IEB + AWFP, 1999-2017.}
\end{figure}

\end{landscape}


\clearpage

\section{Discussion: Further Evidence}
\label{sec:I}
\setcounter{table}{0} 
\setcounter{figure}{0} 

The following section provides more information on the German literature on the own-wage elasticity of employment based on minimum wage variation. In addition, I report conventional minimum wage effects that do not take into account the (potentially) moderating role of labor market concentration.

\paragraph{The Own-Wage Elasticity of Employment.} To make my results and estimates from the literature comparable, it is necessary to interpret employment responses to minimum wages in relation to the magnitude of the underlying bite. For each study in Figure \ref{fig:6}, I calculate the own-wage elasticity of employment based on minimum wage variation by dividing the minimum wage elasticity of employment by the respective minimum wage elasticity of wages. Note that, in the polar case of perfect competition, the own-wage elasticity of employment is equivalent to the own-wage elasticity of labor demand.

For this study, I determine own-wage elasticities of employment at different HHI levels by dividing the minimum wage elasticity of employment from Column (4) in Table \ref{tab:4} by the minimum wage elasticity of wages from Column (4) in Table \ref{tab:3}. The respective standard errors stem from a Bootstrap algorithm with 50 replications. For studies from the literature that do not explicitly report the own-wage elasticity of employment, I divide the reported minimum wage elasticities of employment by the respective minimum wage elasticities of wages by myself (provided that both effects stem from sufficiently comparable specifications). If the standard error of the own-wage elasticity of employment is not reported, I apply the Delta method to reported minimum wage elasticities of wages and employment along with their standard errors (or t statistics) and assume that the covariance between wage and employment effects is zero. I disregard own-wage elasticities of employment that are not based on a significantly positive minimum wage elasticity of wages because, in the absence of a bite, causal minimum wage effects on employment will not manifest. When there are estimates for a range of different subgroups, I take the estimate for the broadest group of workers. If the study reports several post-treatment effects, I select the effect in closest proximity to the timing of the treatment. If available, I collect separate effects by East and West Germany and by unit of observation.

For the international minimum wage literature, similar overviews on the own-wage elasticity of employment are available in \citet{AzarEtAl2024}, \citet{BaileyEtAl2021}, \citet{BrownHamermesh2019}, \citet{Dube2019}, \citet{HarasztosiLindner2019}, \citet{DerenoncourtMontialoux2021}, and \citet{DubeZipperer2024}. In these summaries, the majority of own-wage elasticities of employment based on minimum wage variation do not turn out to be significantly different from zero, mirroring evidence from the German labor market. Specifically, the most recent overview from \citet{DubeZipperer2024} collects 72 minimum wage studies, which are published in peer-reviewed journals, and arrives at a relatively small average own-wage elasticity of employment of -0.14.


\paragraph{Conventional Minimum Wage Effects.} Conventional minimum wage studies generally pool information across different labor markets and, thus, might conceal heterogeneity by market form. By pooling negative employment effects in more competitive markets and positive employment effects in more monopsonistic markets, these studies typically arrive at average employment effects in the midst of both extremes. The exact level of this average effect hinges on the unit of observation underlying the estimation. While regressions at the labor market level assign equal weight to each labor market, regressions at the firm level weight labor markets with many firms (i.e., with low HHI) more strongly. Consequently, the average own-wage elasticity of employment across labor markets should approach zero more closely than the average elasticity across firms (see Appendix Table \ref{tab:I1}).

Table \ref{tab:I1} reports conventional minimum wage effects at the micro and the aggregate level. Specifically, I simplify Equation (\ref{eq:5}) such that it no longer involves an interaction term between minimum wage levels and HHI and estimate the resulting wage and employment effects at the firm and sector-by-market level. In line with the above proposition, the own-wage elasticity of employment at the firm level turns out significantly negative (-2.806) by mainly reflecting emplyoment responses in low-HHI markets. In contrast, the elasticity at the sector-by-market level, which assigns equal weight to all labor markets, is not significantly different from zero (+0.415). In line, \citet{DuetschEtAL2024} review the German minimum wage literature and conclude, that employment effects generally turn out more negative at the firm than at the regional level. For instance, \citet{DustmannEtAl2022} analyze minimum wage effects at both the region and firm level and arrive at a significantly negative own-wage elasticities of employment for firms, but an insignificant close-to-zero elasticity at the level of regions. Vice versa, many of the remaining estimates originate from the aggregate (or worker) level and tend to deliver close-to-zero or insignificant own-wage elasticities of employment.

\clearpage
\vspace*{\fill}

\begin{table}[!ht]
\centering
\scalebox{0.90}{
\begin{threeparttable}
\caption{Conventional Minimum Wage Effects}
\label{tab:I1}
\begin{tabular}{L{3.75cm}C{2.6cm}C{2.6cm}C{0.2cm}C{2.6cm}C{2.6cm}} \hline
&&&&& \\[-0.3cm]
\multirow{6}{*}{} & \multirow{6.4}{*}{\shortstack{(1) \\ \textbf{Micro\vphantom{/}} \\ \textbf{Level\vphantom{/}} \\ Log \\ Mean Daily\vphantom{/} \\ Wages }} & \multirow{6.4}{*}{\shortstack{(2) \\ \textbf{Micro\vphantom{/}} \\ \textbf{Level\vphantom{/}} \\ Log \\ Regular FT\vphantom{/} \\ Employment }}  && \multirow{6.4}{*}{\shortstack{(3) \\ \textbf{Aggregate\vphantom{/}} \\ \textbf{Level\vphantom{/}} \\ Log \\ Mean Daily\vphantom{/} \\ Wages }}   & \multirow{6.4}{*}{\shortstack{(4) \\ \textbf{Aggregate\vphantom{/}} \\ \textbf{Level\vphantom{/}} \\ Log \\ Regular FT\vphantom{/} \\ Employment }} \\
&&&&&      \\
&&&&&      \\
&&&&&      \\
&&&&&      \\
&&&&&      \\[0.3cm] \hline
&&&&& \\[-0.3cm]
\multirow{2.4}{*}{Log Minimum Wage} &  \multirow{2.4}{*}{\shortstack{\hphantom{***}0.086*** \\ (0.019)}}  & \multirow{2.4}{*}{\shortstack{\hphantom{***}-0.215***\hphantom{-} \\ (0.032)}} && \multirow{2.4}{*}{\shortstack{\hphantom{***}0.098*** \\ (0.033)}} & \multirow{2.4}{*}{\shortstack{0.041 \\ (0.123)}}     \\
&&&&&  \\[0.2cm] 
Control Variables  & Yes & Yes  && Yes & Yes   \\
\multirow{2.4}{*}{\shortstack{Fixed Effects}}  & \multirow{2.4}{*}{\shortstack{Firm \\ CZ $\times$ Year}} & \multirow{2.4}{*}{\shortstack{Firm \\ CZ $\times$ Year}}  && \multirow{2.4}{*}{\shortstack{Sector $\times$ Market\\ CZ $\times$ Year}}  & \multirow{2.4}{*}{\shortstack{Sector $\times$ Market\\ CZ $\times$ Year}} \\
&&&&& \\[0.2cm] \hline
&&&&& \\[-0.2cm]
Observations &  2,700,155       & 2,700,155      && 44,417         & ~44,417              \\[0.2cm]
Adjusted R$^2$ &  0.811    & 0.877     &&   0.837  &   ~0.962         \\[0.2cm] \cline{2-3} \cline{5-6}
&&&&& \\[-0.2cm]
Unit of Observation  & \multicolumn{2}{c}{Firm $\times$ Year}  && \multicolumn{2}{c}{Sector $\times$ Market $\times$ Year}   \\[0.2cm] 
\multirow{3.4}{*}{\shortstack[l]{Own-Wage \\ Elasticity of \\ Employment}}  & \multicolumn{2}{c}{\multirow{3.4}{*}{\shortstack{\hphantom{***}-2.806***\hphantom{-} \\ (0.450) \\ p = 0.000}}} && \multicolumn{2}{c}{\multirow{3.4}{*}{\shortstack{0.415 \\ (1.432) \\ p = 0.772}}}   \\
& \multicolumn{2}{c}{} && \multicolumn{2}{c}{} \\
& \multicolumn{2}{c}{} && \multicolumn{2}{c}{} \\[0.2cm] \hline
\end{tabular}
\begin{tablenotes}[para]
\footnotesize\textsc{Note. ---} The table displays fixed effects regressions of the outcome variable (in logs) on log sectoral minimum wage levels. The set of control variables includes log sectoral employment, the sectoral share of firms subject to a collective bargaining agreement, and sector-specific linear time trends. Standard errors (in parentheses) are clustered at the sector-by-federal-state level. The own-wage elasticities of employment are calculated by dividing the minimum wage elasticity of employment from Column (3) or (4) by the respective minimum wage elasticity of wages from Column (1) or (2). The respective standard errors stem from a Bootstrap algorithm with 50 replications. CZ = Commuting Zone. FT = Full-Time. * = p$<$0.10. ** = p$<$0.05. *** = p$<$0.01. Sources: IEB $\plus$ BHP $\plus$ IAB Establishment Panel, 1999-2017.
\end{tablenotes}
\end{threeparttable}
}
\end{table}

\vspace*{\fill}

\begin{landscape}


\begin{table}[!ht]
\centering
\renewcommand{\arraystretch}{1.075}
\scalebox{0.8}{
\begin{threeparttable}
\caption{Own-Wage Elasticity of Employment for the German Labor Market}
\label{tab:I2}
\vspace{-0.1cm}
\begin{tabular}[c]{L{6.5cm}C{5.5cm}C{5.5cm}C{5.5cm}C{5cm}} \hline
\multirow{4.4}{*}{\shortstack{Study}}   & \multirow{4.4}{*}{\shortstack{Reported (or Resulting)\vphantom{/} \\ Own-Wage Elasticity\vphantom{/} \\ of Employment\vphantom{/} }}    & \multirow{4.4}{*}{\shortstack{Reported\vphantom{/} \\ Minimum Wage (Semi-) \vphantom{/} \\ Elasticity of Wages\vphantom{/} }}   & \multirow{4.4}{*}{\shortstack{Reported\vphantom{/} \\ Minimum Wage (Semi-) \vphantom{/} \\ Elasticity of Employment\vphantom{/} }}    & \multirow{4.4}{*}{\shortstack{Coverage\vphantom{/} \\ Unit of Observation\vphantom{/} }}  \\
&&&& \\
&&&& \\
&&&& \\ \hline

\multirow{5.4}{*}{\shortstack[l]{\textbf{Ahlfeldt, Roth \& Seidel \citeyearpar{AhlfeldtEtAl2018}}\vphantom{/}}} &  \multirow{5.4}{*}{\shortstack{0.122 (0.063) \\ Delta Method}}    & \multirow{5.4}{*}{\shortstack{0.0007 (0.0002)\vphantom{/} \\ Table A2\vphantom{/} \\ Column (3)\vphantom{/} \\ Row ``2016 Treatment Effect''\vphantom{/} }} &  \multirow{5.4}{*}{\shortstack{0.0006 (0.0003)\vphantom{/} \\ Table A3\vphantom{/} \\ Column (2)\vphantom{/} \\ Row ``2016 Treatment Effect''\vphantom{/} }} &  \multirow{5.4}{*}{\shortstack{All Sectors\vphantom{/} \\ Region-Level\vphantom{/} }}   \\[-0.2cm]
&&&&  \\
&&&&  \\
&&&&  \\
&&&&  \\

\multirow{5.4}{*}{\shortstack[l]{\textbf{\citet{BoockmannEtAl2011a}}:\vphantom{/} \\ \textbf{West Germany}\vphantom{/} }} &  \multirow{5.4}{*}{\shortstack{ -1.718 (2.476) \\ Delta Method}}    & \multirow{5.4}{*}{\shortstack{1.0998 (0.4507)\vphantom{/} \\ Table 10.6, Panel ``West''\vphantom{/} \\ Column (2)\vphantom{/} \\ Row ``Lohnlücke''\vphantom{/} }} &  \multirow{5.4}{*}{\shortstack{-1.8898 (2.6111)\vphantom{/} \\ Table 10.12, Panel ``West''\vphantom{/} \\ Column (2)\vphantom{/} \\ Row ``Lohnlücke''\vphantom{/} }} &  \multirow{5.4}{*}{\shortstack{Nursing Care\vphantom{/} \\ Firm-Level\vphantom{/} }}   \\[-0.2cm]
&&&&  \\
&&&&  \\
&&&&  \\
&&&&  \\

\multirow{5.4}{*}{\shortstack[l]{\textbf{\citet{BoockmannEtAl2011a}}:\vphantom{/} \\ \textbf{East Germany}\vphantom{/} }} &  \multirow{5.4}{*}{\shortstack{1.504 (1.047) \\ Delta Method}}    & \multirow{5.4}{*}{\shortstack{1.6412 (0.2990)\vphantom{/} \\ Table 10.6, Panel ``Ost''\vphantom{/} \\ Column (2)\vphantom{/} \\ Row ``Lohnlücke''\vphantom{/} }} &  \multirow{5.4}{*}{\shortstack{2.4679 (1.6589)\vphantom{/} \\ Table 10.12, Panel ``Ost''\vphantom{/} \\ Column (2)\vphantom{/} \\ Row ``Lohnlücke''\vphantom{/} }} & \multirow{5.4}{*}{\shortstack{Nursing Care\vphantom{/} \\ Firm-Level\vphantom{/} }}     \\[-0.2cm]
&&&&  \\
&&&&  \\
&&&&  \\
&&&&  \\

\multirow{5.4}{*}{\shortstack[l]{\textbf{\citet{BoockmannEtAl2011b}}:\vphantom{/} \\ \textbf{East Germany}\vphantom{/} }} &  \multirow{5.4}{*}{\shortstack{ 0.500 (1.284) \\ Delta Method}}    & \multirow{5.4}{*}{\shortstack{0.06 (t=7.36)\vphantom{/} \\ Table 7.1 a, Panel ``... (2005)''\vphantom{/} \\ Column 1B \vphantom{/} \\ Row ``Treatment-Effekt''\vphantom{/} }} &  \multirow{5.4}{*}{\shortstack{0.03 (t=-0.39)\vphantom{/} \\ Table 7.5 a, Panel ``Lange Frist'' \vphantom{/} \\ Column 1B \vphantom{/} \\ Row ``Treatment-Effekt''\vphantom{/} }} & \multirow{5.4}{*}{\shortstack{Painting \& Varnishing\vphantom{/} \\ Firm-Level\vphantom{/} }}     \\[-0.2cm]
&&&&  \\
&&&&  \\
&&&&  \\
&&&&  \\

\multirow{5.4}{*}{\shortstack[l]{\textbf{\citet{BoockmannEtAl2011c}}:\vphantom{/} \\ \textbf{2007 Minimum Wage Introduction}\vphantom{/} }} &  \multirow{5.4}{*}{\shortstack{-0.167 (0.198) \\ Delta Method}}    & \multirow{5.4}{*}{\shortstack{0.03 (t=4.66)\vphantom{/} \\ Table 7.3, Panel `` (2008)''\vphantom{/} \\ Column 1B \vphantom{/} \\ Row ``Treatment-Effekt''\vphantom{/} }} &  \multirow{5.4}{*}{\shortstack{-0.005 (t=-0.80)\vphantom{/} \\ Table 7.7a\vphantom{/} \\ Column 1B \vphantom{/} \\ Row ``Interaktionseffekt 1''\vphantom{/} }} & \multirow{5.4}{*}{\shortstack{Electrical Trade\vphantom{/} \\ Worker-Level\vphantom{/} }}     \\[-0.2cm]
&&&&  \\
&&&&  \\
&&&&  \\
&&&&  \\

\multirow{5.4}{*}{\shortstack[l]{\textbf{\citet{BosslerEtAl2022}}:\vphantom{/} \\ \textbf{2015 Minimum Wage Introduction}\vphantom{/} }} &  \multirow{5.4}{*}{\shortstack{ -0.302 (0.199) \\ Delta Method}}    & \multirow{5.4}{*}{\shortstack{0.043 (0.010)\vphantom{/} \\ Table 8\vphantom{/} \\ Column (1) \vphantom{/} \\ Row ``Treatmenteffekte 2015''\vphantom{/} }} &  \multirow{5.4}{*}{\shortstack{-0.013 (0.008)\vphantom{/} \\ Table 8\vphantom{/} \\ Column (1) \vphantom{/} \\ Row ``Treatmenteffekte 2015''\vphantom{/} }} & \multirow{5.4}{*}{\shortstack{All Sectors\vphantom{/} \\ Firm-Level\vphantom{/} }}     \\[-0.2cm]
&&&&  \\
&&&&  \\
&&&&  \\
&&&&  \\[0.2cm] \hline

\end{tabular}
\begin{tablenotes}[para]
\footnotesize\textsc{Note. ---} The figure provides the sources of estimated own-wage elasticities of employment based on minimum wage variation in the German labor market.
When the own-wage elasticity of employment is not explicitly reported, I divide the minimum wage elasticity of employment by the minimum wage elasticity of wages. If the standard error of the own-wage elasticity of employment is not reported, I apply the Delta method to reported coefficients and their standard errors and assume that the covariance between the wage and employment effects is zero. Standard errors (or t statistics) are in parentheses. Source: Own calculations.
\end{tablenotes}
\end{threeparttable}
}
\end{table}

\clearpage
\vspace*{\fill}

\begin{table}[!ht]
\addtocounter{figure}{-1}
\centering
\scalebox{0.8}{
\begin{threeparttable}
\caption{Own-Wage Elasticity of Employment for the German Labor Market (Cont.)}
\vspace{-0.1cm}
\begin{tabular}[c]{L{6.5cm}C{5.5cm}C{5.5cm}C{5.5cm}C{5cm}} \hline
\multirow{4.4}{*}{\shortstack{Study}}   & \multirow{4.4}{*}{\shortstack{Reported (or Resulting)\vphantom{/} \\ Own-Wage Elasticity\vphantom{/} \\ of Employment\vphantom{/} }}    & \multirow{4.4}{*}{\shortstack{Reported\vphantom{/} \\ Minimum Wage (Semi-) \vphantom{/} \\ Elasticity of Wages\vphantom{/} }}   & \multirow{4.4}{*}{\shortstack{Reported\vphantom{/} \\ Minimum Wage (Semi-) \vphantom{/} \\ Elasticity of Employment\vphantom{/} }}    & \multirow{4.4}{*}{\shortstack{Coverage\vphantom{/} \\ Unit of Observation\vphantom{/}  }}  \\
&&&& \\
&&&& \\
&&&& \\ \hline

\multirow{5.4}{*}{\shortstack[l]{\textbf{Bossler, Chittka \& Schank \citeyearpar{BosslerEtAl2024}}\vphantom{/}}} &  \multirow{5.4}{*}{\shortstack{ -0.167 (0.094) \\ Delta Method}}    & \multirow{5.4}{*}{\shortstack{6 (0.24)\vphantom{/} \\ Formula (3)\vphantom{/} \\ S.E.\ from Authors\vphantom{/} }} &  \multirow{5.4}{*}{\shortstack{-1 (0.56)\vphantom{/} \\ Formula (3)\vphantom{/} \\ S.E.\ from Authors\vphantom{/} }} & \multirow{5.4}{*}{\shortstack{All Sectors\vphantom{/} \\ Firm-Level\vphantom{/} }}     \\[-0.2cm]
&&&&  \\
&&&&  \\
&&&&  \\
&&&&  \\

\multirow{5.4}{*}{\shortstack[l]{\textbf{\citet{BosslerGerner2020}:}}} &  \multirow{5.4}{*}{\shortstack{-0.278 (0.212) \\ Table 2, Panel B \\ Column (3)}}    & \multirow{5.4}{*}{} &  \multirow{5.4}{*}{} & \multirow{5.4}{*}{\shortstack{All Sectors\vphantom{/} \\ Firm-Level\vphantom{/} }}     \\[-0.2cm]
&&&&  \\
&&&&  \\
&&&&  \\
&&&&  \\

\multirow{5.4}{*}{\shortstack[l]{\textbf{\citet{BosslerSchank2023}} \vphantom{/} }} &  \multirow{5.4}{*}{\shortstack{ -0.198 (0.204) \\ Delta Method}}    & \multirow{5.4}{*}{\shortstack{0.353 (0.086)\vphantom{/} \\ Table 2\vphantom{/} \\ Column (1) \vphantom{/} \\ Row ``Bite x Year 2015''\vphantom{/} }} &  \multirow{5.4}{*}{\shortstack{-0.070 (0.070)\vphantom{/} \\ Table 3\vphantom{/} \\ Column (1) \vphantom{/} \\ Row ``Bite x Year Post''\vphantom{/} }} & \multirow{5.4}{*}{\shortstack{All Sectors\vphantom{/} \\ Firm-Level\vphantom{/} }}     \\[-0.2cm]
&&&&  \\
&&&&  \\
&&&&  \\
&&&&  \\

\multirow{5.4}{*}{\shortstack[l]{\textbf{\citet{DustmannEtAl2022}}:\vphantom{/} \\ \textbf{Region-Level}\vphantom{/}}}    & \multirow{5.4}{*}{\shortstack{0.03 (0.12) \\ Footnote 27\vphantom{/}}} &  \multirow{5.4}{*}{}     & \multirow{5.4}{*}{} &  \multirow{5.4}{*}{\shortstack{All Sectors\vphantom{/} \\ Region-Level\vphantom{/} }}   \\[-0.2cm]
&&&&  \\
&&&&  \\
&&&&  \\
&&&&  \\

\multirow{5.4}{*}{\shortstack[l]{\textbf{\citet{DustmannEtAl2022}}:\vphantom{/}\\ \textbf{Firm-Level}\vphantom{/}}}   &  \multirow{5.4}{*}{\shortstack{-0.31 (0.04) \\ Online Appendix\vphantom{/} \\ Page 65\vphantom{/}}} &  \multirow{5.4}{*}{}  & \multirow{5.4}{*}{} &  \multirow{5.4}{*}{\shortstack{All Sectors\vphantom{/} \\ Firm-Level\vphantom{/} }}   \\[-0.2cm]
&&&&  \\
&&&&  \\
&&&&  \\
&&&&  \\

\multirow{5.4}{*}{\shortstack[l]{\textbf{\citet{KoenigMoeller2009}:} \vphantom{/} \\ \textbf{West Germany} \vphantom{/} }} &  \multirow{5.4}{*}{\shortstack{2.200 (1.654) \\ Delta Method}}    & \multirow{5.4}{*}{\shortstack{0.005 (0.002)\vphantom{/} \\ Table 5\vphantom{/} \\ Column ``West, Variant 2'' \vphantom{/} \\ Row ``D ... * D Year 1997''\vphantom{/} }} &  \multirow{5.4}{*}{\shortstack{0.011 (0.007)\vphantom{/} \\ Table 6\vphantom{/} \\ Column ``West, Variant 2'' \vphantom{/} \\ Row ``Dummy ... Effect''\vphantom{/} }} & \multirow{5.4}{*}{\shortstack{Main Construction\vphantom{/} \\ Worker-Level\vphantom{/} }}     \\[-0.2cm]
&&&&  \\
&&&&  \\
&&&&  \\
&&&&  \\[0.2cm] \hline
\end{tabular}
\begin{tablenotes}[para]
\footnotesize\textsc{Note. ---} The figure provides the sources of estimated own-wage elasticities of employment based on minimum wage variation in the German labor market.
When the own-wage elasticity of employment is not explicitly reported, I divide the minimum wage elasticity of employment by the minimum wage elasticity of wages. If the standard error of the own-wage elasticity of employment is not reported, I apply the Delta method to reported coefficients and their standard errors and assume that the covariance between the wage and employment effects is zero. Standard errors (or t statistics) are in parentheses. Source: Own calculations.
\end{tablenotes}
\end{threeparttable}
}
\end{table}

\vspace*{\fill}
\clearpage
\vspace*{\fill}

\begin{table}[!ht]
\addtocounter{figure}{-1}
\centering
\scalebox{0.8}{
\begin{threeparttable}
\caption{Own-Wage Elasticity of Employment for the German Labor Market (Cont.)}
\vspace{-0.1cm}
\begin{tabular}[c]{L{6.5cm}C{5.5cm}C{5.5cm}C{5.5cm}C{5cm}} \hline
\multirow{4.4}{*}{\shortstack{Study}}   & \multirow{4.4}{*}{\shortstack{Reported (or Resulting)\vphantom{/} \\ Own-Wage Elasticity\vphantom{/} \\ of Employment\vphantom{/} }}    & \multirow{4.4}{*}{\shortstack{Reported\vphantom{/} \\ Minimum Wage (Semi-) \vphantom{/} \\ Elasticity of Wages\vphantom{/} }}   & \multirow{4.4}{*}{\shortstack{Reported\vphantom{/} \\ Minimum Wage (Semi-) \vphantom{/} \\ Elasticity of Employment\vphantom{/} }}    & \multirow{4.4}{*}{\shortstack{Coverage\vphantom{/} \\ Unit of Observation\vphantom{/}  }}  \\
&&&& \\
&&&& \\
&&&& \\ \hline

\multirow{5.4}{*}{\shortstack[l]{\textbf{\citet{KoenigMoeller2009}:} \vphantom{/} \\ \textbf{East Germany} \vphantom{/} }} &  \multirow{5.4}{*}{\shortstack{-1.818 (1.034) \\ Delta Method}}    & \multirow{5.4}{*}{\shortstack{0.011 (0.003)\vphantom{/} \\ Table 5\vphantom{/} \\ Column ``East, Variant 2'' \vphantom{/} \\ Row ``D ... * D Year 1997''\vphantom{/} }} &  \multirow{5.4}{*}{\shortstack{-0.002 (0.001)\vphantom{/} \\ Table 6\vphantom{/} \\ Column ``East, Variant 2'' \vphantom{/} \\ Row ``Dummy ... Effect''\vphantom{/} }} & \multirow{5.4}{*}{\shortstack{Main Construction\vphantom{/} \\ Worker-Level\vphantom{/} }}     \\[-0.2cm]
&&&&  \\
&&&&  \\
&&&&  \\
&&&&  \\

\multirow{5.4}{*}{\shortstack[l]{\textbf{\citet{Kunaschk2024}} \vphantom{/}}} &  \multirow{5.4}{*}{\shortstack{-0.0177 (0.033) \\ Table 3\vphantom{/} \\ Column (1)\vphantom{/} \\ Row (A)\vphantom{/} }}    & \multirow{5.4}{*}{} &  \multirow{5.4}{*}{} & \multirow{5.4}{*}{\shortstack{Hairdressing\vphantom{/} \\ Region-Level\vphantom{/} }}     \\[-0.2cm]
&&&&  \\
&&&&  \\
&&&&  \\
&&&&  \\

\multirow{5.4}{*}{\shortstack[l]{\textbf{\citet{MoellerEtAl2011}:} \vphantom{/} \\ \textbf{East Germany} \vphantom{/} }} &  \multirow{5.4}{*}{\shortstack{-1.437 (2.104) \\ Delta Method}}    & \multirow{5.4}{*}{\shortstack{0.071 (t=1.65)\vphantom{/} \\ Table 5.16\vphantom{/} \\ Column ``Ostdeutschland, A'' \vphantom{/} \\ Row ``Betroffenheit Einführung''\vphantom{/} }} &  \multirow{5.4}{*}{\shortstack{-0.102 (t=-1.24)\vphantom{/} \\ Table 6.5\vphantom{/} \\ Column ``Ostdeutschland, C'' \vphantom{/} \\ Row ``... 1997 (Einführung)''\vphantom{/} }} & \multirow{5.4}{*}{\shortstack{Main Construction\vphantom{/} \\ Region-Level\vphantom{/} }}     \\[-0.2cm]
&&&&  \\
&&&&  \\
&&&&  \\
&&&&  \\

\multirow{5.4}{*}{\shortstack[l]{\textbf{\citet{Rattenhuber2014}:} \vphantom{/} \\ \textbf{East Germany} \vphantom{/} }} &  \multirow{5.4}{*}{\shortstack{0.025 (0.050) \\ Delta Method}}    & \multirow{5.4}{*}{\shortstack{0.081 (0.028)\vphantom{/} \\ Table 3, Panel ``East Germany''\vphantom{/} \\ Column (1) \vphantom{/} \\ Row ``Blue*Post*Sector''\vphantom{/} }} &  \multirow{5.4}{*}{\shortstack{0.002 (0.004)\vphantom{/} \\ Table 4, Panel ``East Germany''\vphantom{/} \\ Column (1) \vphantom{/} \\ Row ``Blue*Post*Sector''\vphantom{/} }} & \multirow{5.4}{*}{\shortstack{Main Construction\vphantom{/} \\ Region-Level\vphantom{/} }}     \\[-0.2cm]
&&&&  \\
&&&&  \\
&&&&  \\
&&&&  \\

\multirow{5.4}{*}{\shortstack[l]{\textbf{\citet{vomBergeFrings2020}:} \vphantom{/} \\ \textbf{East Germany} \vphantom{/} }} &  \multirow{5.4}{*}{\shortstack{-1.724 (1.834) \\ Delta Method}}    & \multirow{5.4}{*}{\shortstack{0.105 (0.029)\vphantom{/} \\ Table 1\vphantom{/} \\ Column (5) \vphantom{/} \\ Row ``Treatment Effect (East)''\vphantom{/} }} &  \multirow{5.4}{*}{\shortstack{-0.181 (0.186)\vphantom{/} \\ Table 3\vphantom{/} \\ Column (5) \vphantom{/} \\ Row ``Treatment Effect (East)''\vphantom{/} }} & \multirow{5.4}{*}{\shortstack{Main Construction\vphantom{/} \\ Region-Level\vphantom{/} }}     \\[-0.2cm]
&&&&  \\
&&&&  \\
&&&&  \\
&&&&  \\[0.2cm] \hline
\end{tabular}
\begin{tablenotes}[para]
\footnotesize\textsc{Note. ---} The figure provides the sources of estimated own-wage elasticities of employment based on minimum wage variation in the German labor market.
When the own-wage elasticity of employment is not explicitly reported, I divide the minimum wage elasticity of employment by the minimum wage elasticity of wages. If the standard error of the own-wage elasticity of employment is not reported, I apply the Delta method to reported coefficients and their standard errors and assume that the covariance between the wage and employment effects is zero. Standard errors (or t statistics) are in parentheses. Source: Own calculations.
\end{tablenotes}
\end{threeparttable}
}
\end{table}

\vspace*{\fill}

\renewcommand{\arraystretch}{1.5}

\end{landscape}

\clearpage

\clearpage
\addcontentsline{toc}{section}{References} 
\printbibliography[title=References] 

\end{refsection}
\end{appendix}
\end{document}